\documentclass[12pt]{article}
\usepackage{amsmath}
\usepackage{graphicx}
\usepackage{natbib} %comment out if you do not have the package
\usepackage{url} % not crucial - just used below for the URL

\newcommand{\blind}{0}

\usepackage{bm}
\usepackage{amsfonts}
\usepackage{subfigure}
\usepackage{boldline}
\usepackage{changepage}
\usepackage{booktabs}
\usepackage{algorithm2e}
\usepackage{float}
\usepackage[section]{placeins}
\usepackage{xcolor}
\definecolor{darkblue}{rgb}{0.0, 0.0, 0.55}
\definecolor{darkred}{rgb}{0.8, 0.0, 0.0}

\usepackage[breaklinks]{hyperref}

\renewenvironment{abstract}
 {\small
  \begin{center}
  \bfseries \abstractname\vspace{-.5em}\vspace{0pt}
  \end{center}
  \list{}{
    \setlength{\leftmargin}{.4cm}%
    \setlength{\rightmargin}{\leftmargin}%
  }%
\item\relax}
{\endlist}

\begin{document}

\def\spacingset#1{\renewcommand{\baselinestretch}%
{#1}\small\normalsize} \spacingset{1}

\if0\blind
{
  \title{\bf Multipartition model for multiple change point identification}
  \author{Ricardo C. Pedroso\thanks{
    This work was partially funded by Coordenação de Aperfeiçoamento de Pessoal de Ensino Superior (CAPES), Conselho Nacional de Desenvolvimento Científico e Tecnológico (CNPq) and Fundação de Amparo à Pesquisa do Estado de Minas Gerais (FAPEMIG). Additional support for this work was obtained from grants Fondecyt 1180034 and ANID - Millennium Science Initiative Program - NCN17\_059.}\hspace{.2cm}\\
    Departamento de Estatística, Universidade Federal de Minas Gerais\\
    %ricardocunhap@gmail.com\\%[.2in]
    and \\%[.2in]
    Rosangela H. Loschi$^*$\\
    Departamento de Estatística, Universidade Federal de Minas Gerais\\
    %loschi@est.ufmg.br\\%[.2in]
    and \\%[.2in]
    Fernando Andr\'es Quintana$^*$\\
    Departamento de Estad\'{\i}stica, Pontificia Universidad Católica de Chile\\
    %ANID - Millennium Science Initiative Program - Millennium Nucleus\\ Center for the Discovery of Structures in Complex Data%\\
    Millennium Nucleus Center for the Discovery of Structures in Complex Data%\\
    %quintana@mat.uc.cl
    }
  \maketitle
} \fi

\if1\blind
{
  \bigskip
  \bigskip
  \bigskip
  \begin{center}
    {\LARGE\bf Title}
\end{center}
  \medskip
} \fi

%\bigskip
\vspace{-.2in}
\begin{abstract}
Among the main goals in multiple change point problems are the estimation of the number and positions of the change points, as well as the regime structure in the clusters induced by those changes. The product partition model (PPM) is a widely used approach for the detection of multiple change points. The traditional PPM assumes that change points split the set of time points in random clusters that define a partition of the time axis. It is then typically assumed that sampling model parameter values  within each of these blocks are identical. Because changes in different parameters of the observational model may occur at different times, the PPM thus fails to identify the parameters that experienced those changes. A similar problem may occur when detecting changes in multivariate time series. To solve this important limitation, we introduce a multipartition model to detect multiple change points occurring in several parameters at possibly different times. The proposed model assumes that the changes experienced by each parameter generate a different random partition of the time axis, which facilitates identifying which parameters have changed and when they do so.
We discuss a partially collapsed Gibbs sampler scheme to implement
posterior simulation under the proposed model. We apply the proposed model to identify multiple change points in Normal means and variances and evaluate the performance of the proposed model through Monte Carlo simulations and data illustrations. Its performance is compared with some previously proposed approaches for change point problems. These studies show that the proposed model is competitive and enriches the analysis of change point problems.

\end{abstract}

\noindent%
\vfill

\newpage
\spacingset{1.8} % DON'T change the spacing!

\section{Introduction}
\label{secINTRO}

Change point identification is not a new problem; it plays an important
role in many different fields, such as finance, genetics, public health,
historical environmental measurements, and many others. It serves a range
of purposes such as to improve forecasts or to identify the events that
produced the changes, thus guiding future decisions and policies
definition.
Some approaches to handle the case of a single change point include
\cite{chernoffzacks1964}, \cite{smith75} and \cite{mira1996}.
\cite{smith75} assumes independence between the two regimes and that data
within each regime are exchangeable. \cite{mira1996} proposed a more
general non-parametric model that accounts for correlated observations
between and within the regimes. Their model considers a mixture of
products of Dirichlet processes to model the dependence among the
observations within and between the two regimes.

The main goals when addressing multiple change point problems are to
precisely estimate the number and the positions of the changes. Estimation
of the structural parameters within the clusters is also typically of
interest. There is a wide range of models to handle the identification of
multiple change points.
Because the multiple change point identification is a particular case of
clustering analysis in which only contiguous clusters are possible,
approaching this problem by way of the product partition model (PPM),
introduced by \cite{h90}, is an appealing strategy.
The PPM was first applied to detect change points by \cite{bh92}, who
consider that change points define a random partition $\rho$. They also explored
theoretical aspects of PPMs under this setting, assuming that $\rho$
has a product distribution that imposes a Markovian relationship among the
change points. Later, \cite{bh93} applied the PPM to detect multiple
changes in the mean of a sequence of Normal variables with unknown
constant variance. The posterior distribution of partitions is usually
intractable, but simulation-based methods are available for inference,
e.g. as developed in \cite{bh93}. Extensions of the PPM to detect changes
in several structural parameters can be found in \cite{loschi2005p},
\cite{loschi2010} and references therein.

Multiple change point problems have been addressed through many different
methodologies. \cite{chib98} formulated a change point model in terms of a
latent discrete state variable that evolves according to a discrete-time
Markov process and indicates the regime of each particular observation.
%
%\cite{fearnhead2005} proposed a regression model that is a combination of
%independent linear regression models on disjoint segments, such that both
%the number, the position and the parameters of the regressions are to be
%estimated.
 %	
\cite{fearnhead2006} and \cite{fearnhead2007} introduced a direct
simulation algorithm to change point models. \cite{fearnhead2006} presents
efficient recursions to calculate the posterior probabilities of different
numbers of change points and the posterior mean of the structural
parameters, obtaining exact solutions. \cite{fearnhead2007} proposed an
on-line algorithm for the exact filtering of multiple change point
problems. In this case, inference is made incrementally over time, and new
estimates are produced at each time that a new observation is made.
\cite{fearnhead2019} presented a penalized cost approach to change point
detection that shows to be robust to the presence of outliers if a
bi-weight loss function is assumed. \cite{martinez2014} proposed a new way
to model the prior uncertainty about $\rho$, assuming that the cohesion
functions are given by a suitable modification of an exchangeable
partition probability function derived from Pitman's sampling formula.
They also assumed the Ornstein-Uhlenbeck process to model regime behavior.
\cite{garcia2019} proposed a nonparametric extension of the PPM assuming a
random measure to model data within each cluster. This approach does not
impose any specific form for the sampling distribution while allowing for
correlation within regime observations. \cite{monteiro2011} defined another
PPM by assuming that observations within the same cluster have their
distributions indexed by correlated and different parameters. The change
point model introduced by \cite{wyse2011} also imposes a temporal
dependence among the regime data through a Gaussian Markov random field. A
dependence structure among the regimes is considered in the models
introduced by \cite{fearnhead2011} and \cite{ferreira2014}. In both
models, the across-cluster correlation is introduced through the prior
distribution for structural parameters. \cite{nyamundanda2015} introduce a
PPM based model for multiple change point detection in both the mean and
covariance structure of multivariate correlated sequences of Gaussian
data.
\cite{jin2021} proposed a Bayesian hierarchical model to identify changes
in the mean of multivariate sequences in the presence of correlation
between and within the sequences.

The PPM, a main focus of this work, has also been applied for several
other purposes such as outlier detection \citep{quintana2003, QuIgBo2005},
meta-analysis and classification problems \citep{jordan2007},
% to estimate the time-grid in piecewise exponential models \citep{demarqui2008}
spatial \citep{hegarty2008,teixeira2015,page2016} and spatio-temporal
analysis \citep{teixeira2019}.

Despite being a competitive model, if applied to the identification of
multiple change points for sampling models involving two or more
parameters, the PPM  fails to identify the parameter or subset of
parameters associated to the change. In financial data, for example, some
events may produce changes in a return volatility but not in its mean
return. This problem may occur in any situation where interest is on
identifying multiple changes in multiparametric models as well as in
multivariate ones. Under the PPM structure, we only obtain the posterior
distribution of the random partition, that indicates when the structural
changes occurred. However, one or more parameters may experience changes,
and changes in different parameters may occur asynchronously.
This is a long-standing limitation of the PPM. Recently, \cite{peluso2019}
proposed a semiparametric model for detecting multiple change points
occurring in several structural parameters. Extending the \cite{chib98} model,
they assume a specific latent Markovian discrete state
variable for each structural parameter, allowing to identify which
parameter experienced the change. They fix the maximum number of change
points, that can be equal to the number of time points minus one, and carry out
inference using posterior simulation.

Our approach for tackling this problem differs from that in
\cite{peluso2019} in the prior construction. Our main contribution is the
introduction of a multipartition model to detect multiple changes in
sequential data (Section \ref{secBMCP_def}). We refer to it as the Bayesian
multipartition change point model (BMCP). The proposed model assumes that
different parameters may change at different and unknown times and also
experiences an unknown number of change points.  BMCP is a natural
generalization of \cite{bh92}'s PPM, assuming different and independent random
partitions for each structural parameter in the model. Changes in
different parameters are independently driven by different Markovian
processes imposing different product distributions for each partition. The
posterior distributions for these partitions allow us to identify the
instants when the changes occurred and the parameters that experienced
each of these changes.
As the random partition is not an Euclidean vector, a great challenge in
random partition models is to explore the corresponding posterior distributions. Another contribution of this work is thus the derivation of a
computationally efficient algorithm to sample from the joint posterior of
parameters and partitions (Section~\ref{secBMCP_pcg}). In Section
\ref{secBMCPnormal} we offer a detailed discussion for the special case of
change points for means and/or variances in normal data, thus extending
\cite{bh93} and \cite{loschi2005p}. To evaluate the BMCP performance, we
run a Monte Carlo simulation study (Section~\ref{secBMCPnormal_MC}). BMCP
is compared to the models introduced by \cite{bh93}, \cite{loschi2005p}
and \cite{peluso2019} in different scenarios. The BMCP is also fitted to
analyze two data sets, one in finance and the other in genetics
(Section~\ref{secReal}). Some additional comparisons between BMCP and the
other models are provided in the supplementary material.  final discussion
in presented in Section~\ref{secConclusion}.

\section{Model definition}
\label{secBMCP_def}

Consider a sequence of $n$ random variables
$\bm{X}=\left(X_1,\dots,X_n\right)$. Let
$\bm{\theta}_1,\dots,\bm{\theta}_d$ be sequences of $d$ unknown structural
parameters where the $k$th sequence is $\bm{\theta}_k=(\theta_{k,1},\dots,
\theta_{k,n})$ for $k=1,\dots,d$. Let
$f(X_i\mid\theta_{1,i},\dots,\theta_{d,i})$ represent the sampling
distribution of $X_i$, parametrized by $\theta_{1,i},\dots,\theta_{d,i}$,
$i=1,\dots,n$, and assume that $X_1,\dots,X_n$ are conditionally
independent given $\bm{\theta}_1,\dots,\bm{\theta}_d$. Suppose that each
$\bm{\theta}_k$, $k=1,\dots,d$, is affected by an unknown number $N_k$ of
changes, that occur at unknown positions of the sequence. Let $\rho_k$
represents the random partition that splits the set of indexes
$I=\{1,\dots,n\}$ of $\bm{\theta}_k$ into contiguous clusters induced by those changes. The
partition $\rho_k$ may be defined by

\begin{equation} \label{def:rho_k}
	\rho_k = \{i_{k,0},i_{k,1},\dots,i_{k,b_k}\}, \hspace{.1in} 0=i_{k,0}<i_{k,1}<\dots<i_{k,b_k}=n,
\end{equation}
\noindent where the values $i_{k,1},\dots,i_{k,b_k}$ represent the end
points of the contiguous clusters $S_{k,j_k}=\{i_{k,j_k-1}	
+1,\dots,i_{k,j_k}\}$, \hbox{$j_k=1,\dots,b_k$},  \hbox{$k=1,\dots,d$}.
This partition is equally defined by the set of clusters
$\{S_{k,1},\dots,S_{k,b_k}\}$. The first element of each cluster
$S_{k,j_k}$ is called a change point of $\bm{\theta}_k$.
%Thus, we can define the random number of clusters $B_k$ in $\bm{\theta}_k$ as the random variable ranging from $1$ to $n$ that measures the cardinality of $\{S_{k,1},\dots,S_{k,b_k}\}$.

Given $\rho_k=\{S_{k,1},\dots,S_{k,b_k}\}$, we assume that all
observations $i\in S_{k,j_k} $ share the same value for the $k$th
structural parameter. Thus, the vector $\bm{\theta}_k$ can be written as
\begin{equation*}
\bm{\theta}_k = \sum_{j_k=1}^{b_k} \left(\theta^{\star}_{k,j_k} {\bf{1}} \{ 1 \in S_{k,j_k} \}, \dots,  \theta^{\star}_{k,j_k} {\bf{1}} \{ n \in S_{k,j_k} \}\right),
%\label{}
\end{equation*}
\noindent where ${\bf{1}}\{A\}$ denotes the indicator function of event $A$ and $\theta^{\star}_{k,1},\dots,\theta^{\star}_{k,b_k}$ are the cluster parameters such that
\begin{eqnarray*}
	\theta_{k,i}  = \theta^{\star}_{k,j_k}  \;\;\;\text{for}\;\;\; i\in S_{k,j_k},
	\;\;\;i=1,\dots,n,
	\;\;\;j_k=1,\dots,b_k,
	\;\;\;k=1,\dots,d.
\end{eqnarray*}

We assume {\it a priori} that changes in different parameters occur
independently, so that the $d$ random partitions $\rho_1,\dots,\rho_d$ are
independent, with joint prior distribution
\begin{eqnarray}
\label{rhoJoint}
&&P(\rho_1=\{S_{1,1},\dots,S_{1,b_1}\}, \ldots, \rho_d=\{S_{d,1},\dots,S_{d,b_d}\})
\;\;=\;\;\prod_{k=1}^{d}   P(\rho_k=\{S_{k,1},\dots,S_{k,b_k}\}),
\end{eqnarray}
and following \cite{h90},  for each partition $\rho_k$, we assume the
product prior distribution
\begin{equation}
\label{rho_BMCP}
P(\rho_k=\{S_{k,1},\dots,S_{k,b_k}\}) = \frac{\displaystyle\prod_{j_k=1}^{b_k}c_k(S_{k,j_k})}
{\displaystyle\sum\limits_{\rho_k\in\mathcal{P}}\;\,\prod_{S_{k,\ell_k}\in\rho_k}\!\!\!\!\!c_k(S_{k,\ell_k})}, \hspace{.1in} 
\end{equation}
\noindent $k=1,\dots,d,$ where $\mathcal{P}$ represents the set of all possible partitions of $I$ and the cohesions $c_k(S_{k,j_k})$ are positive numbers measuring how strongly we believe the components of $\bm{\theta}_k$ in $S_{k,j_k}$ are to co-cluster {\it a priori}.

It is interesting to point out that the product form for the prior on $\rho_k$ arises naturally under a Markovian structure assumption for the sequence of change points occurring in $\bm{\theta}_k$ \citep{bh92}. Indeed, if the sequence of end points $i_{k,0},\dots,i_{k,b_k}$ at the $k$th structural parameter is a realization of a Markov chain $\{Z_{\ell}^k,\,\ell\in{\mathbb{N}} \}$ in which $Z_{\ell}^k=i_{k,0}=0$ if $\ell=0$ and, for $\ell>0$, $Z_{\ell}^k$ assumes values in the set $\{Z_{\ell-1}^k+1, \ldots,n\}$ if $Z_{\ell-1}^k
\neq n$ and $Z_\ell^k=i_{k,b_k}=n$ if $Z_{\ell-1}^k=n$, then the prior for $\rho_k$ is given by
\begin{eqnarray*}
P(\rho_k=\{i_{k,0},\dots,i_{k,b_k}\}) &=& P(Z_{b_k}^k = i_{k,b_k} \mid Z_{b_k-1}^k = i_{k,b_k-1})
\cdots
P(Z_{1}^k = i_{k,1} \mid Z_{0}^k = i_{k,0}),
\end{eqnarray*}
considering that $p(Z_{0}^k = i_{k,0})=1.$ In this case, the cohesions define the one-step transition probabilities on such a Markov chain. However, the model in (\ref{rho_BMCP}) is more general and may accommodate different dependence structures among change points, which is determined by the choice of $c_k(S_{k,j_k})$.
% For instance, if $c_k(S_{k,j_k})=1$, for all $j_k$,  all partitions of set $I$ related to the structural parameter $\theta_k$ have equal probabilities

Recently, \cite{peluso2019} \citep[see also][]{chib98} proposed a different prior construction that determines change point locations indirectly. This model requires that the maximum number of change points is a known value ${m}$, that may be equal to the maximum possible number of changes $n-1$. To model uncertainty about the change points, a vector of states $E_k=(\epsilon_1^k,\dots,\epsilon_n^k)$ is defined such that $\epsilon_i^k=\ell$, $\ell=1,\dots,{m}+1$, if the  structural parameter at time $t$ belongs to the $\ell$th cluster. An uni-directional Markov process then models the uncertainty about the state variables
$\epsilon_1^k,\dots,\epsilon_n^k$. Change point positions are thus obtained by identifying  cluster components. Fixing the maximum number of changes at a known value ${m\le n-1}$ implies assuming a null probability for realizations of the process with more than ${m}$ change
points, which requires that reliable prior information about ${m}$ should be available. We note here that, {\it a priori}, our proposal \eqref{rho_BMCP} does not limit the number of change points to a pre-specified maximum. {Another important issue is that \citeauthor{peluso2019}'s model has shown to be very sensitive to the choice of $m$, providing very different results when setting different values for $m$ (see Section~\ref{secReal_IR}).}

Given the partitions $\rho_1,\dots,\rho_d$, we assume that (i) the $d$ sequences of structural parameters $\bm{\theta}_1,\dots,\bm{\theta}_d$ are independent and (ii) the cluster parameters
$\theta^{\star}_{k,1},\dots,\theta^{\star}_{k,b_k}$, $k=1,\dots,d$, related to each sequence are independent. Under these assumptions, the joint prior distribution of $\bm{\theta}_1,\dots,\bm{\theta}_d$, given $\rho_1,\dots,\rho_d$, is 
\begin{equation}\label{theta_rho_BMCP}
f(\bm{\theta}_1,\dots,\bm{\theta}_d\mid\rho_1,\dots,\rho_d)
%& = \prod_{k=1}^{d}f(\bm{\theta}_k\mid\rho_1,\dots,\rho_d)
%= \prod_{k=1}^{d}f(\bm{\theta}_k\mid\rho_k)\\[.1in]
 = \prod_{k=1}^{d}\prod_{j_k=1}^{b_k}f_k(\theta^{\star}_{k,j_k}),
\end{equation}

\noindent where $f_k(\theta^{\star}_{k,j_k})$ is the prior distribution for the cluster parameter $\theta^{\star}_{k,j_k}$, $k=1,\dots,d$, $j_k=1,\dots,b_k$.
%To simplify the notation, we set $f_{S_{k,j_k}}(\theta^{\star}_{k,j_k})=f_k(\theta^{\star}_{k,j_k})$.

In the PPM introduced by \cite{bh92}, the partition indirectly induces the clusterization of the sequence of variables $\bm{X}=\left(X_1,\dots,X_n\right)$ by imposing that observations which indexes belong to the same cluster  are identically distributed. It also assumes independence across clusters.
In the proposed model, as the change points in the sequences of parameters $\bm{\theta}_1,\dots,\bm{\theta}_d$ are realizations of independent processes, the random partitions associated with different parameters do not necessarily induce the same number of clusters in the parameters and, even when such numbers are equal, the clusters may be distinct.
However, the partitions $\rho_1,\dots,\rho_d$ will induce a unique partition $\rho^\star$ in the sequence of variables $\bm{X}=\left(X_1,\dots,X_n\right)$. As each partition $\rho_k$ is defined by an ordinate sequence of  end points $0=i_{k,0}<i_{k,1}<\dots<i_{k,b_k}=n$, if we take the union of the end points of all $d$ partitions, we will obtain an ordered sequence of end points
$0=i^\star_{0}<i^\star_{1}<\dots<i^\star_{b^\star}=n$ belonging to $I$ that splits  the sequence of observations $\bm{X}$ in contiguous sets of independent and identically distributed (iid) variables. Thus, partition $\rho^\star$ is defined as

\begin{equation}\label{def:rho_star}
	\rho^\star = \{i^\star_{0},i^\star_{1},\dots,i^\star_{b^\star}\} = \cup_{k=1}^{d}\rho_k,
	\;\;\;\;
	0 = i^\star_{0} < i^\star_{1} < \dots < i^\star_{b^\star} = n.
\end{equation}

\begin{sloppypar}
\noindent Alternatively, $\rho^\star$ may be represented by the set of clusters ${\{S^\star_{1},\dots,S^\star_{b^\star}\}}$, where ${S^\star_{j}=\{i^\star_{j-1}+1,\dots,i^\star_{j}\}}$, $j=1,$ $\dots,$ $b^\star$. Under this representation, each nonempty subset $S_{1,j_1}\cap\dots\cap S_{d,j_d}$,  ${j_k=1,\dots,b_k}$ and
${k=1,\dots,d}$, specifies one of the clusters $S^\star_j\in\rho^\star$, such that $S^\star_j=S_{1,j_1}\cap$ $\dots\cap $ $S_{d,j_d}$. Thus the observations that have an index belonging to $S^\star_{j}$ share the same structural parameters.
%The random number of clusters in $\bm{X}$ induced by $\rho^\star$ may be defined by $B^\star= \#\{S^\star_{1},\dots,S^\star_{b^\star}\}$.
\end{sloppypar}

%\begin{equation*}\label{def:Bstar}
%B^\star=\#\left\{(j_1,\dots,j_d)\,:\, {j_k=1,\dots,b_k},\; {k=1,\dots,d},\; %{S_{1,j_1}\cap\dots\cap S_{d,j_d}}\ne\emptyset\right\}.
%\end{equation*}

%\begin{equation*}\label{def:Bstar}
%	B^\star= \#\{S^\star_{1},\dots,S^\star_{b^\star}\}.
%\end{equation*}

Denote by $\bm{X}_{S^\star_j}$ the subsequence of observations with index belonging to the set $S^\star_j$.
Assume that,  given $(\bm{\theta},\bm{\rho})=(\bm{\theta_1},\dots,\bm{\theta_d},\rho_1,\dots,\rho_d)$, $\bm{X}_{S^\star_1}, \dots, \bm{X}_{S^\star_{b^\star}}$ are independent and that, for all $i\in S^\star_j$, $X_i$ are iid variables with conditional marginal density
$f(X_i\mid\theta^{\star}_{1,j_1},\dots,\theta^{\star}_{d,j_d})$, where ${j_1,\dots,j_d}$ are the indexes defining $S^\star_j$. Under these assumptions, the likelihood function is given by

\begin{equation}\label{likelihood}
\begin{aligned}
f(\bm{X}\mid\bm{\theta}, \bm{\rho})\,
%& = \prod_{S^\star_j\in\rho^\star}
%& = \prod_{j=1}^{b^\star}
%f(\bm{X}_{S^\star_j} \mid \theta^{\star}_{1,j_1},\dots,\theta^{\star}_{d,j_d})\\
%& = \prod_{S^\star_j\in\rho^\star}\;\prod_{i\in S^\star_j}
& = \prod_{j=1}^{b^\star}\;\prod_{i\in S^\star_j}
f(X_i\mid\theta^{\star}_{1,j_1},\dots,\theta^{\star}_{d,j_d}).
\end{aligned}	
\end{equation}

\noindent Considering the likelihood in (\ref{likelihood}) and the prior specifications in (\ref{theta_rho_BMCP}) and (\ref{rho_BMCP}), the joint posterior distribution for $(\bm{\theta},\bm{\rho})$
%parameters $\bm{\theta}_1,\dots,\bm{\theta}_d$ and partitions $\rho_1,\dots,\rho_d$
is given by
\begin{equation}\label{all_X}
f(\bm{\theta}, \bm{\rho} \,\mid\, \bm{X}) \;\propto\; f(\bm{X}\mid\bm{\theta}, \bm{\rho})\;
\prod_{k=1}^{d}\prod_{j_k=1}^{b_k}f_k(\theta^{\star}_{k,j_k})c_k(S_{k,j_k}).
\end{equation}

\subsection{Sampling from the posterior distributions}\label{secBMCP_pcg}

The posterior distribution in (\ref{all_X}) is intractable and some computational procedures must be used to approximate it. We propose a partially collapsed Gibbs sampler \citep{pcg2008} based on a blocking strategy to sample from the joint posterior distribution of the partitions $\bm{\rho} = \{\rho_1,\dots,\rho_d \}$ and parameters $\bm{\theta}=\{\bm{\theta}_1,\dots,\bm{\theta}_d\}$.
To consider a more general context, let $\bm{\delta}$ denote the set of hyperparameters indexing the prior distribution of $(\bm{\theta},\bm{\rho})$.
%
%Let the subscript $(-l)$ indicates the absence of the element indexed by
%$l$ in the respective object
Denote by $\bm{\theta}_{(-\ell)}$ the set $\bm{\theta}$ without vector $\bm{\theta}_\ell$ and  by $\bm{\theta}_{k,(-i)}$ the vector $\bm{\theta}_{k}$ without coordinate $i$, that is,
$\bm{\theta}_{k,(-i)}=(\theta_{k,1},\dots,\theta_{k,i-1},$$\theta_{k,i+1},\dots,$$\theta_{k,n})$.
Assuming that the proposed model holds, and also that given a partition $\rho_k$ the $i$th coordinate of $\bm{\theta}_k$ belongs to cluster $S_{k,J}$, the full conditional posterior distributions for $\rho_k$, $\bm{\theta}_k$ and $\bm{\delta}$ are, respectively,
\begin{eqnarray*}
&&P(\rho_k =\{S_{k,1},\dots,S_{k,b_k} \} \mid \bm{\rho}_{(-k)}, \bm{\theta}, \bm{\delta}, \bm{X}) \;\;\propto\;\;  f(\bm{X}\mid\bm{\theta}, \bm{\rho})  \prod_{j_k=1}^{b_k} f_k(\theta^{\star}_{k,j_k})c_k(S_{k,j_k}),\\
&&f_k( \theta_{k,i} \mid \bm{\rho}, \bm{\theta}_{(-k)}, \bm{\theta}_{k,(-i)}, \bm{\delta}, \bm{X})
\;\;\propto\;\; f_k(\theta^{\star}_{k,J}) \int_{\bm{\Theta} - \{\theta_{k,i}\}} f(\bm{X}\mid\bm{\theta},\bm{\rho}) d\bm{\theta},\\
&&f(\bm{\delta} \mid \bm{\rho}, \bm{\theta}, \bm{X})
\;\;\propto\;\;
f(\bm{\delta}) \prod_{k=1}^{d}\prod_{j_k=1}^{b_k}f_k(\theta^{\star}_{k,j_k})c_k(S_{k,j_k}),	
\end{eqnarray*}
\noindent where $\bm{\Theta}-\{\theta_{k,i}\}$ is the parameter space for the vectors $(\bm{\theta}_1,\dots,\bm{\theta}_{k-1},\bm{\theta}_{k,(-i)}, \bm{\theta}_{k+1},$ $\dots,$ $\bm{\theta}_d)$.

Because $\rho_1,\dots,\rho_d$ are not supported on Euclidean spaces but discrete ones, a major challenge in the proposed multipartition model is to handle their posterior distributions. To sample from the full conditional distributions of $\rho_1,\dots,\rho_d$, we adapt the method
proposed by \cite{bh93} to our multipartition model. Each random partition $\rho_k$ is represented by a fixed dimension random vector $\bm{U}_k=(U_{k,1},\dots,U_{k,n-1})$ where each coordinate $U_{k,i}$, $k=1,\dots,d$, $i=1,\dots,n-1$, indicates whether or not a change point occurred at position $i+1$ of the parameter vector $\bm{\theta}_k$, that is
\[
U_{k,i}=\begin{cases}
\;\; 1 \;\;\;\;\text{if}\;\;\; \theta_{k,i}=\theta_{k,i+1},\\
\;\; 0 \;\;\;\;\text{if}\;\;\; \theta_{k,i}\ne\theta_{k,i+1}.
\end{cases}
\]

Thus, the equivalence $(\bm{X},\bm{\theta}_1,\dots,\bm{\theta}_d,\rho_1,\dots,\rho_d,\bm{\delta})$
 $\Leftrightarrow$
 $(\bm{X},\bm{\theta}_1,\dots,\bm{\theta}_d,\bm{U_1},\dots,\bm{U_d},\bm{\delta})$
is verified. Pseudo-code to sample from the joint posterior distribution $f(\bm{\theta},\bm{\rho},\bm{\delta}\mid\bm{X})$ is presented next, in Algorithm \ref{Pseudocode}, where 
$\mathcal{U}_{(-k)}^{(t)}=$ $(\bm{U}_1^{(t)},\dots,\bm{U}_{k-1}^{(t)},$ $\bm{U}_{k+1}^{(t-1)},$
$\dots,\bm{U}_d^{(t-1)})$, and
$\bm{U}_{k,(-i)}^{(t)}=$  $(U_{k,1}^{(t)},$ $\dots,U_{k,i-1}^{(t)},$  $ U_{k,i+1}^{(t-1)},\dots,$ $U_{k,n-1}^{(t-1)})$
denotes the imputed value of $\bm{U}_{k}$ at iteration $t$, with the $i$th
coordinate removed.

\begin{figure}[ht]
%\begin{adjustwidth}{-.2cm}{-.2cm}
		\centering
		\begin{minipage}{1\linewidth}		
			\begin{algorithm}[H]
				\hrulefill\\
				\KwIn{$\bm{X}$}
				\text{Initialize} $\bm{\delta}^{(0)},\bm{\theta}^{(0)}_1,\dots,\bm{\theta}^{(0)}_d,
				\bm{U}^{(0)}_1,\dots,\bm{U}^{(0)}_d$ ($U^{(0)}_{k,i}$ may be initialized as all 0)\;
				\For{t = 1 {\rm to} T} {
					$\bm{\delta}^{(t)} \sim f(\bm{\delta} \mid \bm{U}^{(t-1)}_1,\dots,\bm{U}^{(t-1)}_d, \bm{\theta}^{(t-1)}_1,\dots,\bm{\theta}^{(t-1)}_d, \bm{X})$\;
					\For{k = 1 {\rm to} d} {
						\For{i = 1 {\rm to} n} {
							$U_{k,i}^{(t)} \sim p(U_{k,i} \mid
							\bm{\delta}^{(t)}, \mathcal{U}_{(-k)}^{(t)}, \bm{U}_{k,(-i)}^{(t)}, \bm{\theta}^{(t)}_1,\dots,\bm{\theta}^{(t)}_{k-1},
							\bm{\theta}^{(t-1)}_{k+1},\dots,\bm{\theta}^{(t-1)}_d,\bm{X})$\;
						}
						\For{$j_k$ = 1 {\rm to} $b_k$} {
							$\theta_{k,j_k}^{\star(t)} \sim f(\theta^{\star}_{k,j_k} \mid
							\bm{\delta}^{(t)},\mathcal{U}_{(-k)}^{(t)}, \bm{U}_{k}^{(t)},
							\bm{\theta}^{(t)}_1,\dots,\bm{\theta}^{(t)}_{k-1},
							\bm{\theta}^{(t-1)}_{k+1},\dots,\bm{\theta}^{(t-1)}_d,
							\bm{X})$\;
						}
						$\bm{\theta}_k^{(t)} = \displaystyle\sum_{j_k=1}^{b_k} (\theta^{\star(t)}_{k,j_k}{\bf{1}}\{1\in S_{k,j_k}\},
						\dots,
						\theta^{\star(t)}_{k,j_k}{\bf{1}}\{n\in S_{k,j_k}\}).$
					}
				}
				\hrulefill
				\vspace{.1in}
				\caption{MCMC scheme to sample from $f(\bm{\theta},\bm{\rho},\bm{\delta}\mid\bm{X})$.}
				\label{Pseudocode}
			\end{algorithm}
		\end{minipage}
%\end{adjustwidth}
\end{figure}

Samples from the posterior distribution of each partition $\rho_k$, $k=1,\dots,d$, are obtained by sampling from the full conditional distribution of $\bm{U}_{k}$. In the $t$th iteration, these samples are obtained considering the following ratio:

\begin{equation}\label{Rt}
R_{k,i}^{(t)} = 
%\frac{p_{k,i}}{1-p_{k,i}} =
\frac{p(U_{k,i}=1\hspace{.05in}\mid\hspace{.05in}
	\mathcal{U}_{(-k)}^{(t)}, \bm{U}_{k,(-i)}^{(t)},
	\bm{\theta}^{(tk)},
	\bm{\delta}^{(t)},\bm{X})}
{p(U_{k,i}=0
	\hspace{.05in}\mid\hspace{.05in}
	\mathcal{U}_{(-k)}^{(t)}, \bm{U}_{k,(-i)}^{(t)},
	\bm{\theta}^{(tk)},
	\bm{\delta}^{(t)},\bm{X})},
\end{equation}
where $\bm{\theta}^{(tk)}= (\bm{\theta}_1^{(t)},\dots,\bm{\theta}_{k-1}^{(t)}, \bm{\theta}_k^{(t-1)}, \dots, \bm{\theta}_d^{(t-1)})$.
The partitions in the numerator and denominator of (\ref{Rt}) only differ at position $i$. Assume that the partition in the numerator is $(U_{k,1},\dots,U_{k,i-1},U_{k,i}=1, U_{k,i+1},\dots,$ $ U_{k,n-1})$ and that a cluster $S_{k,J}$ contains the $i$th element of $I$. The partition in the denominator splits $S_{k,J}$ creating two new clusters. Although all the other clusters are shared by both partitions, these two different partitions induce a different number of distinct coordinates in $\bm{\theta}_k$ in the numerator and denominator in (\ref{Rt}). To sample from the posterior of each $\rho_k$ using  Gibbs sampler,  it is needed to
integrate $\bm{\theta}_k$ out in expression (\ref{Rt}). Such integrals can be done anallytically or using computational approaches.
Under the model assumptions, the probabilities in (\ref{Rt}) are given by
\begin{eqnarray}
	p(U_{k,i}&\mid&\mathcal{U}_{(-k)}^{(t)}, \bm{U}_{k,(-i)}^{(t)},
	\bm{\theta}_{(-k)}, \bm{\delta},\bm{X})
	\propto \displaystyle\prod_{j_k=1}^{b_k}
	f(\bm{X}_{S_{k,j_k}}\mid\bm{\theta}^{\star}_{(-k)})
	c_k(S_{k,j_k}),
\end{eqnarray}
\noindent where
$\bm{\theta}^{\star}=\{\bm{\theta}^{\star}_1,\dots,\bm{\theta}^{\star}_d\}$,
$\bm{\theta}^{\star}_k=(\theta^{\star}_{k,1},\dots,\theta^{\star}_{k,b_k})$
and the likelihood function restricted to cluster $S_{k,j_k}$ is given by
\begin{eqnarray}\label{integrated_likelihood}
	&&f(\bm{X}_{S_{k,j_k}}\mid\bm{\theta}^{\star}_{(-k)})=
	\displaystyle\int
	\Bigg(
	\prod_{S^\star_j\subset S_{k,j_k}}\!\!\!\!\!\!
	f(\bm{X}_{S^\star_j}\mid\theta^{\star}_{1,j_1},\dots,\theta^{\star}_{d,j_d})
	\Bigg)
	f_k(\theta^{\star}_{k,j_k})\,d\theta^{\star}_{k,j_k}. \nonumber
\end{eqnarray}

\noindent For all $i\in S_{k,j_k}$, the new value for coordinate $U_{k,i}$
is given by
\begin{equation}\label{U_ki}
U^{(t)}_{k,i} =  \mathbf{1}\left\{\frac{u}{1-u}\le R_{k,i}^{(t)}\right\},
\end{equation}

\noindent where $u$ is a draw from the $U(0,1)$ distribution and
\begin{eqnarray}\label{RU_ki}
&&R_{k,i}^{(t)} = \frac{f(\bm{X}_{S_{k,j_k}}\mid\bm{\theta}^{\star}_{(-k)})c_k(S_{k,j_k})}
{f(\bm{X}_{S_{ki,j_k}}\mid\bm{\theta}^{\star}_{(-k)})c_{k}(S_{ki,j_k})\;
f(\bm{X}_{S_{ik,j_k}}\mid\bm{\theta}^{\star}_{(-k)})c_{k}(S_{ik,j_k})},\nonumber
\end{eqnarray}
\noindent where $S_{ki,j_k}=\{i_{k,j_k-1}+1,i_{k,j_k-1}+2,\dots,i-1,i\}$
and $S_{ik,j_k}=\{i+1,i+2,\dots,i_{k,j_k}-1,i_{k,j_k}\}$.

\section{The BMCP model for Normal data}\label{secBMCPnormal}

The previous development was general. We now specialize the discussion to the Normal likelihood case where both location and scale parameters are unknown, giving detailed account of all the calculations required to implement Algorithm \ref{Pseudocode} as described earlier. So, we consider the sequence of random variables $\bm{X}=(X_1,\dots,X_n)$ and the sequences of unknown structural parameters $\bm{\mu}=(\mu_1,\dots,\mu_n)$ and $\bm{\sigma}=(\sigma^2_1,\dots,\sigma^2_n)$. Following the general discussion in Section~\ref{secBMCP_def} we assume that $X_i\mid\bm{\mu},\bm{\sigma} \overset{ind}{\sim} N(\mu_i,\sigma^2_i)$, $i=1,\dots,n$. In addition, change points in $\bm{\mu}$ and $\bm{\sigma}$ are assumed to occur independently, at unknown and possibly different instants.
Let  $\rho_1$ and $\rho_2$ be the random partitions of $I$ that induce contiguous clusters in $\bm{\mu}$ and $\bm{\sigma}$, respectively. Denote by $\mu^{\star}_{j_1}$ the common mean into the cluster $S_{1,j_1}$, $j_1=1,\dots,b_1$ and let $\sigma^{2\star}_{j_2}$ be the common variance
for observations into the cluster $S_{2,j_2}$, $j_2=1,\dots,b_2$.
To simplify the notation, let $S_{j_k}=S_{k,j_k}$ and $n_{j_k}\!=\#S_{j_k}$, for $k=1,2$. Also, let $n^\star_j\!=\#S^\star_j$ and $\overline{X}_{S^\star_j}\!=\textstyle\sum_{i\in
S^\star_j}\!X_i/n^\star_j$, for $j=1,\dots,b^\star$.

Given $\rho_1$ and $\rho_2$, assume that (i) the observations $X_i$ for ${i\in S^\star_{j}=S_{j_1}\cap S_{j_2}}$ are independent and identically distributed with
$X_i\mid\mu^{\star}_{j_1},\sigma^{2\star}_{j_2} \overset{iid}{\sim}
N(\mu^{\star}_{j_1},\sigma^{2\star}_{j_2})$ and (ii) observations in
different clusters are independent. Under these assumptions, the
likelihood function is given by
\begin{eqnarray}\label{likelihood_Normal}
&&	f(\bm{X}\mid\bm{\mu},\bm{\sigma},\rho_1,\rho_2)= 
	\prod_{j_1=1}^{b_1}\;\;
	\prod_{j_2\mid S^\star_{j}\ne\emptyset}	
	\left(\frac{1}{2\pi\sigma^{2\star}_{j_2}}\right)^{n^\star_{j}/2}\!\!\!\!\!\!
	\exp\left\{-\!\!\!\sum_{i\in S^\star_{j}}\!\!\! \frac{(X_i-\mu^{\star}_{j_1})^2}{2\sigma^{2\star}_{j_2}}\right\},\nonumber
\end{eqnarray}
\noindent where $\{j_k\mid S^\star_{j}\ne\emptyset\}$ denotes the set of values $j_k$ for which $S^\star_{j}\ne\emptyset$, for $k=1,2$. The double product in \eqref{likelihood_Normal} is equivalent to the single product $\prod_{j=1}^{b^\star}$ over $\rho^\star$ in (\ref{likelihood}) when $d=2$. It is also equivalent to its permuted form ${\prod_{j_2=1}^{b_2}\;\prod_{j_1\mid S^\star_{j}\ne\emptyset}}$.

Given $\rho_1$ and $\rho_2$, we assume that $\bm{\mu}$ and $\bm{\sigma}$ are independent and the structural parameters in different clusters are also independent, with prior distributions
\begin{equation}\label{mu_Normal}
\begin{aligned}
	\mu^{\star}_{j_1} \overset{iid}{\sim} N(\mu_0,\sigma^2_0), & \;\;\;\; j_1=1,...,b_1,\\[.05in]
	\sigma^{2\star}_{j_2} \overset{iid}{\sim}  IG(a/2,d/2),   & \;\;\;\; j_2=1,...,b_2.
\end{aligned}
\end{equation}

For the random partitions $\rho_1$ and $\rho_2$, we assume the independent product partition distributions given in (\ref{rho_BMCP}). The cohesion proposed by \cite{yao84} is considered to quantify how strongly we believe the components of $\bm{\mu}$ and $\bm{\sigma}$ are to co-cluster {\it a priori}. The Yao's cohesion is indexed by a parameter that represents the
probability of a change in the structural parameter to occur at any instant. As $\bm{\mu} $ and $\bm{\sigma}$ can change at different times, to accommodate this feature we assume that these probabilities are $p_1$ and $p_2$, $p_1 \neq p_2$, that model the prior uncertainty about $\rho_1$ and $\rho_2$, respectively. That is, for $k=1,2$, we assume the cohesions
\begin{equation}\label{cohesion_yao}
%\[
c_k(S_{j_k})=\begin{cases}
(1-p_k)^{n_{j_k}-1}p_k \hspace{.2in} \text{if} \hspace{.1in} j_k=1,2,\dots,b_k-1,\\
(1-p_k)^{n_{j_k}-1} \hspace{.33in} \text{if} \hspace{.1in} j_k=b_k.
\end{cases}
%\]
\end{equation}
\noindent Therefore, given $p_k$, the prior distribution for $\rho_k$ is given by 
\begin{equation}\label{rho_p_yao}
P(\rho_k=\{i_{k,0}, i_{k,1}, \cdots, i_{k,b_k}\}\mid p_k) = p_k^{b_k-1}(1-p_k)^{n-b_k}.
\end{equation}
To complete the model specification, we assume {\it a priori} that
\begin{equation}\label{p_Beta}
p_k\overset{ind}{\sim} Beta(\alpha_k,\beta_k),\qquad k=1,2.
\end{equation}

\subsection{On the prior for the number of change points}\label{secBMCPnormal_N}

Assuming the prior distributions in (\ref{rho_p_yao}) and (\ref{p_Beta}), the number of change points $N_k$, related to $\rho_k$, has a $Beta\text{-}Binomial(n-1,\alpha_k,\beta_k)$ prior distribution, such that
\begin{equation}\label{N_yao}
\begin{aligned}
P(N_k=c) & = \dbinom{n-1}{c}
\frac{\Gamma(\alpha_k+\beta_k)\Gamma(\alpha_k+c)\Gamma(n-1+\beta_k-c)}{\Gamma(\alpha_k)\Gamma(\beta_k)\Gamma(\alpha_k+\beta_k+n-1)},\;\;
\end{aligned}
\end{equation}
for $c=0,1,\dots,n-1.$ Thus, mean and variance are given, respectively, by
\begin{equation}\label{EN_yao}
\begin{aligned}
%E(N) & = E(E(N|p)) = E((n-1)p) = (n-1)E(p) = (n-1)\frac{\alpha}{\alpha+\beta}
E(N_k) & = (n-1)\frac{\alpha_k}{\alpha_k+\beta_k},\\[.05in]
Var(N_k) & = (n-1)\frac{\alpha_k\beta_k(\alpha_k+\beta_k+n-1)}{(\alpha_k+\beta_k)^2(\alpha_k+\beta_k+1)}.
\end{aligned}
\end{equation}

If $\alpha_k =\beta_k$, we assume {\it a priori} that around $50\%$ of the observations experienced a change. Considering $\alpha_k=\beta_k=1$ implies the prior assumption that $N_k\sim U\{0,1,\dots,n-1\}$.
%
%If $\alpha_k\rightarrow 0$ and $\beta_k\rightarrow 0$ we have great
%uncertainty about the true number of changes.
%
Also, a prior assumption that $E(N_k)=c$ implies that $\beta_k=\alpha_k L$, where $L=(n-1-c)/c$. In this case,
\begin{equation}\label{VN_yao_lim}
\begin{aligned}
Var(N_k) & = c\hspace{.05in}\frac{L}{1+L}\hspace{.05in}\frac{\alpha_k(1+L)+n-1}{\alpha_k(1+L)+1}.
\end{aligned}
\end{equation}

The derivative of $Var(N_k)$ w.r.t $\alpha_k$ is given by $[cL(2-n)][\alpha_k(1+L)+1]^{-2}$, that is negative for $n\ge3$. We have also that $\lim_{\alpha_k\rightarrow\infty}Var(N_k)=c(n-1-c)/(n-1)$. These results are useful to guide our prior choices of the hyperparameters $\alpha_k$ and $\beta_k$. We express greater uncertainty about $N_k$ by eliciting lower values for $\alpha_k$.

Conditional to $p_1,\dots,p_d$, the number of changes $N^*$ in $\rho^*$ has a ${Binomial(n-1\;,\;p^*)}$ distribution, where $p^*=1-\prod_{k=1}^{d}(1-p_k)$. Therefore,
the expectation and the variance of $N^*$, respectively, are %\footnote{\TB{This result is actually general in the sense that it does not depend on the likelihood but only on the choice of prior cohesions. It could be presented even before discussing likelihood, but it's ok to leave it here too}}
\begin{equation*}\label{EN}
\begin{aligned}
&E(N^*) = (n-1)\left(1-\prod_{k=1}^{d}\frac{\beta_k}{\alpha_k+\beta_k}\right),\\
&Var(N^*)  = (n-1)\prod_{k=1}^{d}\frac{\beta_k}{\alpha_k+\beta_k}
	- (n-1)^2\prod_{k=1}^{d}\left(\frac{\beta_k}{\alpha_k+\beta_k}\right)^2\\
& + (n^2-3n+2)\prod_{k=1}^{d}\left( \left(\frac{\beta_k}{\alpha_k+\beta_k}\right)^2 + \frac{\alpha_k\beta_k}{(\alpha_k+\beta_k)^2(\alpha_k+\beta_k+1)}\right).\\[.05in]
\end{aligned}
\end{equation*}

\subsection{Posterior distribution and sampling}

Considering the likelihood in (\ref{likelihood_Normal}) and the prior distributions in (\ref{mu_Normal}), (\ref{rho_p_yao}) and (\ref{p_Beta}) with known values for the hyperparameters ${(\mu_0,\sigma^2_0,a,d,\alpha_1,\beta_1,\alpha_2,\beta_2)}$, the joint posterior
distribution for the Normal data model becomes 
\begin{eqnarray}\label{all_X_Normal}
&& f(\bm{\mu},\bm{\sigma},\rho_1,\rho_2,p_1,p_2\mid\bm{X})
\propto\Bigg(\prod_{j_1=1}^{b_1}\;\prod_{j_2\mid S^\star_{j}\ne\emptyset}
	\left(\frac{1}{2\pi\sigma^{2\star}_{j_2}}\right)^{n^\star_{j}/2}\!\!\!\!\!\!\!
	\exp\left\{-\!\!\sum_{i\in S^\star_{j}}\!\! \frac{(X_i-\mu^{\star}_{j_1})^2}{2\sigma^{2\star}_{j_2}}\right\}
	\Bigg)\nonumber\\[.1in]
&&\times\;\; \exp \left\{-\frac{1}{2\sigma_0^2} \sum_{j_{1}=1}^{b_1}
	(\mu^{\star}_{j_1}-\mu_0)^2 \right\}
\times\;\Bigg( \prod_{j_2=1}^{b_2}\left(\frac{1}{\sigma^{2\star}_{j_2}}\right)^{({d}+2)/2}
	\!\!\!\!\!\!\!\!
	\exp\left\{ -\frac{a}{2\sigma^{2\star}_{j_2}} \right\}
	\Bigg)\nonumber\\[.1in]
&& \times\;\;p_1^{\alpha_1+b_1-2}(1-p_1)^{n+\beta_1-b_1-1}
\times\;\;p_2^{\alpha_2+b_2-2}(1-p_2)^{n+\beta_2-b_2-1}.
\end{eqnarray}

The posterior distribution in (\ref{all_X_Normal}) has no known closed form. We sample from it through the partially colapsed Gibbs sampler scheme given in Algorithm \ref{Pseudocode}.
%To simplify notation, let ${j_k\mid S^\star_{j}=\{j_k\mid S^\star_{j}\ne\emptyset\}}$ for $k=1,2$.
%
To sample from the full conditional distribution of $\rho_1$ using the ratio $R_{1,i}^{(t)}$ in (\ref{RU_ki}), we consider the specific case of (\ref{integrated_likelihood}) to $\bm{X}_{S_{j_1}}$ given $\bm{\sigma}$ for each cluster $S_{j_1}$, $j_1=1,\dots,b_1$. This distribution is free of $\bm{\mu}$. It is given by
\begin{equation}\label{datafactor_mu_calc}
\begin{aligned}
& {f(\bm{X}_{S_{j_1}}\!\mid\bm{\sigma})}
	= \displaystyle\int
	\Bigg(
	\prod_{j_2\mid S^\star_{j}\ne\emptyset}
	f(\bm{X}_{S^\star_{j}}\!\mid\mu^{\star}_{j_1},\sigma^{2\star}_{j_2})
	\Bigg)
	f_1(\mu^{\star}_{j_1})
	\;d\mu^{\star}_{j_1}\nonumber \\[.1in]
&\;\;\;\;\;=  \;\left(\frac{1}{(2\pi)^{n_{j_1}}\sigma^2_0\bm{Q_1}} \right)^{1/2}
	\Bigg(
	\prod_{j_2\mid S^\star_{j}\ne\emptyset}\left(\frac{1}{\sigma^{2\star}_{j_2}}\right)^{n^\star_{j}/2}
	\Bigg)
	\;\;\times\;\; \exp\left\{-\frac{1}{2}\left[\left(
	\sum_{j_2\mid S^\star_{j}\ne\emptyset}
	\frac{\displaystyle\sum_{i\in S^\star_{j}}X_i^2}{\sigma^{2\star}_{j_2}}
	\right)
	+\frac{\mu^2_0}{\sigma^2_0}
	-\frac{\bm{Q_2^2}}{\bm{Q_1}}
	\right]\right\},
\end{aligned}
\end{equation}
\noindent where $\bm{Q_1}=\sum_{j_2\mid
S^\star_{j}\ne\emptyset}n^\star_{j}(\sigma^{2\star}_{j_2})^{-1}+(\sigma^2_0)^{-1}$
and $\bm{Q_2}=\sum_{j_2\mid S^\star_{j}\ne\emptyset}
{n^\star_{j}\overline{X}_{S^\star_{j}}}(\sigma^{2\star}_{j_2})^{-1} +
{\mu_0}(\sigma^2_0)^{-1}$.
Similarly, to sample from the full conditional distribution of $\rho_2$ we consider the distribution of $\bm{X}_{S_{j_2}}$ given $\bm{\mu}$ for each cluster $S_{j_2}$, $j_2=1,\dots,b_2$, that is free of $\bm{\sigma}$. It is given by
\begin{equation}\label{datafactor_s2_calc}
\begin{split}
	& {f(\bm{X}_{S_{j_2}}\!\mid\bm{\mu})}
	= \int
	\Bigg(
	\prod_{j_1\mid S^\star_{j}\ne\emptyset}
	f(\bm{X}_{S^\star_{j}}\!\mid\mu^{\star}_{j_1},\sigma^{2\star}_{j_2})
	\Bigg)
	f_2(\sigma^{2\star}_{j_2})
	\;d\sigma^{2\star}_{j_2}\\[.051in]
	&\;\;=(2\pi)^{-n_{j_2}/2}\;\frac{(a/2)^{d/2}\Gamma(n_{j_2}+d/2)}{\Gamma(d/2)}
	\;\left(
	\frac{\sum_{j_1\mid S^\star_{j}\ne\emptyset}\sum_{i\in S^\star_{j}}(X_i-\mu^{\star}_{j_1})^2+a}{2}\right)^{-(n_{j_2}+d)/2}.
\end{split}
\end{equation}

To sample from the posterior distribution of $\mu_i$, $i\in S_{j_1}$, consider the Normal distribution

\begin{equation}\label{mu_all_BMCP}
\mu^{\star}_{j_1}\mid\bm{\sigma},\rho_1,\rho_2,\bm{X}\sim N\left(\frac{\bm{Q_2}}{\bm{Q_1}}\;,\;\bm{Q_1}^{-1}\right),
\end{equation}

\noindent with $\bm{Q_1}$ and $\bm{Q_2}$ as defined in (\ref{datafactor_mu_calc}), and sample from $\sigma^2_i$, $i\in S_{j_2}$, through the Inverse-Gamma distribution

\begin{eqnarray}\label{s2_all_BMCP}
&&\sigma^{2\star}_{j_2}\mid\bm{\mu},\rho_1,\rho_2,\bm{X}
\sim IG\Bigg(
%\Bigg(
\sum_{j_1\mid S^\star_{j}\ne\emptyset}\;
\sum_{i\in S^\star_{j}}
(X_i-\mu^{\star}_{j_1})^2/2
%\Bigg)
+a/2\;,\;(n_{j_2}+d)/2
\Bigg).\nonumber
\end{eqnarray}

\noindent Finally, the full conditional distribution of $p_k$, $k=1,2$, is given by
\begin{equation}\label{p1_all_BMCP}
\begin{aligned}
p_k\mid\rho_k,\bm{X} &\sim Beta(\alpha_k+b_k-1,n+\beta_k-b_k).
\end{aligned}
\end{equation}

Performance of the BMCP model introduced in this section is analyzed in next section through simulation studies.

\subsection{Monte Carlo simulation study}\label{secBMCPnormal_MC}

We ran a Monte Carlo simulation study to evaluate the performance of the proposed BMCP model when identifying multiple change points in Normal means and variances. We compare the proposed BMCP with the models BH93 \citep{bh93}, LCIA05 \citep{loschi2005p} and DPM19 \citep{peluso2019}.

BH93 and LCIA05 models consider a single partition $\rho$ to identify the changes. BH93 is proposed to identify changes only in the mean, under constant variance. It assumes that\linebreak
${X_i\mid\mu_i,\sigma^2\overset{ind}{\sim}N(\mu_i,\sigma^2)}$, $i=1,\dots,n$ and that, {\it a priori}, ${\mu^{\star}_j\overset{iid}{\sim}N(\mu_0,\sigma_0^2/n_j)}$, where $n_j=\#S_j$, $j=1,\dots,b$. We consider the same prior specifications proposed in \cite{bh93} for $\sigma^2$, $\mu_0$ and $\sigma_0^2$. LCIA05 identifies changes in $\bm{\mu}$ or $\bm{\sigma}$, but does not specify which parameter has changed. To analyze the data using LCIA05 the following prior distributions for the cluster parameters are assumed: ${\mu^{\star}_j\mid\sigma^{2\star}_j\overset{iid}{\sim}N(m,v\sigma^{2\star}_j)}$
and ${\sigma^{2\star}_j\overset{iid}{\sim}IG(a/2,d/2)}$ where ${(m,v,a,d)=(0,2,0.1,2.1)}$.
In the simulation study, the changes estimated by the LCIA05 and BH93 models should be compared to the true $\rho^\star$ and $\rho_1$, respectively.
For the BMCP model, we assume the prior distributions  given in (\ref{mu_Normal}), with hyperparameters $(\mu_0,\sigma^2_0,a,d)=(0,100,0.1,2.1)$, which is a reasonably flat prior for both $\mu^{\star}_{j_1}$ and $\sigma^{2\star}_{j_2}$.
For all the partition models (BH93, LCIA05 and BMCP), we consider the Yao's cohesion to model the prior uncertainty about the random partitions. The $Beta(1,1)$ is assumed as the prior distributions for parameters $p_1$ and $p_2$ in BMCP and for parameter $p$ in the LCIA05. In BH93, we assume $p\sim U(0,0.05)$.

The DPM19 model identifies changes in $\bm{\mu}$ and $\bm{\sigma}$ separately through random discrete state vectors that indicate the regime of each parameter at each instant. We refer to these state vectors by $E_1$ and $E_2$, respectively. We consider the DPM19 prior specifications
exactly as proposed in the original paper: the $Beta(1,1)$ prior distribution for the time-dependent probability of regime change; for the regime parameters, we assume Dirichlet process prior distributions with concentration parameters $M_1\overset{D}{=}M_2\sim Ga(0.05,0.0001)$ and base distributions $N(\mu_0,\sigma_0^2)$ for the regime means and
$IG(a,d)$ for the regime variances, where $(\mu_0,\sigma_0^2,a,d)=(0,1,1,1)$. We fix the maximum number of changes $m_1$ and $m_2$ as the true number of changes in $\bm{\mu}$ and $\bm{\sigma}$, respectively. This is a highly informative choice.

Three different scenarios are considered and $400$ data sets are generated in each case. For each data set, samples of the posterior distribution are obtained through the proposed MCMC scheme. In the case of the BMCP, LCIA05 and BH93 models, $20,000$ samples were generated after a warm-up period of $30,000$ iterations. These models were implemented in \texttt{C++} language and integrated to \texttt{R} through the \texttt{Rcpp} package \citep{rcpp2011}. For the DPM19 model, $2,000$ samples were generated after a warm-up period of $3,000$ iterations. The specification of small values for $m_1$ and $m_2$ considerably reduces the parameter space of the DPM19 model, when compared to the parameter space of the PPM based models. Therefore, a small number of iterations is required for the DPM19 sampling procedure. The \texttt{R} code of the DPM19 is available at \hbox{\url{https://github.com/stefanopel/DPM-change-point}}.\footnote{The authors thank Stefano Peluso for making code to fit the DPM19 model freely available.}

\subsubsection{Scenario 1: Changes in the mean and constant variance}\label{sec_scene1}

\begin{figure}[!htb]
%\begin{adjustwidth}{-.4cm}{-.4cm}
	%\centering
	\subfigure[][BMCP]{
		\includegraphics[width=3.5cm, height=3cm, trim=0 .5cm 0 0]{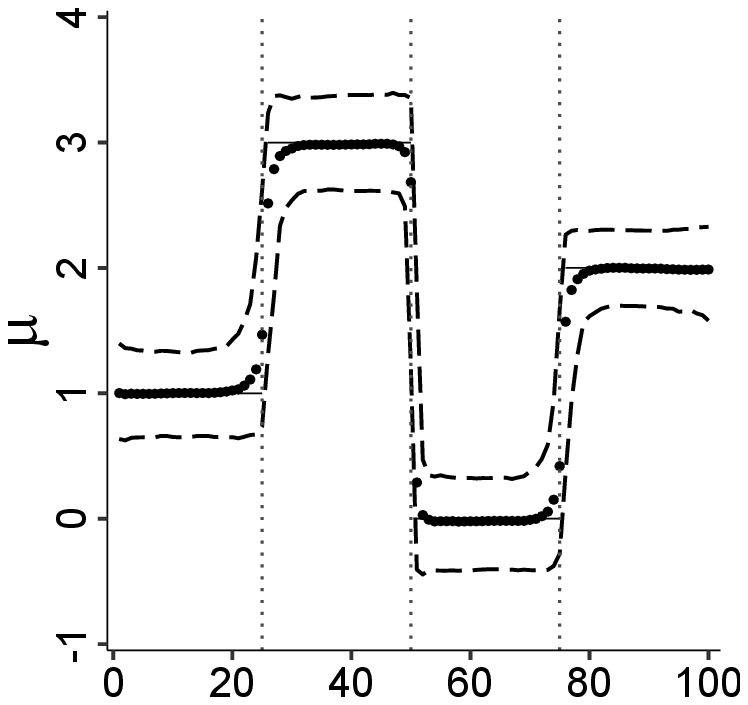}
		\label{fig:scene1_mu_LP20}}%\hspace{.01in}
	\subfigure[][DPM19]{
		\includegraphics[width=3.5cm, height=3cm, trim=0 .5cm 0 0]{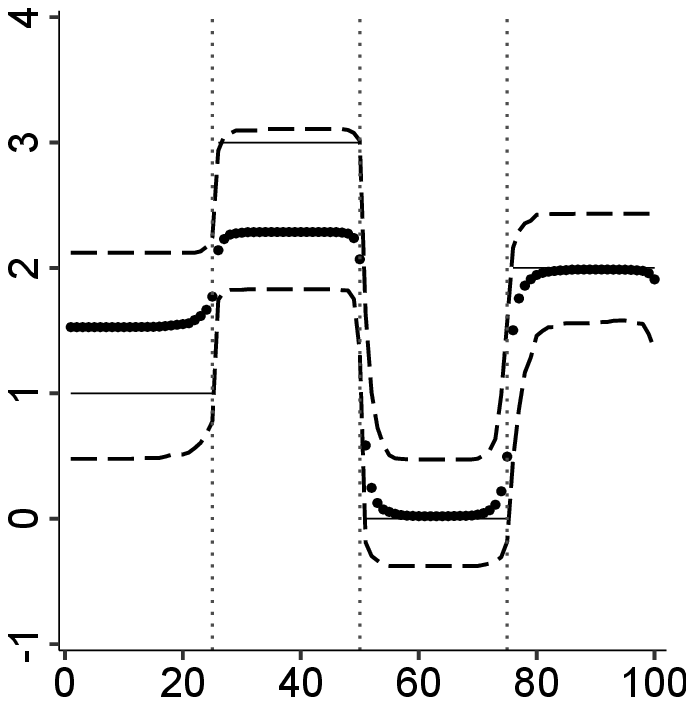}
		\label{fig:scene1_mu_P18}}%\hspace{.01in}
	\subfigure[][LCIA05]{
		\includegraphics[width=3.5cm, height=3cm, trim=0 .5cm 0 0]{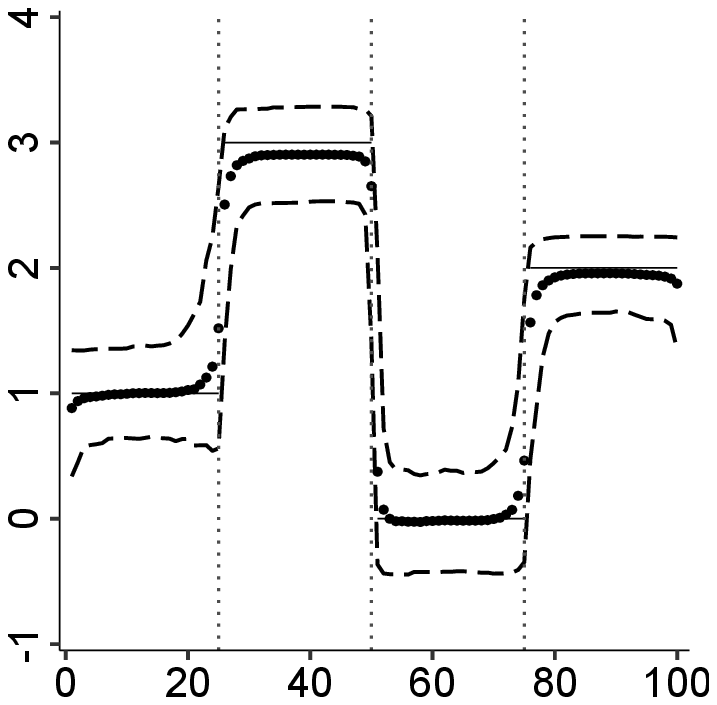}
		\label{fig:scene1_mu_LC02}}%\hspace{.01in}
	\subfigure[][BH93]{
		\includegraphics[width=3.5cm, height=3cm, trim=0 .5cm 0 0]{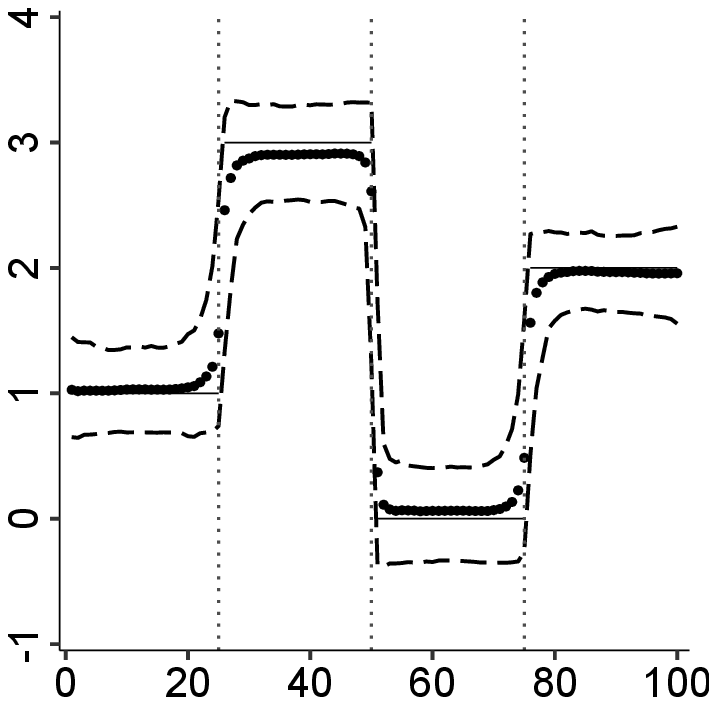}
		\label{fig:scene1_mu_BH93}}
	\\
	\subfigure[][BMCP]{
		\includegraphics[width=3.5cm, height=3cm, trim=0 .5cm 0 0]{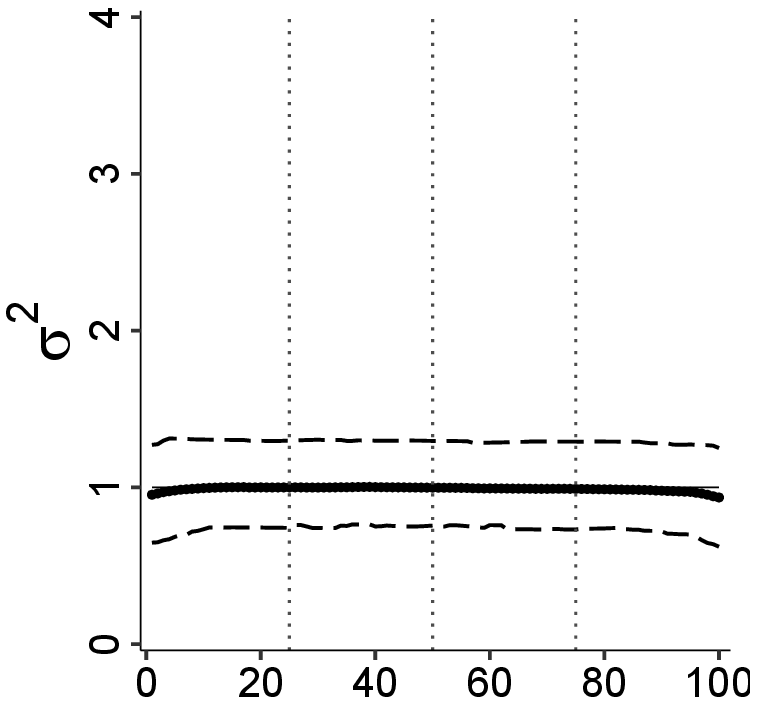}
		\label{fig:scene1_s2_LP20}}%\hspace{.01in}
	\subfigure[][DPM19]{
		\includegraphics[width=3.5cm, height=3cm, trim=0 .5cm 0 0]{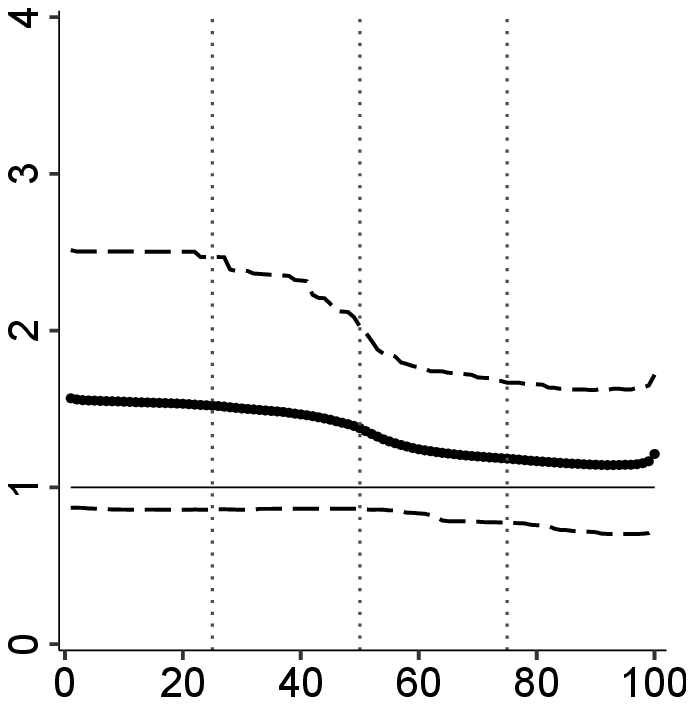}
		\label{fig:scene1_s2_P18}}%\hspace{.01in}
	\subfigure[][LCIA05]{
		\includegraphics[width=3.5cm, height=3cm, trim=0 .5cm 0 0]{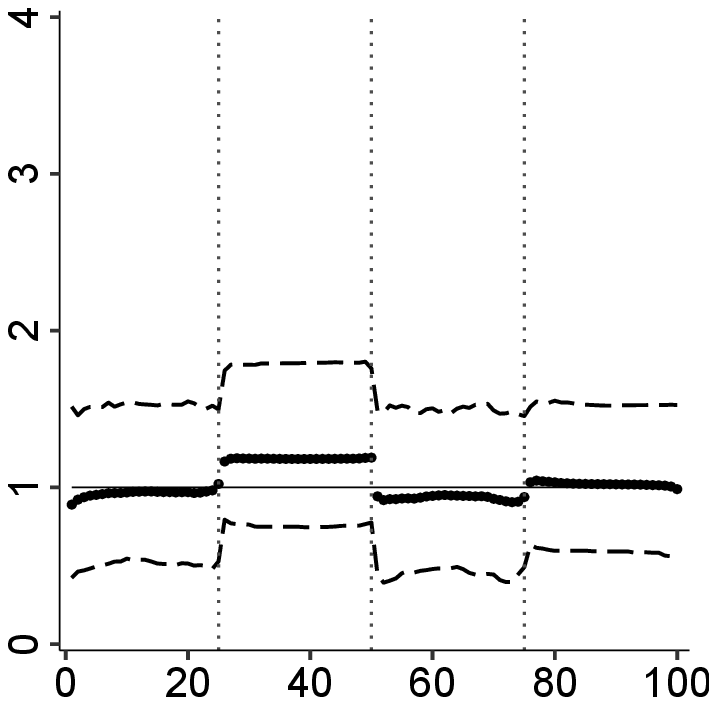}
		\label{fig:scene1_s2_LC02}}%\hspace{.01in}
	\subfigure[][BH93]{
		\includegraphics[width=3.5cm, height=3cm, trim=0 .5cm 0 0]{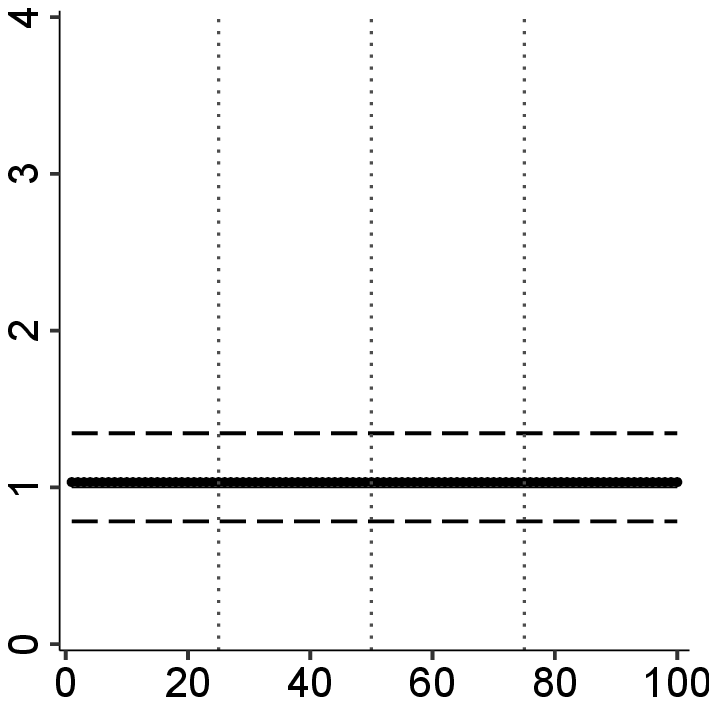}
		\label{fig:scene1_s2_BH93}}
%\end{adjustwidth}
\caption{Average of the product estimates (black dots) for the means (top) and variances (bottom) at each instant and the $5\%$ and $95\%$ quantiles of such estimates, based on the Monte Carlo replications, under BMCP (a,e), DPM19 (b,f), LCIA05 (c,g) and BH93 (d,h) models, for Scenario 1. The true mean and variance values are indicated by the solid gray horizontal lines. The vertical dotted gray lines indicate the true endpoints in $\rho_1$.}
\label{fig:scene1_PE}
\end{figure}

In Scenario 1 we assume homoscedasticity and three equally spaced changes in the mean along data sequences of size $n=100$. The constant variance equals one. The mean changes are determined by the partition {$\rho_1=\{0,25,50,75,100\}$} and the cluster parameters are $\bm{\mu}^\star=(1,3,$ $0,2)$. This scenario follows the BH93 model assumptions of mean changes and constant variance. The average of the product estimates given in Figure \ref{fig:scene1_PE} shows that the three PPM based models provide reasonable estimates for the means at each instant. The BMCP provides less biased estimates for the means and identifies the homoscedastic behavior in the data, which is not detected by LCIA05 model. The product estimates under the LCIA05 model indicate changes in the variance (that do not exist in the data) at the same instants of the changes in the mean. Although BMCP and BH93 models produce good estimates for the constant variance, the product estimates for the means provided by BH93 model are a bit worse, being more biased in some clusters (e.g., the means are underestimated in the $2nd$ cluster and overestimated in the $3rd$ cluster). This is unexpected as the configuration for this scenario favors the BH93 model. The DPM19 model provides the most biased mean and variance estimates among the four models, with higher errors in the $1st$ and $2nd$ clusters.

%\vspace{.2in}

Considering the posterior most likely partition to estimate the change point positions, Table \ref{tab:scene1_mod} shows that the true partition $\rho_1$ was the most estimated for all the PPM based models. Moreover, it is also noticeable that, for those data sets in which the posterior mode is not equal to the true partition, the estimated partitions are very close to the true $\rho_1$, differing only by one point. The BMCP model correctly identified the true partition $\rho_2=\{0,100\}$ for almost all data sets, indicating no changes in the variance. DPM19 indicates no changes in the variance for most of the data sets, but the most estimated $E_1$ vector ignores the first change in the mean.

\begin{table}[!htb]
%\begin{adjustwidth}{-.5cm}{}
\centering
\footnotesize
\begingroup
\renewcommand{\arraystretch}{.6} % Default value: 1
\begin{tabular}{p{0.3\textwidth}r}
% 	\\[-1ex]
%	\hline \\[-1.8ex]
	\clineB{1-2}{2} \\[-1.8ex]
	BMCP ${Mo(\rho_1|\bm{X})}$ & \# \\[-.5ex]
%	\hline \\[-1.8ex]
	\cmidrule(r{.15cm}l{.15cm}){1-2}
	$\{0,25,50,75,100\}$ & 128 \\
	$\{0,24,50,75,100\}$ &  30 \\
	$\{0,26,50,75,100\}$ &  29 \\
%	$\{0,25,50,74,100\}$ &  27 \\
%	$\{0,25,50,76,100\}$ &  24 \\
	\hline \\[-1.8ex]
\end{tabular}\hspace{.1in}
\begin{tabular}{p{0.3\textwidth}r}
%	\\[-1ex]
%	\hline \\[-1.8ex]
	\clineB{1-2}{2} \\[-1.8ex]
	BMCP ${Mo(\rho_2|\bm{X})}$ & \# \\[-.5ex]
%	\hline \\[-1.8ex]
	\cmidrule(r{.15cm}l{.15cm}){1-2}
	$\{0,100\}$    & 393 \\
	$\{0,88,100\}$ &   2 \\
	$\{0,4,100\}$  &   1 \\
%	$\{0,5,100\}$  &   1 \\
%	$\{0,9,100\}$  &   1 \\
	\hline \\[-1.8ex]
\end{tabular}
\begin{tabular}{p{0.3\textwidth}r}
%	\\[-1ex]
%	\hline \\[-1.8ex]
	\clineB{1-2}{2} \\[-1.8ex]
	DPM19 ${Mo(E_1|\bm{X})}$ & \# \\[-.5ex]
%	\hline \\[-1.8ex]
	\cmidrule(r{.15cm}l{.15cm}){1-2}
	$\{0,50,75,100\}$    & 122 \\
	$\{0,25,50,75,100\}$ &  48 \\
	$\{0,51,75,100\}$    &  31 \\
%	$\{0,50,76,100\}$    &  23 \\
%	$\{0,50,74,100\}$    &  22 \\
	\hline \\[-1.8ex]
\end{tabular}\hspace{.1in}
\begin{tabular}{p{0.3\textwidth}r}
%	\\[-1ex]
%	\hline \\[-1.8ex]
	\clineB{1-2}{2} \\[-1.8ex]
	DPM19 ${Mo(E_2|\bm{X})}$ & \# \\[-.5ex]
%	\hline \\[-1.8ex]
	\cmidrule(r{.15cm}l{.15cm}){1-2}
	$\{0,100\}$    & 353 \\
	$\{0,50,100\}$ &   5 \\
	$\{0,53,100\}$ &   4 \\
%	$\{0,45,100\}$ &   2 \\
%	$\{0,48,100\}$ &   2 \\
	\hline \\[-1.8ex]
\end{tabular}
\begin{tabular}{p{0.3\textwidth}r}
%	\\[-1ex]
%	\hline \\[-1.8ex]
	\clineB{1-2}{2} \\[-1.8ex]
	LCIA05 ${Mo(\rho|\bm{X})}$ & \# \\[-.5ex]
%	\hline \\[-1.8ex]
	\cmidrule(r{.15cm}l{.15cm}){1-2}
	$\{0,25,50,75,100\}$ & 121 \\
	$\{0,24,50,75,100\}$ &  26 \\
	$\{0,25,50,74,100\}$ &  25 \\
%	$\{0,26,50,75,100\}$ &  19 \\
%	$\{0,25,50,76,100\}$ &  18 \\
	\hline \\[-1.8ex]
\end{tabular}\hspace{.1in}
\begin{tabular}{p{0.3\textwidth}r}
%	\\[-1ex]
%	\hline \\[-1.8ex]
	\clineB{1-2}{2} \\[-1.8ex]
	BH93 ${Mo(\rho|\bm{X})}$ & \# \\[-.5ex]
%	\hline \\[-1.8ex]
	\cmidrule(r{.15cm}l{.15cm}){1-2}
	$\{0,25,50,75,100\}$ & 130 \\
	$\{0,24,50,75,100\}$ &  26 \\
	$\{0,25,50,74,100\}$ &  25 \\
%	$\{0,26,50,75,100\}$ &  25 \\
%	$\{0,25,50,76,100\}$ &  22 \\
	\hline \\[-1.8ex]
% 	\clineB{1-2}{2} \\[-1.8ex]
\end{tabular}
\endgroup
\caption{Top posterior modes of $\rho_1$  and $\rho_2$ (BMCP), $E_1$ and $E_2$ (DPM19) and $\rho$ (LCIA05, BH93) estimated for each of the 400 data sets of Scenario 1.} \label{tab:scene1_mod}
%\end{adjustwidth}
\end{table}

\begin{figure}[!htb]
%\begin{adjustwidth}{-.4cm}{-.4cm}
\centering
\subfigure[][BMCP ($\rho_1$)]{
	\includegraphics[width=4cm, height=3cm, trim=0 2cm 0 0]{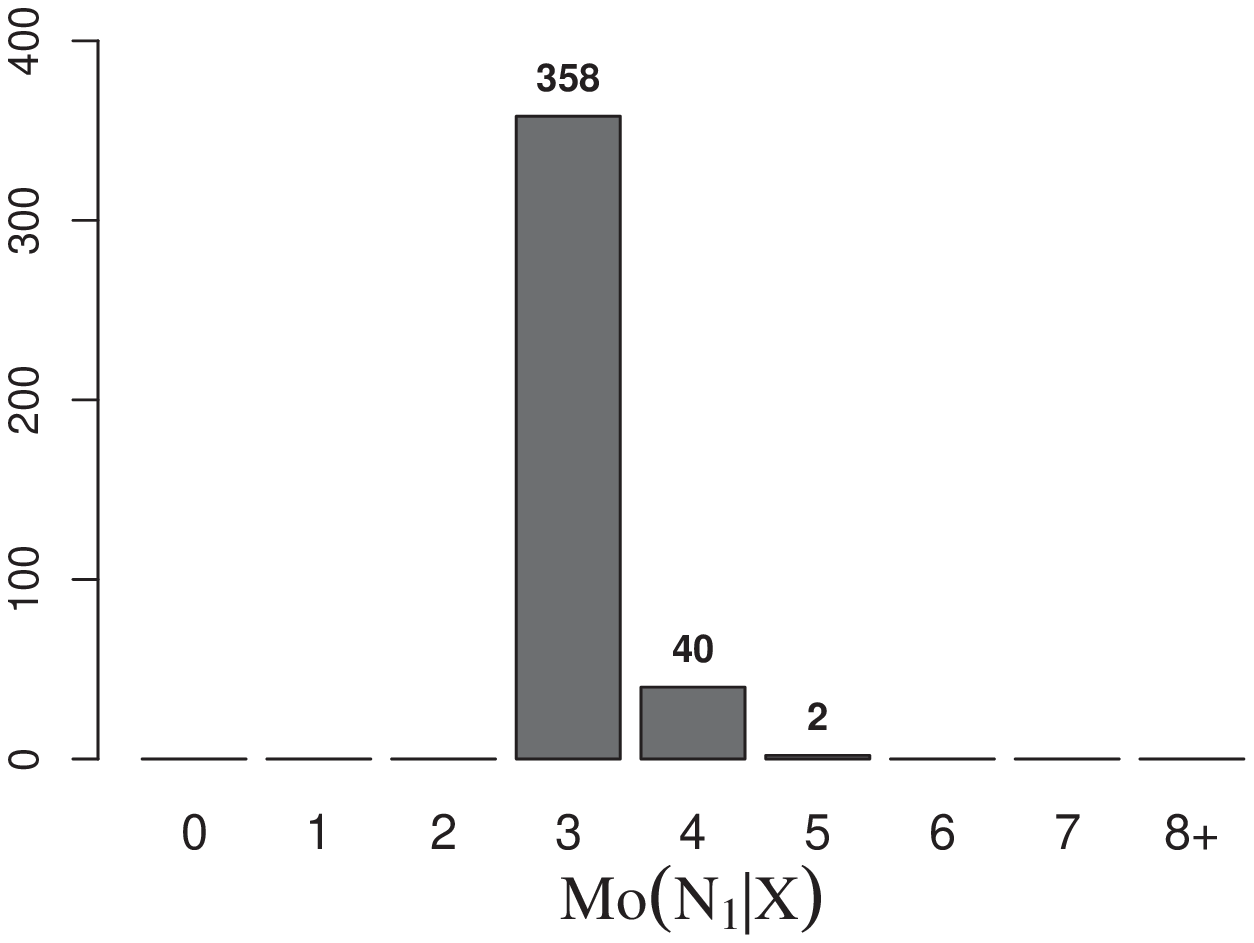}
	\label{fig:scene1_Nmode_LP20_mu}}
\subfigure[][BMCP ($\rho_2$)]{
	\includegraphics[width=4cm, height=3cm, trim=0 2cm 0 0]{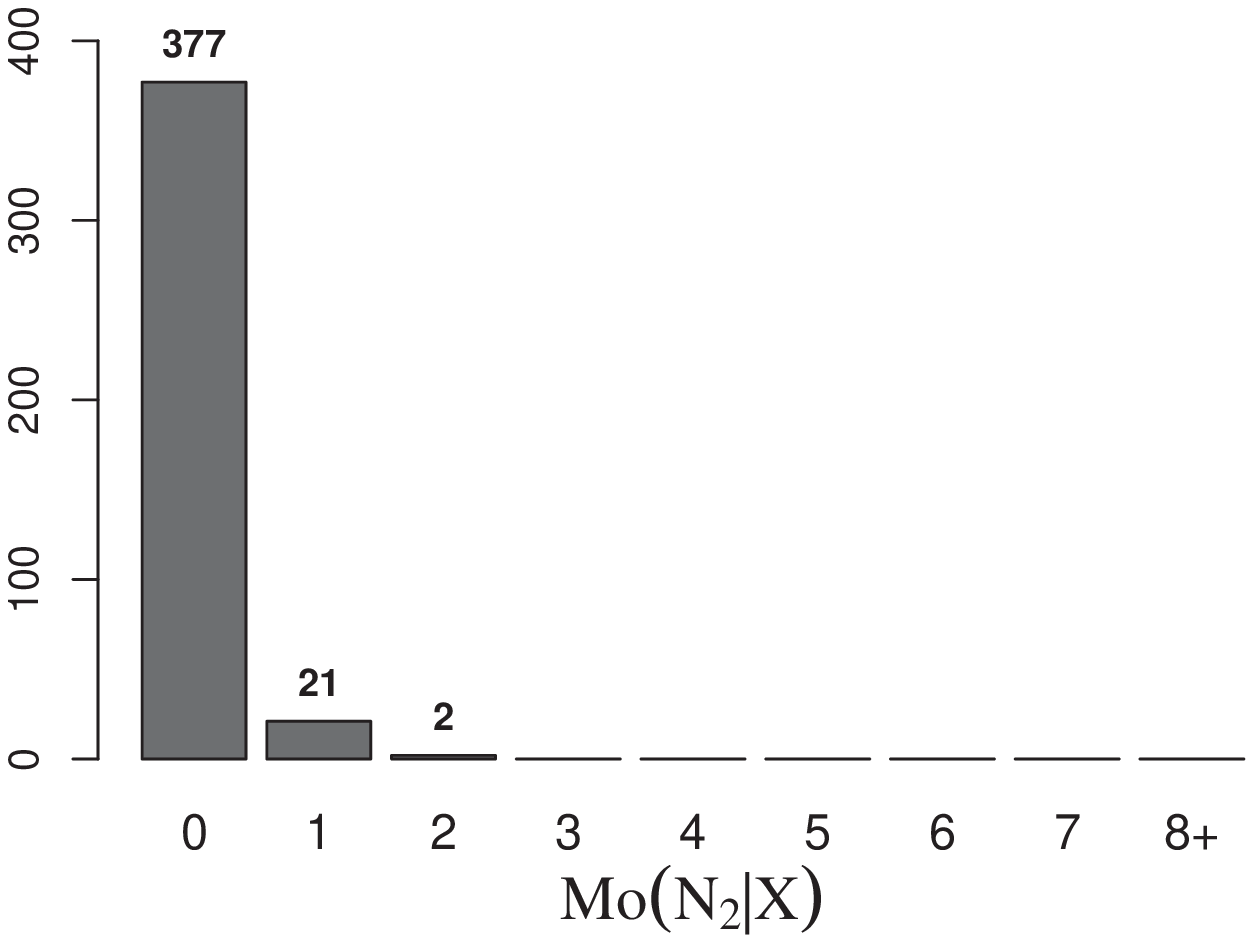}
	\label{fig:scene1_Nmode_LP20_s2}}
\subfigure[][LCIA05 ($\rho$)]{
	\includegraphics[width=4cm, height=3cm, trim=0 2cm 0 0]{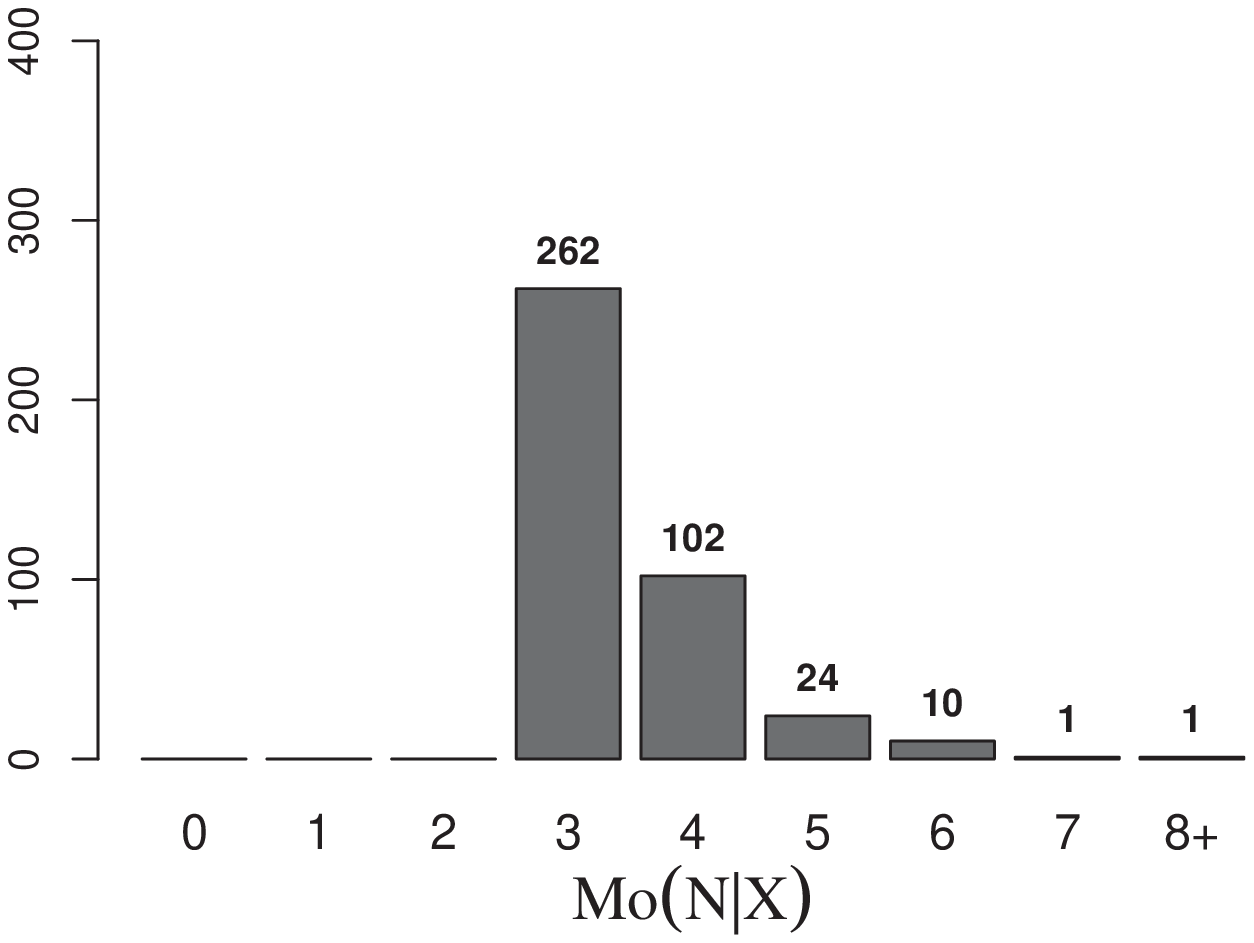}
	\label{fig:scene1_Nmode_LC02}}\\[-.2in]
\subfigure[][DPM19 ($E_1$)]{
	\includegraphics[width=4cm, height=3cm, trim=0 2cm 0 0]{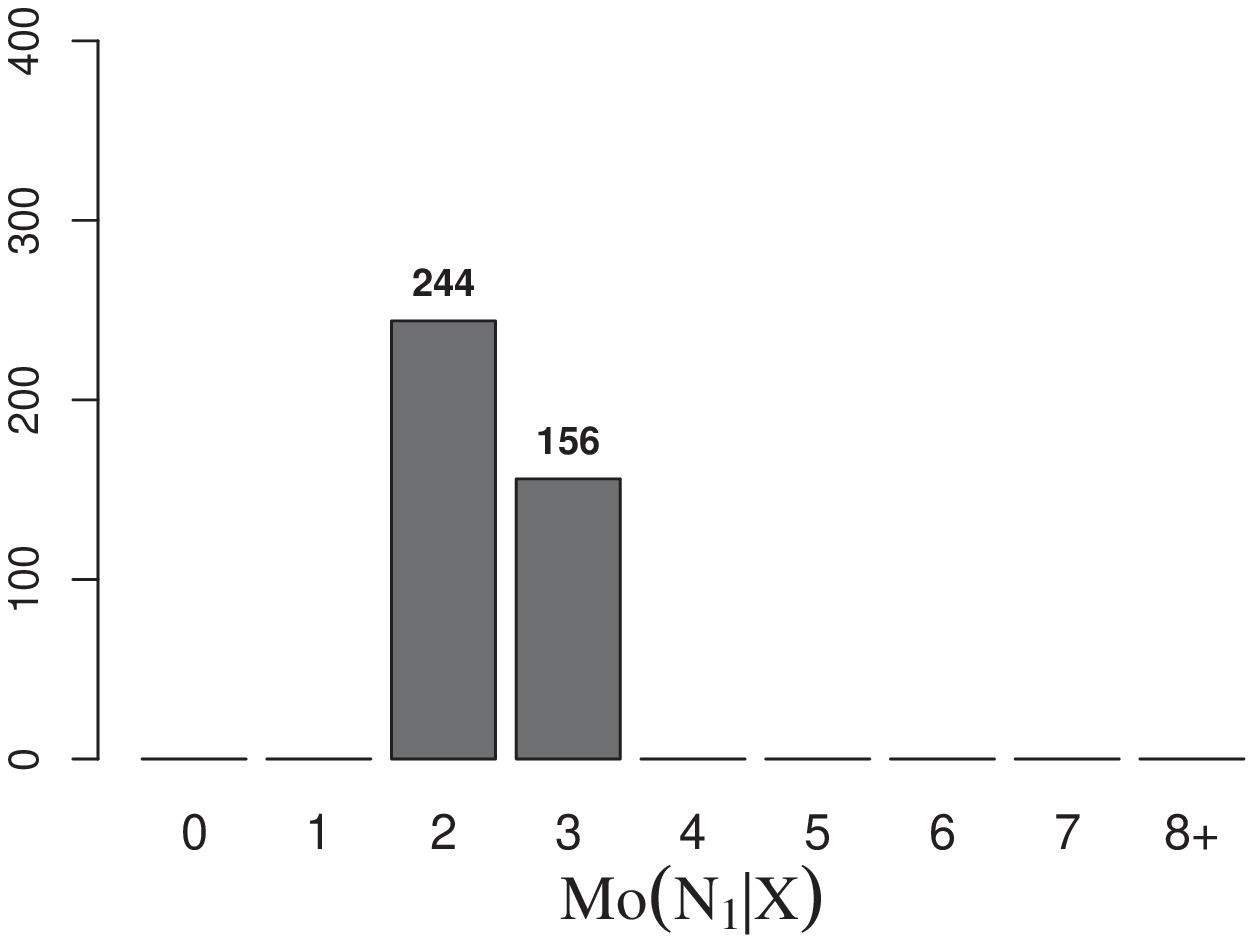}
	\label{fig:scene1_Nmode_P18_mu}}
\subfigure[][DPM19 ($E_2$)]{
	\includegraphics[width=4cm, height=3cm, trim=0 2cm 0 0]{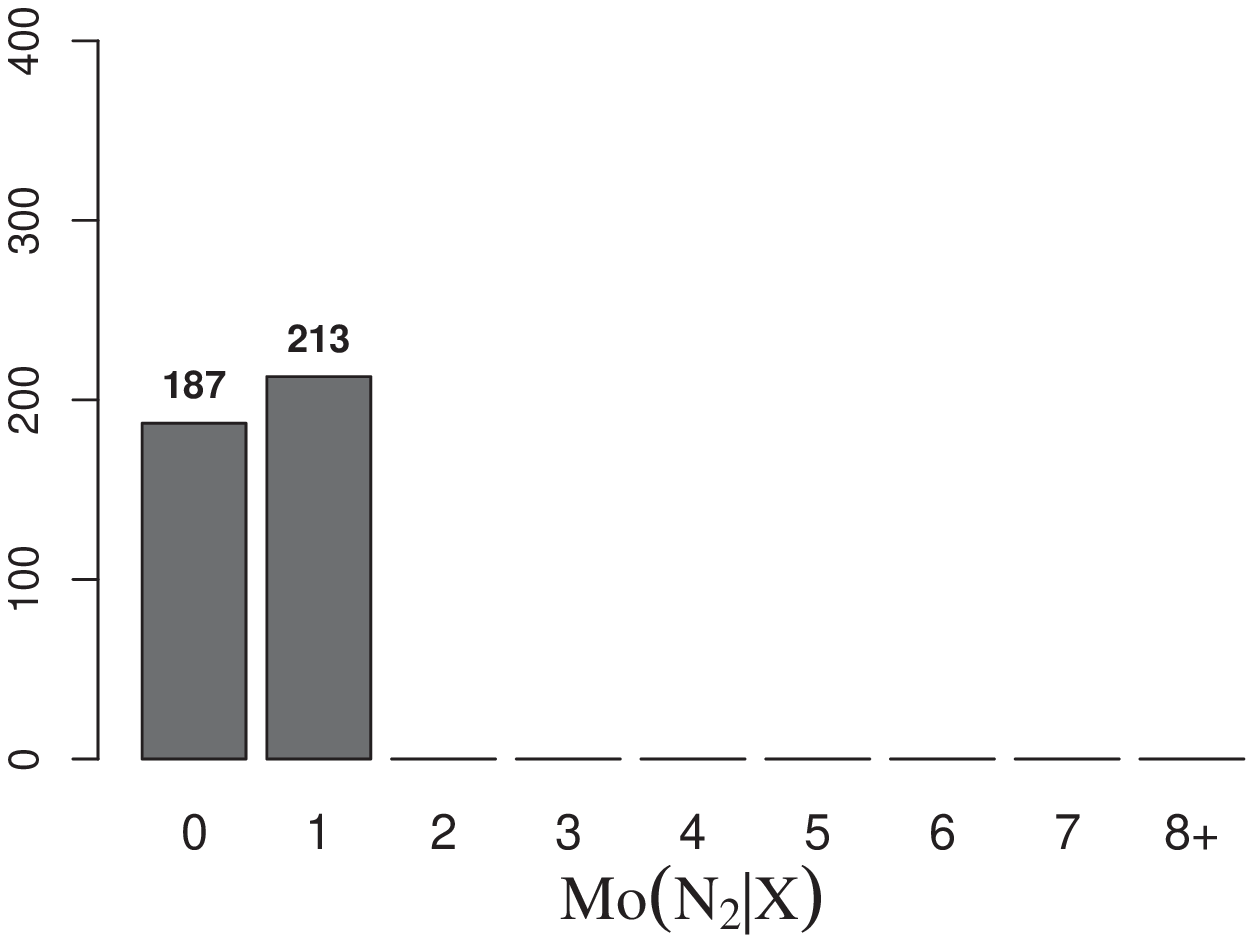}
	\label{fig:scene1_Nmode_P18_s2}}
\subfigure[][BH93 ($\rho$)]{
	\includegraphics[width=4cm, height=3cm, trim=0 2cm 0 0]{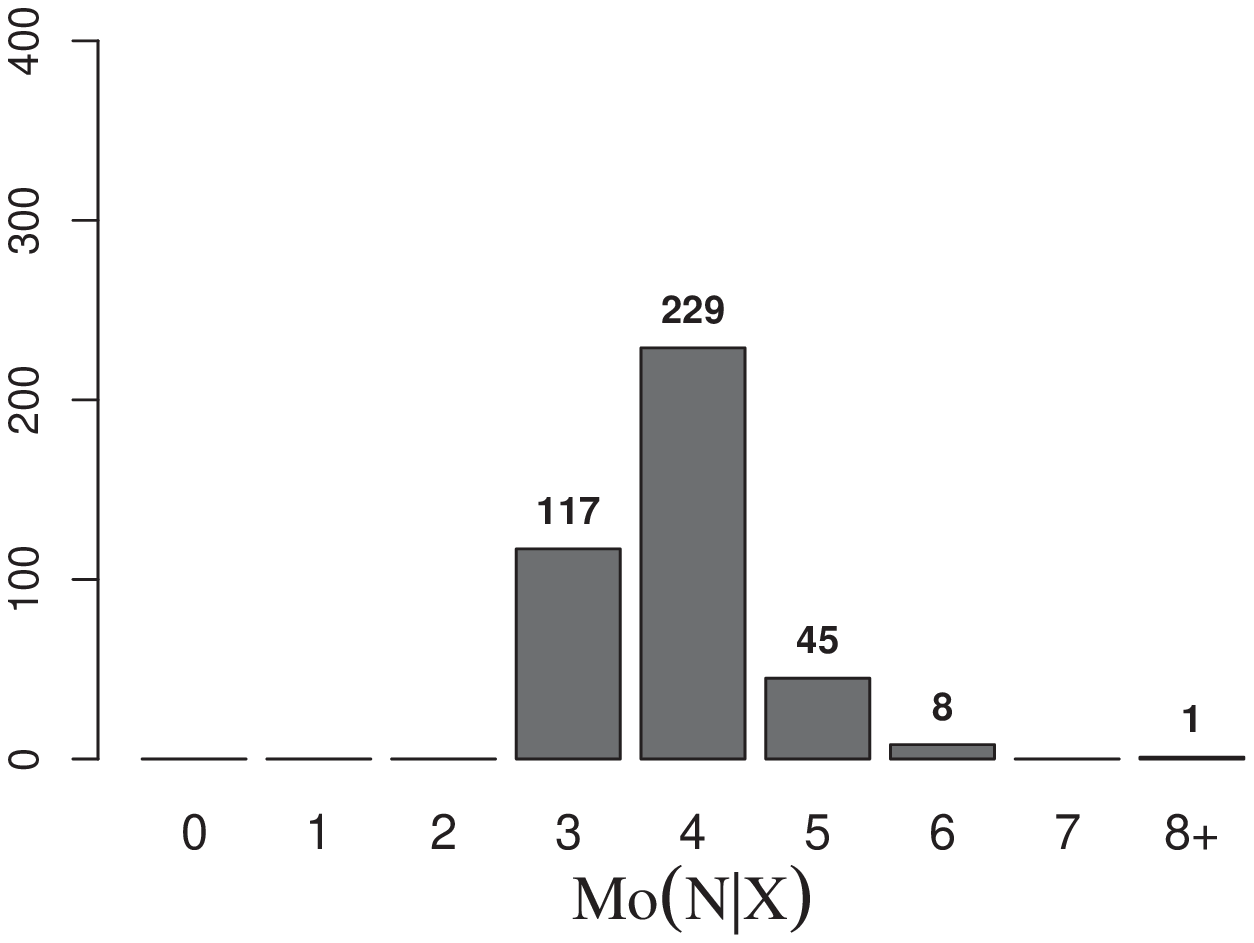}
	\label{fig:scene1_Nmode_BH93}}
%	\end{adjustwidth}
\caption{Counts distribution of the posterior modes of the number of changes, estimated in each of the 400 replications, for models BMCP (a,b), LCIA05 (c), DPM19 (d,e) and BH93 (f), for
Scenario 1.} \label{fig:scene1_Nmode}
\end{figure}

\begin{figure}[!htb]
%\begin{adjustwidth}{-.4cm}{-.4cm}
\centering
\subfigure[][BMCP ($\rho_1$)]{
	\includegraphics[width=6cm, height=4cm, trim=0 2.5cm 0 0]{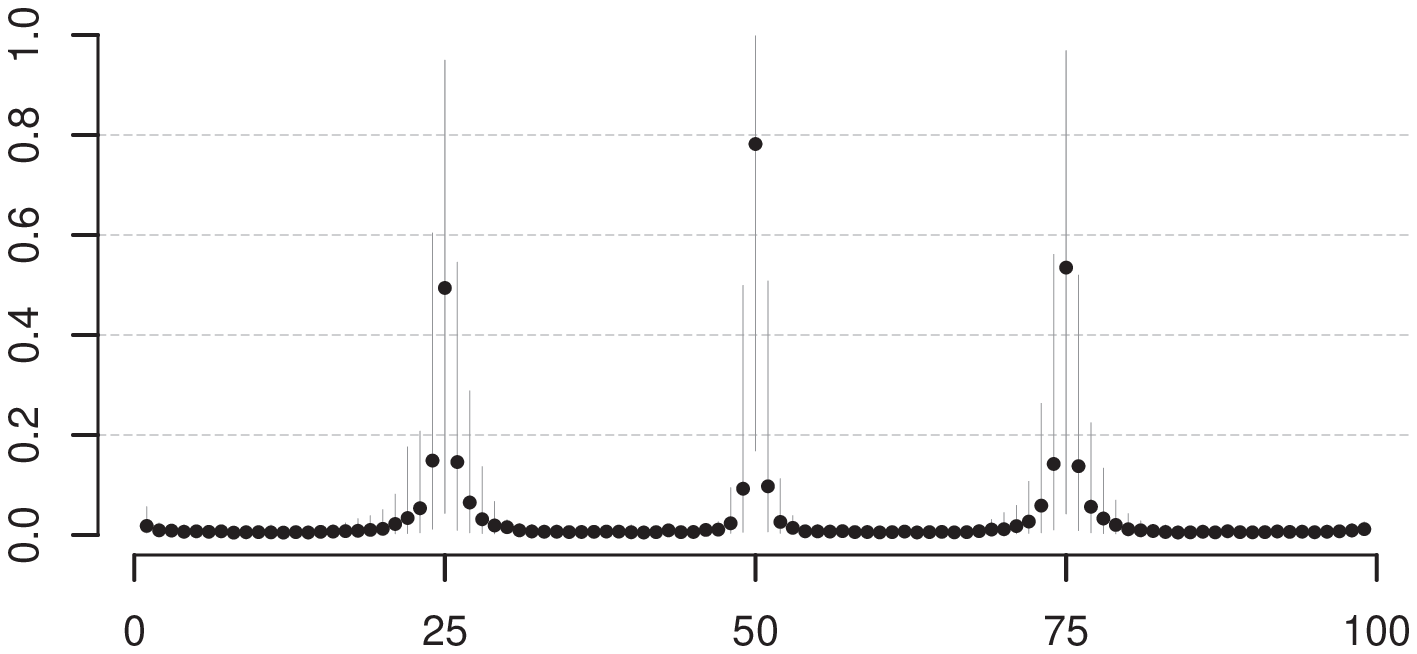}
	\label{fig:scene1_prob_IC_LP20_mu}}%
\subfigure[][DPM19 ($E_1$)]{
	\includegraphics[width=6cm, height=4cm, trim=0 2.5cm 0 0]{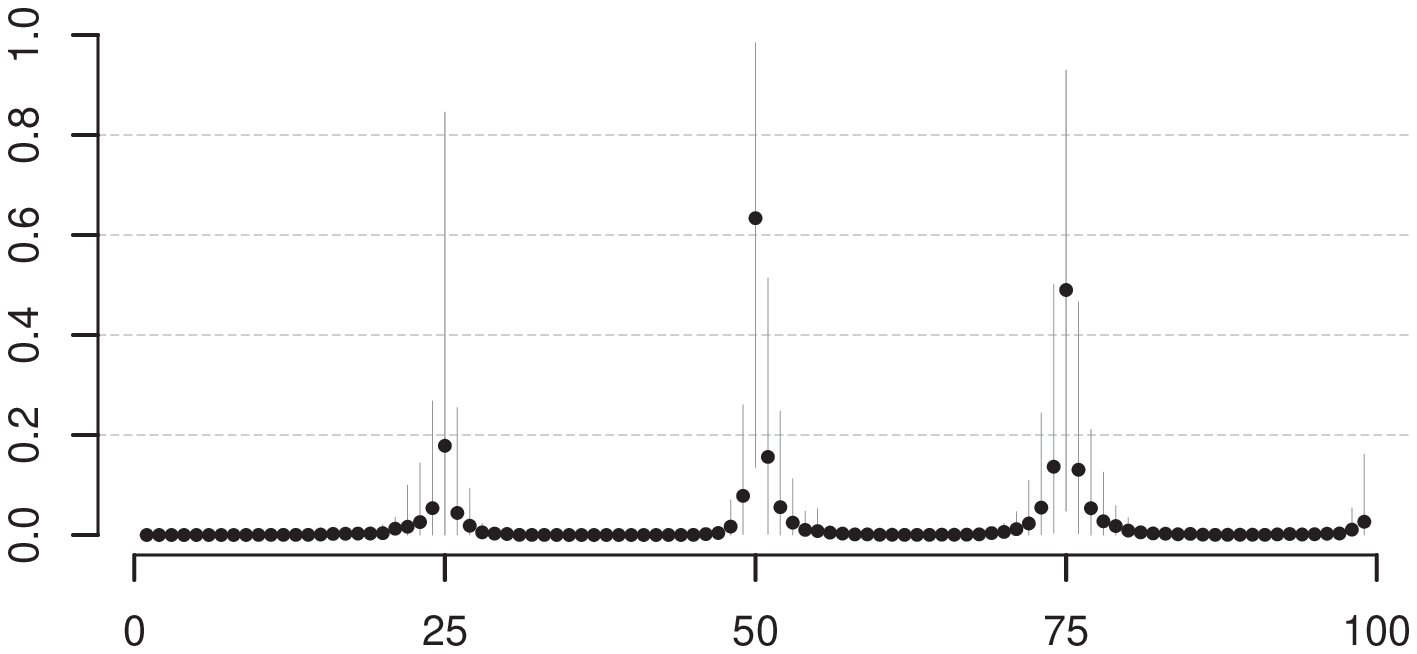}
	\label{fig:scene1_prob_IC_P18_mu}}\\[-.35in]
\subfigure[][BMCP ($\rho_2$)]{
	\includegraphics[width=6cm, height=4cm, trim=0 2.5cm 0 0]{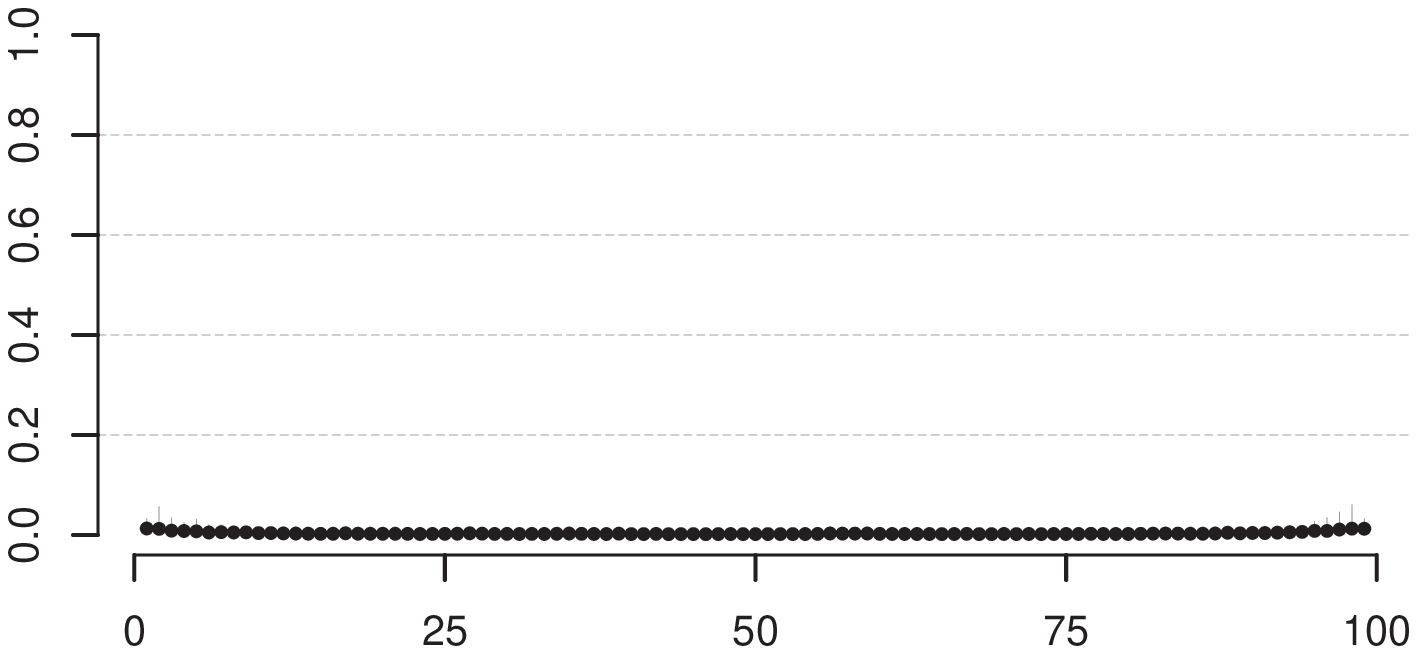}
	\label{fig:scene1_prob_IC_LP20_s2}}%
\subfigure[][DPM19 ($E_2$)]{
	\includegraphics[width=6cm, height=4cm, trim=0 2.5cm 0 0]{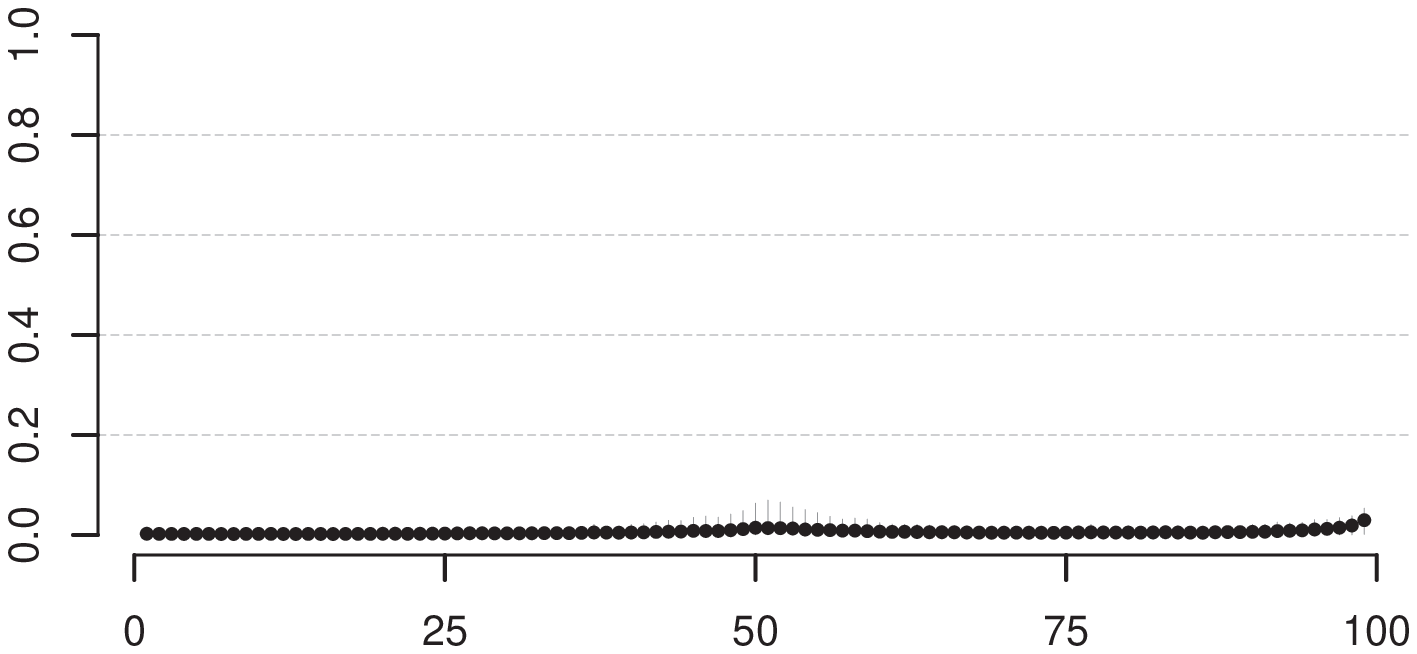}
	\label{fig:scene1_prob_IC_P18_s2}}\\[-.35in]
\subfigure[][LCIA05 ($\rho$)]{
	\includegraphics[width=6cm, height=4cm, trim=0 2.5cm 0 0]{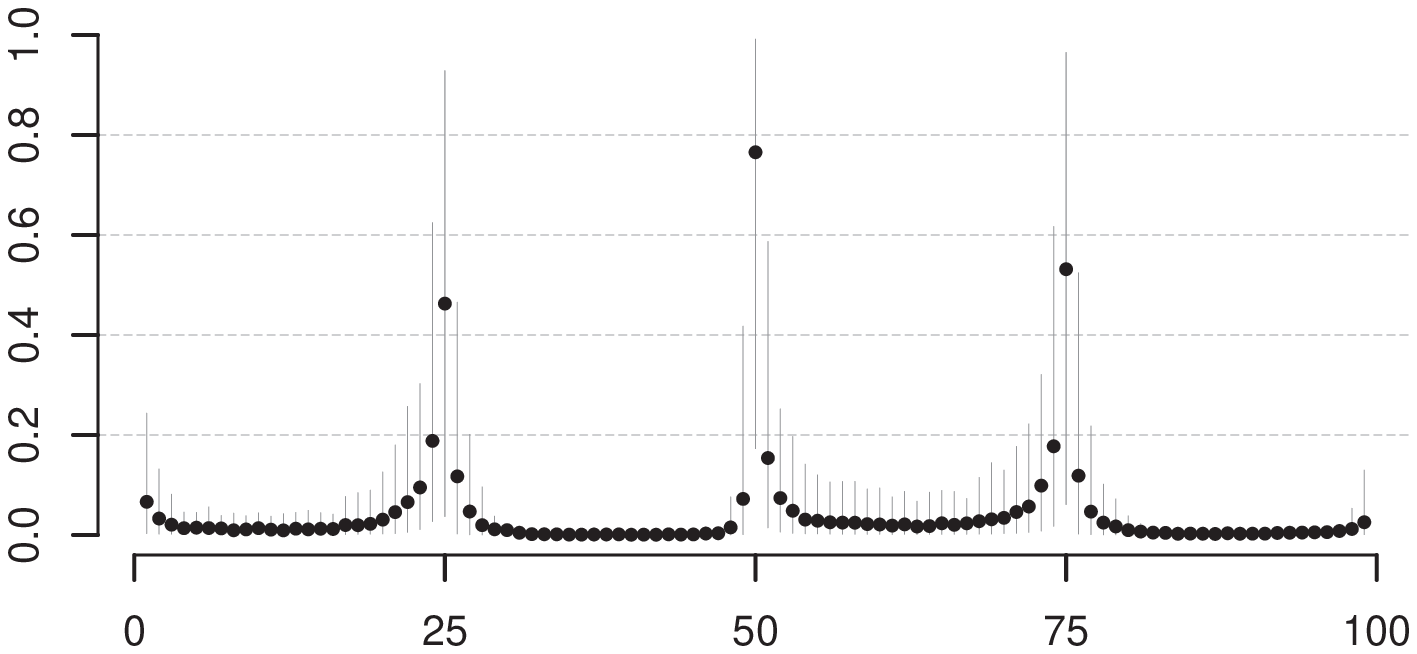}
	\label{fig:scene1_prob_IC_LC02}}%
\subfigure[][BH93 ($\rho$)]{
	\includegraphics[width=6cm, height=4cm, trim=0 2.5cm 0 0]{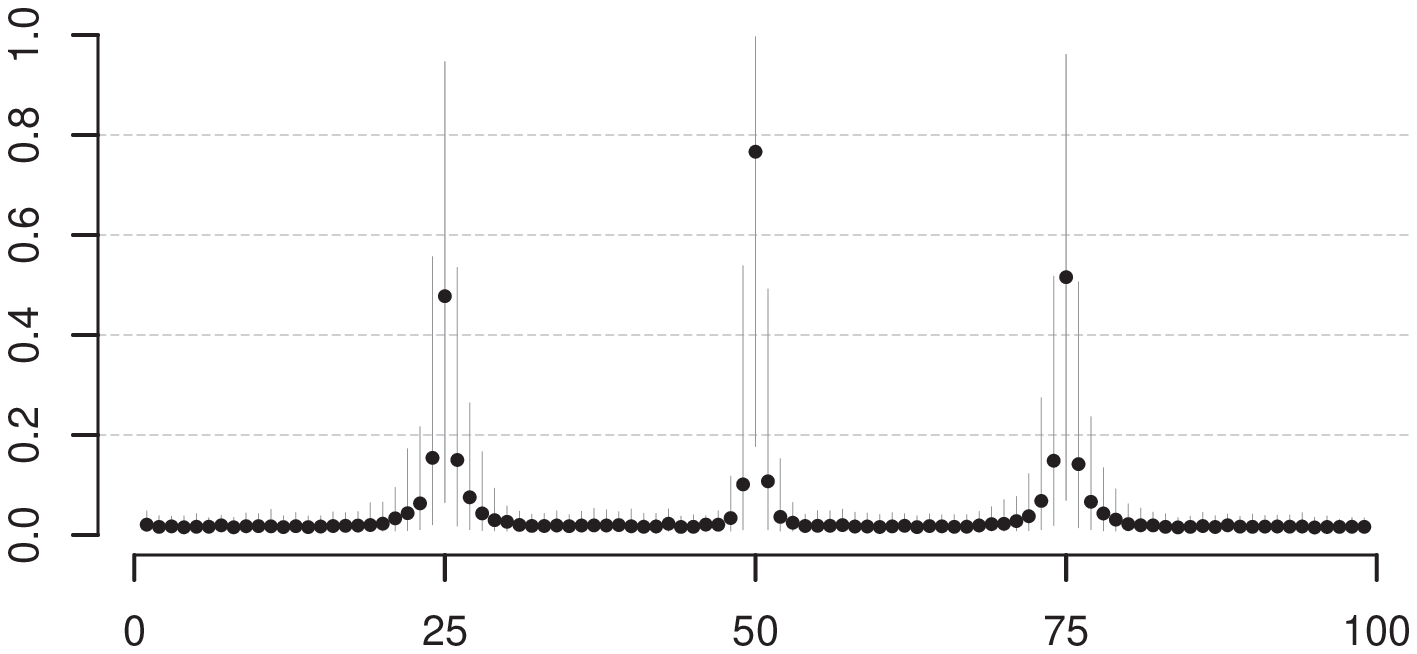}
	\label{fig:scene1_prob_IC_BH93}}
%\end{adjustwidth}
\caption{Average of the posterior probabilities of each instant to be an end point (black dots) for each partition and the $5\%$ and $95\%$ quantiles range of such probabilities based on the Monte Carlo replications, for the BMCP (a,c), DPM19 (b,d), LCIA05 (e) and BH93 (f) models, for Scenario 1.}
\label{fig:scene1_prob_IC}
\end{figure}

Due to the high dimension of the parametric space of a random partition ($2^{n-1}$ different elements), the posterior distribution tends to be too flat in many situations. In these cases, the probabilities of each possible partition are very small and tends to be very close for many different partitions, making difficult the decision about the exact change points. The probabilities of each instant being a change point, obtained from the posterior of the random partition, as proposed by \cite{loschi2005p}, is a good auxiliary tool to detect change points. It is simple to adapt this definition to determine the probability of each point to be an end point of a cluster, as it is presented next along the text. The average of these probabilities over all data sets, for each model, are displayed in Figure \ref{fig:scene1_prob_IC}. They show that the BMCP model correctly identifies the change points in the mean and the absence of changes in the variance. The probabilities for the true changes in the mean are, on average, above $0.5$ while they are below $0.2$ for all the other instants. In the same way, DPM19 also identifies the mean changes except for the first change. Concerning the changes in the variance, such average probabilities are very close of zero at all instants. Models LCIA05 and BH93 also correctly identify the change points in $\bm{\mu}$. However, the LCIA05 model does not identify which parameter experienced these changes. Despite BH93 model has a good performance, it is based on the strong assumption that we know {\it a priori} that the variance is constant along the sequence.
Figure \ref{fig:scene1_Nmode} shows that the BMCP model precisely identifies the true number of change points in the mean ($N_1=3$) and no changes in the variance ($N_2=0$) for around $90\%$ and $94\%$ of the data sets, respectively. Under the LCIA05, the true number of changes ($N=3$) is correctly identified by $66\%$ of the data sets, while it is overestimated under the BH93 model for around $71\%$. The DPM19 underestimates by one the number of changes in the mean and overestimates by one the number of changes in the variance.

\vspace{-.3in}

%\newpage

\FloatBarrier
\subsubsection{Scenario 2: constant mean and changes in the variance}\label{sec_scene2}

\begin{figure}[!htb]
%\begin{adjustwidth}{-.4cm}{-.4cm}
\centering
\subfigure[][BMCP]{
	\includegraphics[width=3.5cm, height=3cm, trim=0 .5cm 0 0]{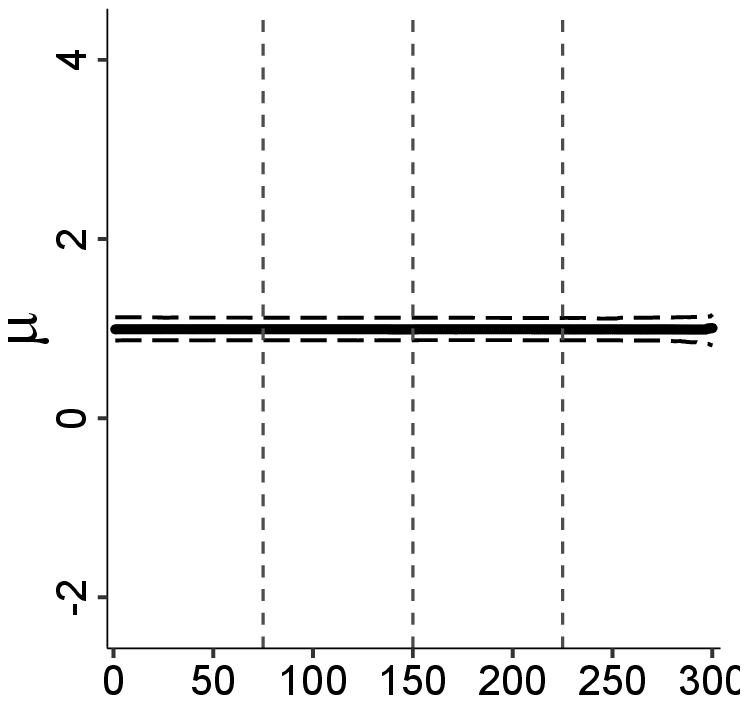}
	\label{fig:scene2_mu_LP20}}%\hspace{.01in}
\subfigure[][DPM19]{
	\includegraphics[width=3.5cm, height=3cm, trim=0 .5cm 0 0]{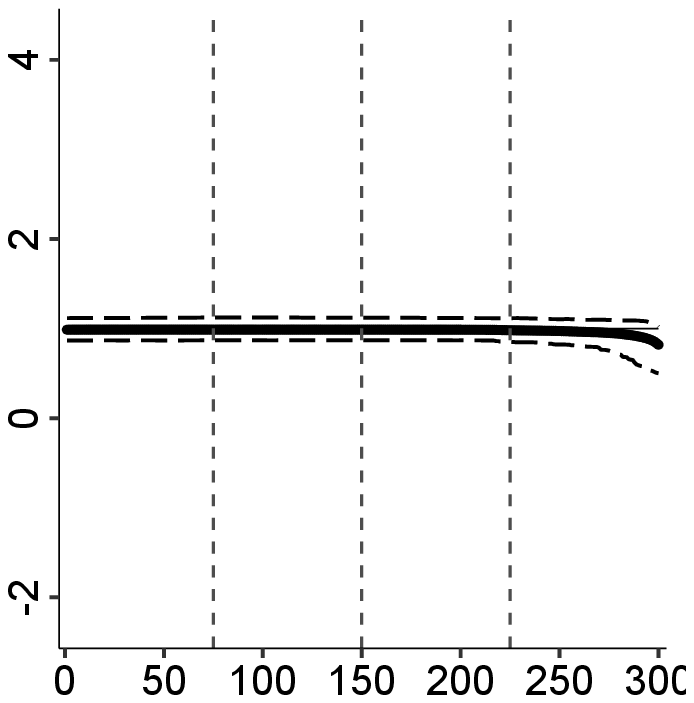}
	\label{fig:scene2_mu_P18}}%\hspace{.01in}
\subfigure[][LCIA05]{
	\includegraphics[width=3.5cm, height=3cm, trim=0 .5cm 0 0]{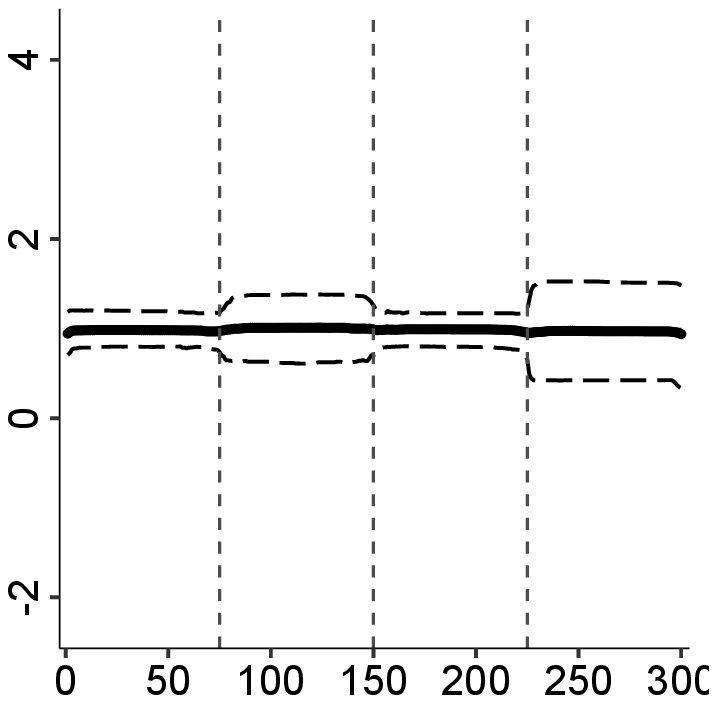}
	\label{fig:scene2_mu_LC02}}%\hspace{.01in}
\subfigure[][BH93]{
	\includegraphics[width=3.5cm, height=3cm, trim=0 .5cm 0 0]{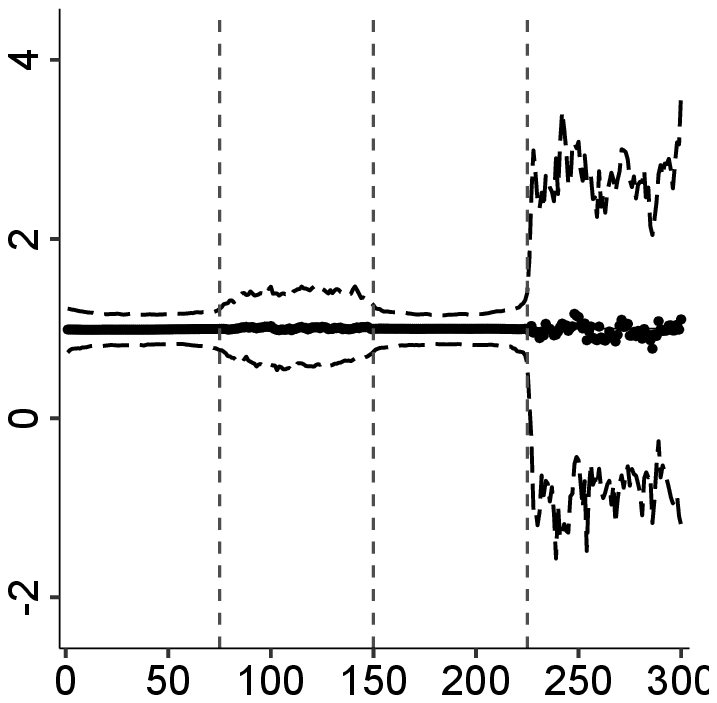}
	\label{fig:scene2_mu_BH93}}
\\
\subfigure[][BMCP]{
	\includegraphics[width=3.5cm, height=3cm, trim=0 .5cm 0 0]{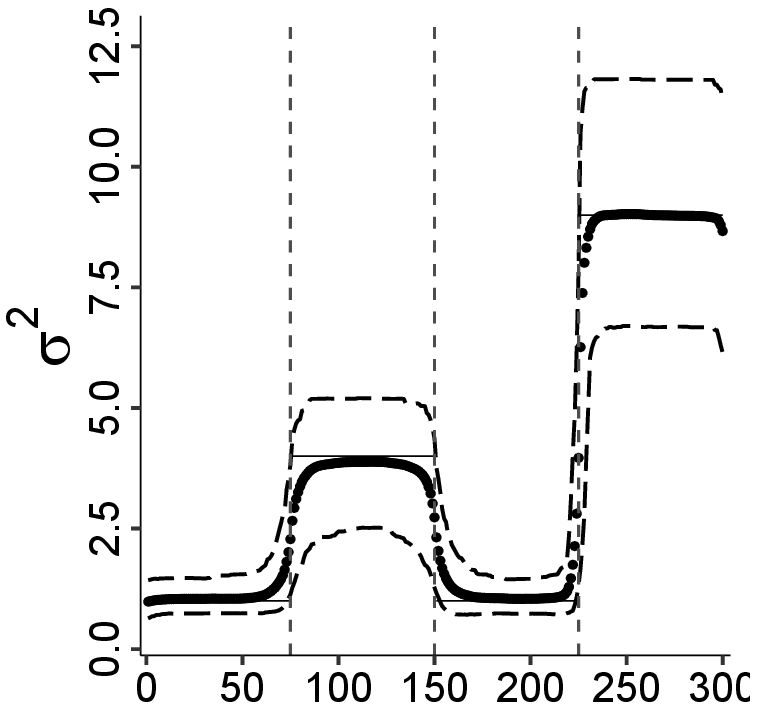}
	\label{fig:scene2_s2_LP20}}%\hspace{.01in}
\subfigure[][DPM19]{
	\includegraphics[width=3.5cm, height=3cm, trim=0 .5cm 0 0]{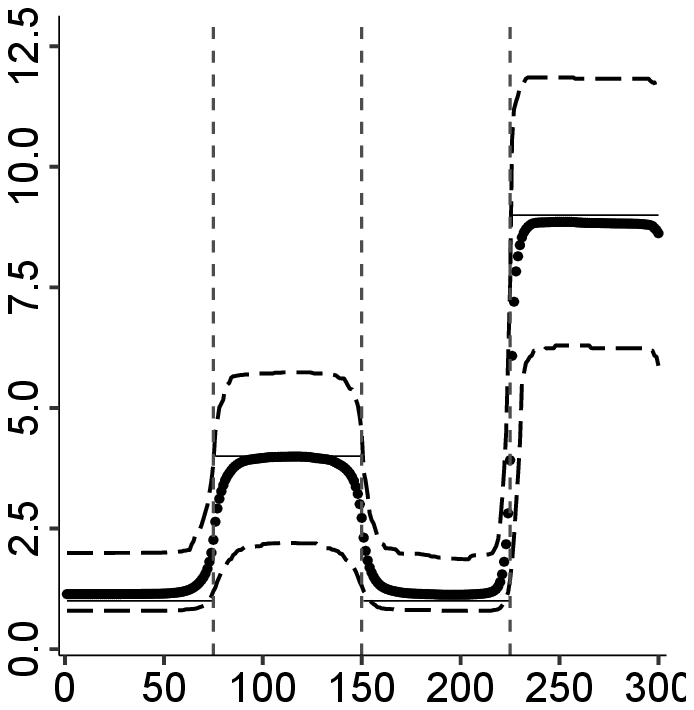}
	\label{fig:scene2_s2_P18}}%\hspace{.01in}
\subfigure[][LCIA05]{
	\includegraphics[width=3.5cm, height=3cm, trim=0 .5cm 0 0]{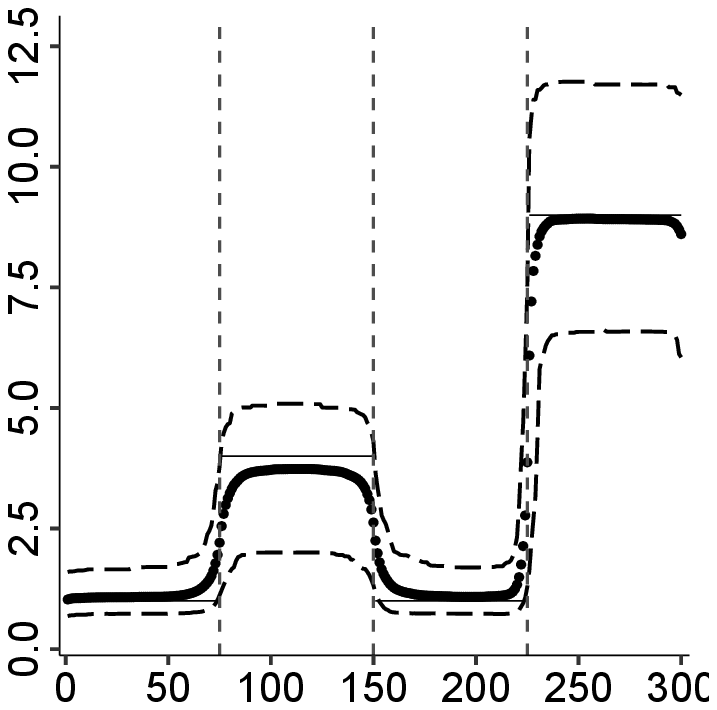}
	\label{fig:scene2_s2_LC02}}%\hspace{.01in}
\subfigure[][BH93]{
	\includegraphics[width=3.5cm, height=3cm, trim=0 .5cm 0 0]{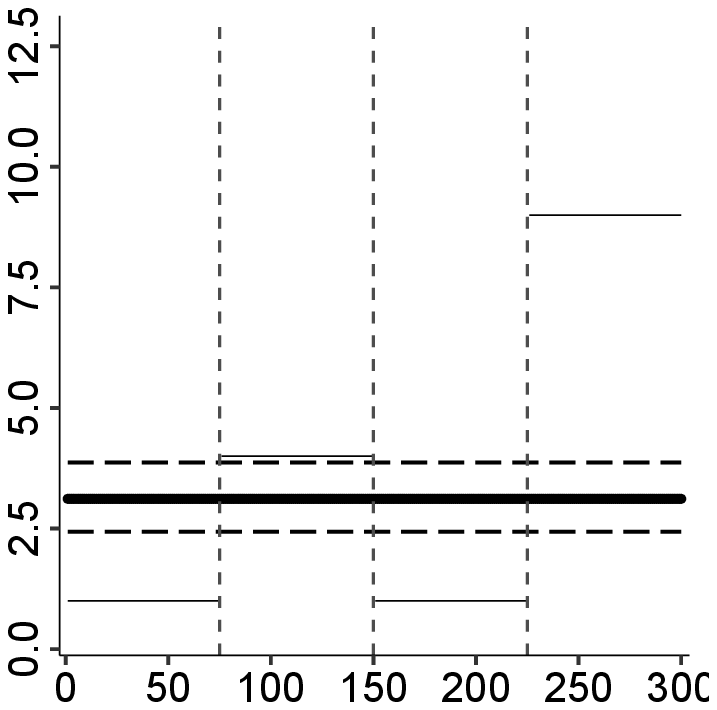}
	\label{fig:scene2_s2_BH93}}
%\end{adjustwidth}
\caption{Average of the product estimates (black dots) for the means (top) and variances (bottom) in each instant and the $5\%$ and $95\%$ quantiles of such estimates based on the Monte Carlo replications, under BMCP (a,e), DPM19 (b,f), LCIA05 (c,g) and BH93 (d,h) models for Scenario 2. The true mean and variance values are indicated by the solid gray horizontal lines.
The vertical dashed gray lines indicate the true end points in $\rho_2$.}
\label{fig:scene2_PE}
\end{figure}

Under this scenario data sets are generated with constant mean and three equally spaced changes in  the variance along sequences of size $n=300$. The mean is constant and equal to one. The variance changes are at instants defined by the partition {$\rho_2=\{0,75,150,225,300\}$} and the
cluster parameters $\bm{\sigma}^\star=(1,4,1,9)$. Figure \ref{fig:scene2_PE} shows that the BMCP, DPM19 and LCIA05 models provide product estimates for the means that are very close to the true value. Under the LCIA05 model, there is a larger variation in such estimates, mainly in the $2nd$ and $4th$ clusters, indicating that for some of the data sets this model estimated different means into the different clusters. It is also observed for the BH93 model, under which the product estimates for the means are more biased, which suggests non-constant mean in the $2nd$ and $4th$ clusters. The BMCP, DPM19 and LCIA05 models provide reasonable estimates for the variance, but these estimates tend to be more biased around the true changes. The BH93 model provides an unsuitable estimation of the variance, as a consequence of the constant variance assumption of this model.

The results in Table \ref{tab:scene2_mod} show that both models, BMCP and DPM19 provide good estimates of $\rho_1$, identifying no change points in the mean for almost all data sets. The reported most estimated partitions for $\rho_2$ are very close to the truth under the BMCP and DPM19 models. The most estimated partition for $\rho_2$ indicates only one of the three changes, the highest one. This agrees with our past experience, namely, changes in the variance are harder to detect than changes in the mean. The LCIA05 had a similar performance to the BMCP model, with many partitions estimated with only one change that is equal or close to the true highest change in the variance. The BH93 model indicates the non-existence of changes in the mean for more than $50\%$ of the data sets.

\begin{table}[!htb]
%\begin{adjustwidth}{-.5cm}{}
\centering
\footnotesize
\begingroup
\renewcommand{\arraystretch}{.6} % Default value: 1
\begin{tabular}{p{0.3\textwidth}r}
	\clineB{1-2}{2} \\[-1.8ex]
	BMCP ${Mo(\rho_1|\bm{X})}$ & \# \\[-.5ex]
	\cmidrule(r{.15cm}l{.15cm}){1-2}
	$\{0,300\}$     & 399 \\
	$\{0,297,300\}$ &   1 \\
	& \\
%	& \\
%	& \\
	\hline \\[-1.8ex]
\end{tabular}\hspace{.1in}
\begin{tabular}{p{0.3\textwidth}r}
	\clineB{1-2}{2} \\[-1.8ex]
	BMCP ${Mo(\rho_2|\bm{X})}$ & \# \\[-.5ex]
	\cmidrule(r{.15cm}l{.15cm}){1-2}
	$\{0,225,300\}$        & $\,\,\,16$ \\
	$\{0,75,149,225,300\}$ &  9 \\
	$\{0,76,150,225,300\}$ &  7 \\
%	$\{0,226,300\}$        &  7 \\
%	$\{0,75,148,225,300\}$ &  6 \\
	\hline \\[-1.8ex]
\end{tabular}\\
\begin{tabular}{p{0.3\textwidth}r}
	\clineB{1-2}{2} \\[-1.8ex]
	DPM19 ${Mo(E_1|\bm{X})}$ & \# \\[-.5ex]
	\cmidrule(r{.15cm}l{.15cm}){1-2}
	$\{0,300\}$ & 400 \\
	& \\
	& \\
%	& \\
%	& \\
	\hline \\[-1.8ex]
\end{tabular}\hspace{.1in}
\begin{tabular}{p{0.3\textwidth}r}
	\clineB{1-2}{2} \\[-1.8ex]
	DPM19 ${Mo(E_2|\bm{X})}$ & \# \\[-.5ex]
	\cmidrule(r{.15cm}l{.15cm}){1-2}
	$\{0,225,300\}$        & $\,\,\,11$ \\
	$\{0,75,149,225,300\}$ &  8 \\
	$\{0,75,150,225,300\}$ &  7 \\
%	$\{0,76,150,225,300\}$ &  7 \\ 	
%	$\{0,226,300\}$        &  5 \\
	\hline \\[-1.8ex]
\end{tabular}\\
\begin{tabular}{p{0.3\textwidth}r}
	\clineB{1-2}{2} \\[-1.8ex]
	LCIA05 ${Mo(\rho|\bm{X})}$ & \# \\[-.5ex]
	\cmidrule(r{.15cm}l{.15cm}){1-2}
	$\{0,225,300\}$        & $\,\,\,25$ \\
	$\{0,226,300\}$        & 22 \\
	$\{0,229,300\}$        &  7 \\
%	$\{0,75,149,225,300\}$ &  6 \\
%	$\{0,76,150,225,300\}$ &  5 \\
	\hline \\[-1.8ex]
\end{tabular}\hspace{.1in}
\begin{tabular}{p{0.3\textwidth}r}
	\clineB{1-2}{2} \\[-1.8ex]
	BH93 ${Mo(\rho|\bm{X})}$ & \# \\[-.5ex]
	\cmidrule(r{.15cm}l{.15cm}){1-2}
	$\{0,300\}$         & 232 \\
	$\{0,297,300\}$     &   4 \\
	$\{0,288,300\}$     &   3 \\
%	$\{0,230,232,300\}$ &   2 \\
%	$\{0,247,250,300\}$ &   2 \\
	\hline \\[-1.8ex]
\end{tabular}
\endgroup
\caption{Top posterior modes of $\rho_1$ and $\rho_2$ (BMCP), $E_1$ and $E_2$ (DPM19) and $\rho$ (LCIA05, BH93) estimated for each of the $400$ data sets of Scenario 2.} \label{tab:scene2_mod}
%\end{adjustwidth}
\end{table}

\begin{figure}[!htbp]
%\begin{adjustwidth}{-.4cm}{-.4cm}
\centering
\subfigure[][BMCP ($\rho_1$)]{
	\includegraphics[width=4cm, height=3cm, trim=0 2cm 0 0]{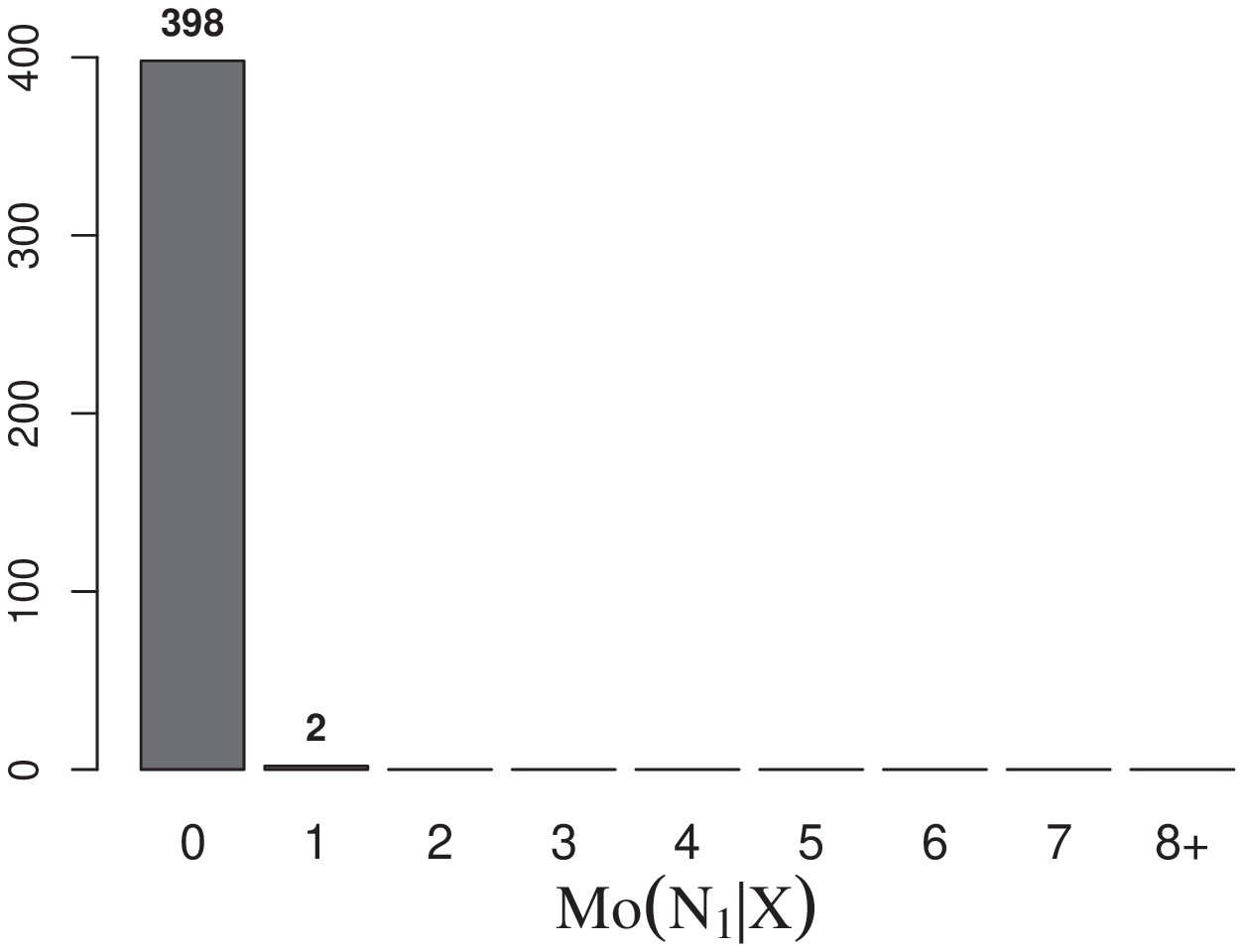}
	\label{fig:scene2_Nmode_LP20_mu}}
\subfigure[][BMCP ($\rho_2$)]{
	\includegraphics[width=4cm, height=3cm, trim=0 2cm 0 0]{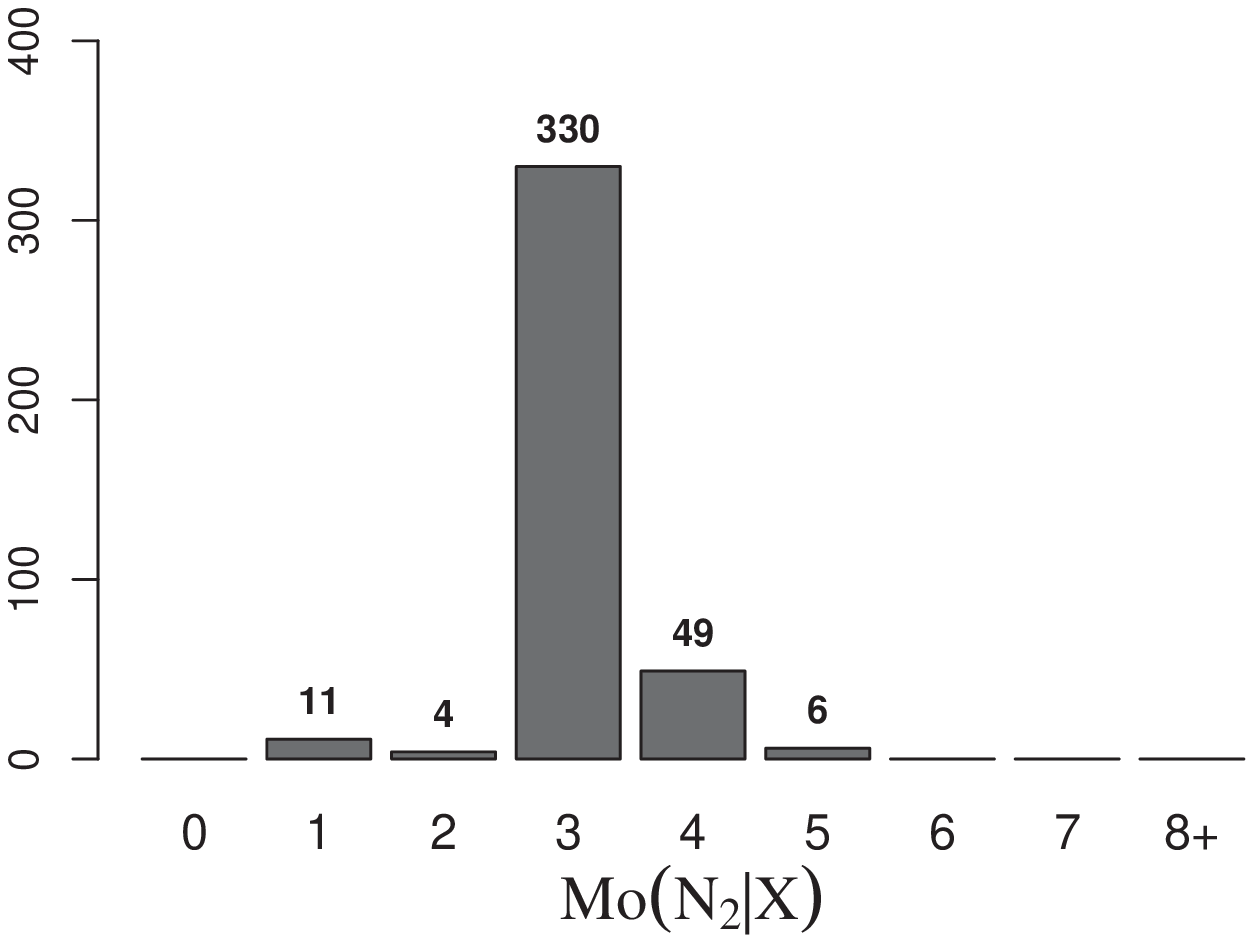}
	\label{fig:scene2_Nmode_LP20_s2}}
\subfigure[][LCIA05 ($\rho$)]{
	\includegraphics[width=4cm, height=3cm, trim=0 2cm 0 0]{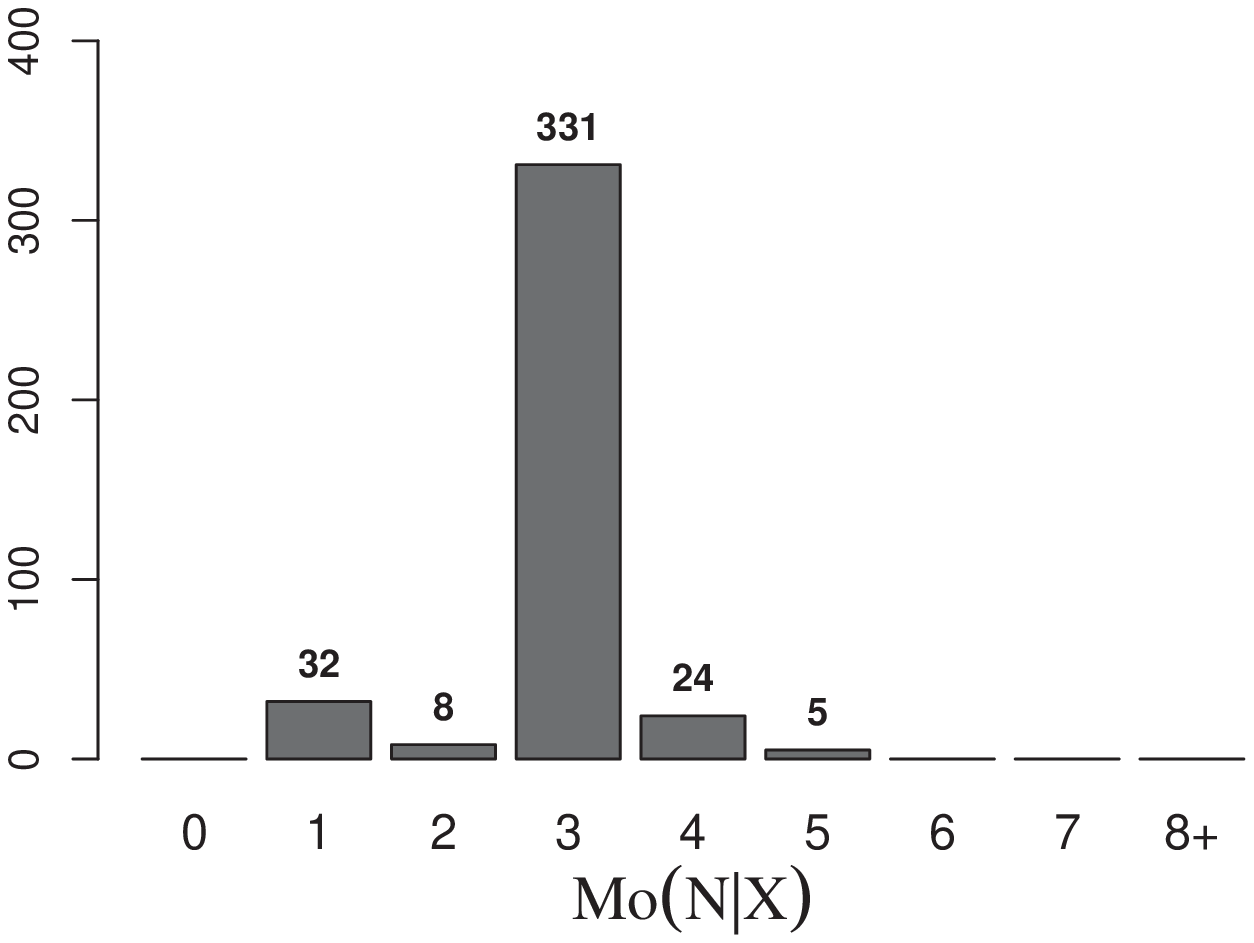}
	\label{fig:scene2_Nmode_LC02}}\\[-.2in]
\subfigure[][DPM19 ($E_1$)]{
	\includegraphics[width=4cm, height=3cm, trim=0 2cm 0 0]{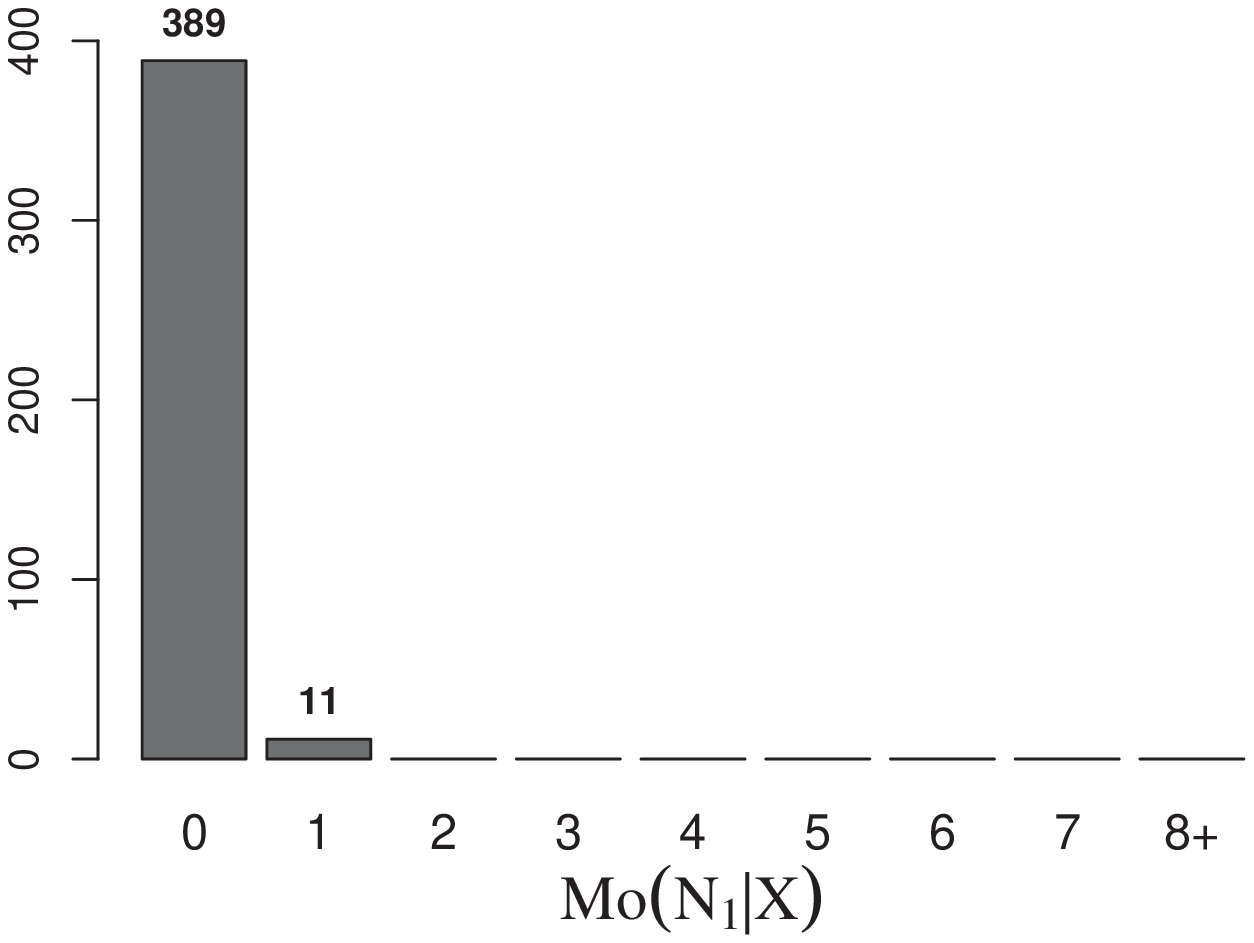}
	\label{fig:scene2_Nmode_P18_mu}}
\subfigure[][DPM19 ($E_2$)]{
	\includegraphics[width=4cm, height=3cm, trim=0 2cm 0 0]{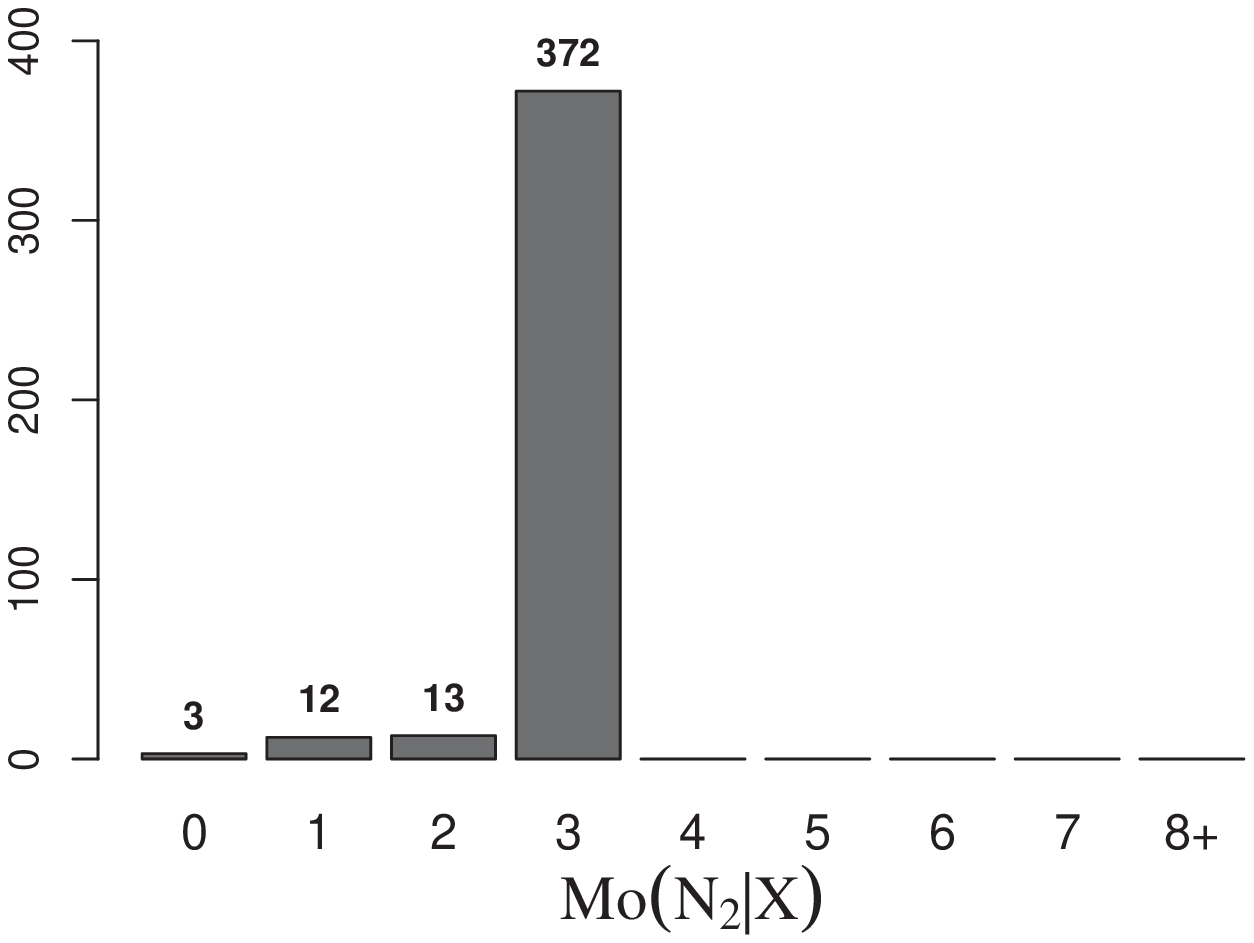}
	\label{fig:scene2_Nmode_P18_s2}}
\subfigure[][BH93 ($\rho$)]{
	\includegraphics[width=4cm, height=3cm, trim=0 2cm 0 0]{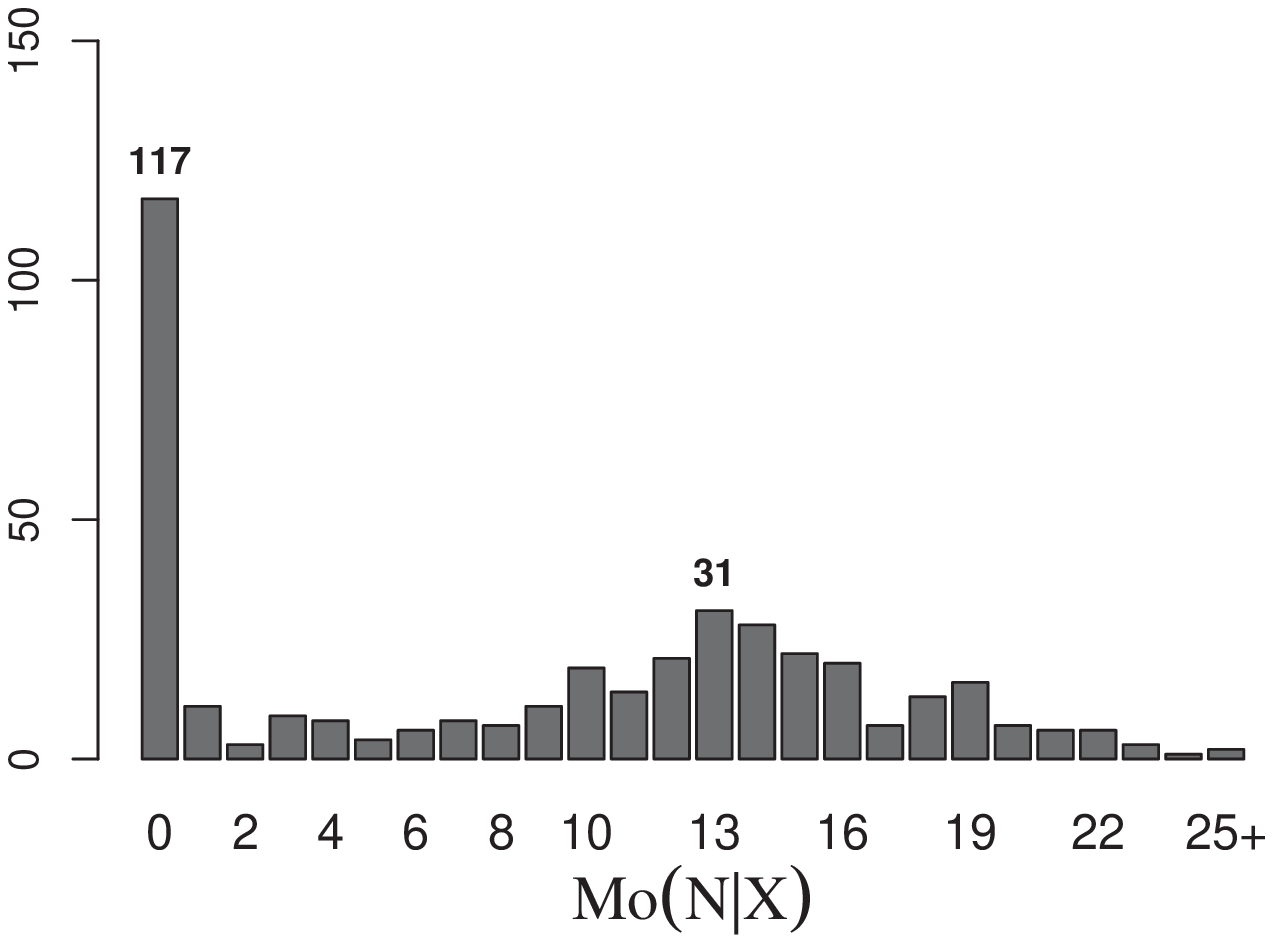}
	\label{fig:scene2_Nmode_BH93}}
%\end{adjustwidth}
\caption{Counts distribution of the posterior modes of the number of changes estimated in each of the $400$ replications, for models BMCP (a,b), LCIA05 (c), DPM19 (d,e) and BH93 (f), for
Scenario 2.} \label{fig:scene2_Nmode}
\end{figure}

Figure \ref{fig:scene2_Nmode} shows that the absence of changes in the mean and the true number of changes in the variance are correctly indicated for almost all data sets under the BMCP and DPM19 models. The total number of changes is correctly identified by the LCIA05 model in $83\%$ of the replications. Under the BH93 model, the number of mean changes is overestimated for the most of the data sets, as a consequence of its constant variance assumption applied to data sequences with non-constant variance.

\begin{figure}[!htb]
%\begin{adjustwidth}{-.4cm}{-.4cm}
\centering
\subfigure[][BMCP ($\rho_1$)]{
	\includegraphics[width=6cm, height=4cm, trim=0 2.5cm 0 0]{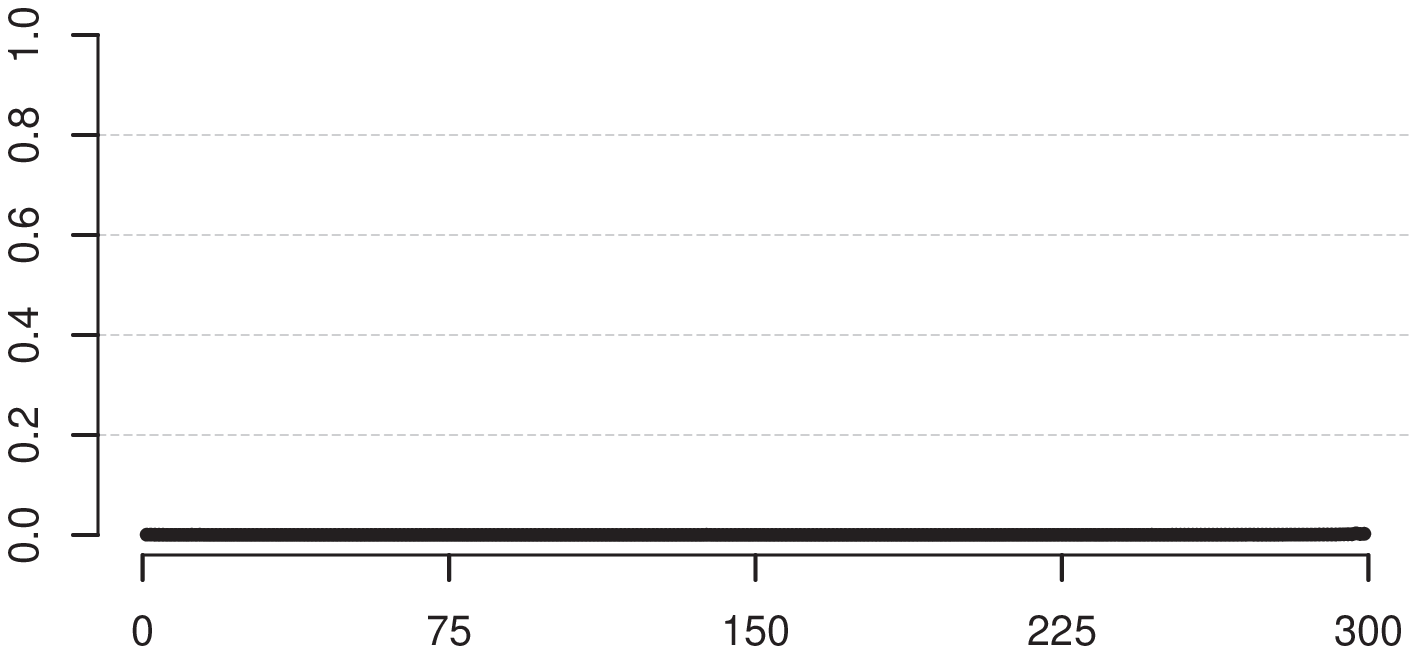}
	\label{fig:scene2_prob_IC_LP20_mu}}
\subfigure[][DPM19 ($E_1$)]{
	\includegraphics[width=6cm, height=4cm, trim=0 2.5cm 0 0]{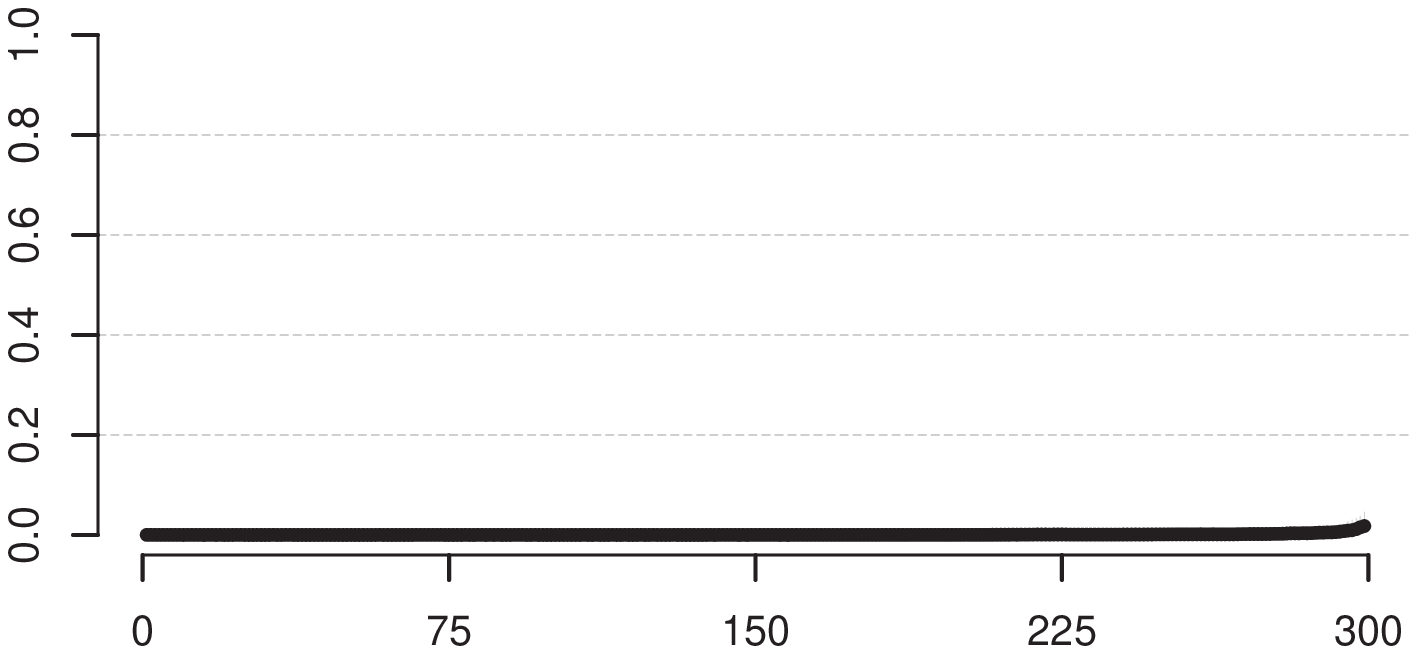}
	\label{fig:scene2_prob_IC_P18_mu}}\\[-.35in]
\subfigure[][BMCP ($\rho_2$)]{
	\includegraphics[width=6cm, height=4cm, trim=0 2.5cm 0 0]{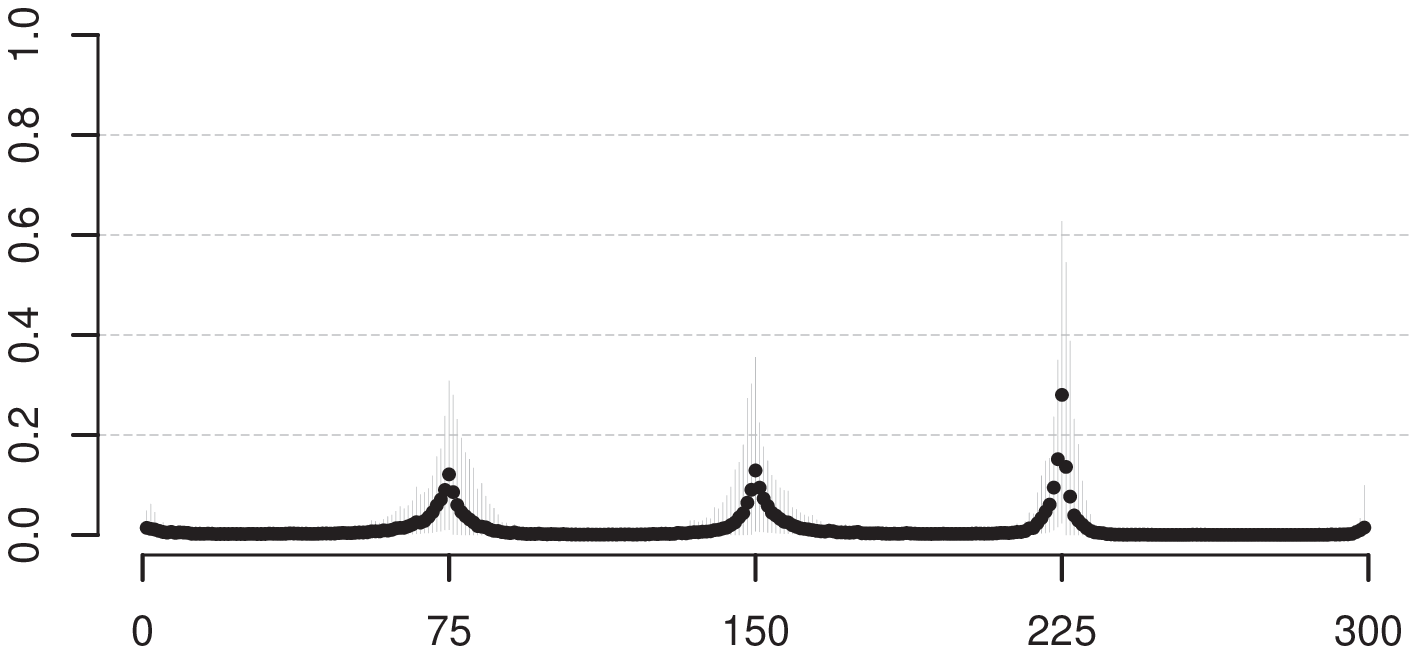}
	\label{fig:scene2_prob_IC_LP20_s2}}
\subfigure[][DPM19 ($E_2$)]{
	\includegraphics[width=6cm, height=4cm, trim=0 2.5cm 0 0]{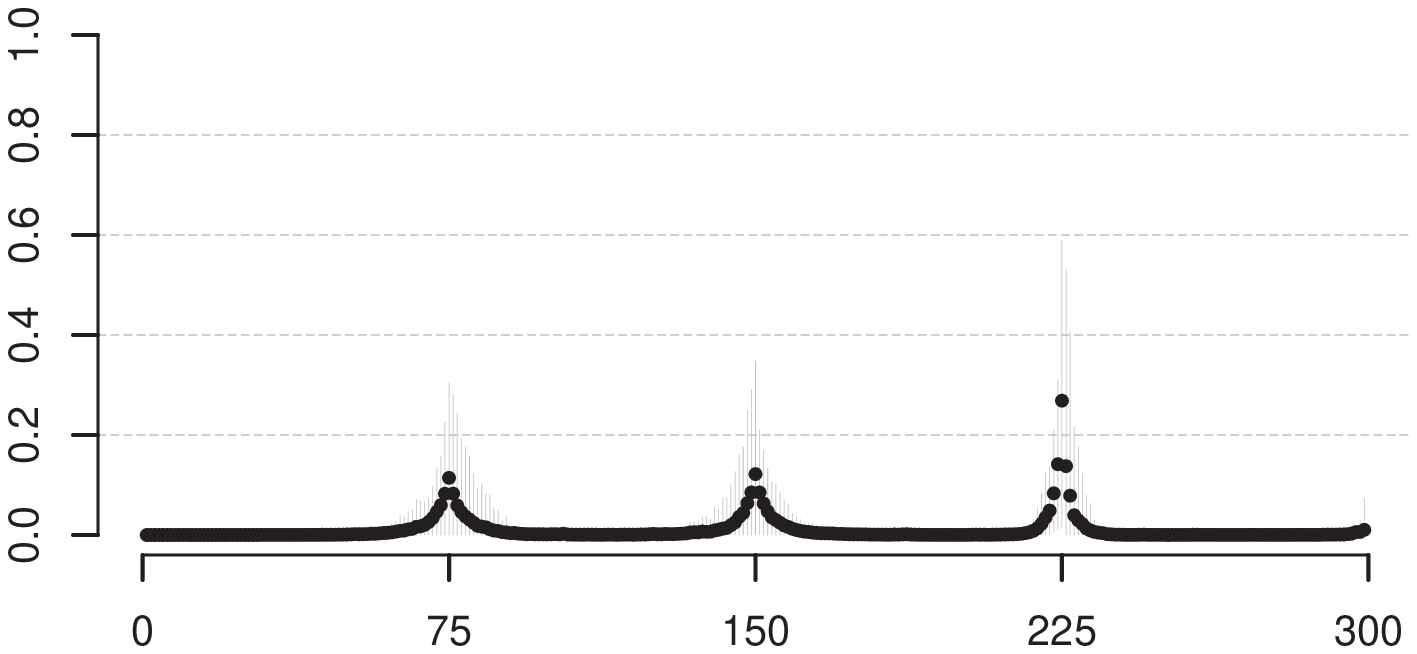}
	\label{fig:scene2_prob_IC_P18_s2}}\\[-.35in]
\subfigure[][LCIA05 ($\rho$)]{
	\includegraphics[width=6cm, height=4cm, trim=0 2.5cm 0 0]{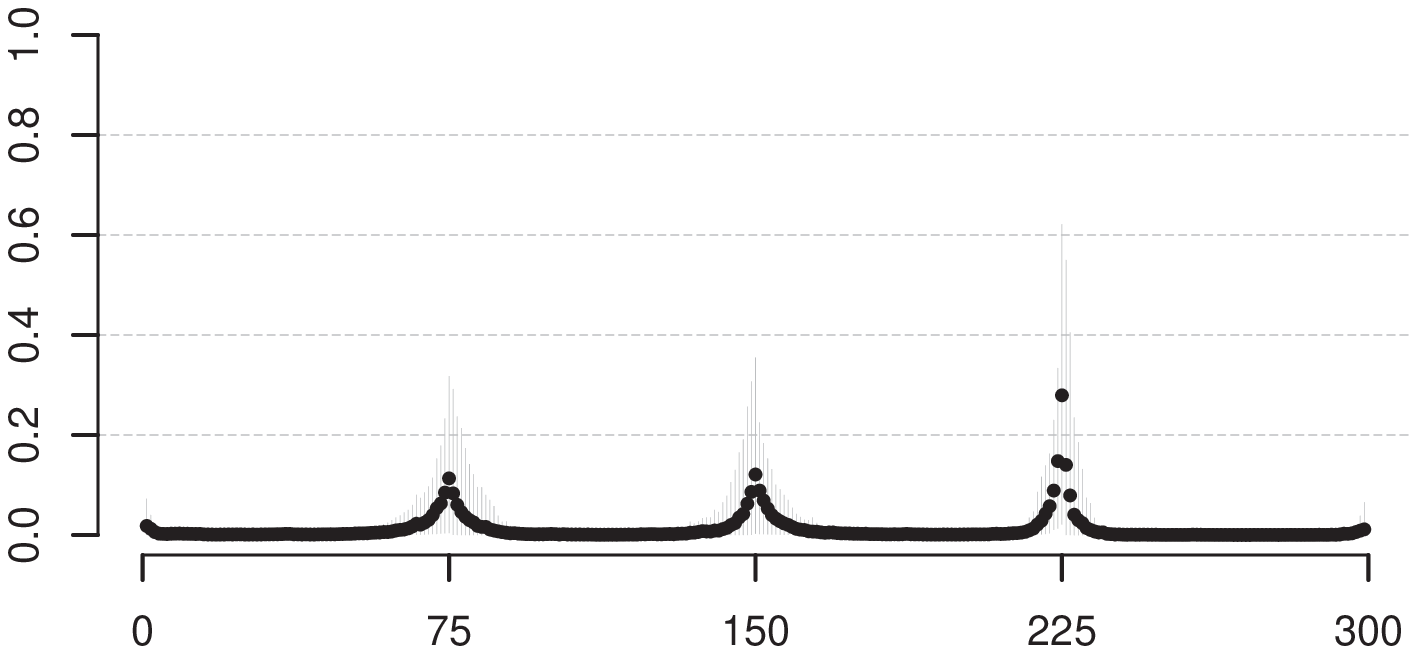}
	\label{fig:scene2_prob_IC_LC02}}
\subfigure[][BH93 ($\rho$)]{
	\includegraphics[width=6cm, height=4cm, trim=0 2.5cm 0 0]{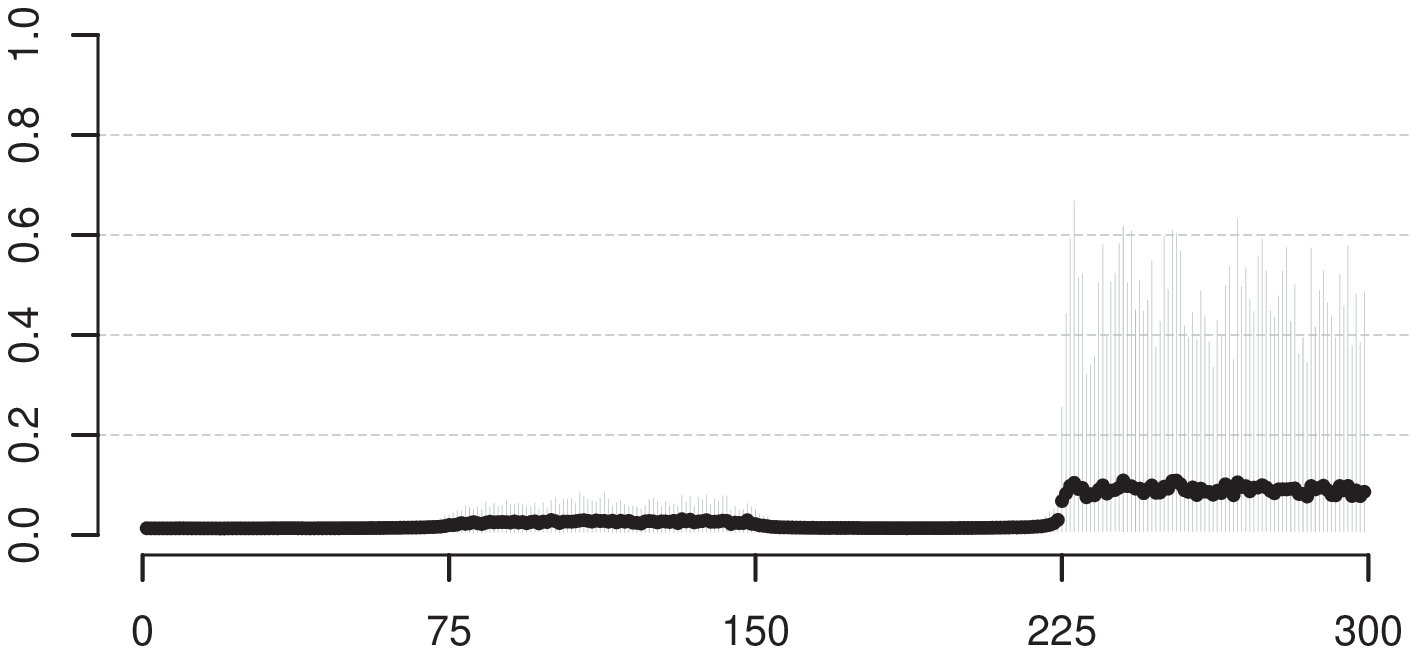}
	\label{fig:scene2_prob_IC_BH93}}
%\end{adjustwidth}
\caption{Average of the posterior probabilities of each instant to be an end point 
(black dots) for each partition and the $5\%$ and $95\%$ quantiles range of such probabilities based on the Monte Carlo replications, for models BMCP (a,c), DPM19 (b,d), LCIA05 (e) and BH93 (f) for Scenario 2.}
\label{fig:scene2_prob_IC}
\end{figure}

The average posterior probabilities displayed in Figure \ref{fig:scene2_prob_IC} improve the previous estimation of the positions of the changes in $\rho_1$ and $\rho_2$ under the proposed model. The probability of a change in the mean in each instant is very close to zero for all instants and all data sets under both BMCP and DPM19 models. Figures \ref{fig:scene2_prob_IC_LP20_s2},  \ref{fig:scene2_prob_IC_P18_s2} and \ref{fig:scene2_prob_IC_LC02} show that, in comparison with the other instants, on average, those in which the true changes occurred are much more likely. According to this criterion, Figure \ref{fig:scene2_prob_IC_BH93} shows that BH93 model has a poor performance, indicating that, on average, all the instants in the $4th$ cluster have a higher probability of being change points in the mean. This behavior is related to the overestimated number of changes in the mean shown in Figure \ref{fig:scene2_Nmode_BH93}. In addition to the LCIA05 results, the BMCP and DPM19 models specify that the changes occurred in the variance.

\FloatBarrier
\subsubsection{Scenario 3: changes in the mean and variance at different times}\label{sec_scene3}

\begin{figure}[!htbp]
%\begin{adjustwidth}{-.4cm}{-.4cm}
\centering
\subfigure[][BMCP]{
	\includegraphics[width=3.5cm,height=3cm, trim=0 .5cm 0 0]{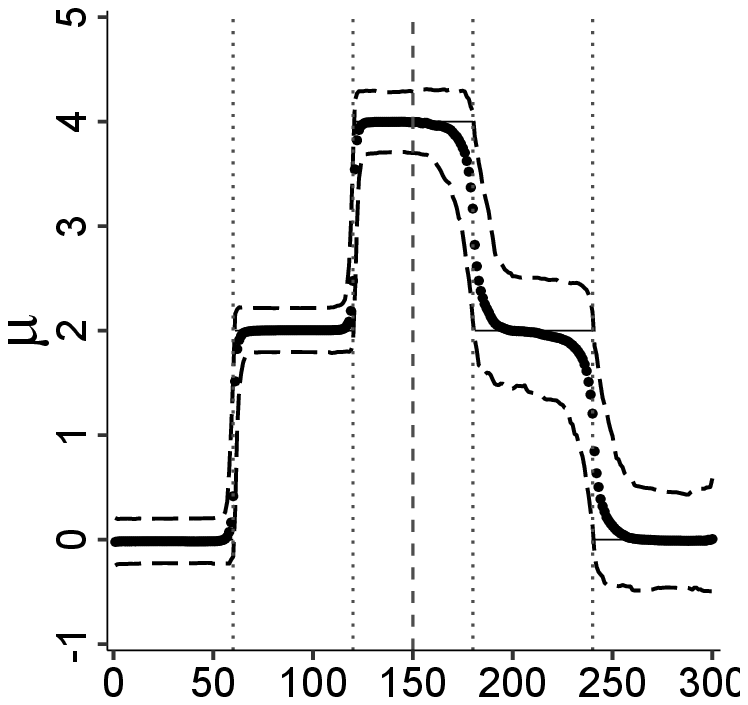}
	\label{fig:scene3_mu_LP20}}%\hspace{.01in}
\subfigure[][DPM19]{
	\includegraphics[width=3.5cm,height=3cm, trim=0 .5cm 0 0]{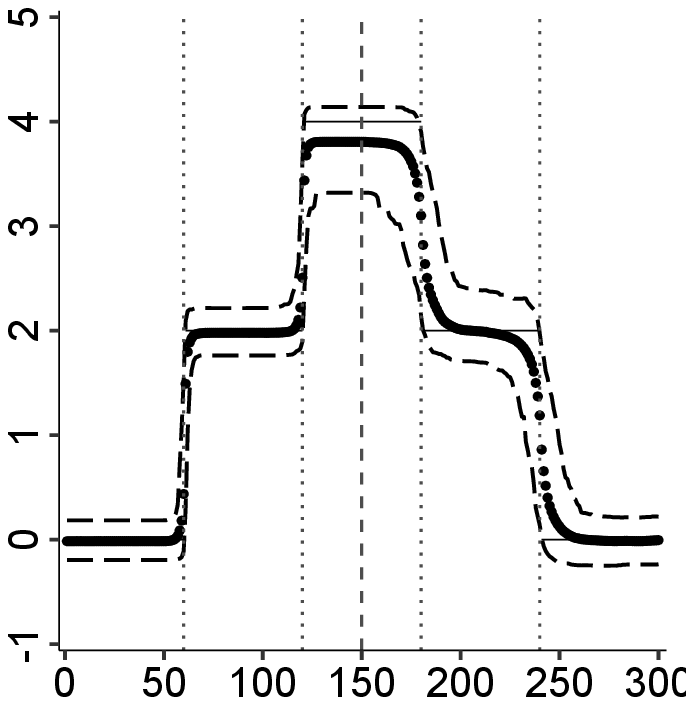}
	\label{fig:scene3_mu_P18}}%\hspace{.01in}
\subfigure[][LCIA05]{
	\includegraphics[width=3.5cm,height=3cm, trim=0 .5cm 0 0]{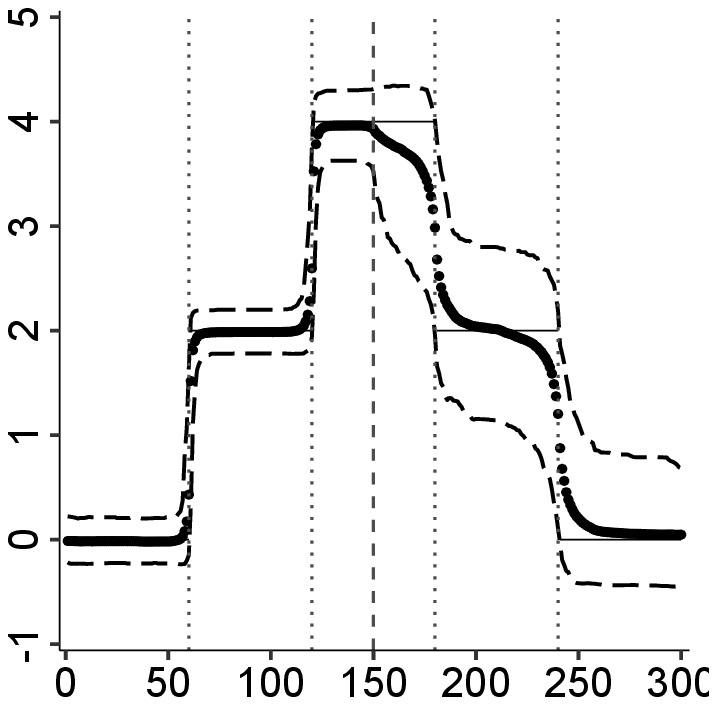}
	\label{fig:scene3_mu_LC02}}%\hspace{.01in}
\subfigure[][BH93]{
	\includegraphics[width=3.5cm,height=3cm, trim=0 .5cm 0 0]{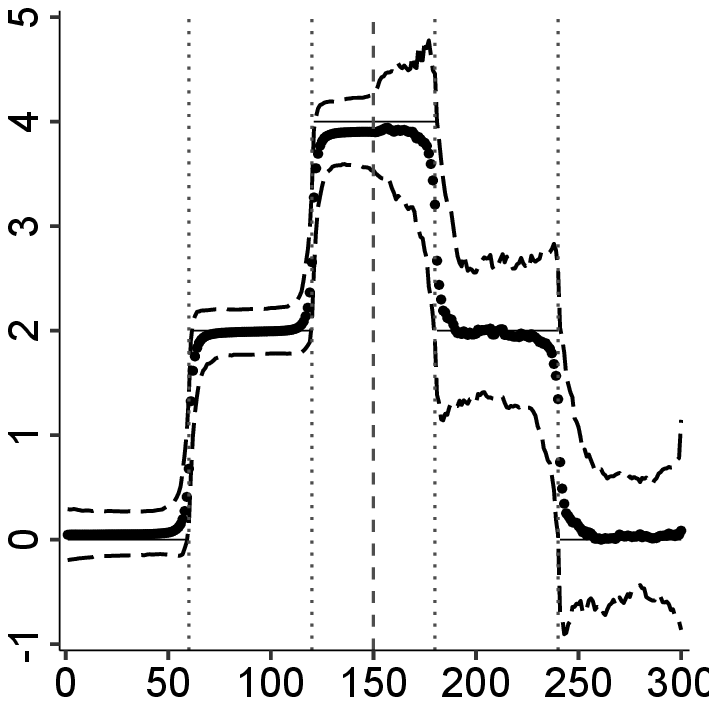}
	\label{fig:scene3_mu_BH93}}
\\
\subfigure[][BMCP]{
	\includegraphics[width=3.5cm,height=3cm, trim=0 .5cm 0 0]{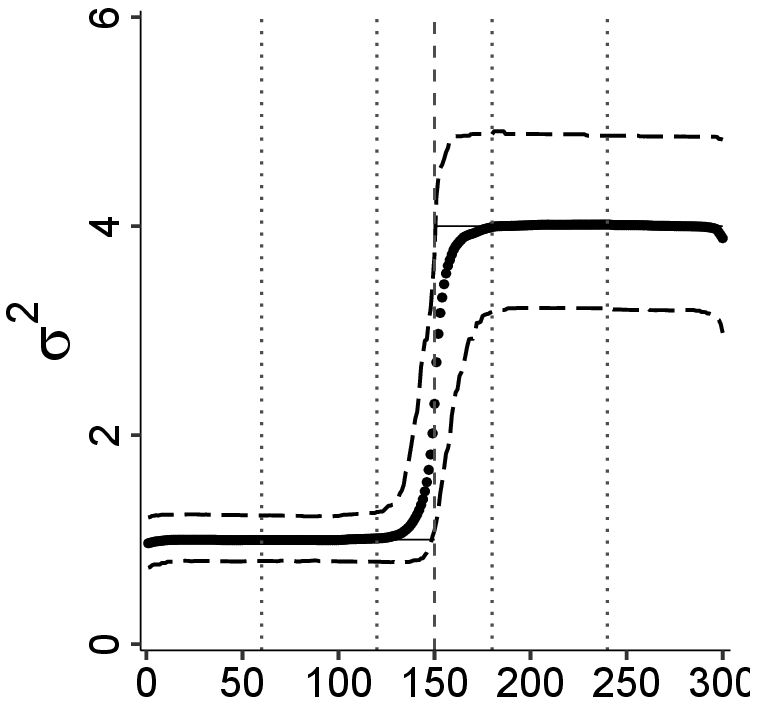}
	\label{fig:scene3_s2_LP20}}%\hspace{.01in}
\subfigure[][DPM19]{
	\includegraphics[width=3.5cm,height=3cm, trim=0 .5cm 0 0]{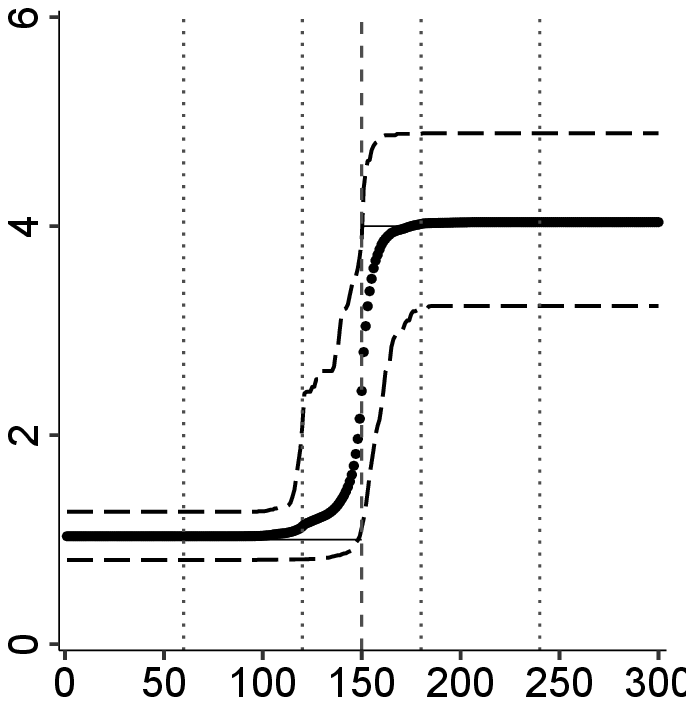}
	\label{fig:scene3_s2_P18}}%\hspace{.01in}
\subfigure[][LCIA05]{
	\includegraphics[width=3.5cm,height=3cm, trim=0 .5cm 0 0]{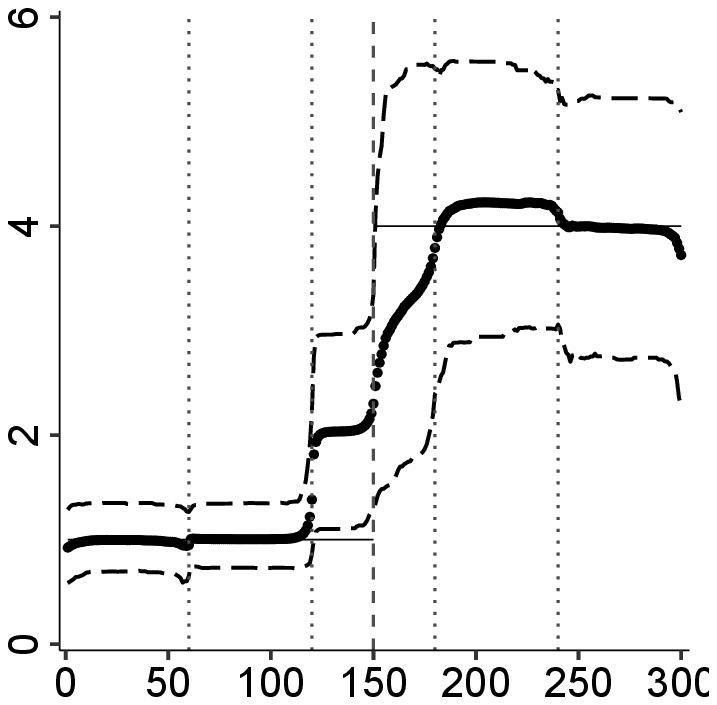}
	\label{fig:scene3_s2_LC02}}%\hspace{.01in}
\subfigure[][BH93]{
	\includegraphics[width=3.5cm,height=3cm, trim=0 .5cm 0 0]{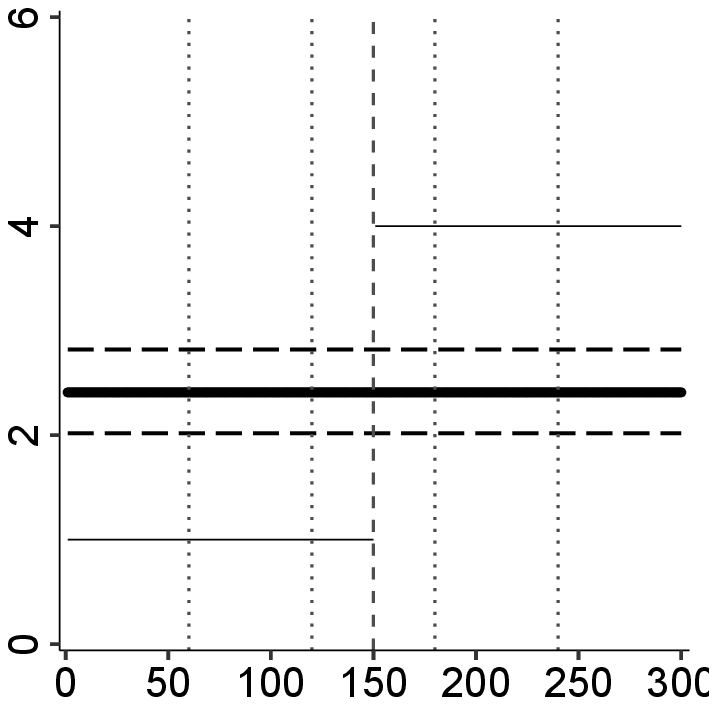}
	\label{fig:scene3_s2_BH93}}
%\end{adjustwidth}
\caption{Average of the product estimates (black dots) for the means (top) and variances (bottom) in each instant and the $5\%$ and $95\%$ quantiles of such estimates based on the Monte Carlo replications, under BMCP (a,e), DPM19 (b,f), LCIA05 (c,g) and BH93 (d,h) models for Scenario 3. The true mean and variance values are indicated by the solid gray horizontal lines.
The vertical gray dotted and dashed lines indicate the true end points in $\bm{\mu}$ and $\bm{\sigma}$, respectively.} \label{fig:scene3_PE}
\end{figure}

In this scenario, we assume that four changes in the mean and one change in the variance occur at different times, inducing a total of six clusters in data sequences of size $n=300$. Changes in the mean and variance are given by the partitions $\rho_1=\{0,60,120,180,240,300\}$ and
$\rho_2=\{0,150,300\}$, respectively, and the cluster parameters are $\bm{\sigma}^\star=(1,4)$ and ${\bm{\mu}^\star=(0,2,4,2,0)}$. Figure \ref{fig:scene3_PE} shows that the four models provide reasonable estimates for the means. These estimates are more biased around the true changes and it is more evident after instant $150$, when the variance experiences an increase. For all the models the  estimates for the means become less accurate after the change in the variance. However, in the LCIA05 and BH93 models, the loss of accuracy is more severe. The BMCP model provides the most accurate estimates for the means (Figure \ref{fig:scene3_mu_LP20}) and for the variances (Figure \ref{fig:scene3_s2_LP20}).
As shown in Figure \ref{fig:scene3_s2_LC02}, the product estimates for the variance provided by model LCIA05 are clearly affected by changes in the mean. For example, the product estimates for the variance present an expressive change at position $120$, indicating the presence of a change
point in this parameter and position, but such change does not truly exist. A similar but less noticeable bias can be seen in Figure \ref{fig:scene3_s2_P18}. BMCP and DPM19 have similar performance but DPM19 produced more biased estimates in the third cluster for the means and for the variance between observations $100$ and $150$.

\begin{figure}[!htbp]
%\begin{adjustwidth}{-.4cm}{-.4cm}
\centering
\subfigure[][BMCP ($\rho_1$)]{
	\includegraphics[width=4cm, height=3cm, trim=0 2cm 0 0]{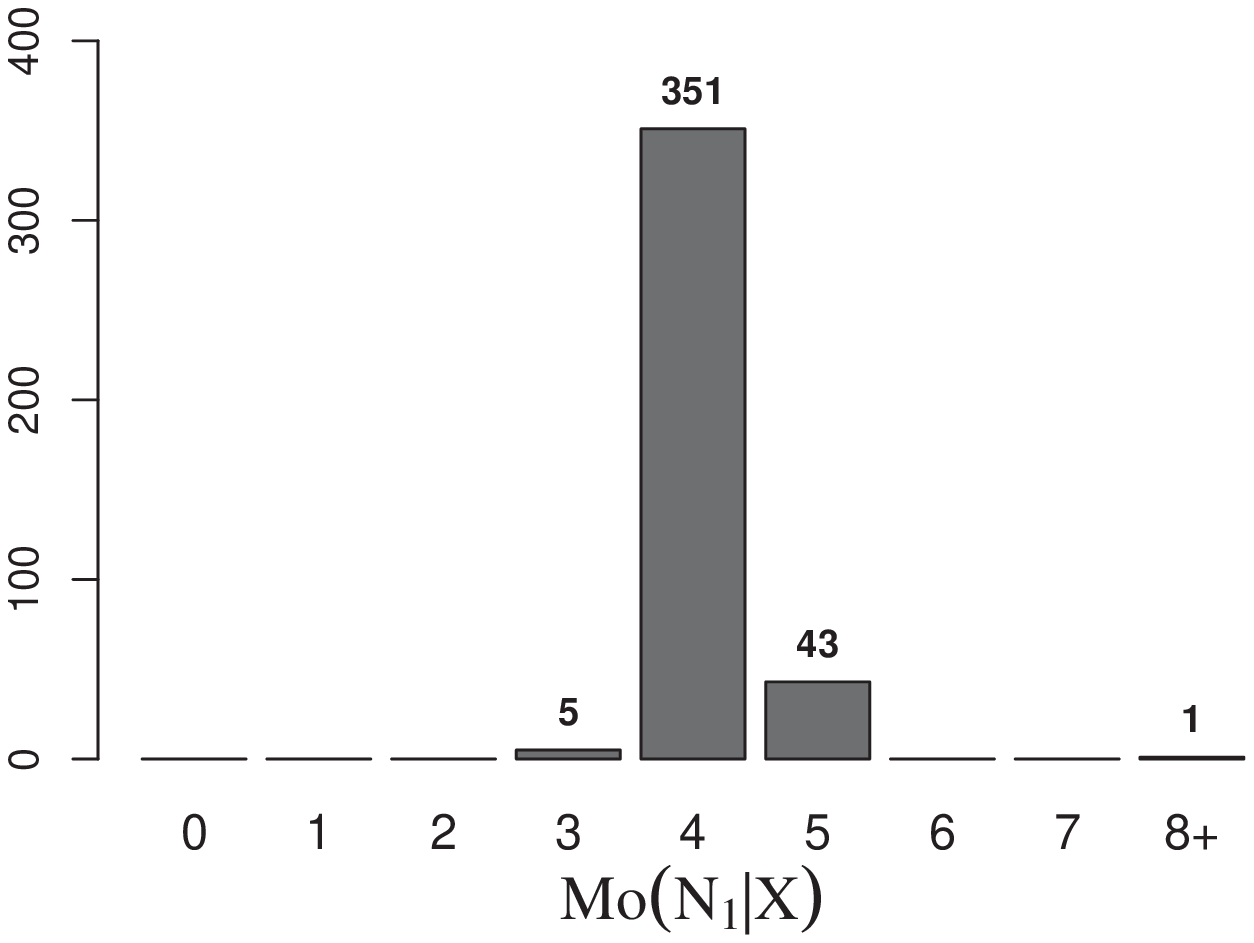}
	\label{fig:scene3_Nmode_LP20_mu}}
\subfigure[][BMCP ($\rho_2$)]{
	\includegraphics[width=4cm, height=3cm, trim=0 2cm 0 0]{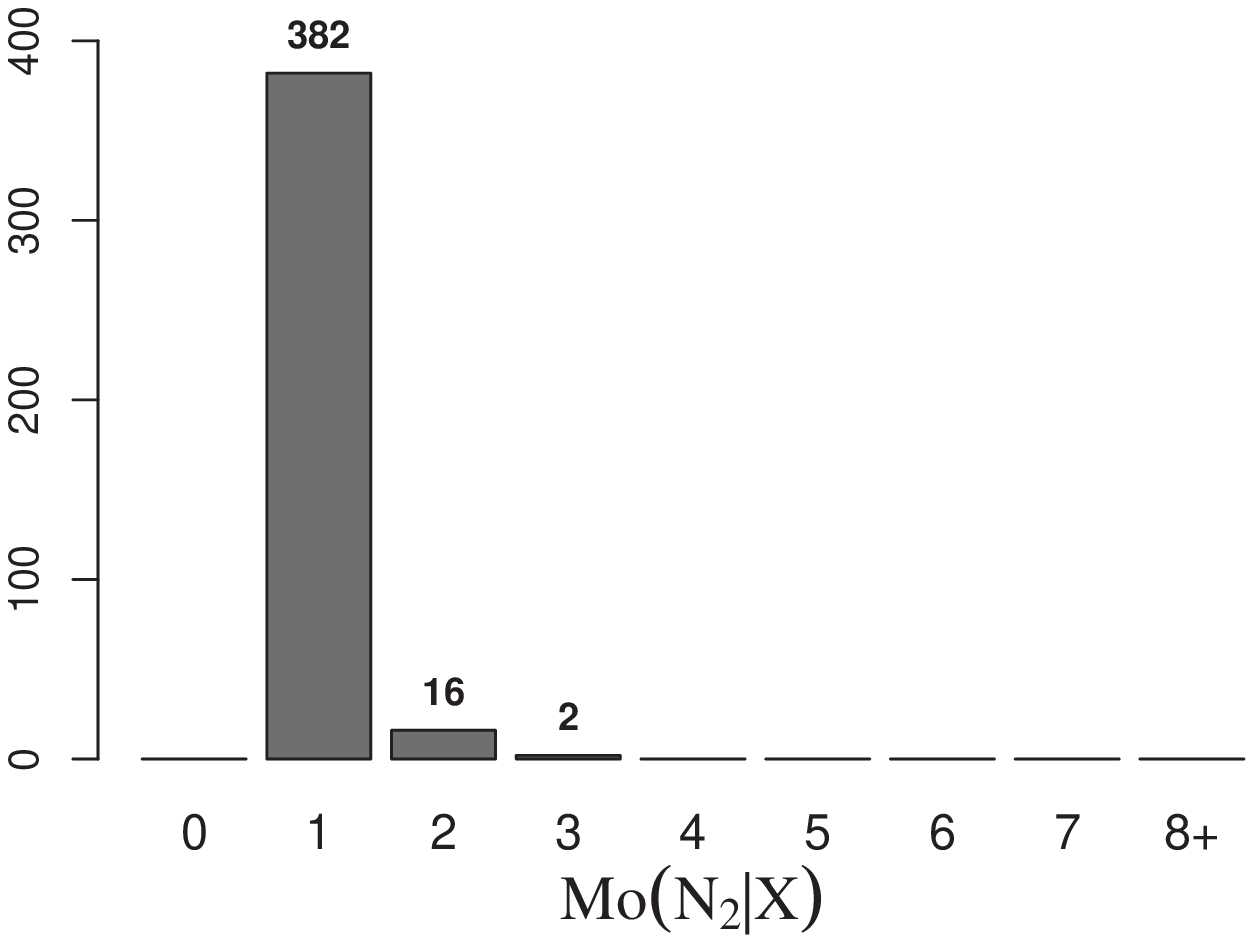}
	\label{fig:scene3_Nmode_LP20_s2}}
\subfigure[][LCIA05 ($\rho$)]{
	\includegraphics[width=4cm, height=3cm, trim=0 2cm 0 0]{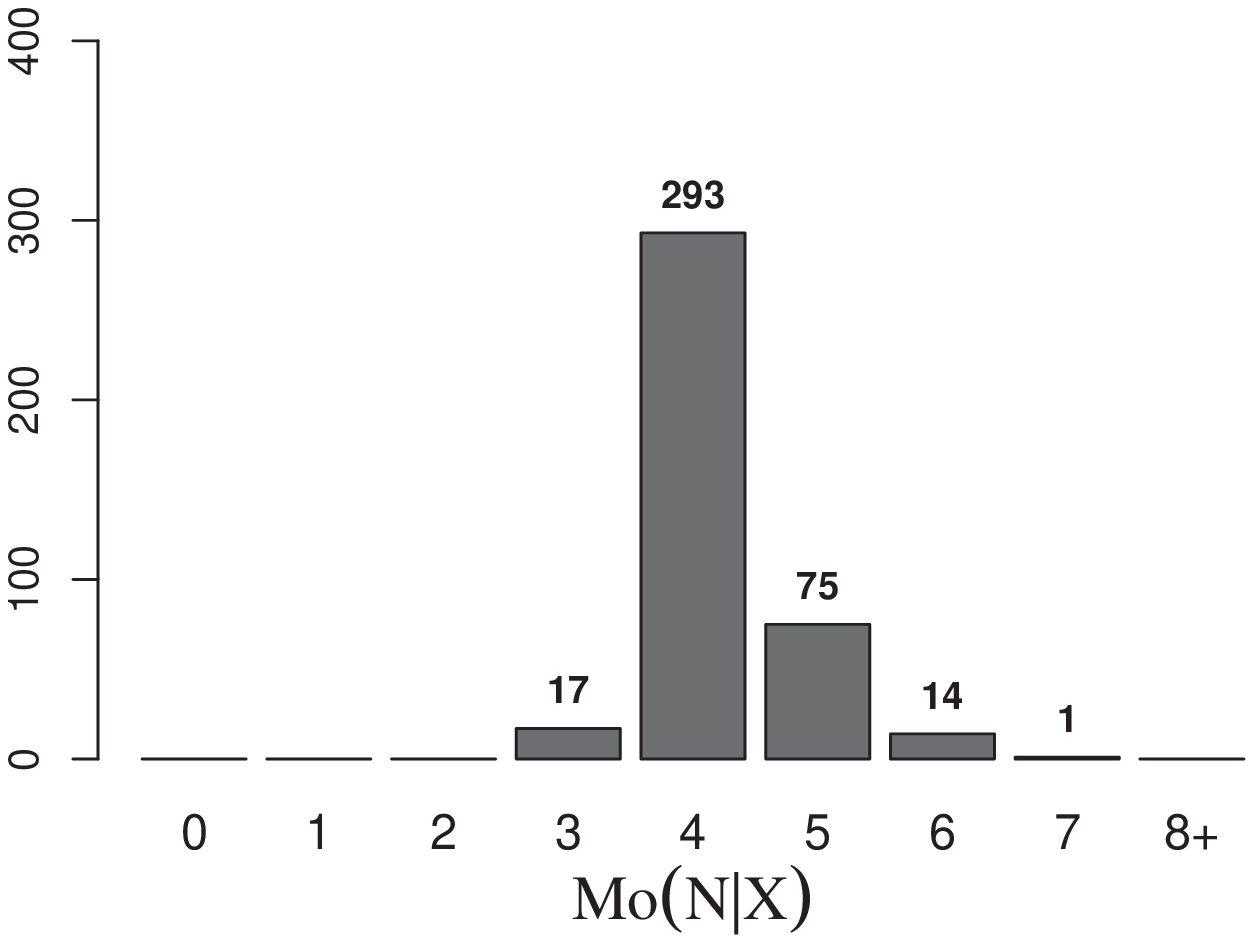}
	\label{fig:scene3_Nmode_LC02}}\\[-.2in]
\subfigure[][DPM19 ($E_1$)]{
	\includegraphics[width=4cm, height=3cm, trim=0 2cm 0 0]{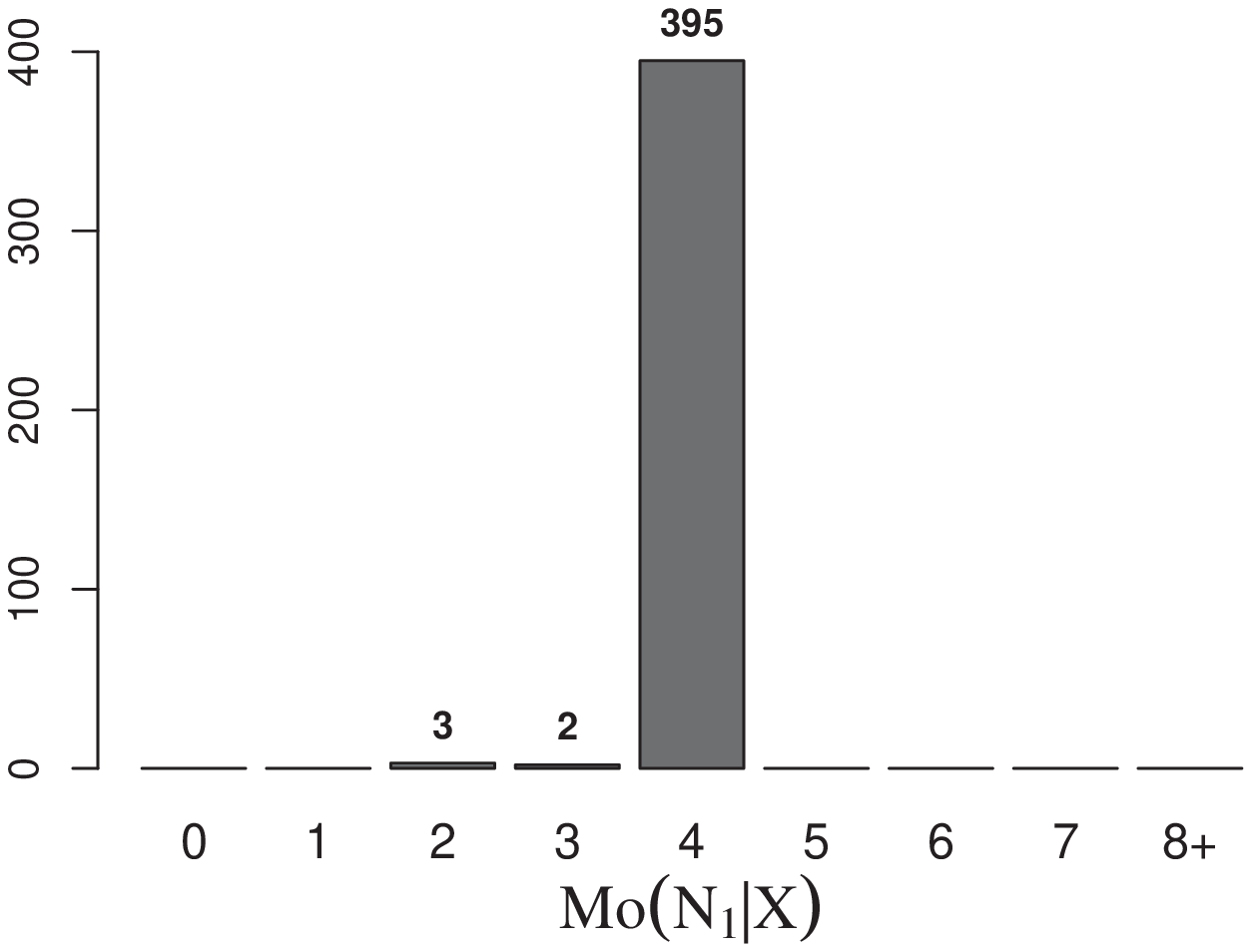}
	\label{fig:scene3_Nmode_P18_mu}}
\subfigure[][DPM19 ($E_2$)]{
	\includegraphics[width=4cm, height=3cm, trim=0 2cm 0 0]{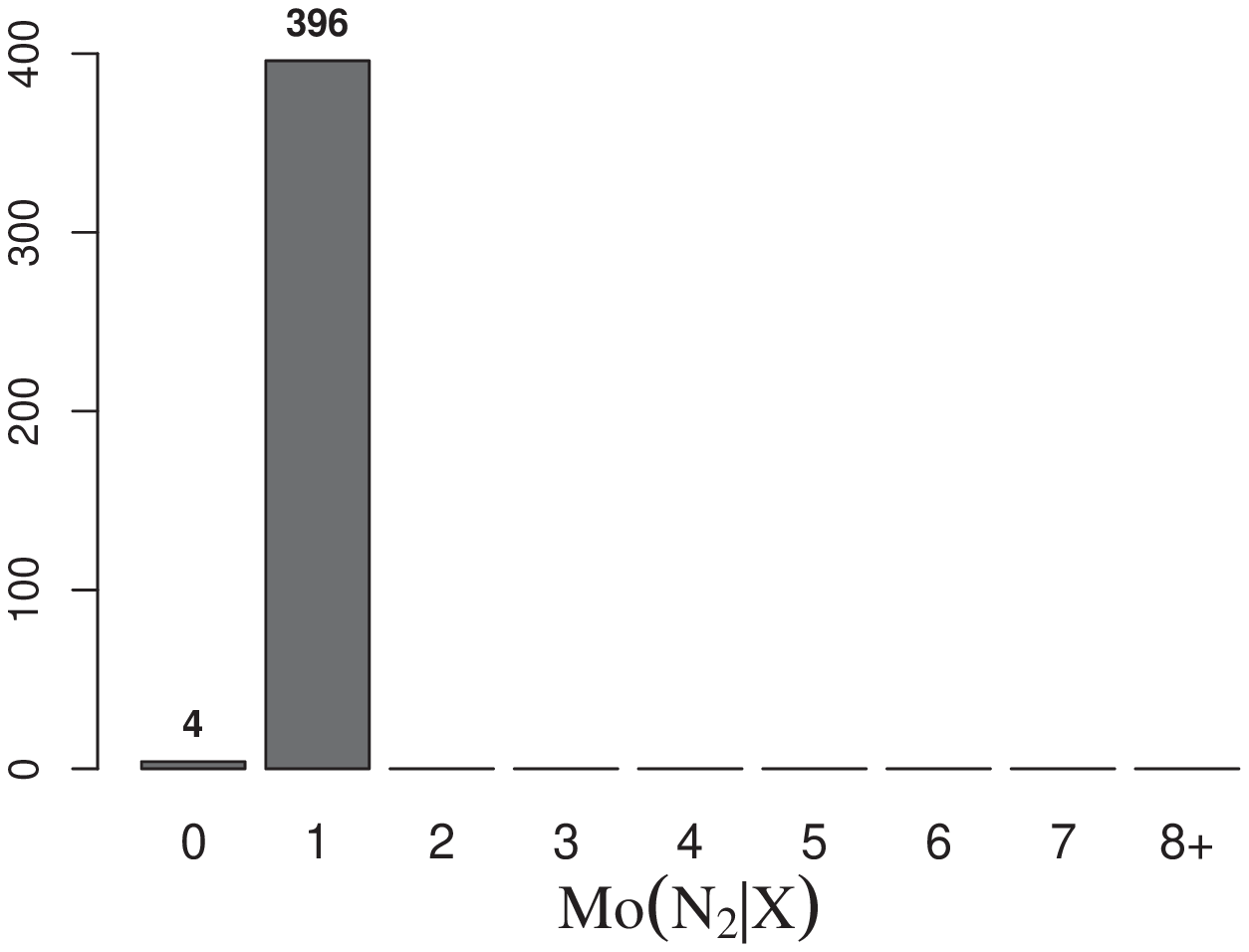}
	\label{fig:scene3_Nmode_P18_s2}}
\subfigure[][BH93 ($\rho$)]{
	\includegraphics[width=4cm, height=3cm, trim=0 2cm 0 0]{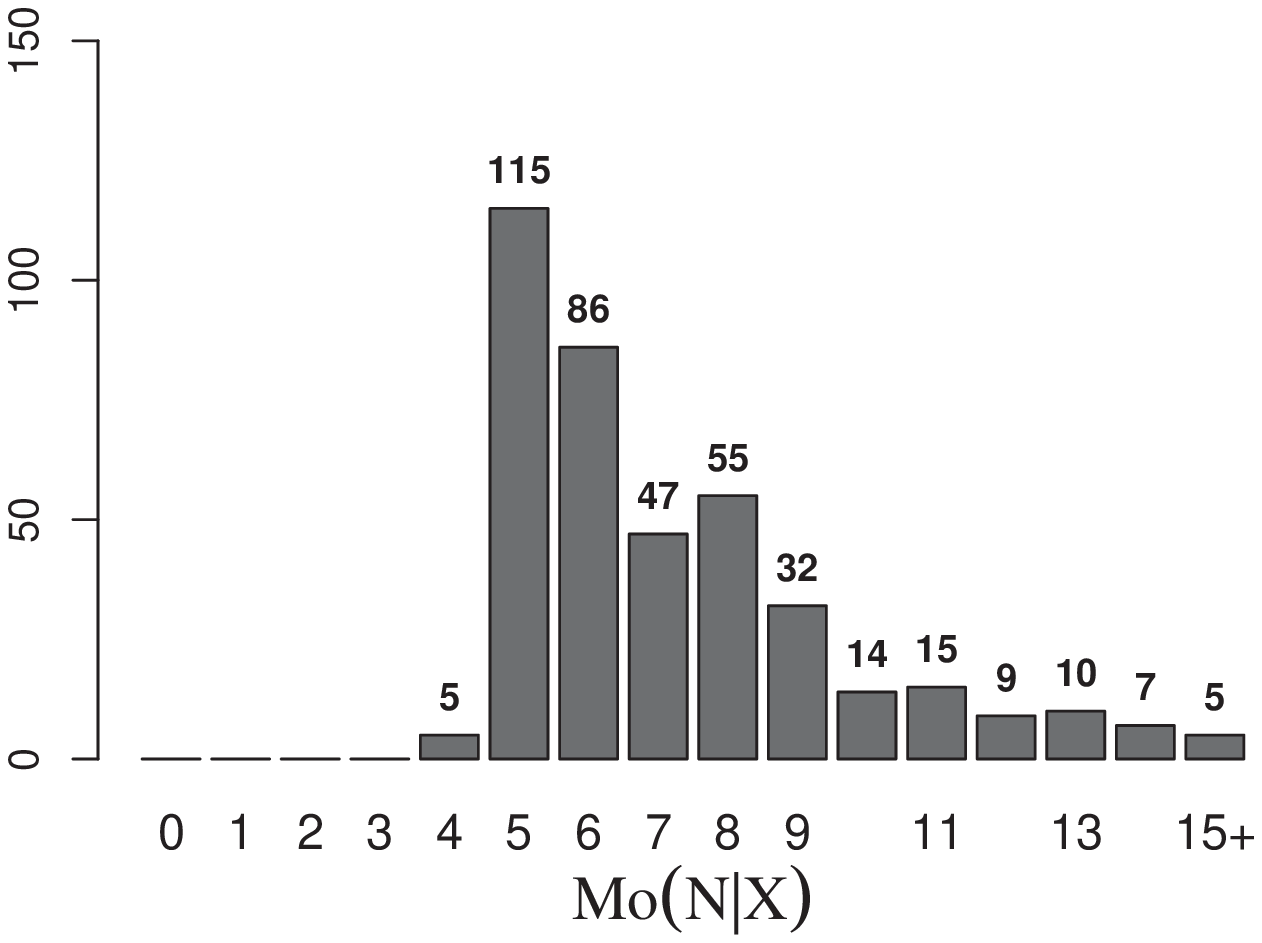}
	\label{fig:scene3_Nmode_BH93}}
%\end{adjustwidth}
\caption{Counts distribution of the posterior modes of the number of changes estimated in each of the 400 replications, for models BMCP (a,b), LCIA05 (c), DPM19 (d,e) and BH93 (f), for Scenario 3.} \label{fig:scene3_Nmode}
\end{figure}

For more than $85\%$ of the data sets (Figure \ref{fig:scene3_Nmode}), both the BMCP and DPM19 models correctly estimate the number of change points in the mean as well as in the variance. The LCIA05 underestimated the true number of changes ($N=5$) by one in $75\%$ of the data sets. For model BH93, the number of mean changes is overestimated for almost all data sets,  reflecting the expected poor performance of this model under the current scenario, due to its constant variance assumption.

Table \ref{tab:scene3_mod} shows competitive performances between the BMCP and DPM19 models to identify the true partitions in the data. The LCIA05 model does not identify the variance change in the reported partitions. Under the LCIA05 model the most likely partitions under the posteriori, in the sense of being most frequently selected in our Monte Carlo study, indicate the occurrence of only the four changes in the mean. Under the BMCP and DPM19 models, the estimated partitions that differ from the true $\rho_1$ or $\rho_2$ are very close to the true ones, in the sense that differences occur only by one or two elements or positions. Assuming the  BMCP and DPM19, for the majority of the data sets, the most likely partitions for the variance correctly indicate the existence of one change point. Not all of these partitions precisely identify its correct position, but even so, they indicate an instant that is very close to the true change point.

\begin{table}[!htb]
%\begin{adjustwidth}{-.5cm}{}
\centering
\footnotesize
\begingroup
\renewcommand{\arraystretch}{.6} % Default value: 1
\begin{tabular}{p{0.3\textwidth}r}
	\clineB{1-2}{2} \\[-1.8ex]
	BMCP ${Mo(\rho_1|\bm{X})}$ & \# \\[-.5ex]
	\cmidrule(r{.15cm}l{.15cm}){1-2}
	$\{0,60,120,180,240,300\}$ & 10 \\
	$\{0,60,120,179,240,300\}$ &  5 \\
	$\{0,60,120,179,241,300\}$ &  5 \\
%	$\{0,60,120,181,240,300\}$ &  5 \\
%	$\{0,60,120,177,242,300\}$ &  3 \\
	\hline \\[-1.8ex]
\end{tabular}\hspace{.1in}
\begin{tabular}{p{0.3\textwidth}r}
	\clineB{1-2}{2} \\[-1.8ex]
	BMCP ${Mo(\rho_2|\bm{X})}$ & \# \\[-.5ex]
	\cmidrule(r{.15cm}l{.15cm}){1-2}
	$\{0,150,300\}$ & 88 \\
	$\{0,151,300\}$ & 58 \\
	$\{0,152,300\}$ & 41 \\
%	$\{0,149,300\}$ & 27 \\
%	$\{0,154,300\}$ & 25 \\
	\hline \\[-1.8ex]
\end{tabular}\\
\begin{tabular}{p{0.3\textwidth}r}
	\clineB{1-2}{2} \\[-1.8ex]
	DPM19 ${Mo(E_1|\bm{X})}$ & \# \\[-.5ex]
	\cmidrule(r{.15cm}l{.15cm}){1-2}
	$\{0,60,120,180,240,300\}$ & 16 \\
	$\{0,60,120,181,240,300\}$ &  7 \\
	$\{0,60,240,300\}$         &  7 \\
%	$\{0,60,120,179,241,300\}$ &  6 \\
%	$\{0,60,120,180,239,300\}$ &  5 \\
	\hline \\[-1.8ex]
\end{tabular}\hspace{.1in}
\begin{tabular}{p{0.3\textwidth}r}
	\clineB{1-2}{2} \\[-1.8ex]
	DPM19 ${Mo(E_2|\bm{X})}$ & \# \\[-.5ex]
	\cmidrule(r{.15cm}l{.15cm}){1-2}
	$\{0,150,300\}$ & 93 \\
	$\{0,151,300\}$ & 51 \\
	$\{0,152,300\}$ & 38 \\
%	$\{0,149,300\}$ & 26 \\
%	$\{0,153,300\}$ & 23 \\
	\hline \\[-1.8ex]
\end{tabular}\\
\begin{tabular}{p{0.3\textwidth}r}
	\clineB{1-2}{2} \\[-1.8ex]
	LCIA05 ${Mo(\rho|\bm{X})}$ & \# \\[-.5ex]
	\cmidrule(r{.15cm}l{.15cm}){1-2}
	$\{0,60,120,180,240,300\}$ & 6 \\
	$\{0,60,120,179,241,300\}$ & 5 \\
	$\{0,60,120,150,241,300\}$ & 4 \\
%	$\{0,60,120,179,240,300\}$ & 4 \\
%	$\{0,60,120,151,300\}$     & 3 \\
	\hline \\[-1.8ex]
\end{tabular}\hspace{.1in}
\begin{tabular}{p{0.3\textwidth}r}
	\clineB{1-2}{2} \\[-1.8ex]
	BH93 ${Mo(\rho|\bm{X})}$ & \# \\[-.5ex]
	\cmidrule(r{.15cm}l{.15cm}){1-2}
	$\{0,60,120,180,240,300\}$ & 13 \\
	$\{0,60,120,180,239,300\}$ &  5 \\
	$\{0,60,120,179,240,300\}$ &  4 \\
%	$\{0,60,120,181,240,300\}$ &  4 \\
%	$\{0,60,120,182,239,300\}$ &  4 \\
	\hline \\[-1.8ex]
\end{tabular}
\endgroup
\caption{Top posterior modes of $\rho_1$ and $\rho_2$ (BMCP), $E_1$ and $E_2$ (DPM19) and $\rho$ (LCIA05, BH93) estimated for each of the 400 data sets of Scenario 3.}
\label{tab:scene3_mod}
%\end{adjustwidth}
\end{table}

\begin{figure}[!htb]
%\begin{adjustwidth}{-.4cm}{-.4cm}
\centering
\subfigure[][BMCP ($\rho_1$)]{
	\includegraphics[width=6cm, height=4cm, trim=0 2.5cm 0 0]{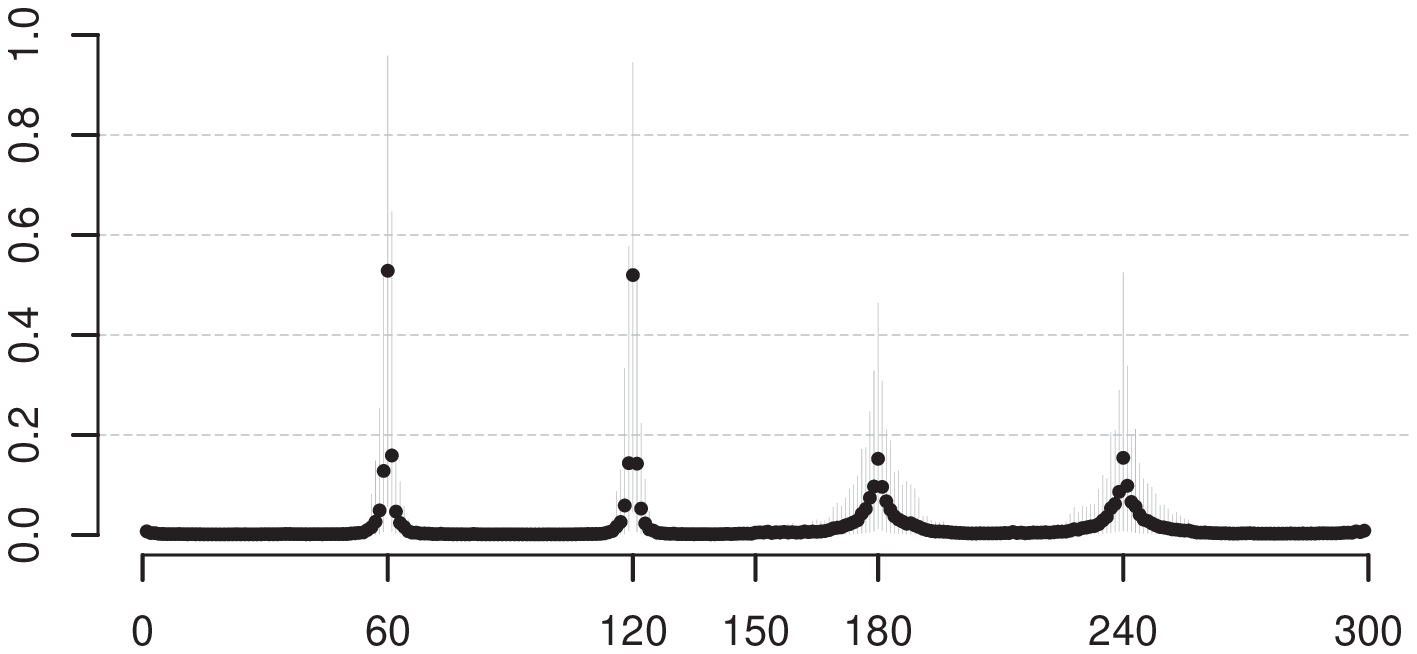}
	\label{fig:scene3_prob_IC_LP20_mu}}
\subfigure[][DPM19 ($E_1$)]{
	\includegraphics[width=6cm, height=4cm, trim=0 2.5cm 0 0]{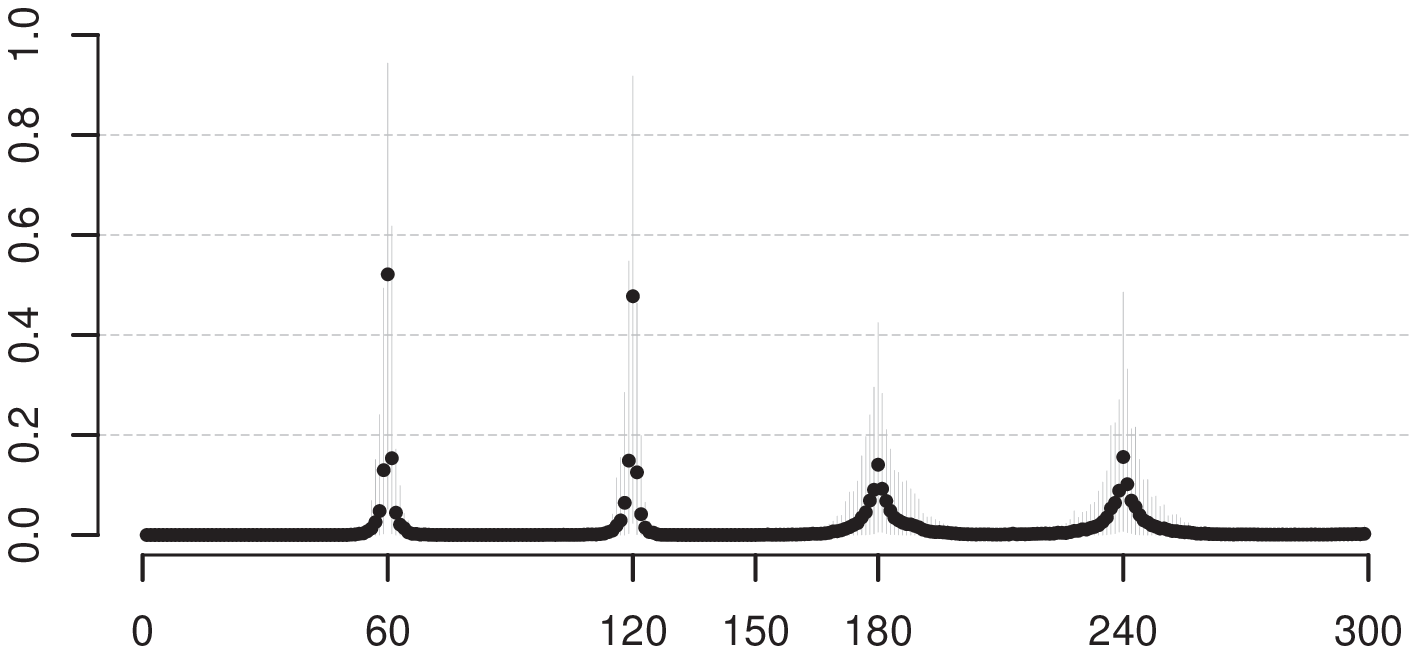}
	\label{fig:scene3_prob_IC_P18_mu}}\\[-.35in]
\subfigure[][BMCP ($\rho_2$)]{
	\includegraphics[width=6cm, height=4cm, trim=0 2.5cm 0 0]{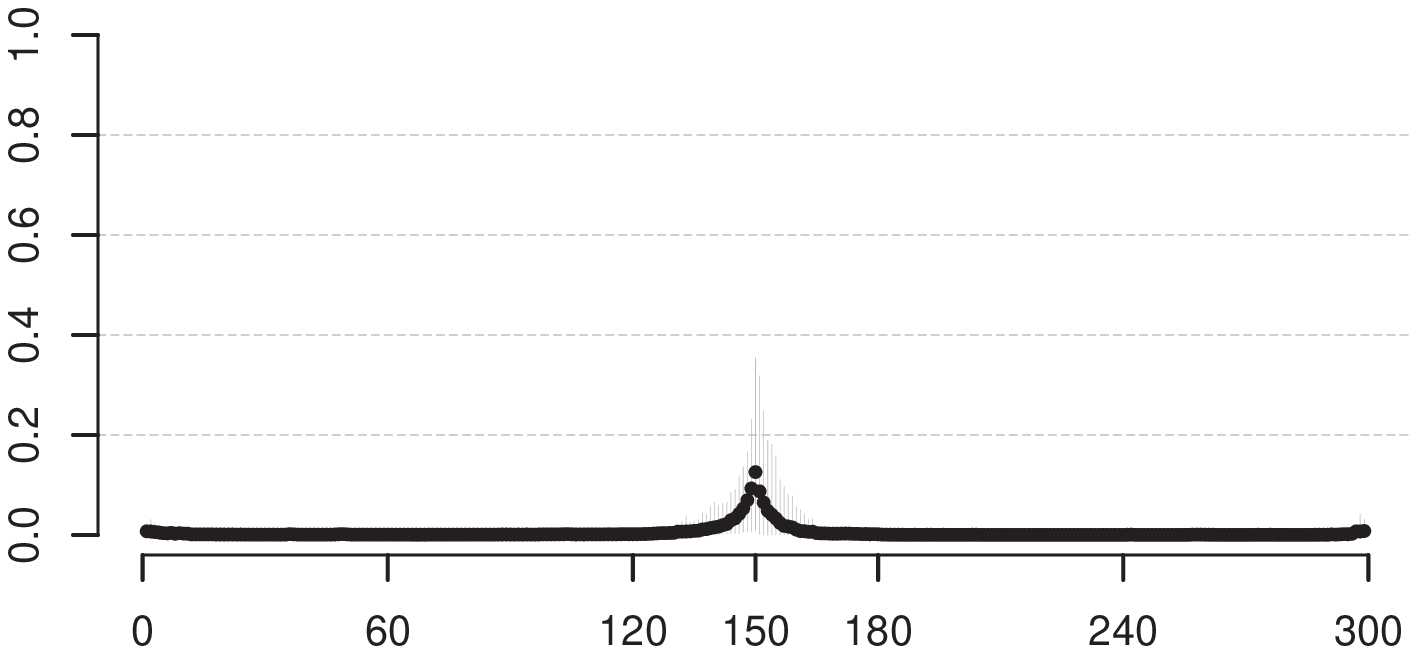}
	\label{fig:scene3_prob_IC_LP20_s2}}
\subfigure[][DPM19 ($E_2$)]{
	\includegraphics[width=6cm, height=4cm, trim=0 2.5cm 0 0]{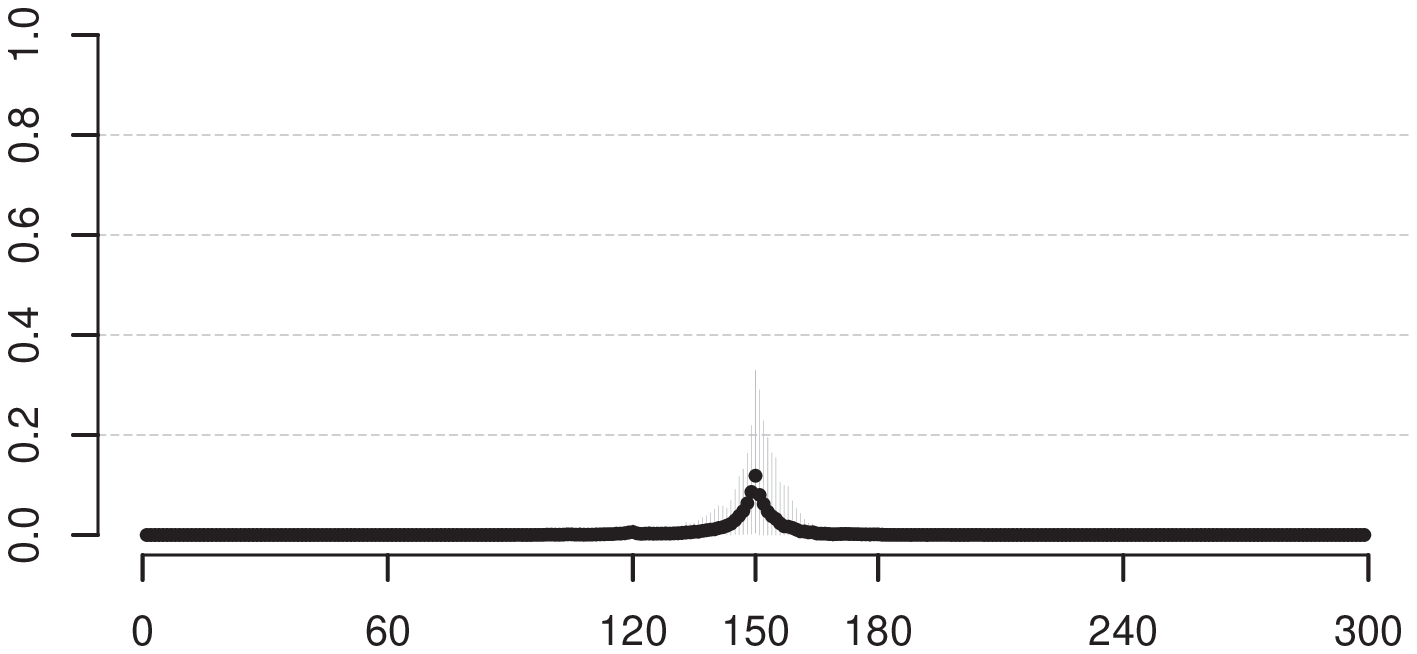}
	\label{fig:scene3_prob_IC_P18_s2}}\\[-.35in]
\subfigure[][LCIA05 ($\rho$)]{
	\includegraphics[width=6cm, height=4cm, trim=0 2.5cm 0 0]{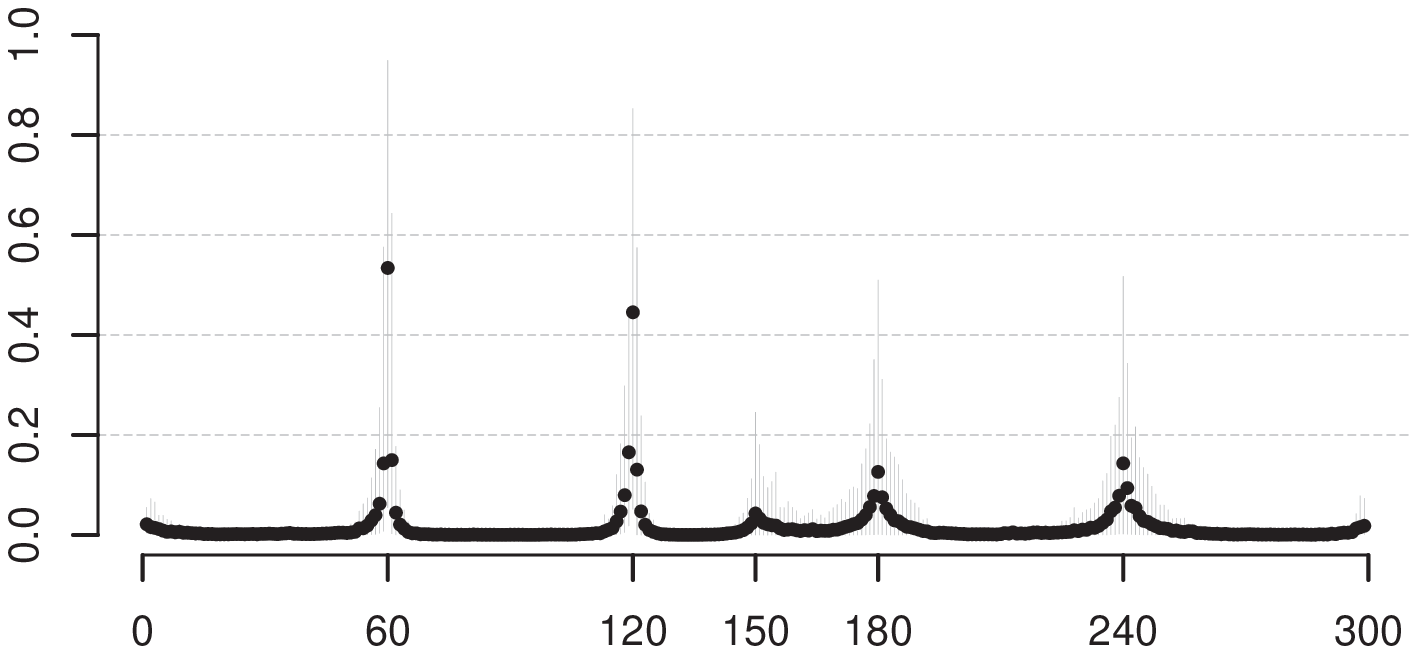}
	\label{fig:scene3_prob_IC_LC02}}
\subfigure[][BH93 ($\rho$)]{
	\includegraphics[width=6cm, height=4cm, trim=0 2.5cm 0 0]{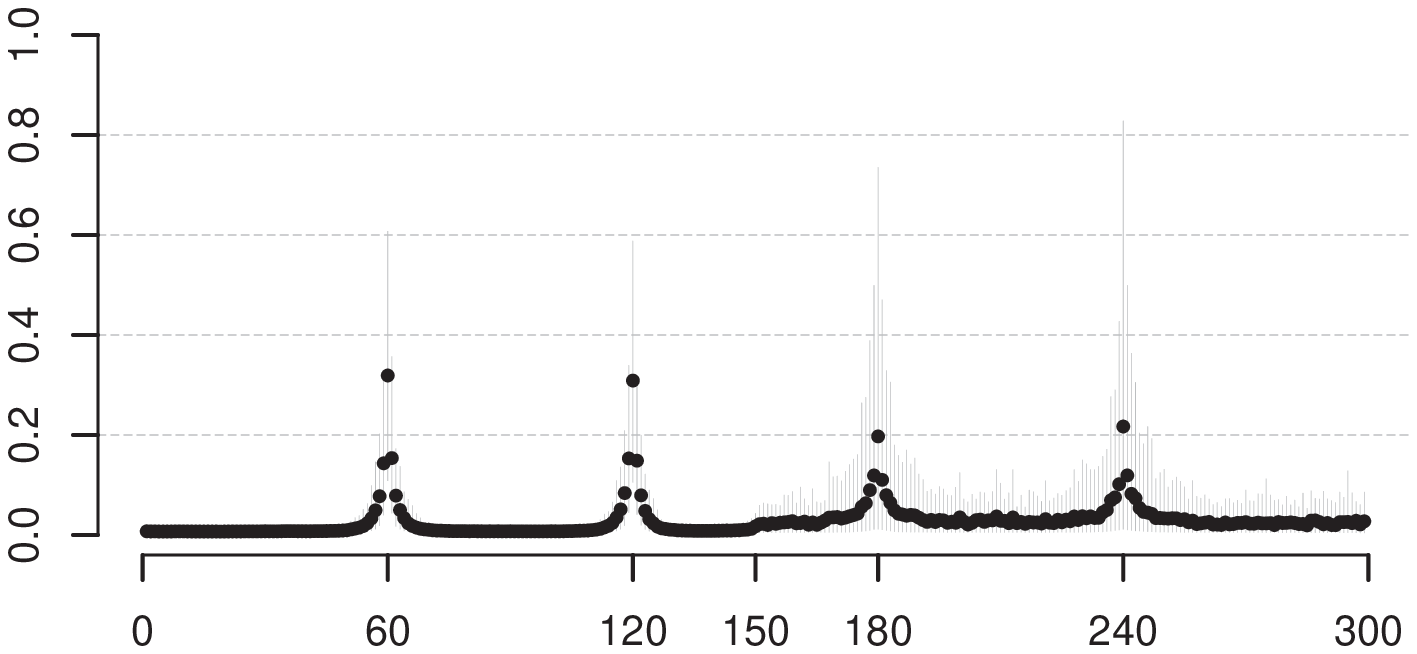}
	\label{fig:scene3_prob_IC_BH93}}
%\end{adjustwidth}
\caption{Average of the posterior probabilities of each instant to be an end point (black dots) for each partition and the $5\%$ and $95\%$ quantiles range of such probabilities based on the Monte Carlo replications, for models BMCP (a,c), DPM19 (b,d), LCIA05 (e) and BH93 (f) for Scenario 3.} 
\label{fig:scene3_prob_IC}
\end{figure}

As in Scenario 2, the average of posterior probabilities at the true change points are greater than at other instants (see Figure \ref{fig:scene3_prob_IC}). BMCP and DPM19 models indicate, with similar accuracy, the instants at which the changes took place in the mean as well as when they occurred for the variance. Unlike what is observed under the other models, the probabilities provided by the BH93 model tend to be greater than zero for all instants after the change in the variance.

In summary, the results in this section show that the BMCP model achieves the goal of identifying changes in the mean and variance separately. It provided better results than LCIA05 and BH93 in the case of changes in the mean and variance occurring at different instants and it also had a very good performance in those scenarios where only one structural parameter has changed along the sequence.

When compared to DPM19, the BMCP model provides highly competitive results. Considering that the DPM19  requires that the maximum number of change points in each parameter should be previously specified, which is not required for the BMCP model, together with the fact that we set these
values equal to the true number of changes, favoring the DPM19, the results show that BMCP is a competitive model for change point analysis. BMCP provides very accurate estimates for the structural parameters as well as for the number and positions of the change points, effectively
overcoming the DMP19 performance in several of the scenarios explored here.
Additional simulation study results evaluating sensitivity of the BMCP and DPM19 models to hyperparameter specifications can be found in the accompanying supplementary material file.

%Additional simulation study results evaluating sensitivity of the BMCP and DPM19 models to hyperparameter specifications can be found in the accompanying supplementary material file.

\FloatBarrier
\section{Case studies}\label{secReal}
In this section, we illustrate the use of the proposed model analyzing datasets from two fields where change point analysis plays an important role: financial (Section \ref{secReal_IR}) and genetic (Section \ref{secReal_GC}) datasets.

\subsection{Case study 1: US ex-post real interest rate:}\label{secReal_IR}

\begin{figure}[!htbp]
	\centering
	\includegraphics[width=9cm, height=4cm, trim=0 .5cm 0 0]{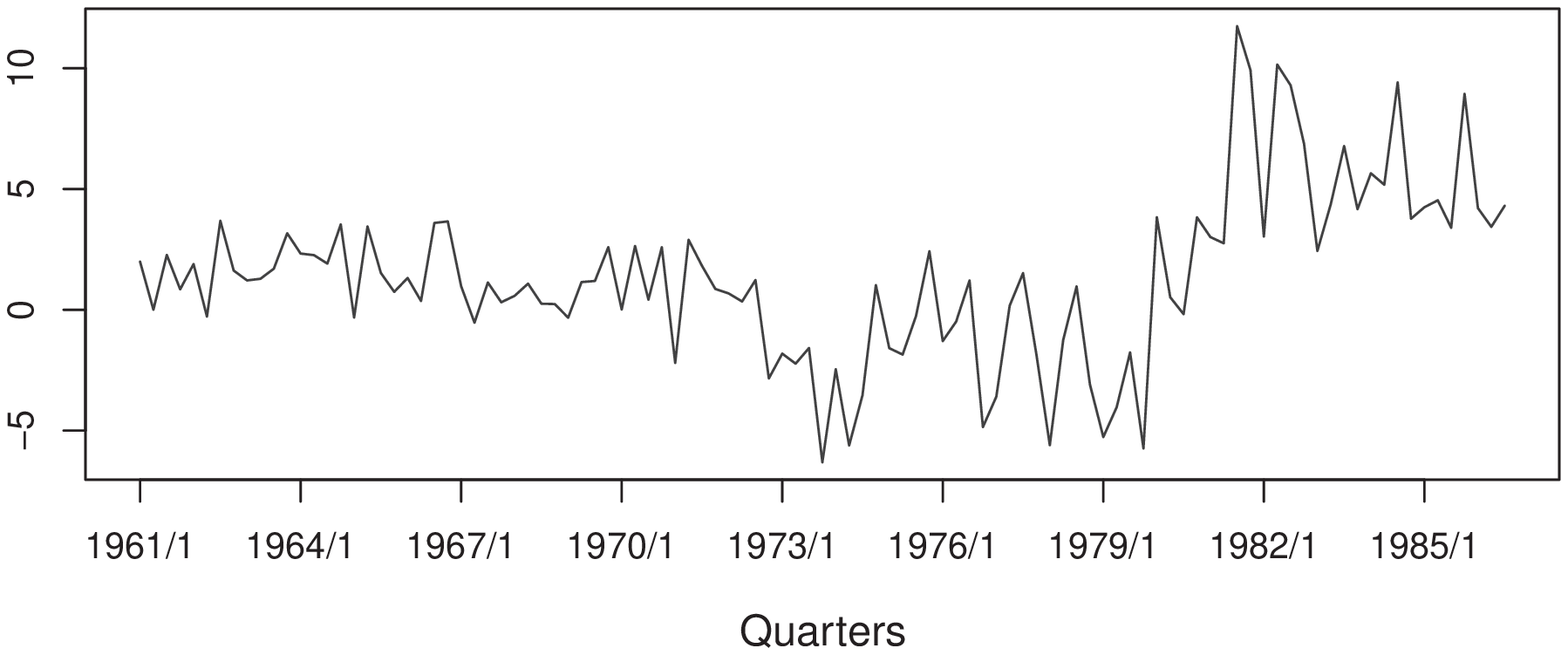}
	\caption{US ex-post real interest rate on a quarterly basis, from 1961/1 to 1986/3.}
	\label{fig:IR_data}
\end{figure}

We apply BMCP, DPM19, LCIA05 and BH93 to analyze the time series of the US ex-post real interest rate, available in the \texttt{R} package \texttt{bcp} \citep{erdman2007}. The data, displayed in Figure \ref{fig:IR_data}, correspond to the sequence of $n=103$ quaterly treasury bill rate deflated by the Consumer Price Index (CPI) inflation rate, denoted by $\bm{X}$, from the first quarter of 1961 to the third quarter of 1986. The presence of regime changes in this data was previously analyzed by \cite{garcia1996} and \cite{bai2003}.

We assume that at quarterly $i$, $X_i\mid\mu_i,\sigma_i^2 \sim N(\mu_i,\sigma_i^2)$, $i=1,\dots,103$. For all models, the prior specifications are the ones considered in the simulation study. To fit DPM19 we fixed the maximum number of changes for both parameters in $m_1=m_2=10$. This number was defined based on the results reported by \cite{garcia1996} and \cite{bai2003}, which indicate up to three changes in the mean and one change in the variance. 
{ Different choices for $m_1$ and $m_2$ are considered (see the supplementary material)  shown that DPM19 is truly sensitive to the specifications of $m_1$ and $m_2$. For this dataset, by assuming  $m_1$ and $m_2$ around the time series size, DPM19 showed to be ineffective for identifying possible changes pointing out most instants as change points with high probability.}
The MCMC procedures took $26$, $715$, $11$ and $14$ seconds to run $50,000$ iterations under the BMCP, DPM19, LCIA05 and BH93 models, respectively. We discarded the first $30,000$ iterations as the warm-up period.

\begin{figure}[!htbp]
    %\begin{adjustwidth}{-.4cm}{-.4cm}
		\centering
		\subfigure[][BMCP]{
			\includegraphics[width=3.5cm, height=3cm, trim=0 .5cm 0 0]{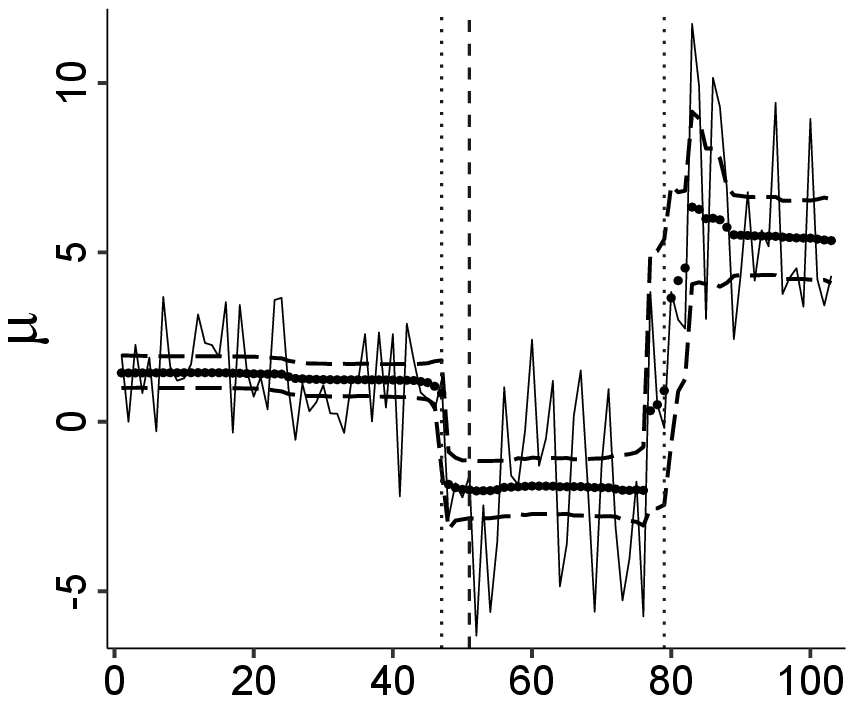}
			\label{fig:IR_mu_LP20}}
		\subfigure[][DPM19]{
			\includegraphics[width=3.5cm, height=3cm, trim=0 .5cm 0 0]{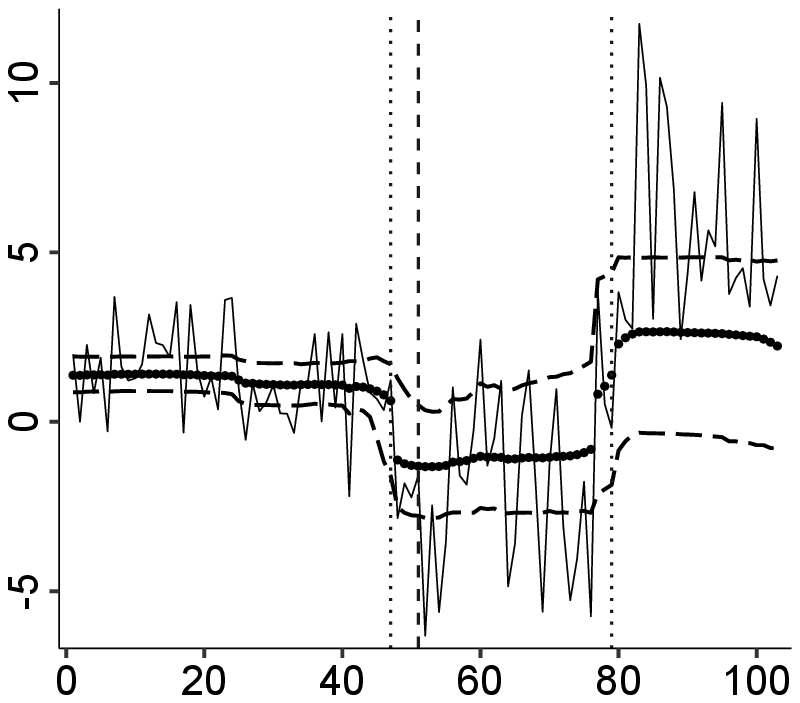}
			\label{fig:IR_mu_P18}}
		\subfigure[][LCIA05]{
			\includegraphics[width=3.5cm, height=3cm, trim=0 .5cm 0 0]{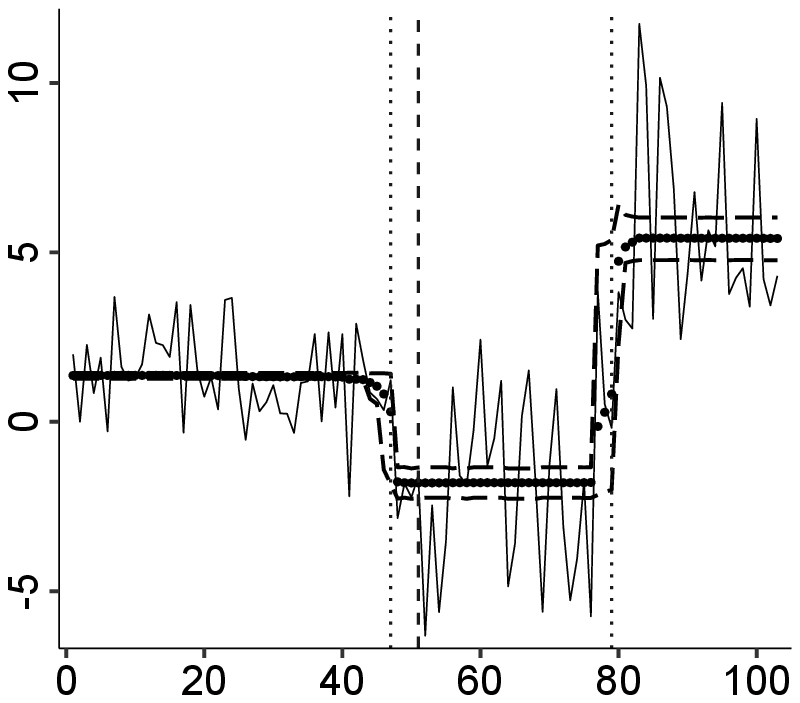}
			\label{fig:IR_mu_LC02}}%\hspace{-.1in}
		\subfigure[][BH93]{
			\includegraphics[width=3.5cm, height=3cm, trim=0 .5cm 0 0]{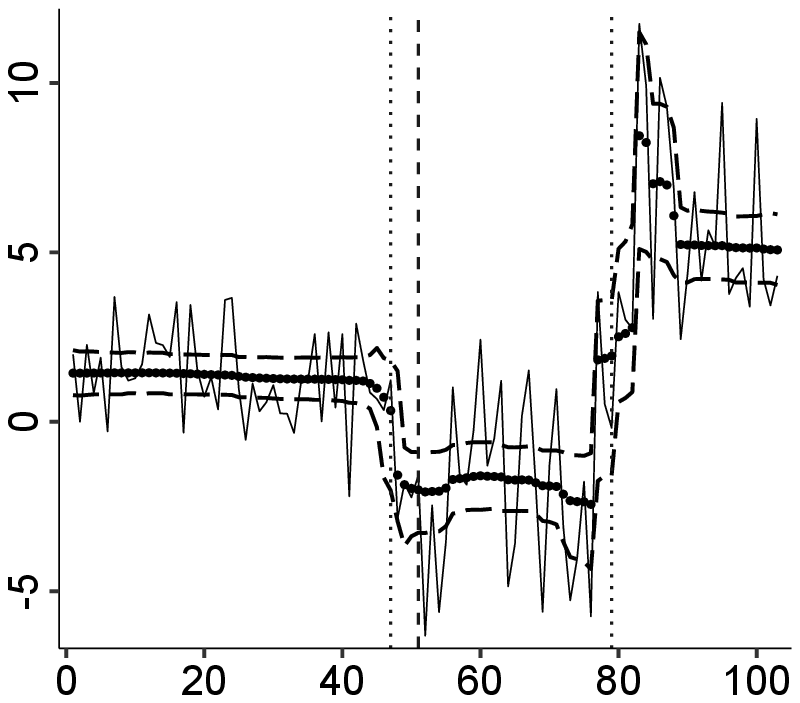}
			\label{fig:IR_mu_BH93}}
		\\
		\subfigure[][BMCP]{
			\includegraphics[width=3.5cm, height=3cm, trim=0 .5cm 0 0]{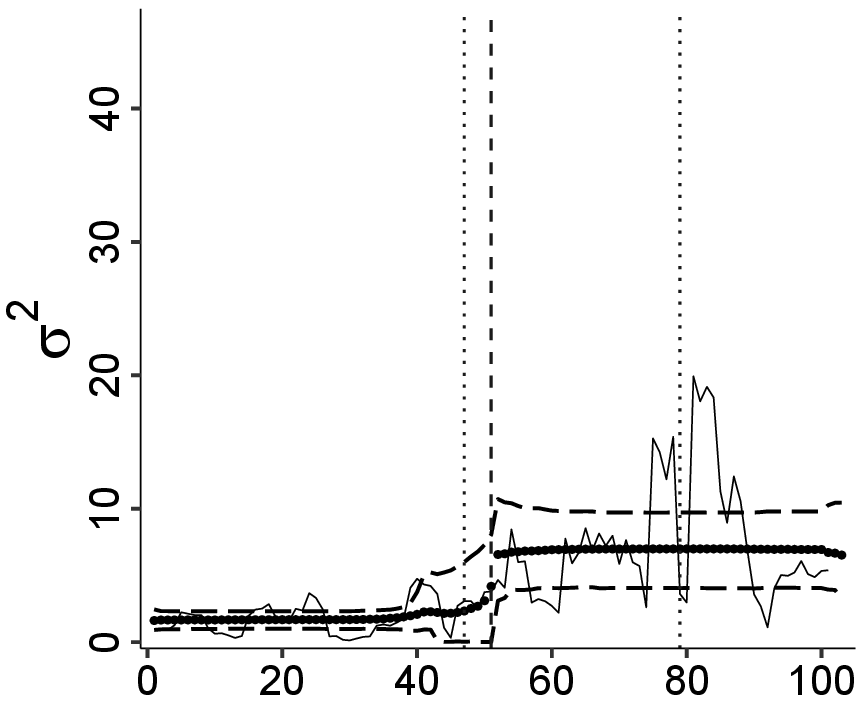}
			\label{fig:IR_s2_LP20}}%\hspace{-.1in}
		\subfigure[][DPM19]{
			\includegraphics[width=3.5cm, height=3cm, trim=0 .5cm 0 0]{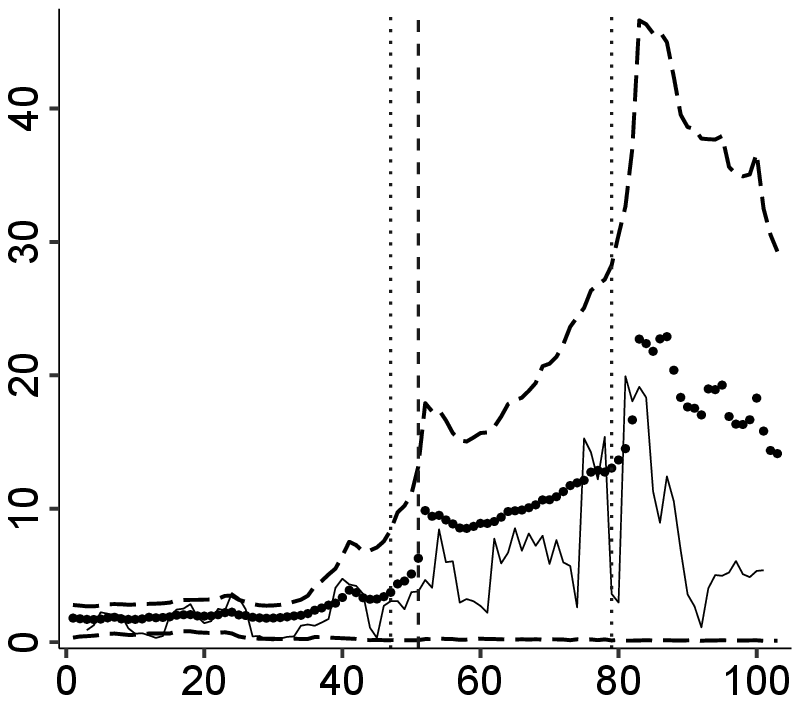}
			\label{fig:IR_s2_P18}}%\hspace{-.1in}
		\subfigure[][LCIA05]{
			\includegraphics[width=3.5cm, height=3cm, trim=0 .5cm 0 0]{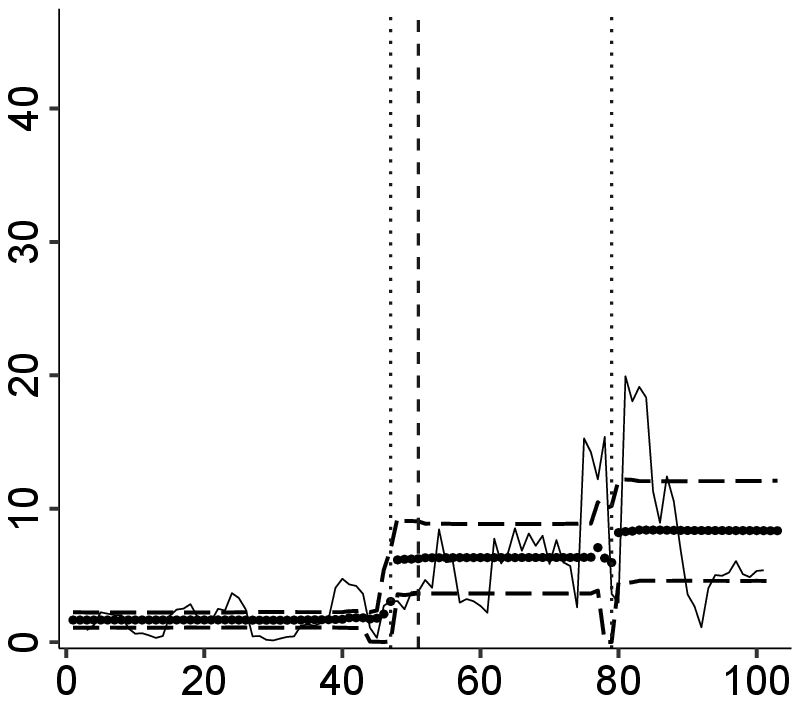}
			\label{fig:IR_s2_LC02}}%\hspace{-.1in}
		\subfigure[][BH93]{
			\includegraphics[width=3.5cm, height=3cm, trim=0 .5cm 0 0]{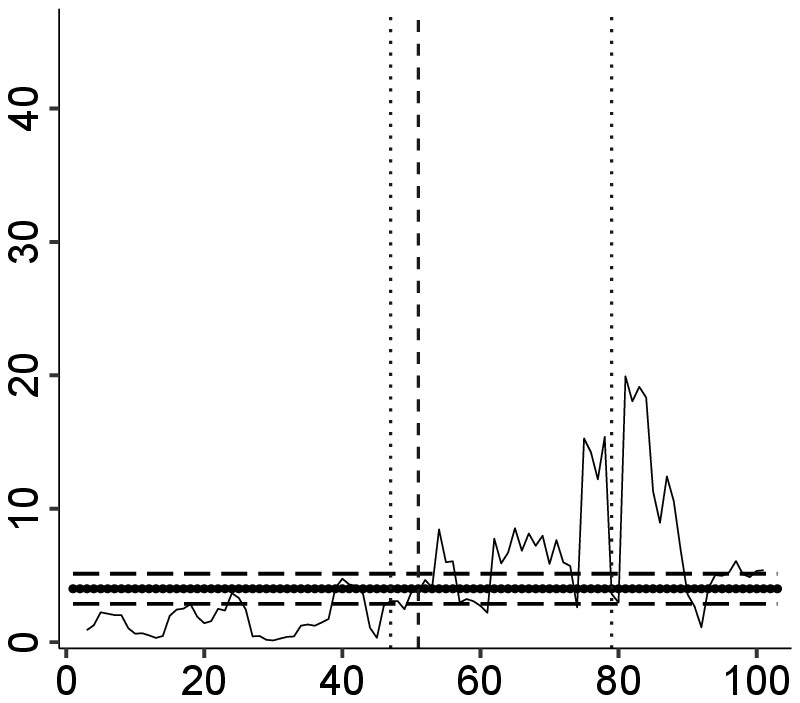}
			\label{fig:IR_s2_BH93}}
    %\end{adjustwidth}
\caption{Parameter estimates (black dots) and $90\%$ highest posterior density intervals (dashed lines) for the means (top) and variances (bottom) under BMCP (a,e), DPM19 (b,f), LCIA05 (c,g) and BH93 (d,h) models for the US ex-post Real Interest Rate dataset. The black solid lines represent the observed data (top) and the moving sample variance calculated over ranges of length 5 (bottom). The vertical dotted and dashed lines indicate the changes in $\bm{\mu}$ and $\bm{\sigma}$, respectively, according to the most likely partitions $\rho_1$ and $\rho_2$ as estimated by the BMCP model.}
\label{fig:IR_PE}
\end{figure}

Figure \ref{fig:IR_PE} shows the parameter estimates for the means (a,c,e,g) and variances (b,d,f,h) along the time. The vertical lines indicate the posterior modes of $\rho_1$ (dotted line) and $\rho_2$ (dashed line) under the BMCP model. All models provided similar point estimates for the means, except the DPM19 for which posterior means after instant $79$ were below the data. Besides, under DPM19, there is more posterior uncertainty about the means as the HPD intervals have a broad range. All models indicate strong changes in the mean occurring around the $47$ and $79$ quarters. The product estimates for the means under the BH93 model are less smooth after the $51st$ quarter, the instant at which both models, BMCP and DPM19 detected a change point in the variance. A similar behavior for the mean estimates under BH93 was observed in  clusters with higher variance in Scenarios 2 and 3 in our simulation study. The estimates for the variance under the BMCP, DPM19 and LCIA05 indicate a change around quarter $51$. The LCIA05 model also indicates that the variance changes after the second mean change. Under this model, the product estimates for the variance are affected by the two changes in the mean, similar to what is observed in the simulation study for Scenario 3 (Section \ref{sec_scene3}), where the mean and the variance change at different times. As for the posterior estimates of the means, DPM19 also presents the highest posterior uncertainty for the variance estimates.

\begin{figure}[!htbp]
\centering
\subfigure[][BMCP ($\rho_1$)]{
	\includegraphics[width=4cm, height=3cm,trim=0 2cm 0 0]{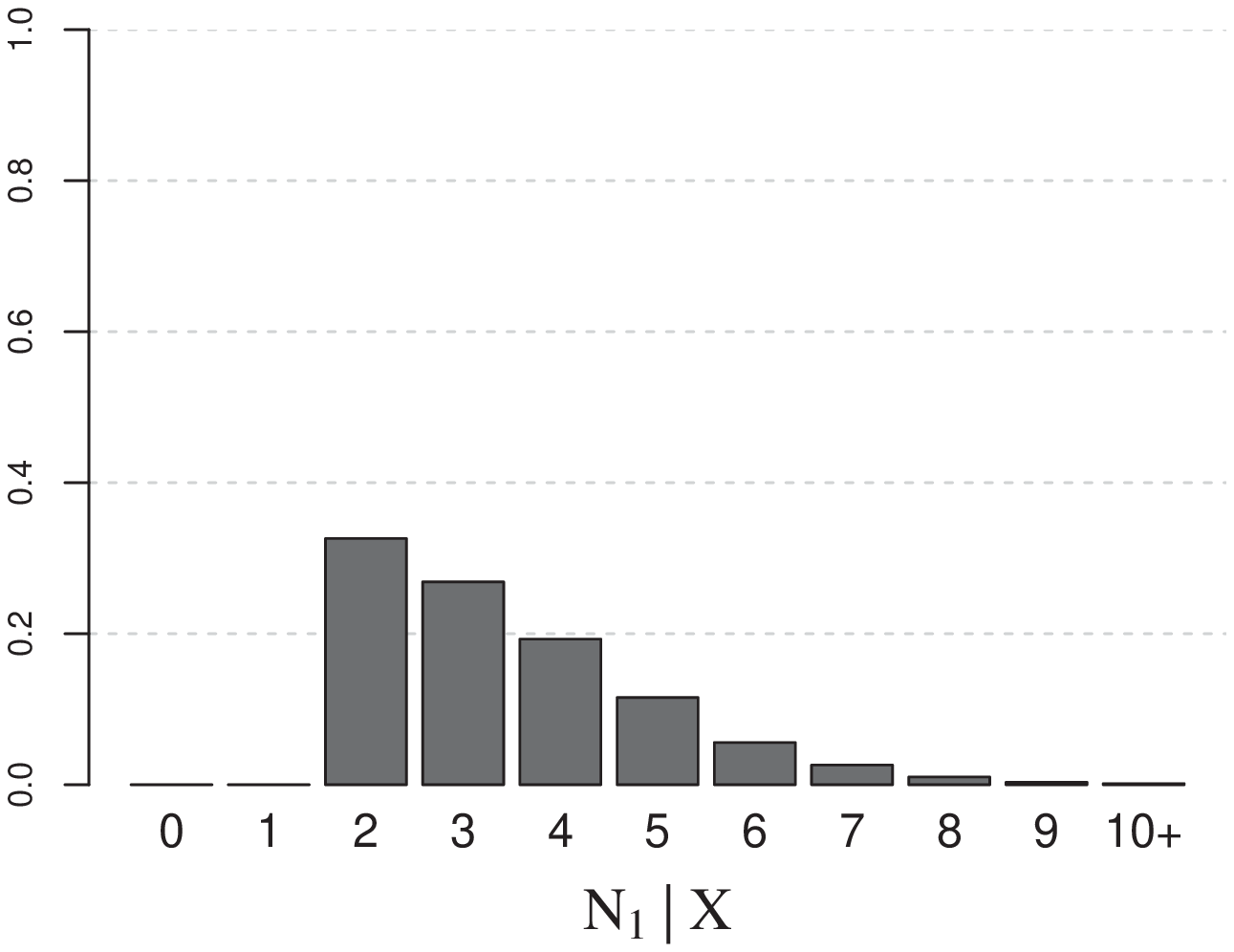}
	\label{fig:IR_N_LP20_mu}}\hspace{-.2in}
\subfigure[][BMCP ($\rho_2$)]{
	\includegraphics[width=4cm, height=3cm, trim=0 2cm 0 0]{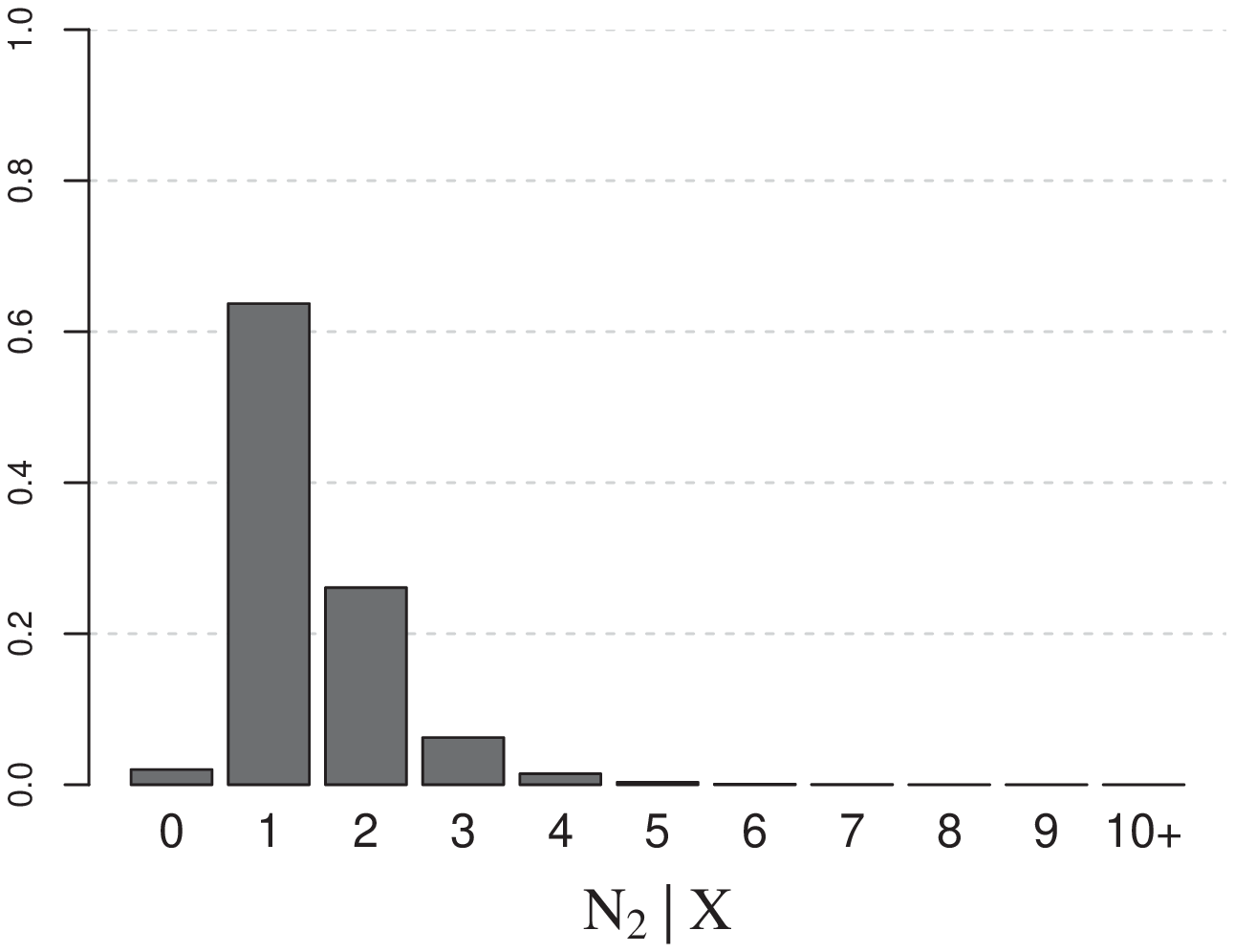}
	\label{fig:IR_N_LP20_s2}}\hspace{-.2in}
\subfigure[][LCIA05 ($\rho$)]{
	\includegraphics[width=4cm, height=3cm,trim=0 2cm 0 0]{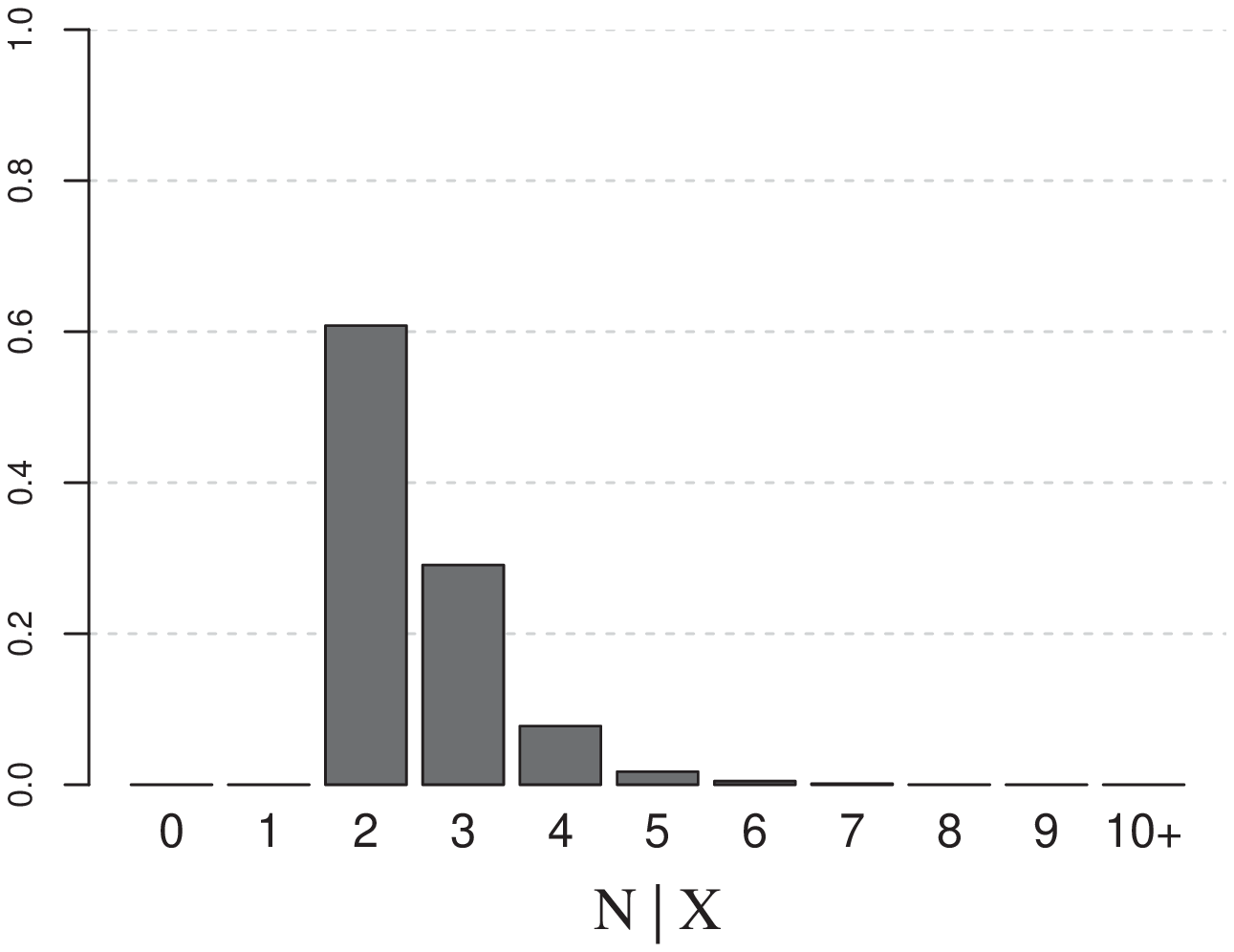}
	\label{fig:IR_N_LC02}}\hspace{-.2in}\\
\subfigure[][DPM19 ($E_1$)]{
	\includegraphics[width=4cm, height=3cm,trim=0 2cm 0 0]{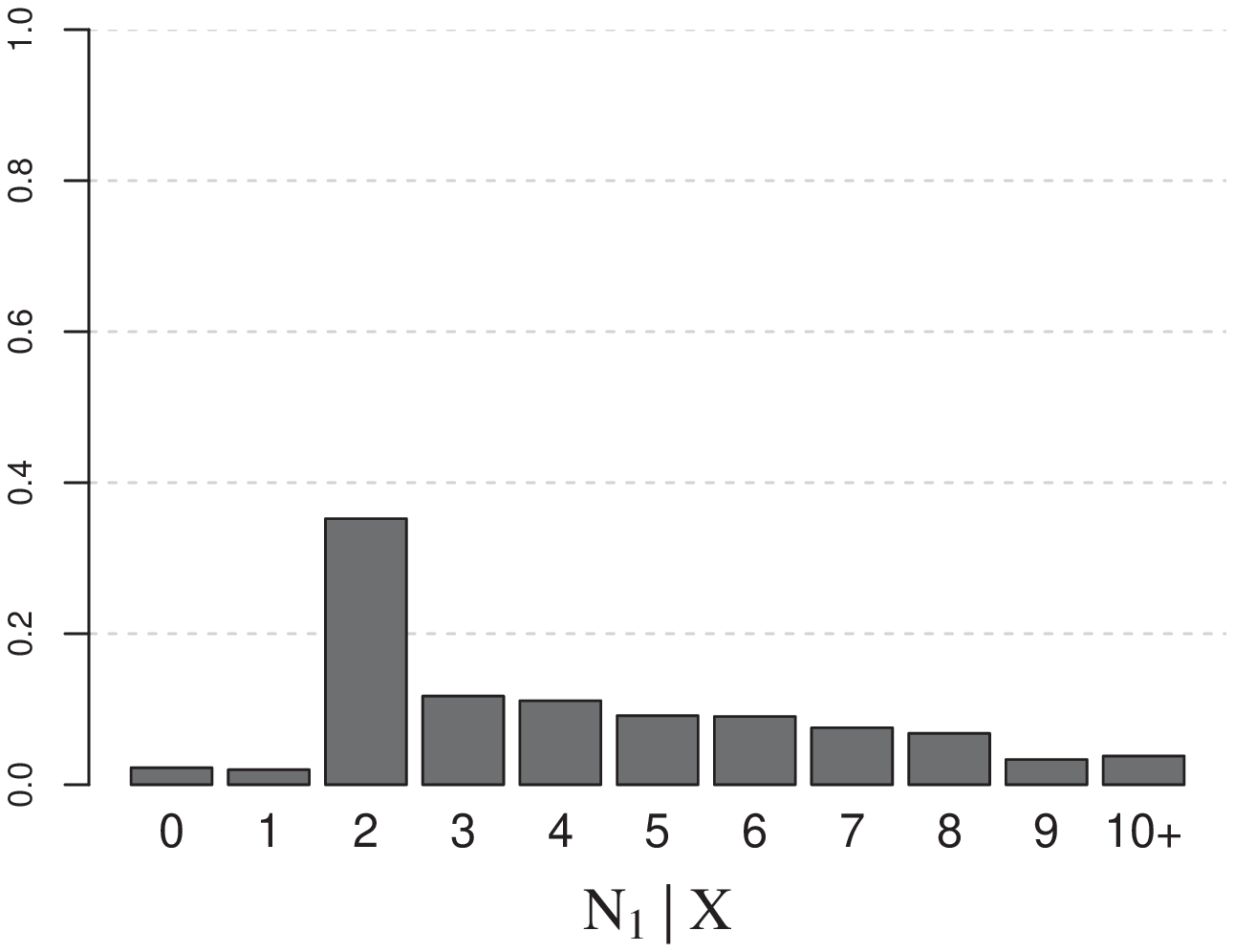}
	\label{fig:IR_N_P18_mu}}\hspace{-.2in}
\subfigure[][DPM19 ($E_2$)]{
	\includegraphics[width=4cm, height=3cm, trim=0 2cm 0 0]{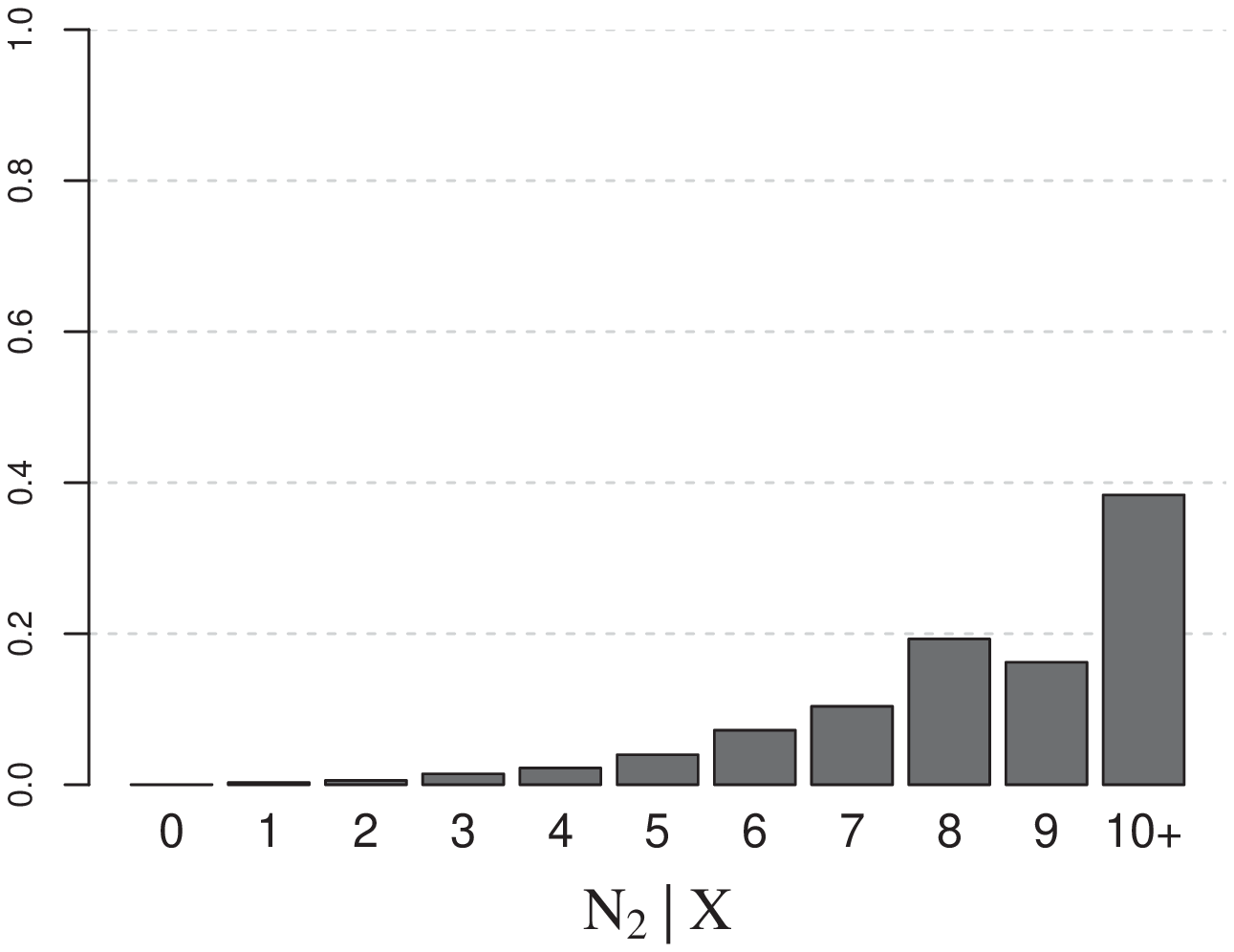}
	\label{fig:IR_N_P18_s2}}\hspace{-.2in}
\subfigure[][BH93 ($\rho$)]{
	\includegraphics[width=4cm, height=3cm,trim=0 2cm 0 0]{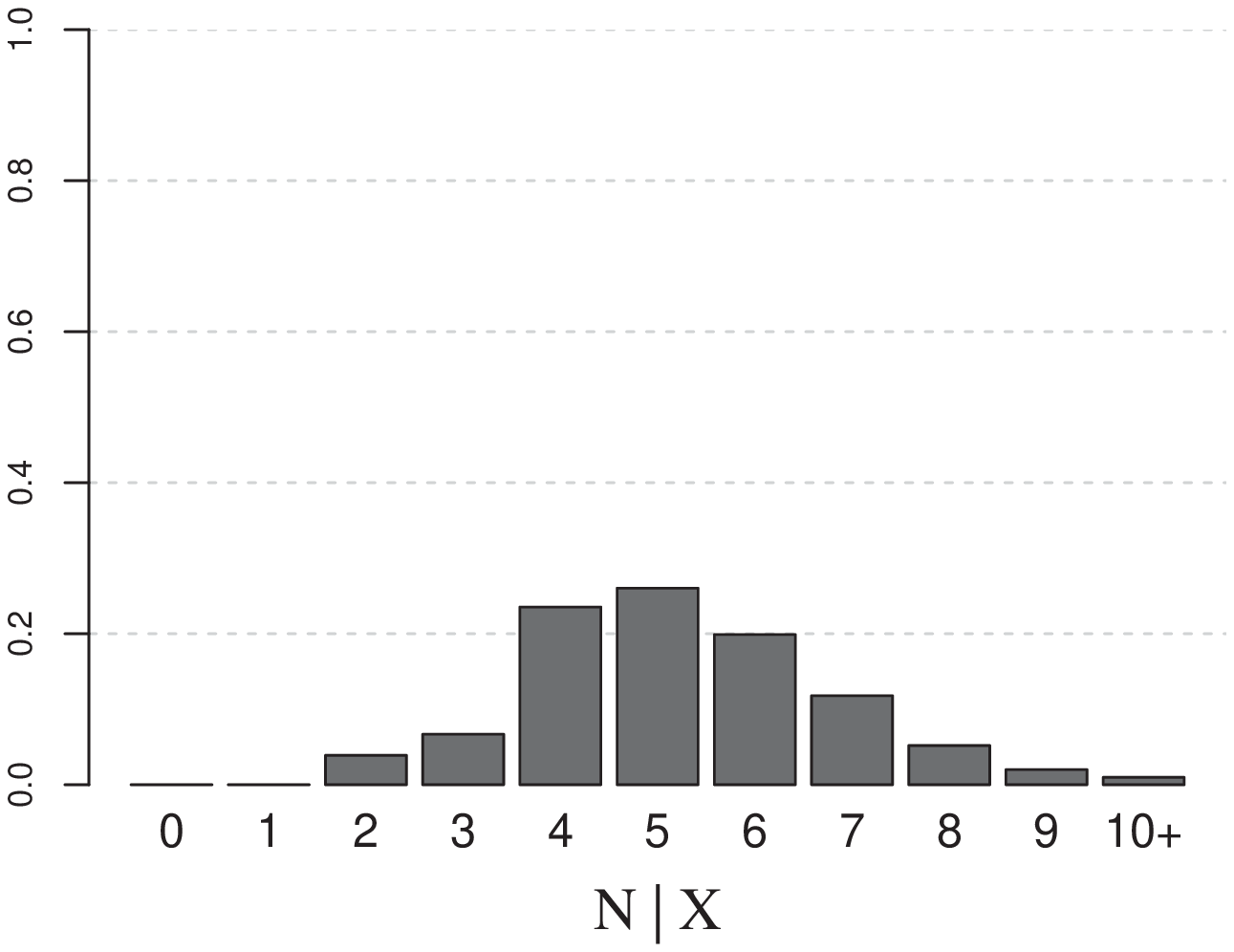}
	\label{fig:IR_N_BH93}}
\caption{Posterior distribution for the number of changes, BMCP (a,b), LCIA05 (c), DPM19 (d,e) and BH93 (f), Case study 1.}
\label{fig:IR_N}
%\end{figure}
%
%\begin{figure}[!htbp]
%\begin{adjustwidth}{-.4cm}{-.4cm}
\centering		
\subfigure[][BMCP ($\rho_1$)]{
	\includegraphics[width=4cm, height=3.0cm, trim=0 2cm 0 0]{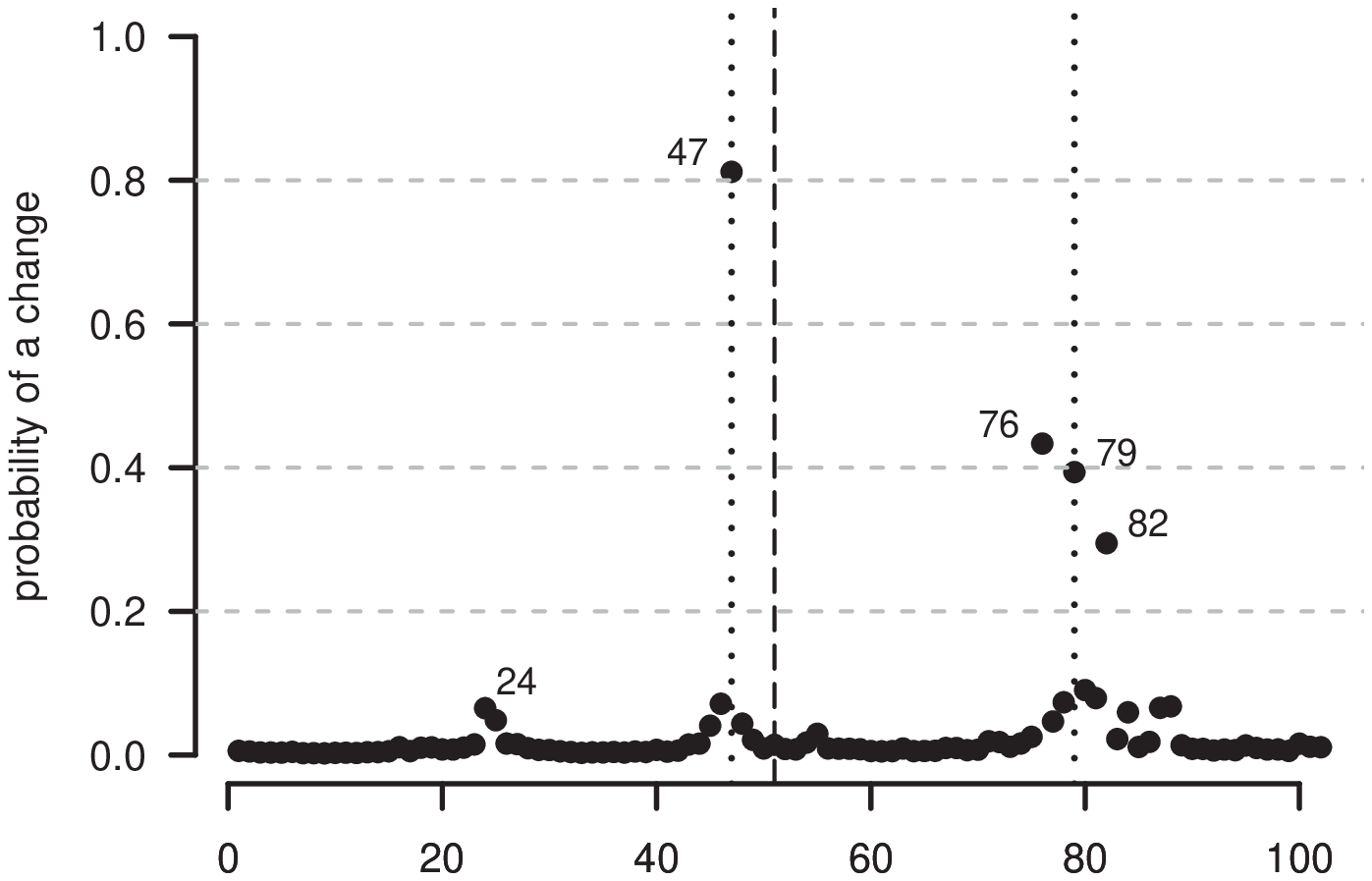}
	\label{fig:IR_prob_LP20_mu}}\hspace{-.2in}
\subfigure[][BMCP ($\rho_2$)]{
	\includegraphics[width=4cm, height=3.0cm, trim=0 2cm 0 0]{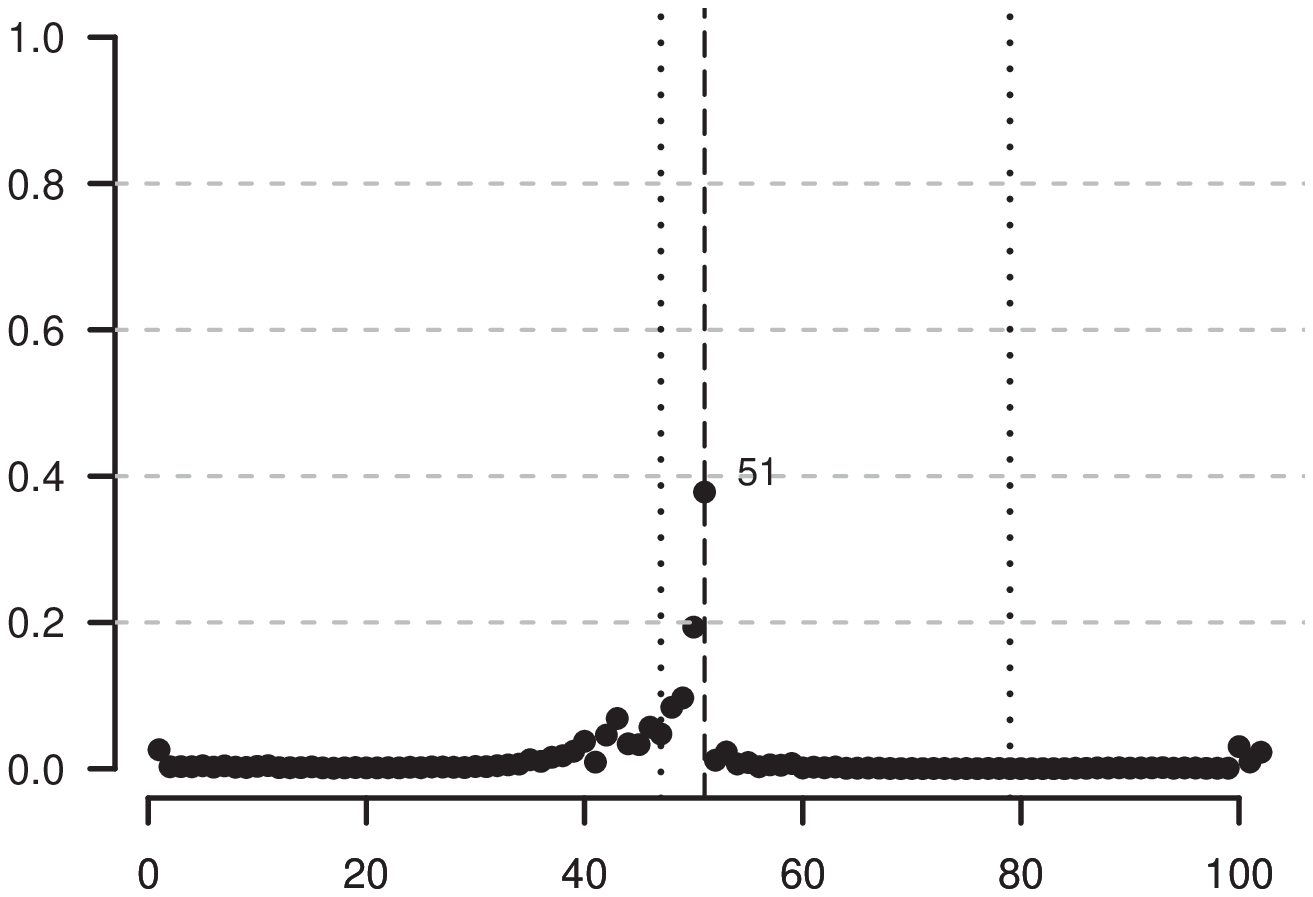}
	\label{fig:IR_prob_LP20_s2}}\hspace{-.2in}
\subfigure[][LCIA05 ($\rho$)]{
	\includegraphics[width=4cm, height=3.0cm, trim=0 2cm 0 0]{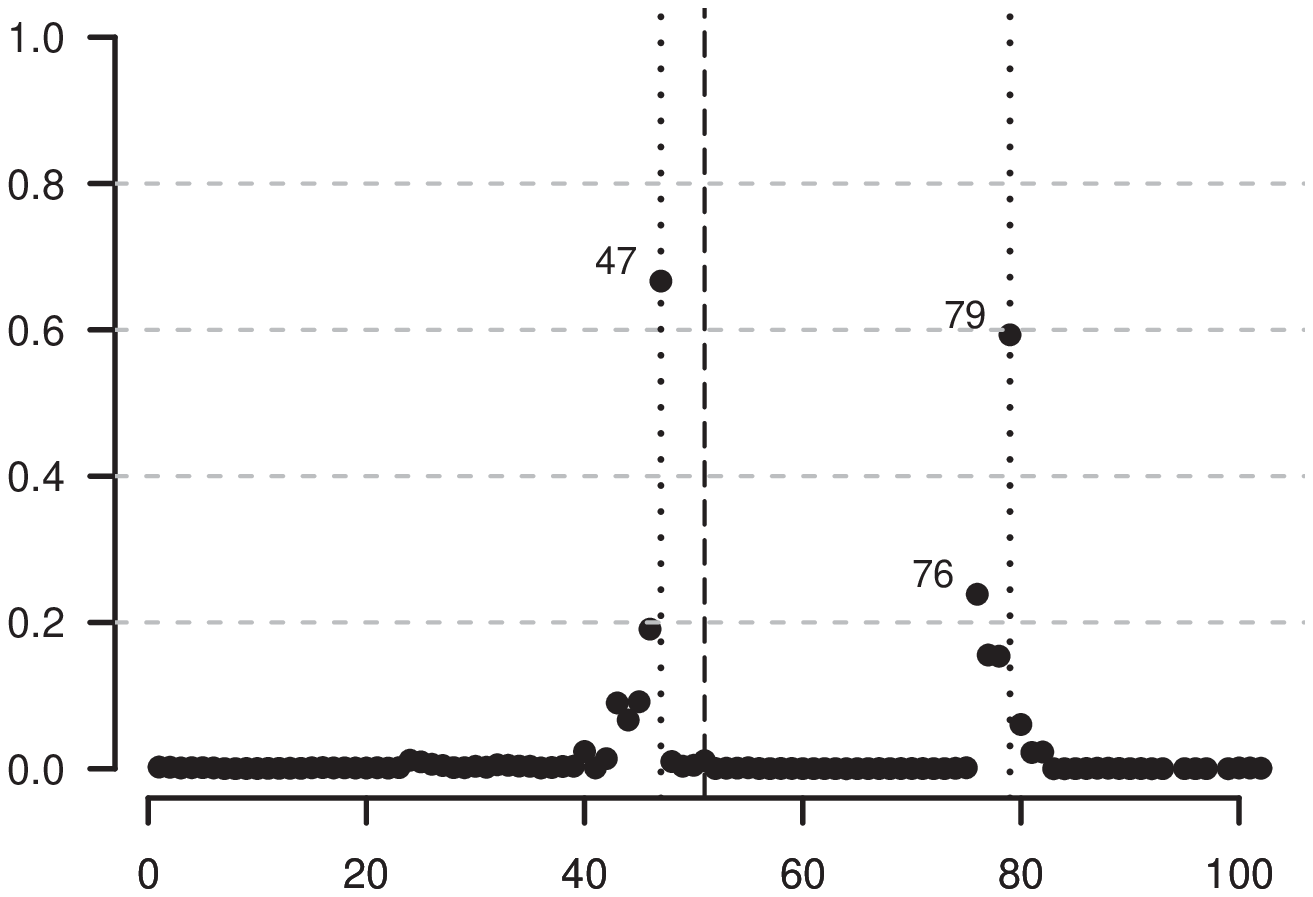}
	\label{fig:IR_prob_LC02}}\hspace{-.2in}\\
\subfigure[][DPM19 ($E_1$)]{
	\includegraphics[width=4cm, height=3.0cm, trim=0 2cm 0 0]{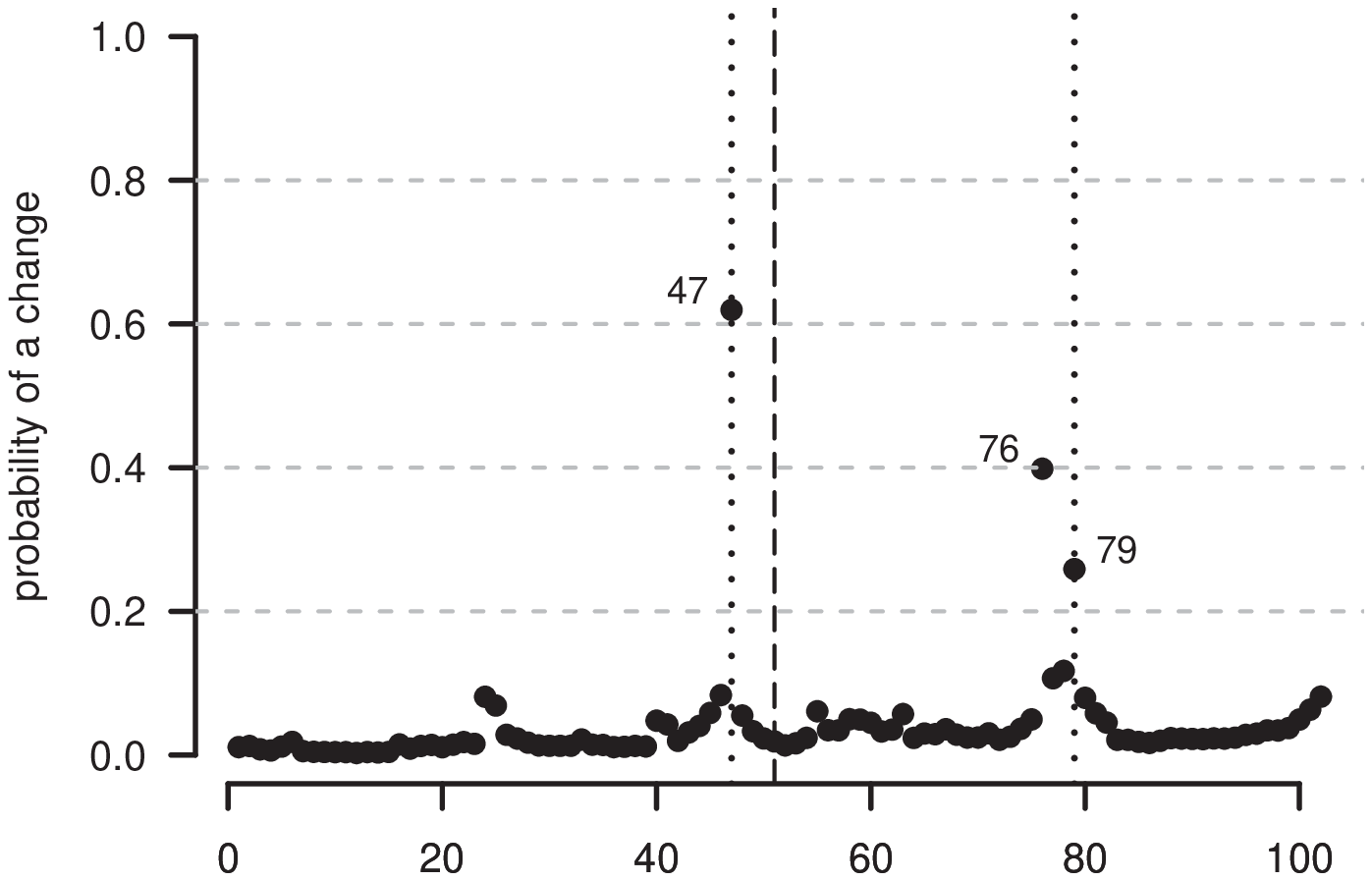}
	\label{fig:IR_prob_P18_mu}}\hspace{-.2in}
\subfigure[][DPM19 ($E_2$)]{
	\includegraphics[width=4cm, height=3.0cm, trim=0 2cm 0 0]{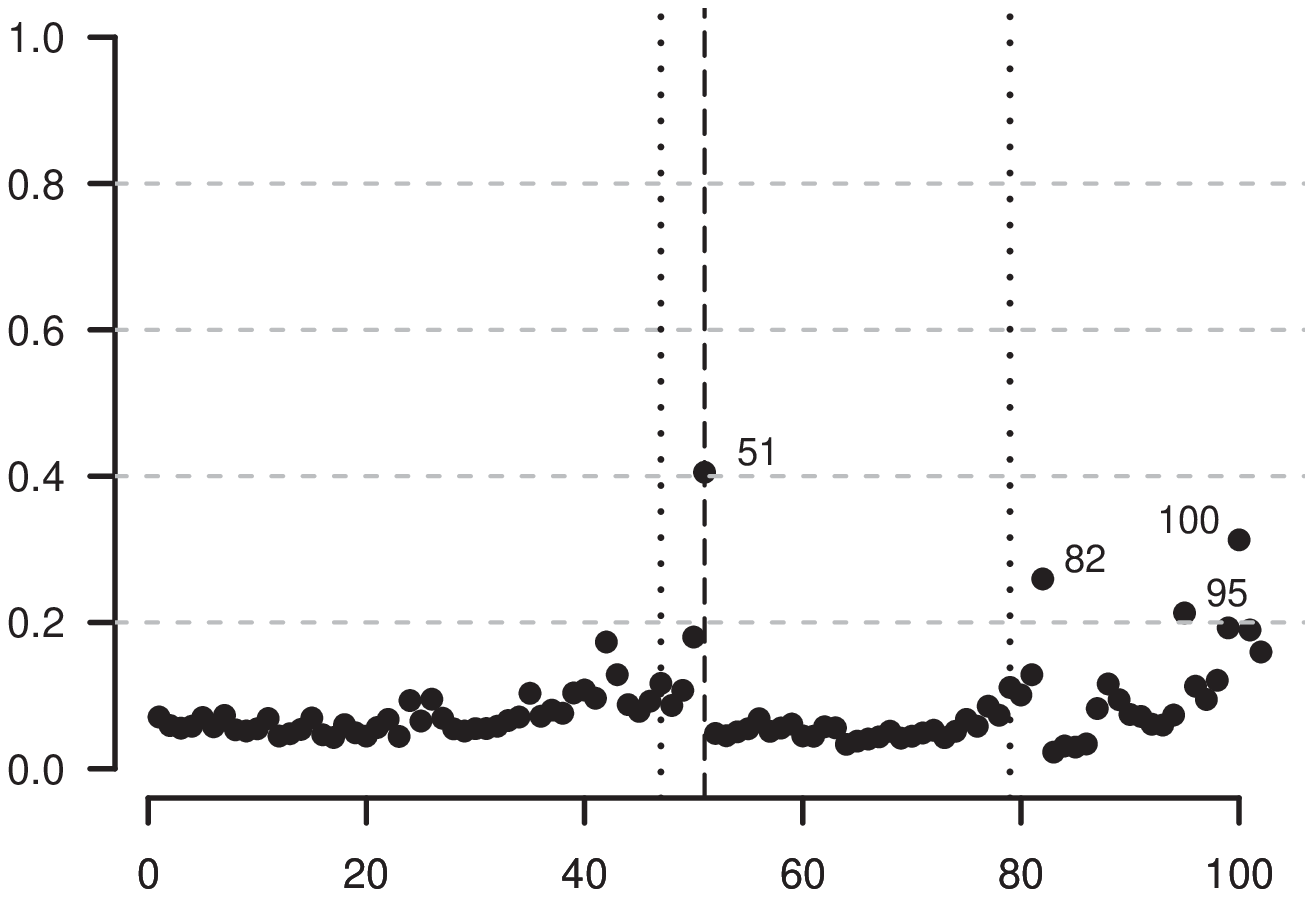}
	\label{fig:IR_prob_P18_s2}}\hspace{-.2in}
\subfigure[][BH93 ($\rho$)]{
	\includegraphics[width=4cm, height=3.0cm, trim=0 2cm 0 0]{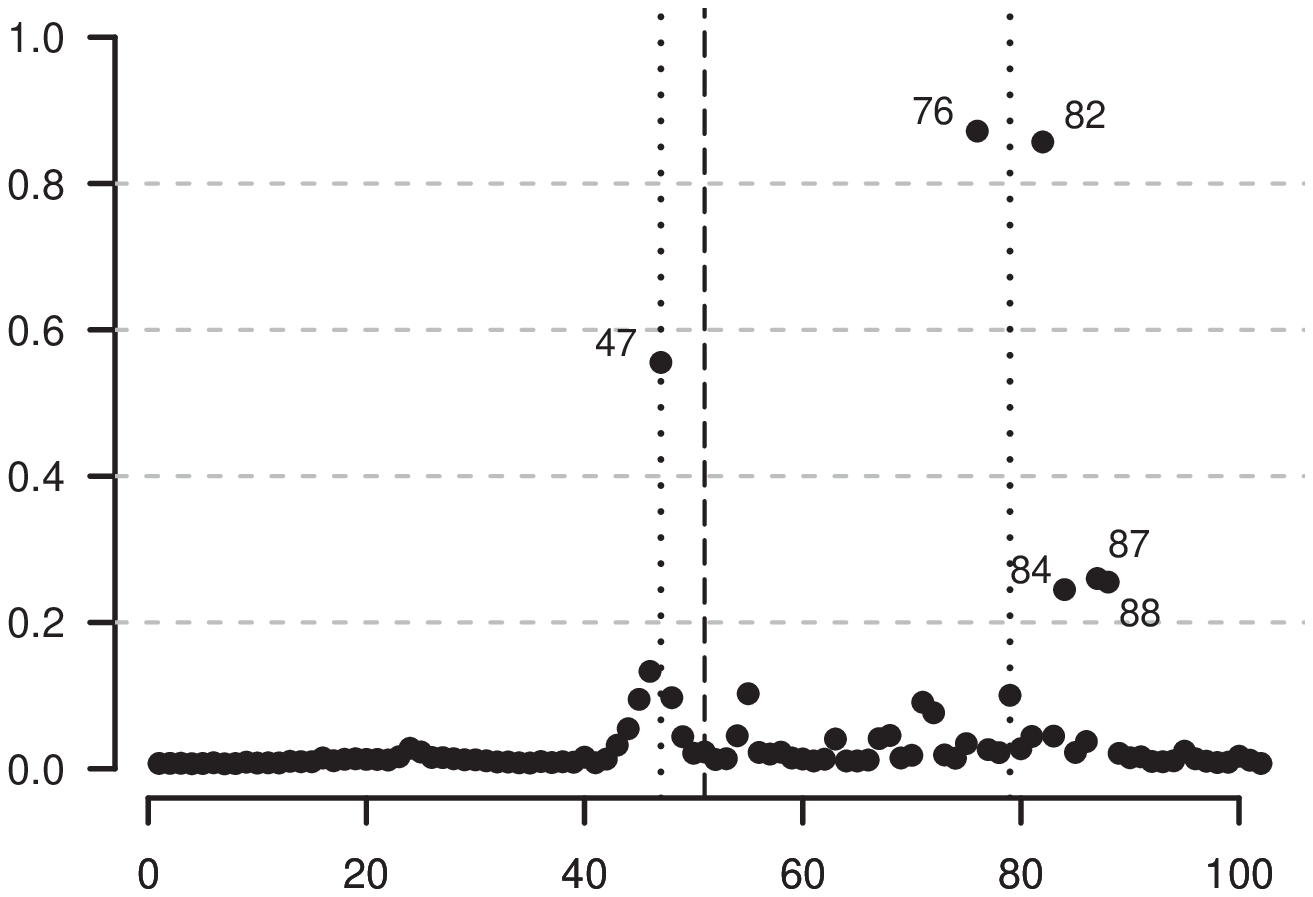}
	\label{fig:IR_prob_BH93}}
%\end{adjustwidth}
\caption{Posterior probabilities that each instant is an end point
(black bullets) under BMCP (a,b), LCIA05 (c), DPM19 (d,e), and BH93 (f), Case study 1. The vertical dotted and dashed lines are the
changes in $\bm{\mu}$ and $\bm{\sigma}$, respectively, given by the
most likely partitions $\rho_1$ and $\rho_2$ estimated by the BMCP.}
\label{fig:IR_prob}
\end{figure}

The posterior modes of the number of changes (Figure~\ref{fig:IR_N}) under BMCP and DPM19  indicated two changes in the mean. For the variance, these models provided strongly different estimates. While BMCP indicated only one change in the variance,  DPM19 indicated ten changes, which is the maximum  number of changes assumed {\it a priori} to fit this model. The BMCP and DPM19 indicate the quarters $47$, $76$ and $79$ as change points in the mean (Figure \ref{fig:IR_prob} (a,c)) and the quarter $51$ as a change point in the variance (Figure \ref{fig:IR_prob} (b,d)), with posterior probabilities much higher than for the other quarters. Although not identical, inference obtained from the posterior distributions for the partitions induce similar findings (see Table \ref{tab:IR}). Under the BMCP model, the posterior most likely partitions for the mean and the variance are $\rho_1=\{0,47,79,103\}$ and $\rho_2=\{0,51,103\}$, respectively. With higher posterior probabilities, the quarters $47$, $76$ and $79$ are also pointed out as change points by model LCIA05 (Figure \ref{fig:IR_prob_LC02}) and the quarters $47$, $76$ and $82$ are indicated as change points in the mean by model BH93 (Figure \ref{fig:IR_prob_BH93}). The posterior for $\rho$ under LCIA05 model only detects the changes that the proposed model indicates as change points in the mean (see Table \ref{tab:IR} and Figure \ref{fig:IR_prob_LC02}).

Estimates provided by BMCP and DPM19 are similar to the ones reported in \cite{garcia1996}. Considering a Markov switching model, \cite{garcia1996} concluded that a better fit for these data is obtained if three different means and two different variances are considered. Changes in the mean were identified in 1972/3 and 1980/1 that correspond to quarters $47$ and $77$.
They also found that the $2nd$ and $3rd$ clusters share the same variance, which is different from the variance in the $1st$ cluster. This finding is in agreement with results obtained fitting  BMCP and DPM19 models which pointed to a unique change in the variance right after the first change in the mean. Results presented by \cite{bai2003} differ from that given by BMCP and \cite{garcia1996}, who detected one more change point in the mean at quarter $24$.

\begin{table}[H]%[!htbp]
%\begin{adjustwidth}{-1cm}{}
\centering
\footnotesize
\begingroup
\renewcommand{\arraystretch}{.6} % Default value: 1
\begin{tabular}{p{0.25\textwidth}p{0.075\textwidth}}
	\clineB{1-2}{2} \\[-1.8ex]
	BMCP & $p(\rho_1|\bm{X})$ \\[-.5ex]
	\cmidrule(r{.15cm}l{.15cm}){1-2}
	$\{0,47,79,103\}$ & 0.1441 \\
	$\{0,47,76,103\}$ & 0.0602 \\
	\hline \\[-1.8ex]
\end{tabular}\hspace{.1in}
\begin{tabular}{p{0.25\textwidth}p{0.075\textwidth}}
	\clineB{1-2}{2} \\[-1.8ex]
	BMCP & $p(\rho_2|\bm{X})$ \\[-.5ex]
	\cmidrule(r{.15cm}l{.15cm}){1-2}
	$\{0,51,103\}$ & 0.2054 \\
	$\{0,50,103\}$ & 0.1038 \\
	\hline \\[-1.8ex]
\end{tabular}\\
\begin{tabular}{p{0.25\textwidth}p{0.075\textwidth}}
	\clineB{1-2}{2} \\[-1.8ex]
	DPM19 & $p(E_1|\bm{X})$ \\[-.5ex]
	\cmidrule(r{.15cm}l{.15cm}){1-2}
	$\{0,47,76,103\}$ & 0.1141 \\
	$\{0,47,79,103\}$ & 0.0656 \\
	\hline \\[-1.8ex]
\end{tabular}\hspace{.1in}
\begin{tabular}{p{0.25\textwidth}p{0.075\textwidth}}
	\clineB{1-2}{2} \\[-1.8ex]
	DPM19 & $p(E_2|\bm{X})$ \\[-.5ex]
	\cmidrule(r{.15cm}l{.15cm}){1-2}
	$\{0,82,103\}$ & 0.0005 \\
	$\{0,51,103\}$ & 0.0004 \\
	\hline \\[-1.8ex]
\end{tabular}\\
\begin{tabular}{p{0.25\textwidth}p{0.075\textwidth}}
	\clineB{1-2}{2} \\[-1.8ex]
	LCIA05 & $p(\rho|\bm{X})$ \\[-.5ex]
	\cmidrule(r{.15cm}l{.15cm}){1-2}
	$\{0,47,79,103\}$ & 0.2005 \\
	$\{0,47,76,103\}$ & 0.1262 \\
	\hline \\[-1.8ex]
\end{tabular}\hspace{.1in}
\begin{tabular}{p{0.25\textwidth}p{0.075\textwidth}}
	\clineB{1-2}{2} \\[-1.8ex]
	BH93 & $p(\rho|\bm{X})$ \\[-.5ex]
	\cmidrule(r{.15cm}l{.15cm}){1-2}
	$\{0,47,76,82,87,103\}$ & 0.0309 \\
	$\{0,47,76,82,88,103\}$ & 0.0296 \\
	\hline \\[-1.8ex]
\end{tabular}
\endgroup
\caption[Most likely partitions for case study 1]{Top most likely $\rho_1$ and $\rho_2$ (BMCP), $E_1$ and $E_2$ (DPM19) and $\rho$ (LCIA05, BH93) based on the posterior probabilities, for case study 1.}
\label{tab:IR}
%\end{adjustwidth}
\end{table}

\FloatBarrier
\subsection{Case study 2: guanine-cytosine content}\label{secReal_GC}

We reanalyze the guanine-cytosine dataset \texttt{HC1}, available in the \texttt{R} package \texttt{changepoint} \citep{killick2016} using BMCP. The analysis using DPM19, LCIA05, and BH93 may be found in the supplementary material. 

This dataset was originally accessed from the National Center for Biotechnology Information and represents a sequence of guanine-cytosine content in
contiguous equal size intervals over a specific range of the human chromosome 1.  Genome segmentation is a relevant strategy to identify connections between genomes and phenotypes \citep{algama2014}. In our analysis, we consider the first $n=2,000$ available observations.
This dataset was previously analyzed by \cite{jewell2021} assuming a change-in-mean model but with a common variance. We fit the proposed model considering {\it a priori} that   $(\mu_0,\sigma_0^2,a,d)=(0,10^6,0.02,0.02)$ and $p_1\overset{D}{=}p_2\sim Beta(1,1)$. The MCMC procedure took $5$ hours to generate $50,000$ samples of the joint posterior distribution of the parameters and partitions, after a discarded warm-up period of $50,000$ iterations. %The results for the other models are available in the supplementary material.

\begin{figure}[!htbp]
\centering		
\subfigure[][]{
	\includegraphics[width=5cm, height=4.5cm, trim=0 2cm 0 0]{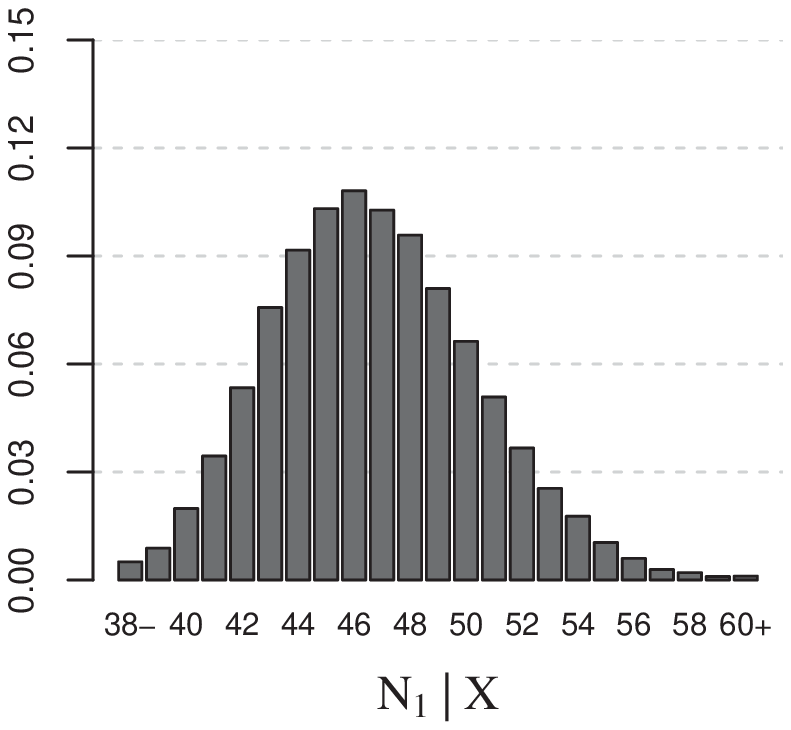}
	\label{fig:GC_N_LP20_mu}}\hspace{-.1in}
\subfigure[][]{
	\includegraphics[width=5cm, height=4.5cm, trim=0 2cm 0 0]{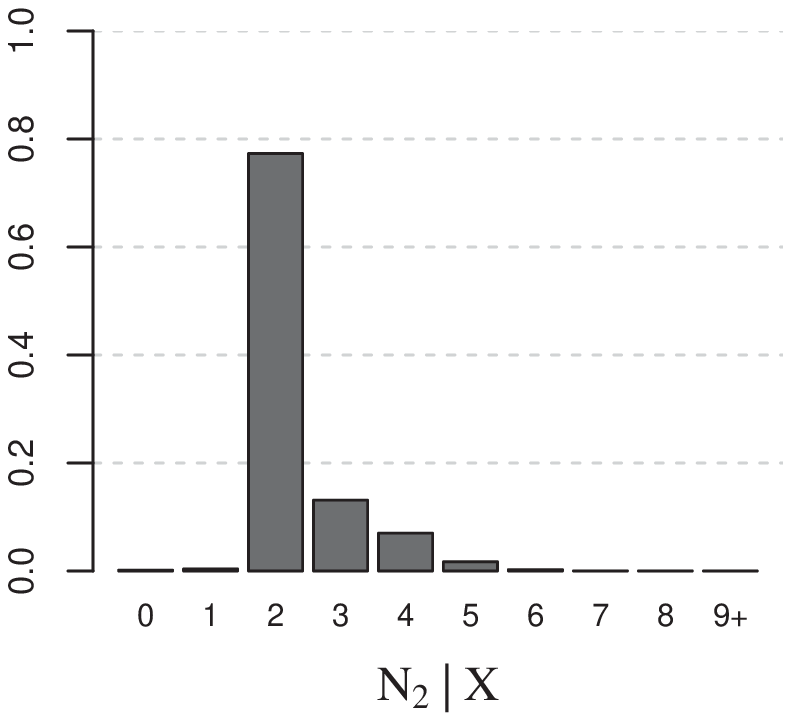}
	\label{fig:GC_N_LP20_s2}}
\vspace{-.2in}
\caption{Posterior distribution of the number of changes in $\bm{\mu}$ (a) and $\bm{\sigma}$ (b), for the \texttt{HC1} data.}
\label{fig:GC_N}
\end{figure}

\begin{figure}[!htbp]
\begin{adjustwidth}{-.4cm}{-.4cm}
 	\centering
    \vspace{-.2in}
	\subfigure[][]{
		\includegraphics[width=12.5cm,height=4.5cm, trim=0 2.5cm 0 0]{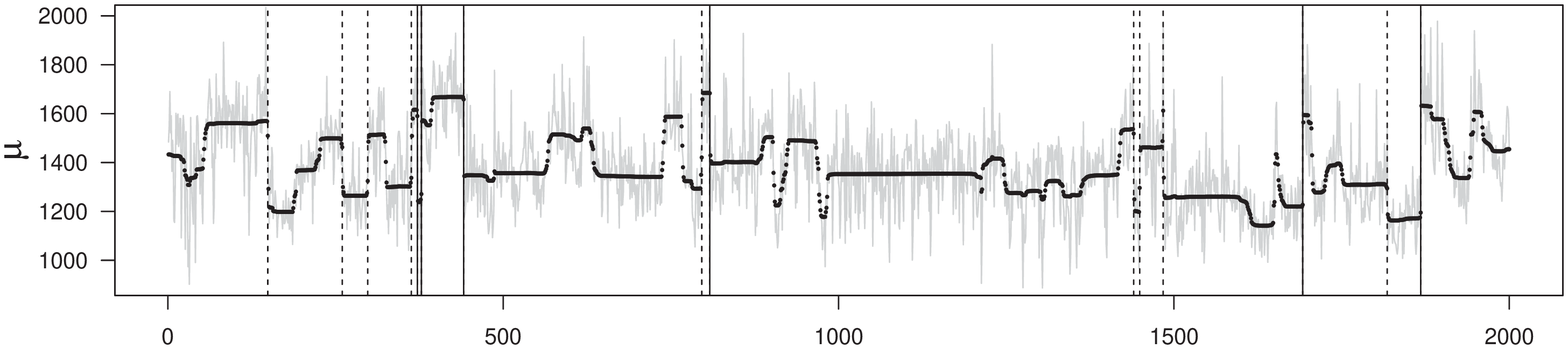}
		\label{fig:GC_mu_LP20}}\\[-.2in]
	\subfigure[][]{
		\includegraphics[width=12.5cm,height=4.5cm, trim=0 2.5cm 0 0]{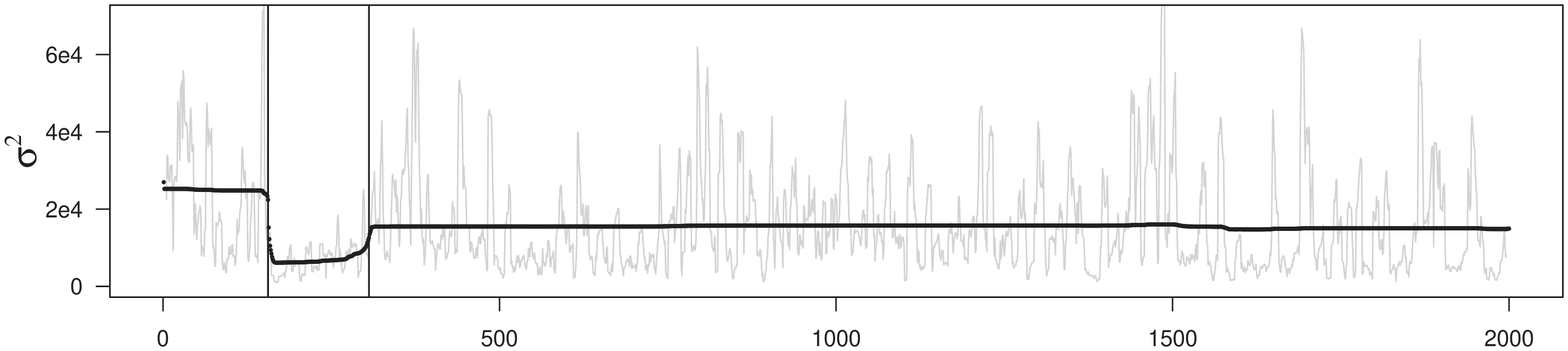}
		\label{fig:GC_s2_LP20}}
\end{adjustwidth}
\caption{Product estimates (black dots) for $\bm{\mu}$ (a) and $\bm{\sigma}$ (b), for the \texttt{HC1} dataset. The gray lines represent the observed data in (a) and the moving sample variance, calculated over ranges of length nine, in (b). The vertical lines in (a) represent the positions with probability of being a change greater than 0.5 (dashed lines) and greater than 0.8 (solid lines). The vertical solid lines in (b) represent the estimated variance changes according to the most likely partition $\rho_2=\{0,156,306,2000\}$.}
\label{fig:GC_PE}
\end{figure}

Contrary to what has been considered in the literature, the posterior distributions displayed in Figure \ref{fig:GC_N_LP20_s2} provides strong evidence that a model  assuming  a common variance may be inappropriate to analyze these data. The most likely number of changes in the variance
is $N_2=2$ and the partition is $\rho_2=\{0,156,306,2000\}$ with posterior probability $0.03$. The product estimates in Figure \ref{fig:GC_PE} show that before position $156$ the variance is higher than for the other   and also the  lowest variance is experienced between positions $156$ and $306$.
%\end{sloppypar}

\begin{figure}[!htbp]
\centering
\begin{adjustwidth}{-.4cm}{-.4cm}
	\centering	
	\subfigure[][]{
		\includegraphics[width=8.5cm,height=4cm, trim=0 2.5cm 0 0]{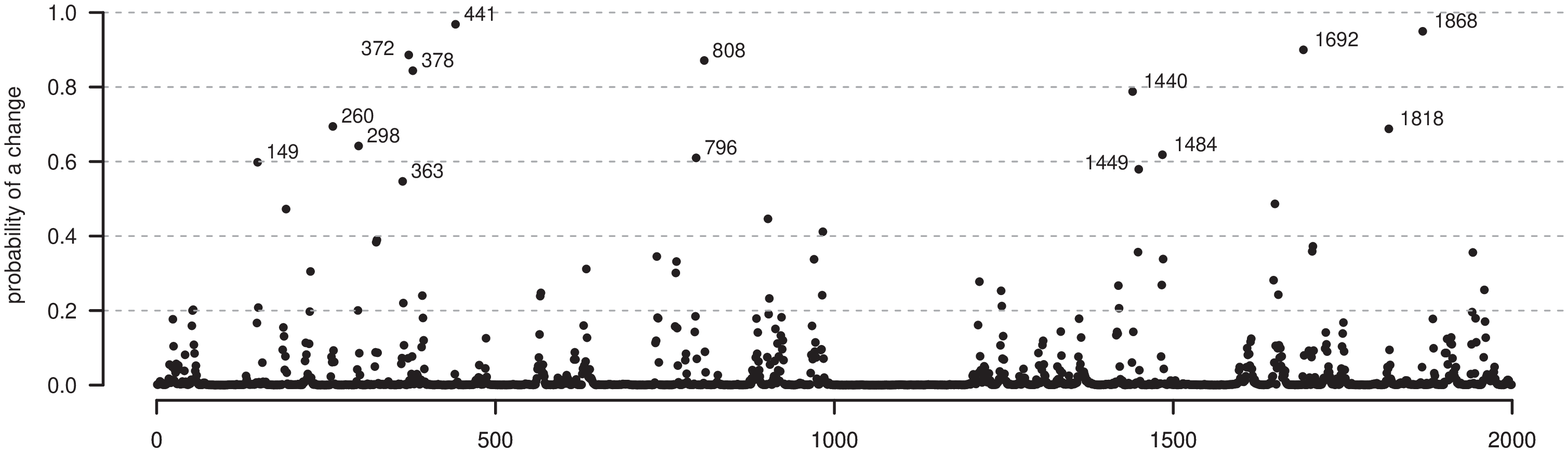}
		\label{fig:GC_prob_LP20_mu}}%\\[-.2in]
	\subfigure[][]{
		\includegraphics[width=8.5cm,height=4cm, trim=0 2.5cm 0 0]{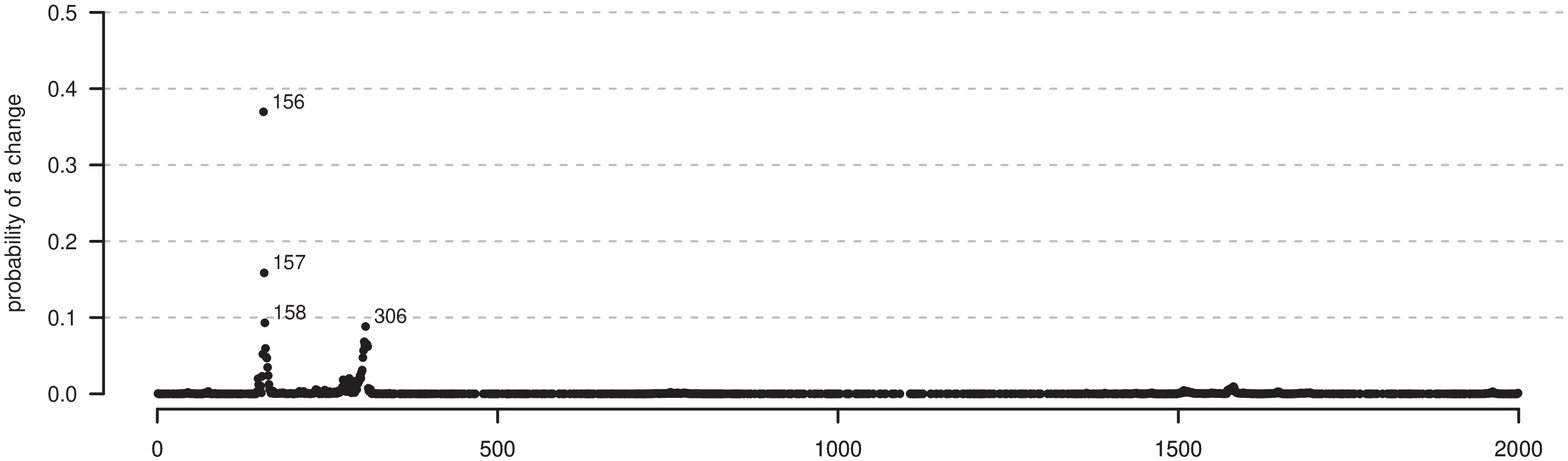}
		\label{fig:GC_prob_LP20_s2}}
	\vspace{-.3in}
	\caption{Posterior probability of each position to be a change in $\bm{\mu}$ (a) and $\bm{\sigma}$ (b), for the \texttt{HC1} dataset. The labeled positions are those with probability greater than 0.5 in (a) and greater than 0.08 in (b).}
	\label{fig:GC_prob}
\end{adjustwidth}
\end{figure}

The \texttt{HC1} data experience a high number of changes in the mean (posterior mode of $N_1$ is $46$) as shown in Figures \ref{fig:GC_N_LP20_mu} and \ref{fig:GC_prob_LP20_mu}. The posterior distribution of $\rho_1$ assigned small probabilities to the sampled partitions making this distribution an inefficient tool to identify the change point positions. Considering the posterior probability of a change (Figure \ref{fig:GC_prob_LP20_mu}) and defining an ad-hoc threshold of $0.5$, we identify $15$ changes in the means. These positions, represented by the vertical lines in Figure \ref{fig:GC_mu_LP20}, coincide with some of the main changes observed in the product estimates for the means and are close to the changes identified by \cite{jewell2021}. DPM19 indicates a small number of changes in the mean and no changes in the variance while LCIA05 provides very different estimates for both the mean and variance of this \texttt{HC1} dataset.

\section{Conclusions}\label{secConclusion}

One of the greatest limitations of the traditional PPM when used to deal with multiple change
point identification is its inability to identify which structural parameter experienced each change. We proposed a multipartition model that provides a reasonable answer to this issue, allowing us to identify the positions and the number of changes that occurred, as well as which
parameters have changed along the observed data sequence. We illustrated the applicability of the proposed model by considering the identification of changes in the mean and variance of Normal data.
In partition models, it is usually challenging to sample from the posterior distribution of the random partition. To deal with this problem, we proposed an efficient partially collapsed Gibbs sampler based on a blocking strategy, which facilitates posterior simulation from the joint
posterior distribution of parameters and partitions.

The simulation studies showed that the proposed model is an efficient approach to identify when and which parameters changed along the sequence. Its performance is as good as that of the original PPM \citep{bh93}, when only the mean changes over time. The proposed model is also competitive when compared to the model described in \cite{peluso2019}, providing better or similar results even considering less prior information about the true number of changes.
Analyzing the guanine-cytosine dataset, the proposed model indicates some changes in the variance, which shows that the assumptions of constant variance considered in some previous analyses may not be appropriate.

Despite its good performance, the proposed model assumes independence among the partitions which may be an unrealistic assumption in many practical situations. To obtain a more general model some type of correlation among the partitions should be considered. This is an interesting topic for future research.

\bibliography{refs}   % name your BibTeX data base
\bibliographystyle{apalike}

\FloatBarrier
\newpage
\begin{center}
{\large\bf SUPPLEMENTARY MATERIAL}
\end{center}

%\onecolumn
%\appendix

\setcounter{section}{19} % 19 -> "S"
\renewcommand{\thesection}{\Alph{section}}
%\section*{Supplementary material}
\section*{} %Ricardo to arXiv
\subsection{A further look at BMCP and DPM19}\label{supp1_peluso}

{Our goal is to evaluate the influence of the prior specification for hyperparameters $(\mu_0, \sigma_0, a, d)$ in the posterior inference provided by the BMCP and DPM19 models. Data sequences of size $n=400$ are  independently generated from Normal distributions, exactly as planned in \cite{peluso2019}.  Three changes in the mean and three changes in the variance occurring at different times are considered, inducing a total of seven clusters in the data. The partitions in the mean and variance are ${\rho_1=\{0,100,200,300,400\}}$ and ${\rho_2=\{0,122,223,325,400\}}$, respectively. The cluster parameters are  ${\bm{\mu}=(0.5,1,0.25,0.75)}$ and ${\bm{\sigma}=(0.3^2,0.6^2,0.15^2,0.45^2)}$. We consider $400$ replications of the data. }

To analyze the data fitting,  following \cite{peluso2019}, for the DPM19 we fix the maximum number of changes in $\bm{\mu}$ and $\bm{\sigma}$ as the respective true number of changes, that is, $m_1=m_2=3$. The DPM19 presented a poor performance whenever greater values for $m_1$ and $m_2$ were assumed thus the results are omitted. The four prior specifications for the hyperparameters  $(\mu_0,\sigma_0^2,a,d)$ shown in Table \ref{tab:scene4_hyper} are considered. The other hyperparameters in DPM19 are specified as described in Section \ref{secBMCPnormal_MC}. For the proposed model, we assume $p_1 \overset{D}{=}p_2 \sim Beta(1,1)$.
\begin{table}[!htb]
\centering
\footnotesize
\begingroup
\renewcommand{\arraystretch}{.6} % Default value: 1
\begin{tabular}{crrrrr}
	\\[-1ex]
	\hline \\[-1.8ex] 
	Prior Specifications & $\mu_0$ & $\sigma_0^2$ & $a$ & $d$ \\ 
	\hline \\[-1.8ex]  
	C1 & 0 & 100 & 0.05 & 1.05 \\ 
	C2 & 0 &   1 & 1    & 1 \\ 
	C3 & 0 & 100 & 1 & 1 \\ 
	C4 & 0 &   1 & 0.05 & 1.05 \\ 
	\hline \\[-1.8ex] 
\end{tabular}
\endgroup
\caption{Hyperparameter values considered in each configuration C1, C2, C3 and C4.}
\label{tab:scene4_hyper}
\end{table}	

\begin{figure}[!htbp]
\centering
% bmcp20
\subfigure[][BMCP (C1)]{
	\includegraphics[width=3.5cm, height=3cm, trim=0 1cm 0 0]{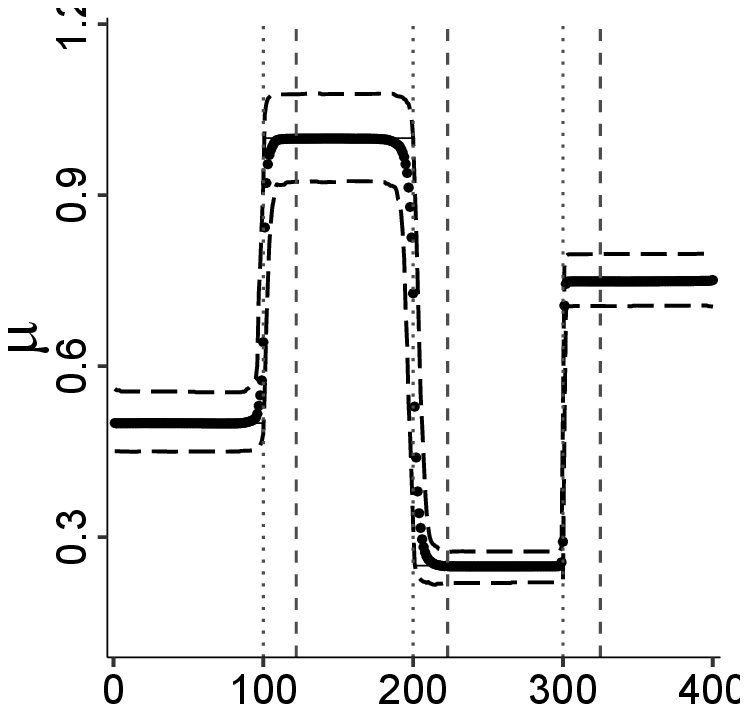}
	\label{fig:scene4_mu_LP20_c1}}%\hspace{.01in}
\subfigure[][BMCP (C2)]{
	\includegraphics[width=3.5cm, height=3cm, trim=0 1cm 0 0]{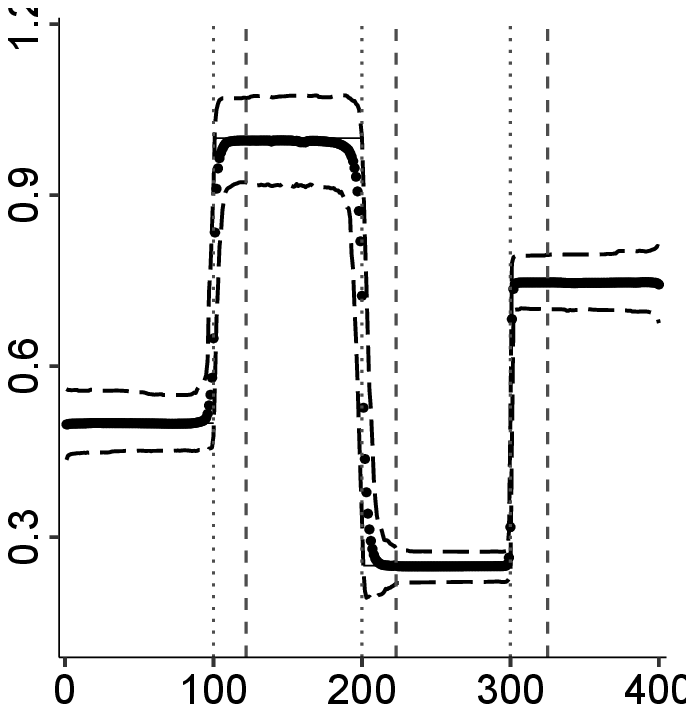}
	\label{fig:scene4_mu_LP20_c2}}%\hspace{.01in}
\subfigure[][BMCP (C3)]{
	\includegraphics[width=3.5cm, height=3cm, trim=0 1cm 0 0]{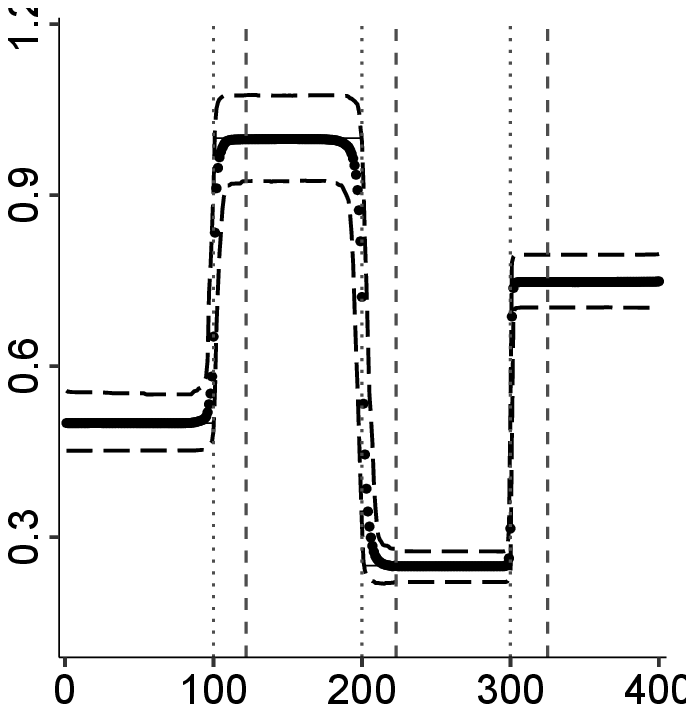}
	\label{fig:scene4_mu_LP20_c3}}%\hspace{.01in}
\subfigure[][BMCP (C4)]{
	\includegraphics[width=3.5cm, height=3cm, trim=0 1cm 0 0]{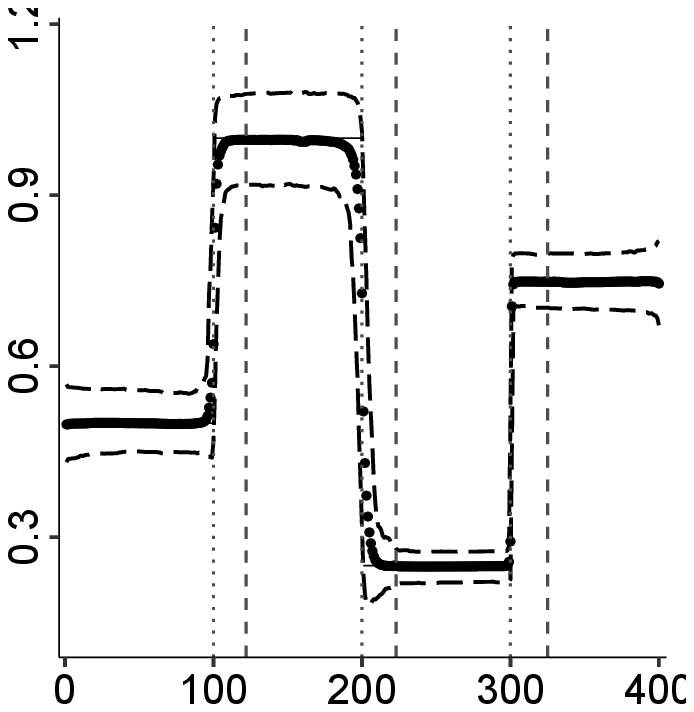}
	\label{fig:scene4_mu_LP20_c4}}\\[.2cm]
% dpm19
\subfigure[][DPM19 (C1)]{
	\includegraphics[width=3.5cm, height=3cm, trim=0 1cm 0 0]{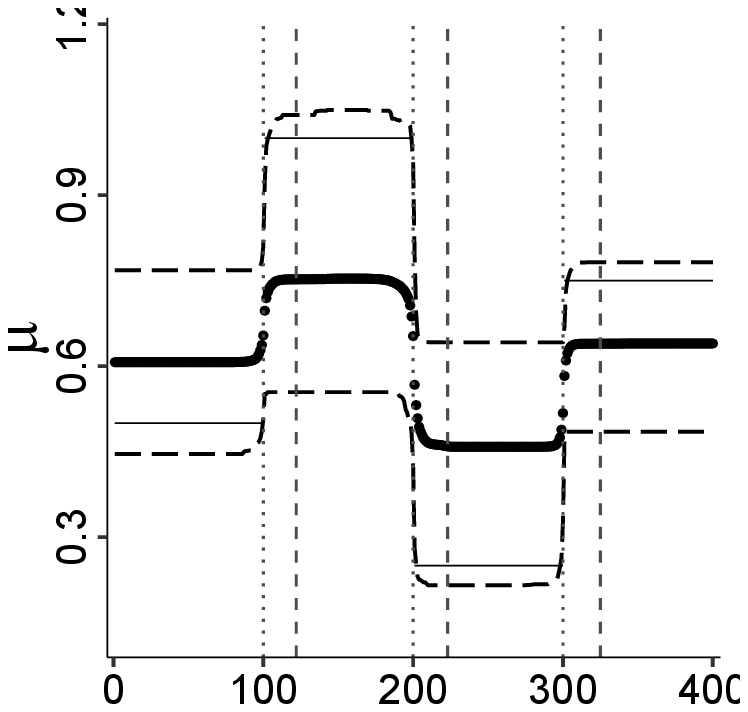}
	\label{fig:scene4_mu_P18_c1}}%\hspace{.01in}
\subfigure[][DPM19 (C2)]{
	\includegraphics[width=3.5cm, height=3cm, trim=0 1cm 0 0]{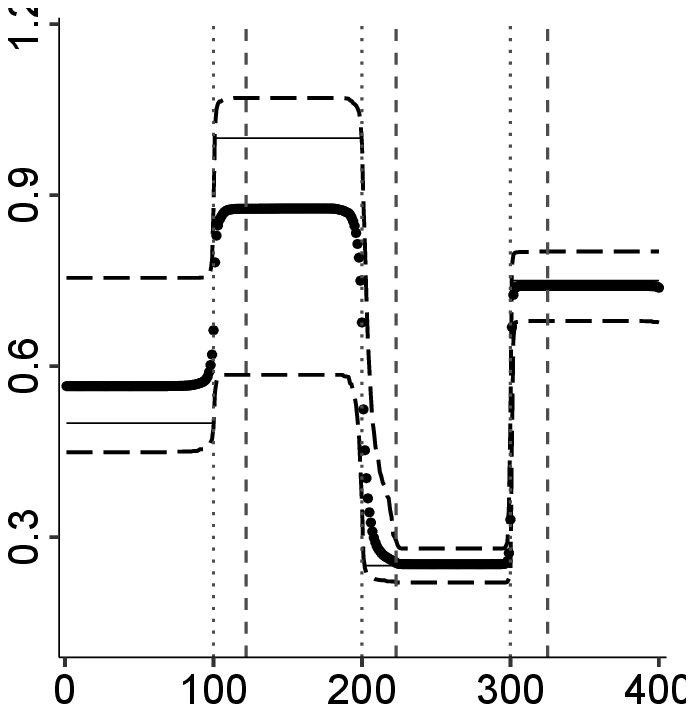}
	\label{fig:scene4_mu_P18_c2}}%\hspace{.01in}
\subfigure[][DPM19 (C3)]{
	\includegraphics[width=3.5cm, height=3cm, trim=0 1cm 0 0]{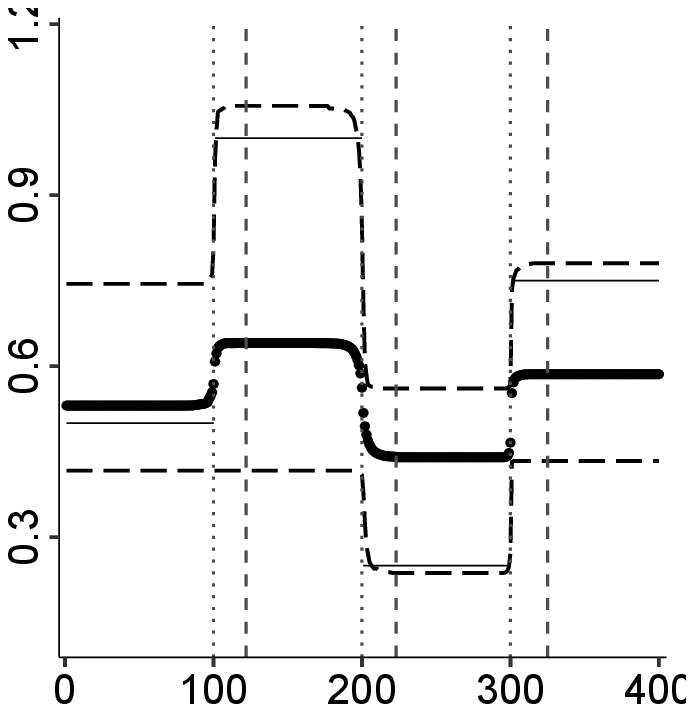}
	\label{fig:scene4_mu_P18_c3}}%\hspace{.01in}
\subfigure[][DPM19 (C4)]{
	\includegraphics[width=3.5cm, height=3cm, trim=0 1cm 0 0]{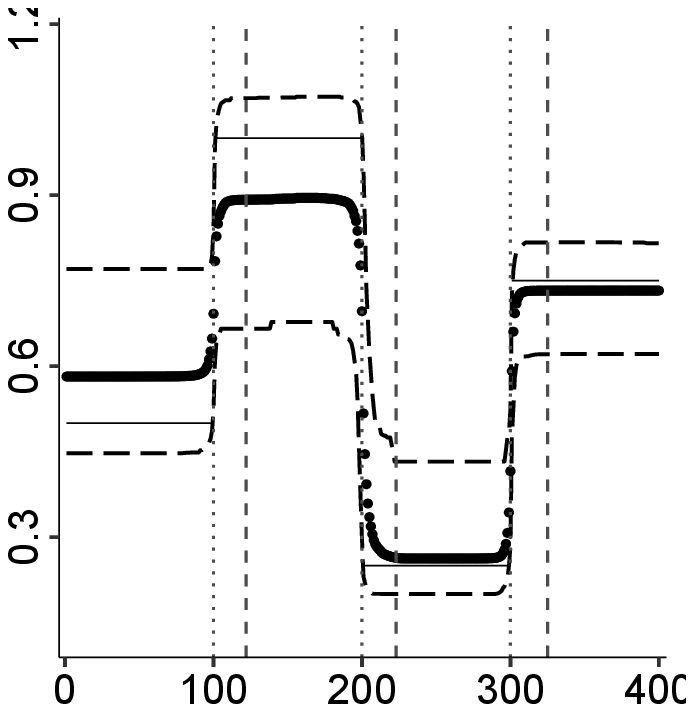}
	\label{fig:scene4_mu_P18_c4}}\\
\caption {Average of the mean estimates (black dots) at each instant and the $5\%$ and $95\%$ quantiles of such estimates based on the Monte Carlo replications, under BMCP (a-d) and DPM19 (e-h) models. The true mean values are indicated by the solid gray horizontal lines. The vertical gray dotted and dashed lines indicate the true end points in $\rho_1$ and $\rho_2$, respectively.}
\label{fig:scene4_PE_mu}
\end{figure}

\begin{figure}[!htbp]
\centering
% bmcp20
\subfigure[][BMCP (C1)]{
	\includegraphics[width=3.5cm, height=3cm, trim=0 1cm 0 0]{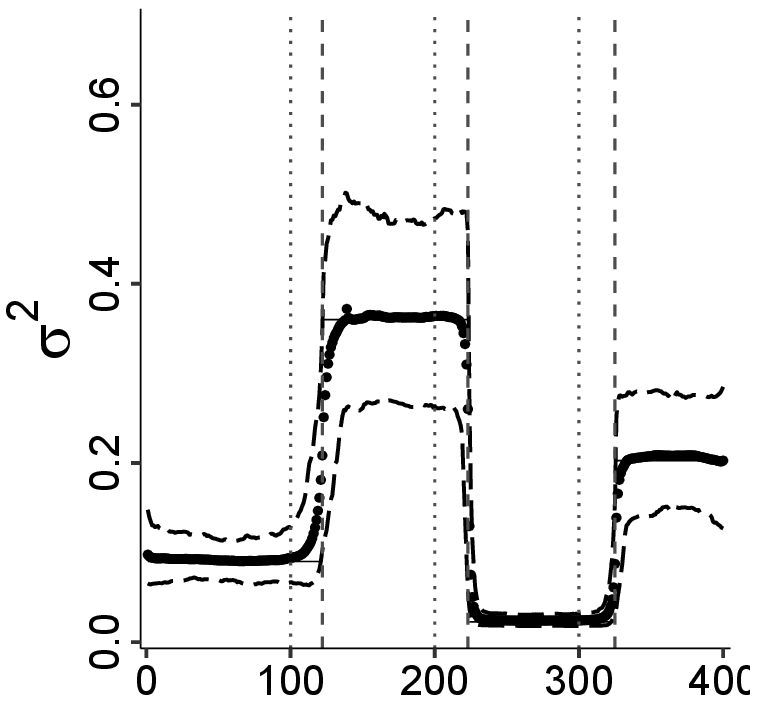}
	\label{fig:scene4_s2_LP20_c1}}%\hspace{.01in}
\subfigure[][BMCP (C2)]{
	\includegraphics[width=3.5cm, height=3cm, trim=0 1cm 0 0]{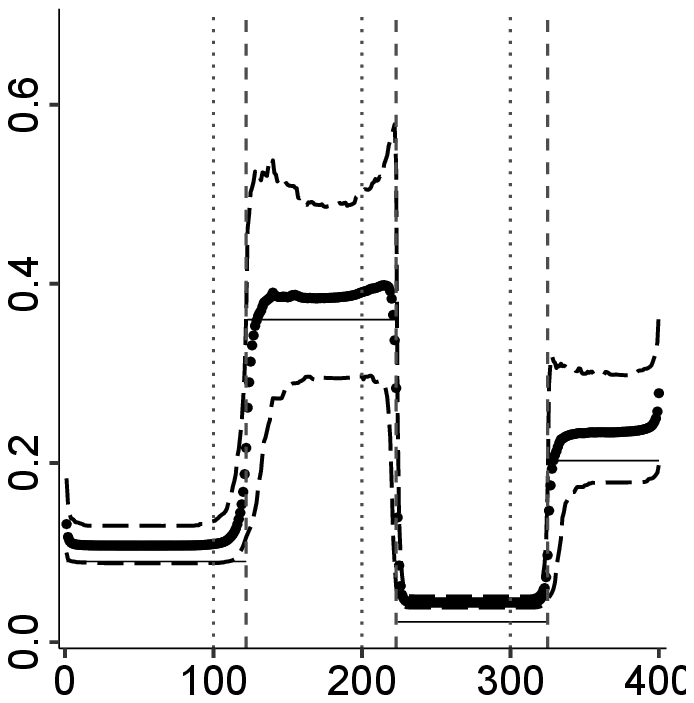}
	\label{fig:scene4_s2_LP20_c2}}%\hspace{.01in}
\subfigure[][BMCP (C3)]{
	\includegraphics[width=3.5cm, height=3cm, trim=0 1cm 0 0]{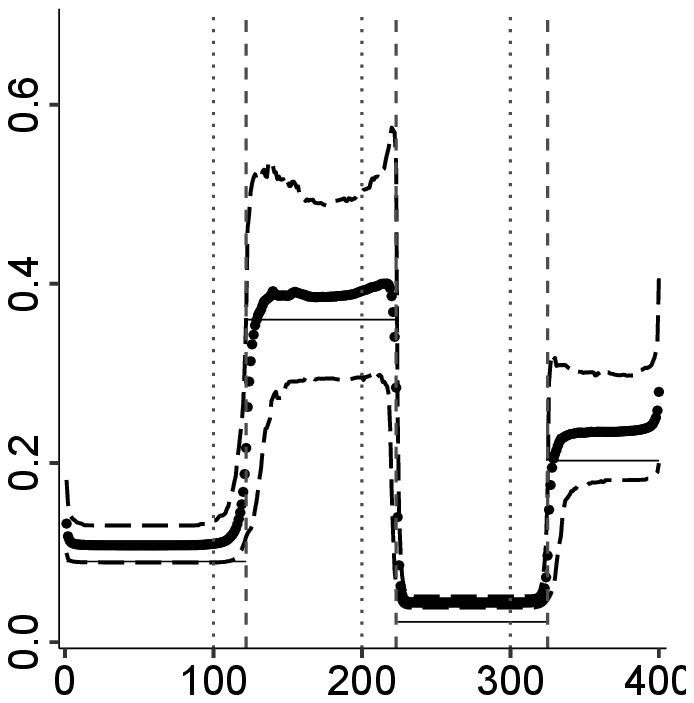}
	\label{fig:scene4_s2_LP20_c3}}%\hspace{.01in}
\subfigure[][BMCP (C4)]{
	\includegraphics[width=3.5cm, height=3cm, trim=0 1cm 0 0]{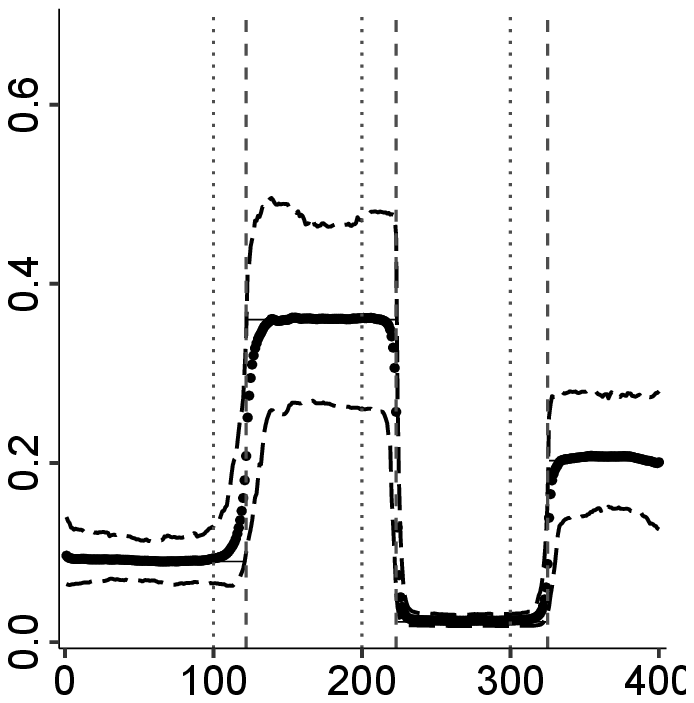}
	\label{fig:scene4_s2_LP20_c4}}\\[.2cm]
% dpm19
\subfigure[][DPM19 (C1)]{
	\includegraphics[width=3.5cm, height=3cm, trim=0 1cm 0 0]{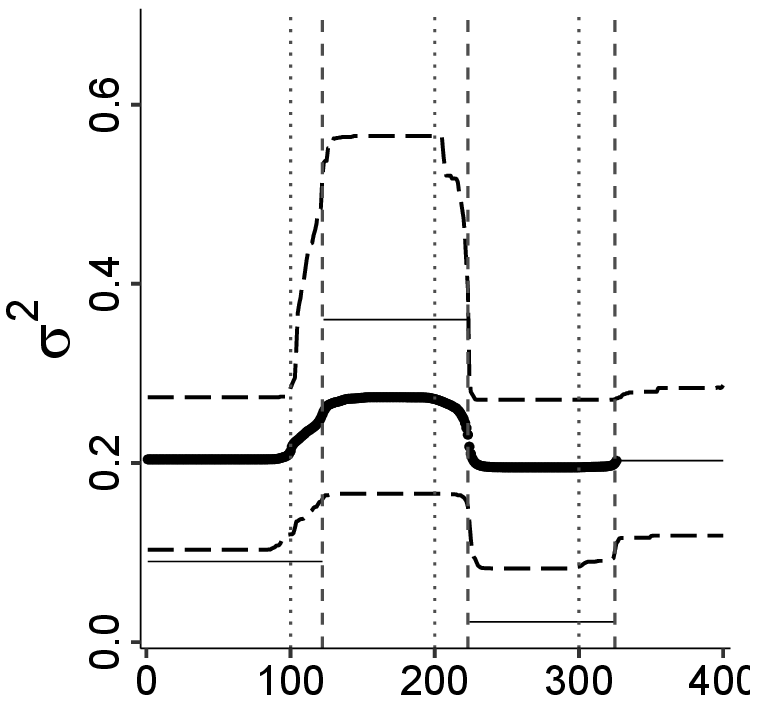}
	\label{fig:scene4_s2_P18_c1}}%\hspace{.01in}
\subfigure[][DPM19 (C2)]{
	\includegraphics[width=3.5cm, height=3cm, trim=0 1cm 0 0]{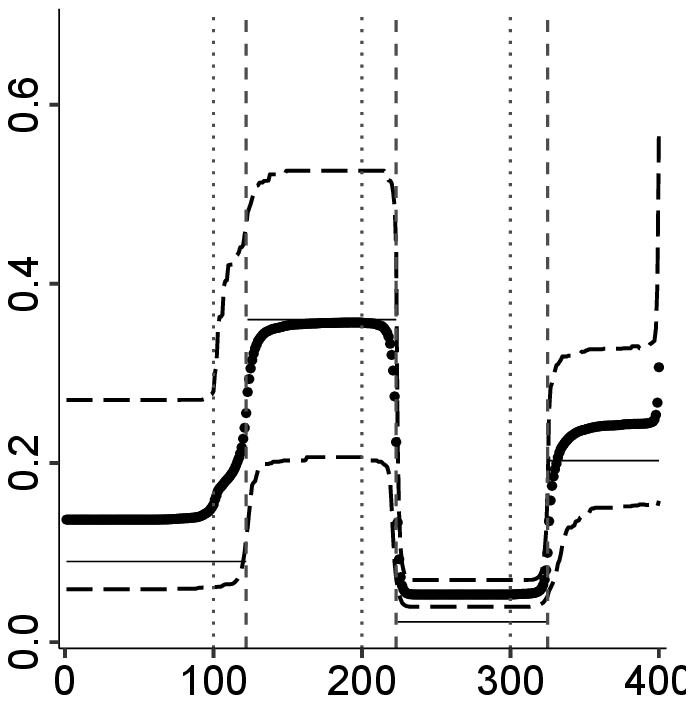}
	\label{fig:scene4_s2_P18_c2}}%\hspace{.01in}
\subfigure[][DPM19 (C3)]{
	\includegraphics[width=3.5cm, height=3cm, trim=0 1cm 0 0]{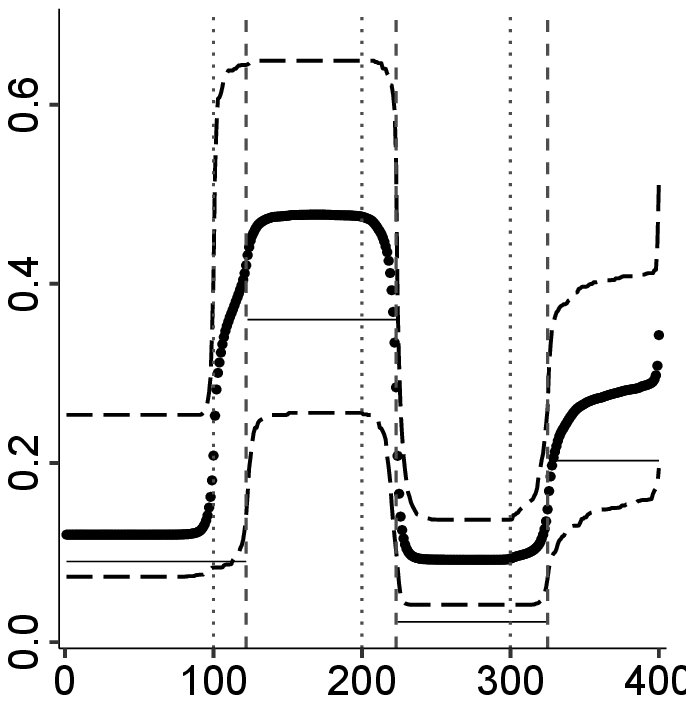}
	\label{fig:scene4_s2_P18_c3}}%\hspace{.01in}
\subfigure[][DPM19 (C4)]{
	\includegraphics[width=3.5cm, height=3cm, trim=0 1cm 0 0]{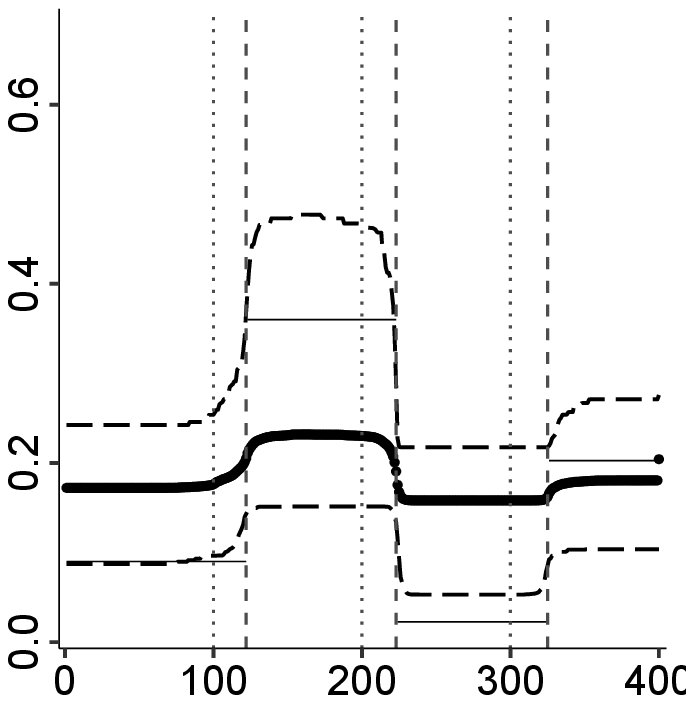}
	\label{fig:scene4_s2_P18_c4}}\\
\caption {Average of the variance estimates (black dots) at each instant and the $5\%$ and $95\%$ quantiles of such estimates based on the Monte Carlo replications, under BMCP (a-d) and DPM19 (e-h) models. The true variance values are indicated by the solid gray horizontal lines. The vertical gray dotted and dashed lines indicate the true end points in $\rho_1$ and $\rho_2$, respectively. { (In (e), the estimates for the variance after instant 326 are extremely high thus they are not shown.)}}
\label{fig:scene4_PE_s2}
\end{figure}

Figure \ref{fig:scene4_PE_mu} shows that, under all prior specifications, the BMCP  provides accurate estimates for $\bm{\mu}$ in all clusters while the DPM19  produced biased estimates in the $1st$ and $2nd$ clusters. DPM19 model presents an improvement in its performance in  scenarios which consider more informative prior distributions for the hyperparameters and for $\mu$ (C2 and C4). Regarding the posterior estimates for the variance, Figure \ref{fig:scene4_PE_s2} shows that BMCP outperforms the DPM19 in all cases. The proposed model provides very precise estimates for the variance in scenarios C1 and C4 but overestimates it in all  clusters in scenarios C2 and C3.  However, biases produced by DPM19 are much higher in all cases and the variances in the 1st and 3rd  clusters are overestimated while for the 2nd  and 4th clusters they are over or underestimated depending on the prior specifications. Additionally, under DPM19, the posterior estimates for the variances in the $1st$ cluster are affected by the 1st change experienced by the mean occurred at the position $100$.

\begin{figure}[!htbp]
\centering
% bmcp20
\subfigure[][BMCP (C1)]{
	\includegraphics[width=3.5cm, height=3.2cm, trim=0 2cm 0 0]{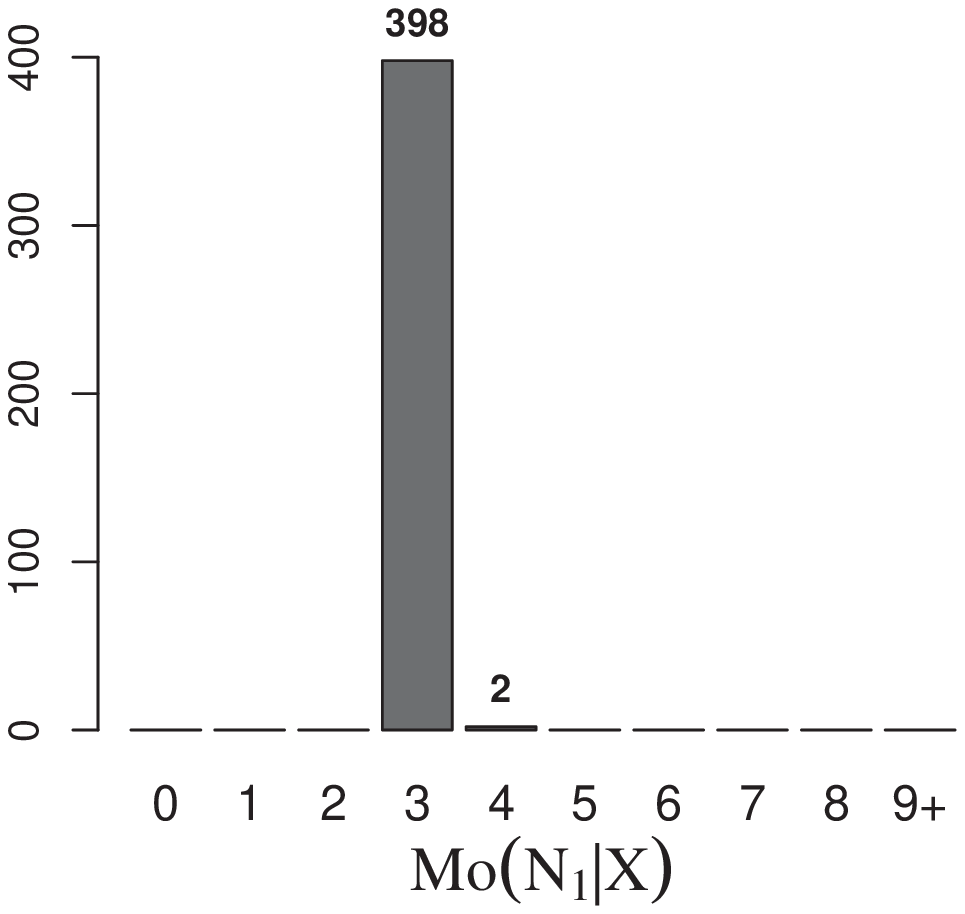}
	\label{fig:scene4_Nmode_LP20_mu_c1}}\hspace{-.2in}
\subfigure[][BMCP (C2)]{
	\includegraphics[width=3.5cm, height=3.2cm, trim=0 2cm 0 0]{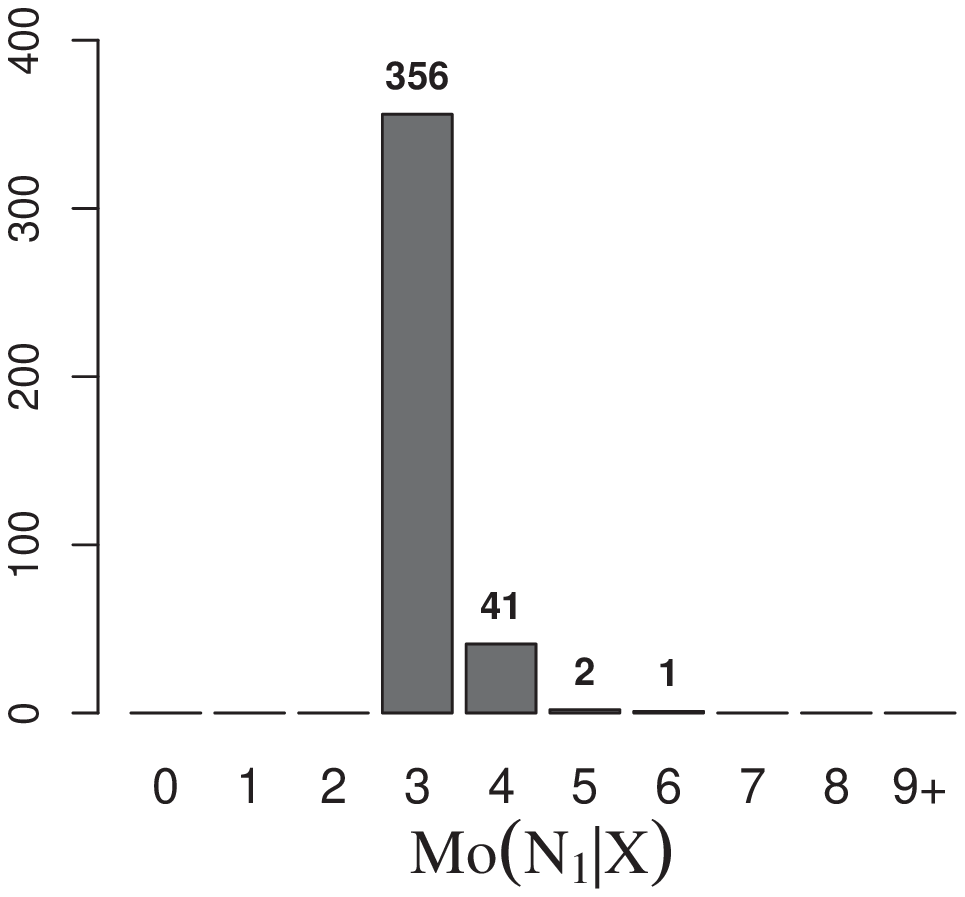}
	\label{fig:scene4_Nmode_LP20_mu_c2}}\hspace{-.2in}
\subfigure[][BMCP (C3)]{
	\includegraphics[width=3.5cm, height=3.2cm, trim=0 2cm 0 0]{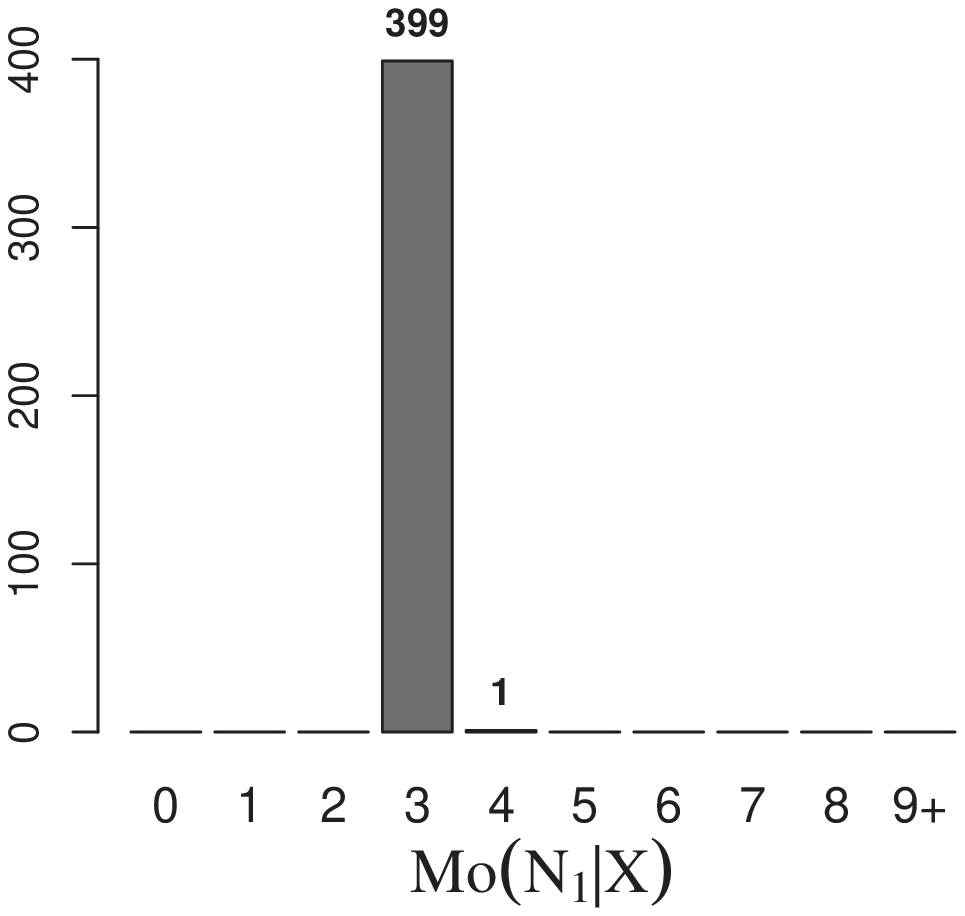}
	\label{fig:scene4_Nmode_LP20_mu_c3}}\hspace{-.2in}
\subfigure[][BMCP (C4)]{
	\includegraphics[width=3.5cm, height=3.2cm, trim=0 2cm 0 0]{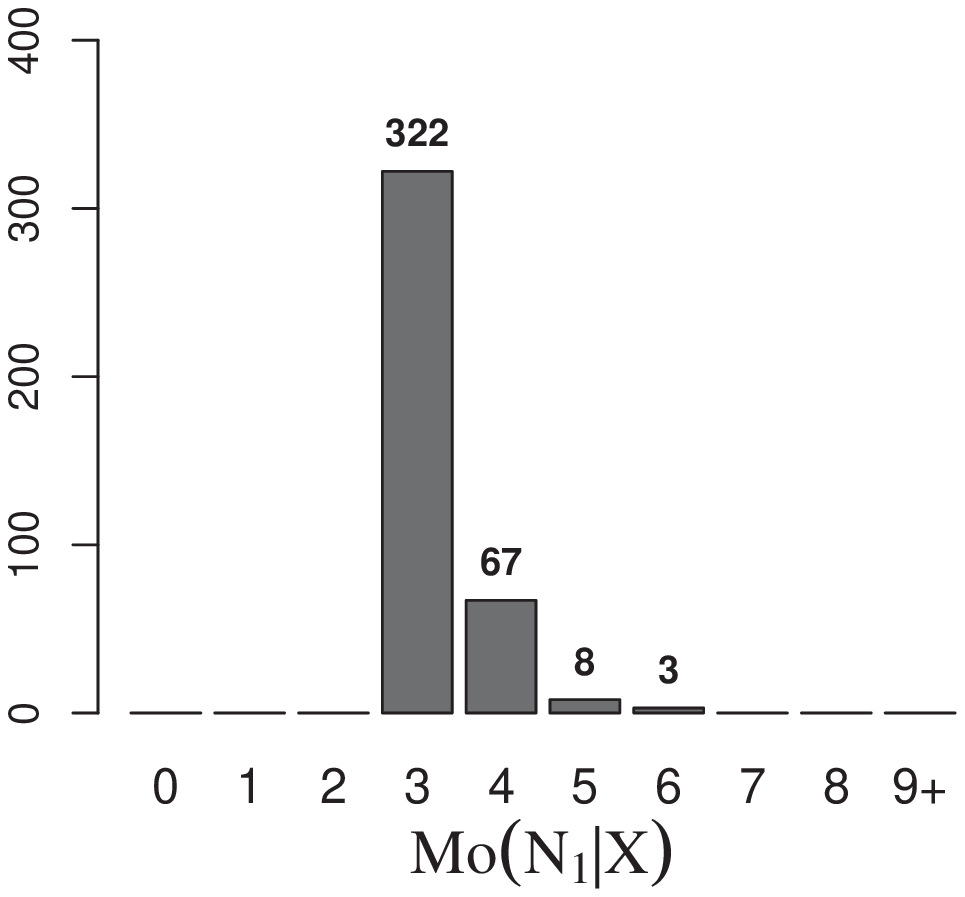}
	\label{fig:scene4_Nmode_LP20_mu_c4}}\\[-.1in]
% dpm19
\subfigure[][DPM19 (C1)]{
	\includegraphics[width=3.5cm, height=3.2cm, trim=0 2cm 0 0]{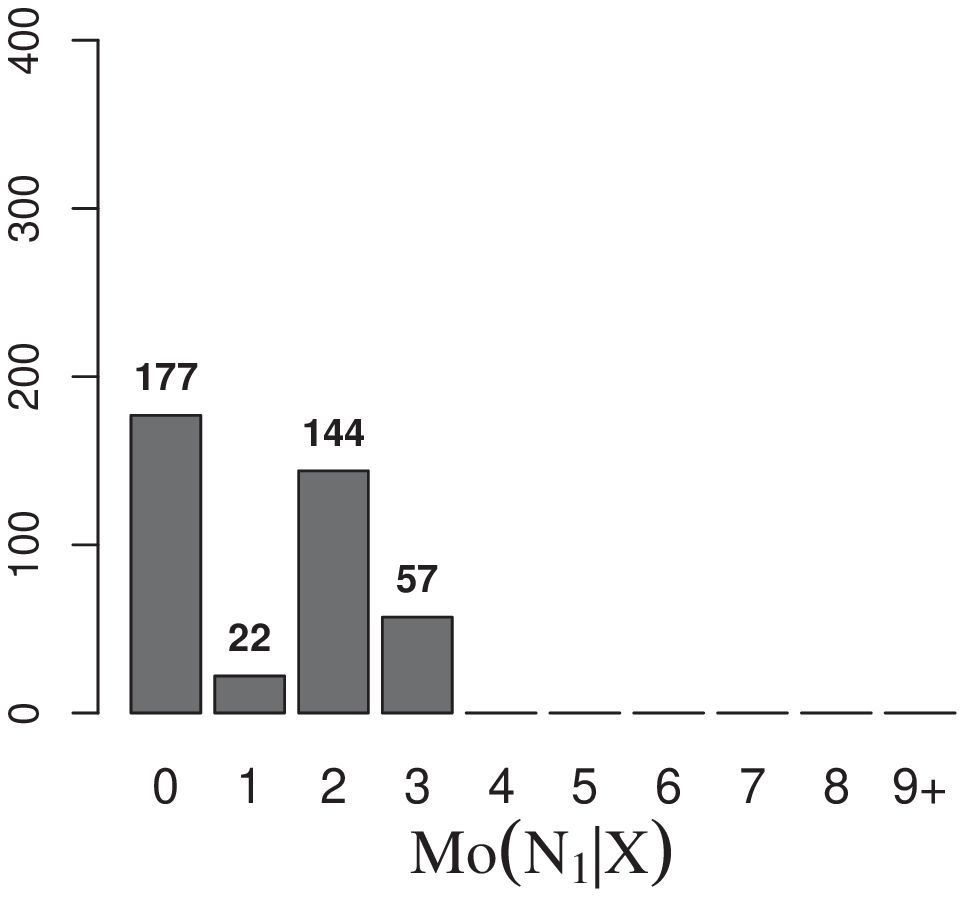}
	\label{fig:scene4_Nmode_P18_mu_c1}}\hspace{-.2in}
\subfigure[][DPM19 (C2)]{
	\includegraphics[width=3.5cm, height=3.2cm, trim=0 2cm 0 0]{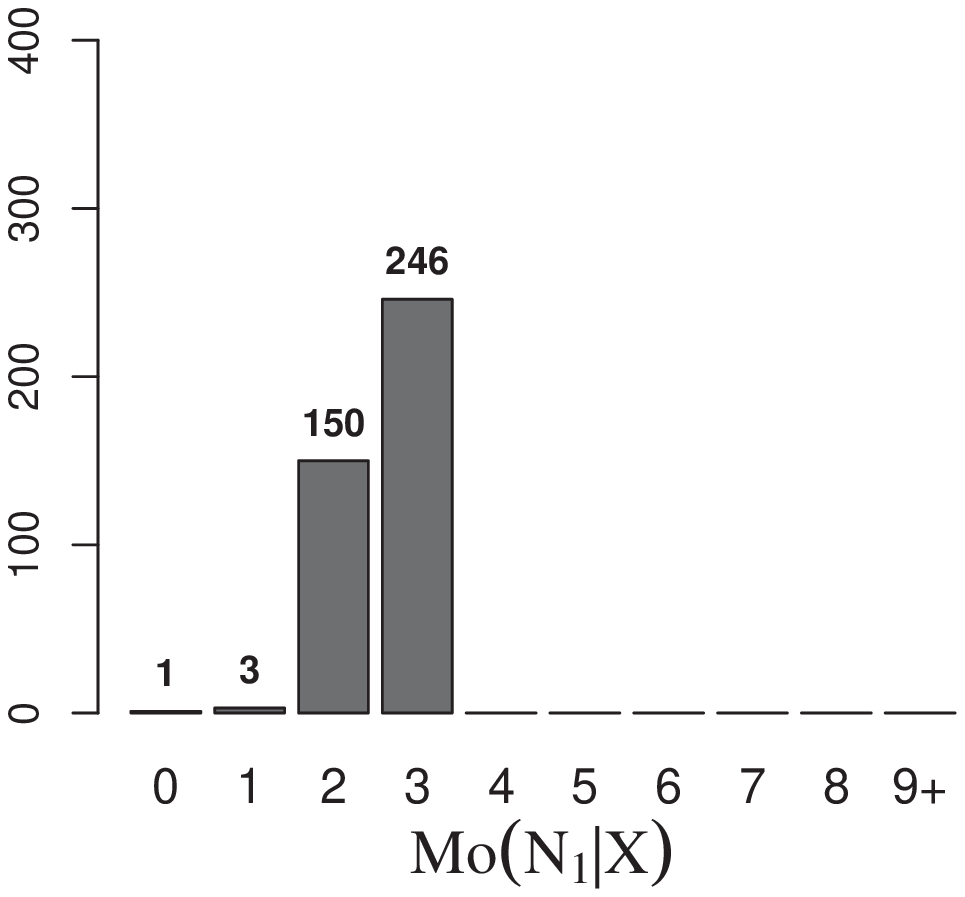}
	\label{fig:scene4_Nmode_P18_mu_c2}}\hspace{-.2in}
\subfigure[][DPM19 (C3)]{
	\includegraphics[width=3.5cm, height=3.2cm, trim=0 2cm 0 0]{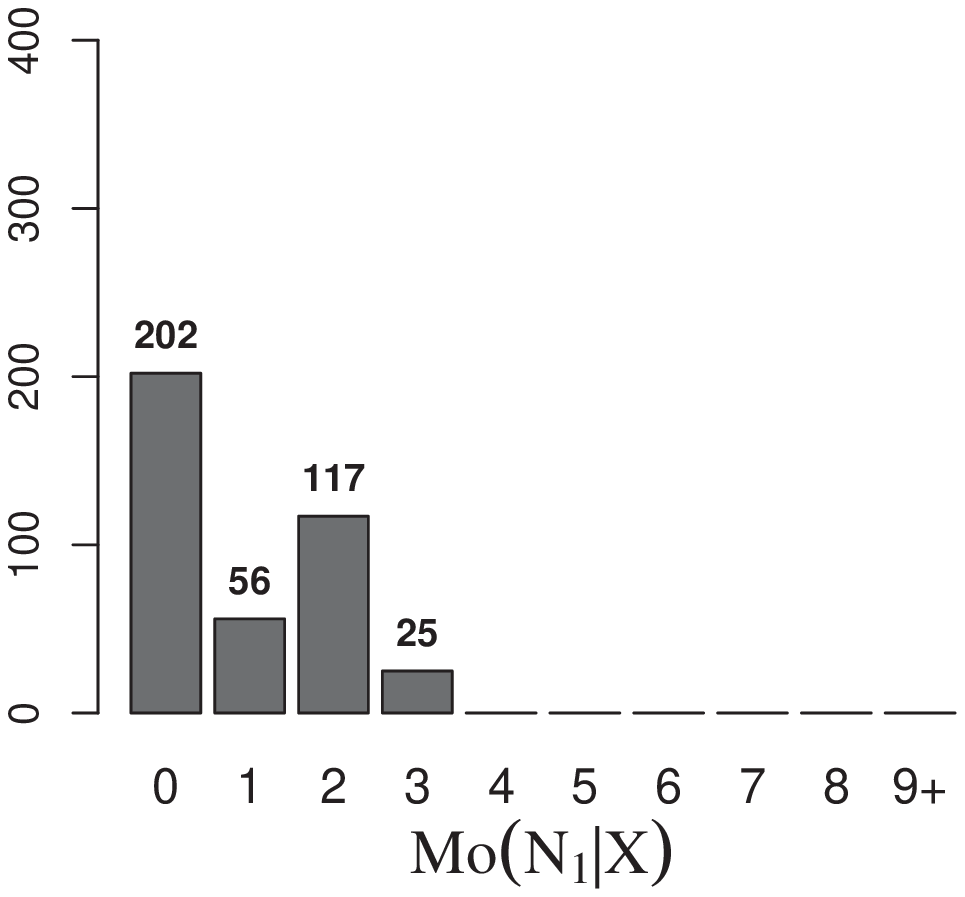}
	\label{fig:scene4_Nmode_P18_mu_c3}}\hspace{-.2in}
\subfigure[][DPM19 (C4)]{
	\includegraphics[width=3.5cm, height=3.2cm, trim=0 2cm 0 0]{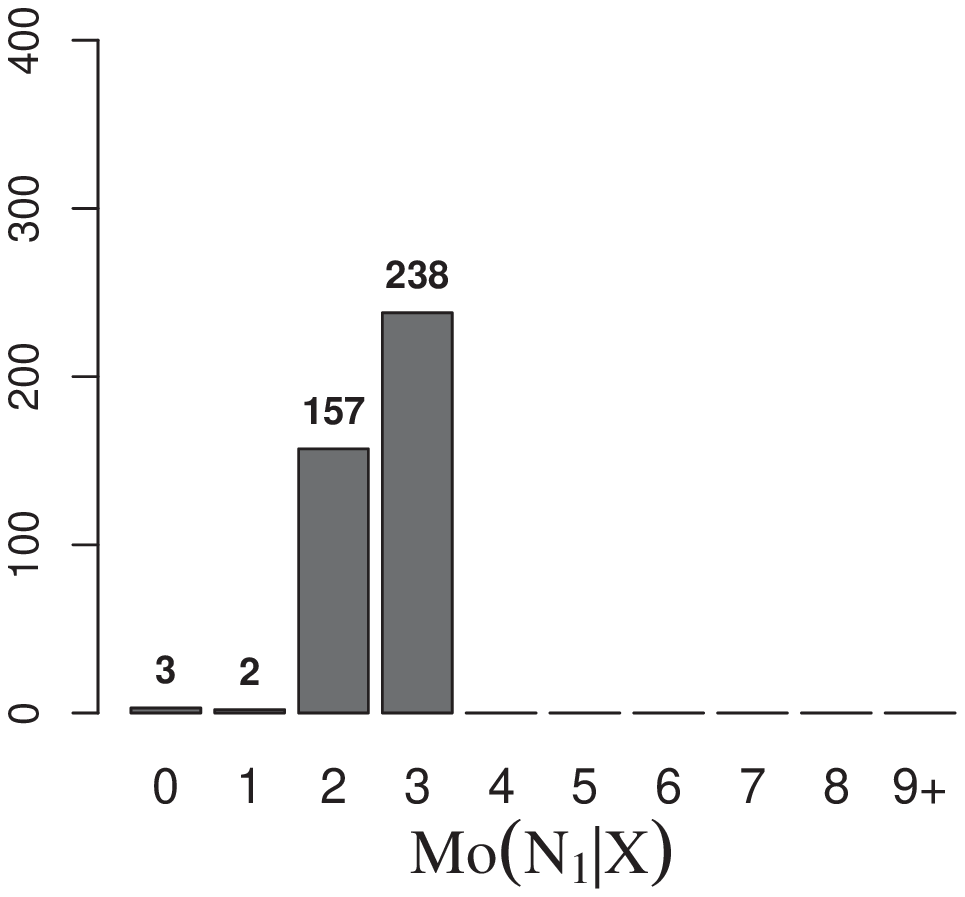}
	\label{fig:scene4_Nmode_P18_mu_c4}}\\
\caption{Counts distribution of the posterior modes of the number of changes in the mean ($N_1$) estimated for each Monte Carlo replication.}
\label{fig:scene4_Nmode_mu}
\end{figure}

The estimated partitions (Table \ref{tab:scene4_mod_rho1}) and the estimated number of change points (Figure \ref{fig:scene4_Nmode_mu}) for the means $\bm{\mu}$ also show the best performance of the proposed model if compared to DPM19 under all prior specifications. 


\begin{figure}[!htbp]
\centering
% bmcp20
\subfigure[][BMCP (C1)]{
	\includegraphics[width=3.5cm, height=3.2cm, trim=0 2cm 0 0]{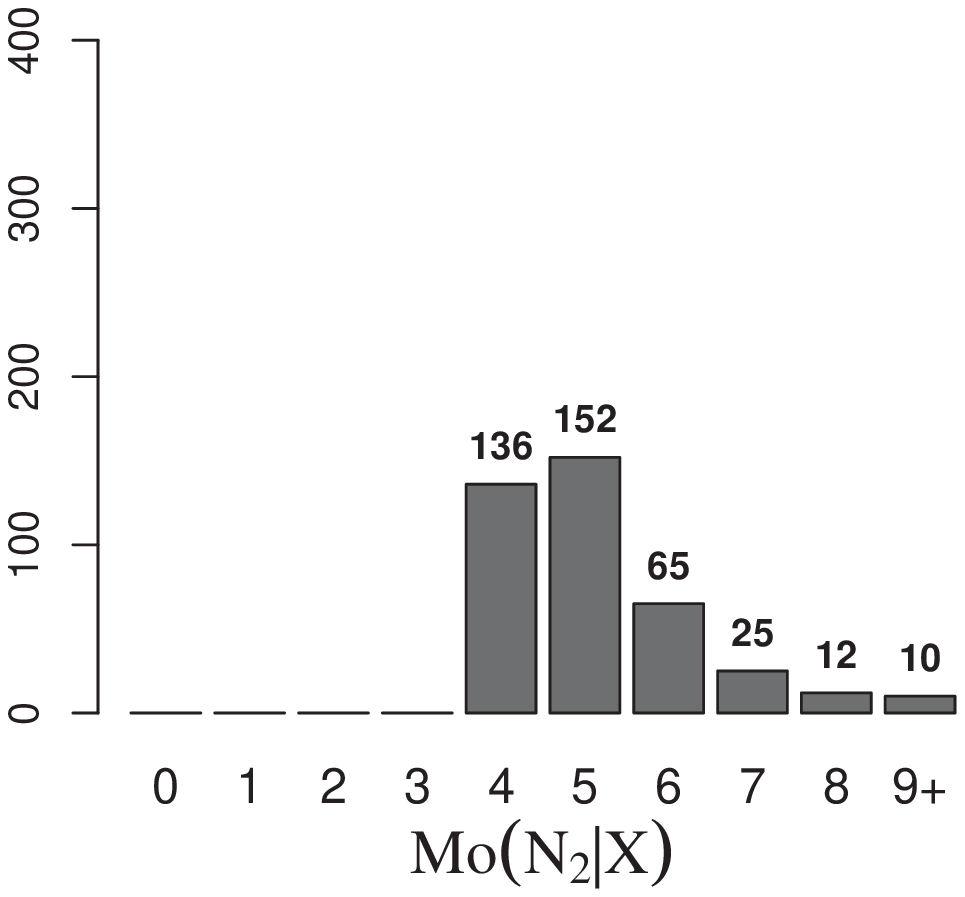}
	\label{fig:scene4_Nmode_LP20_s2_c1}}\hspace{-.2in}
\subfigure[][BMCP (C2)]{
	\includegraphics[width=3.5cm, height=3.2cm, trim=0 2cm 0 0]{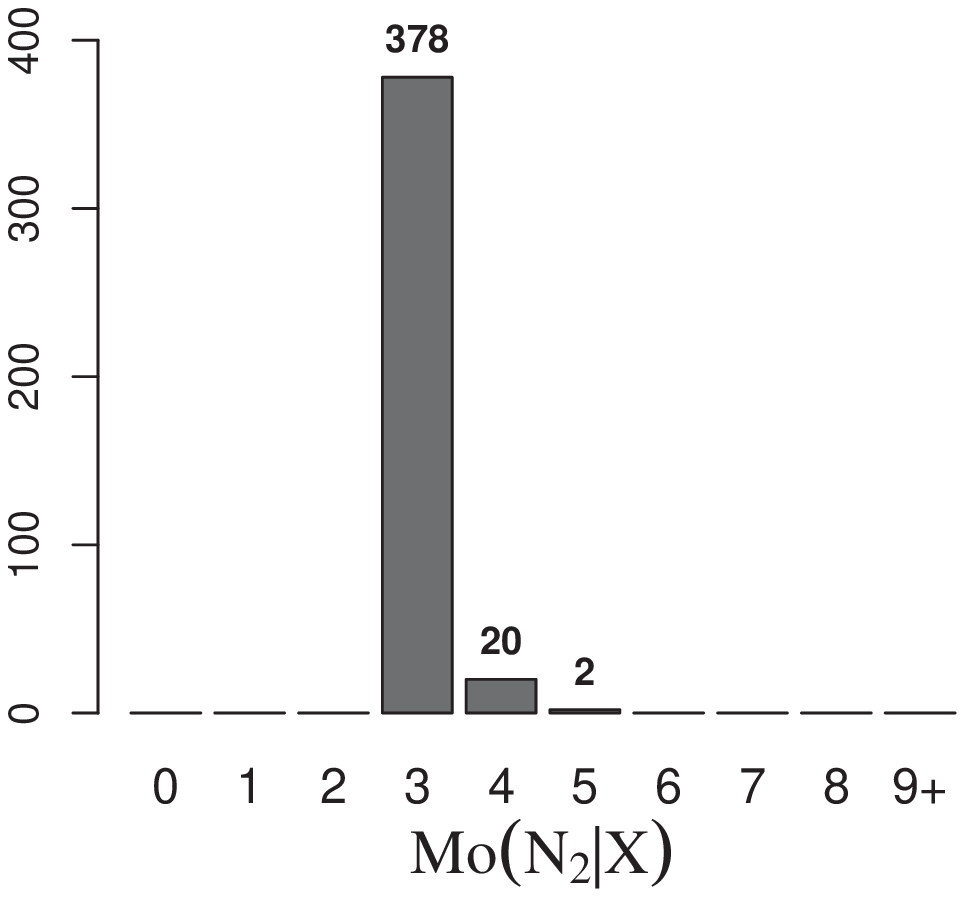}
	\label{fig:scene4_Nmode_LP20_s2_c2}}\hspace{-.2in}
\subfigure[][BMCP (C3)]{
	\includegraphics[width=3.5cm, height=3.2cm, trim=0 2cm 0 0]{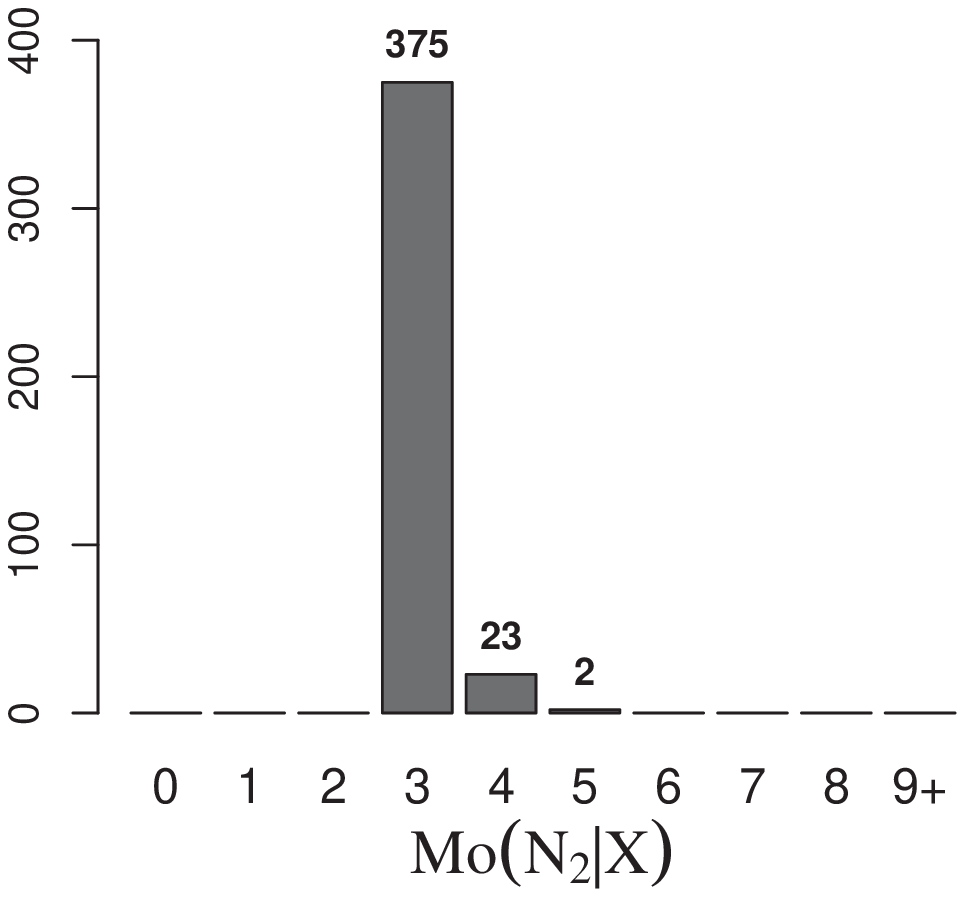}
	\label{fig:scene4_Nmode_LP20_s2_c3}}\hspace{-.2in}
\subfigure[][BMCP (C4)]{
	\includegraphics[width=3.5cm, height=3.2cm, trim=0 2cm 0 0]{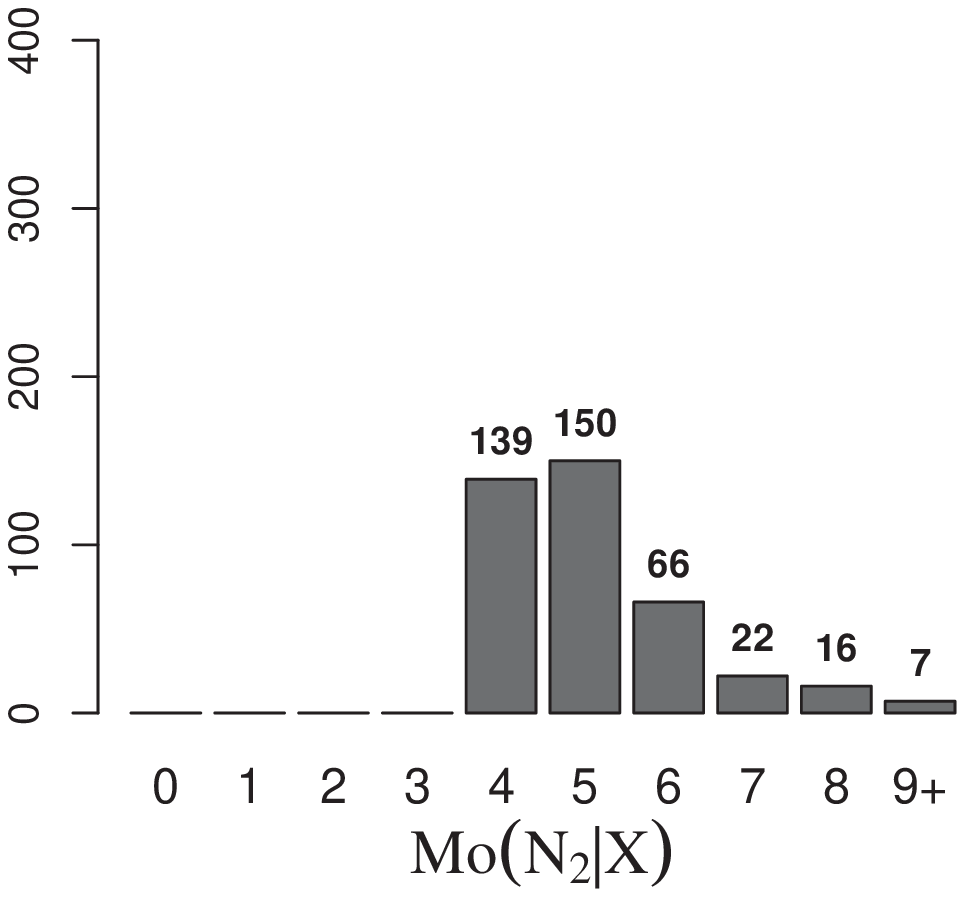}
	\label{fig:scene4_Nmode_LP20_s2_c4}}\\[-.1in]
% dpm19
\subfigure[][DPM19 (C1)]{
	\includegraphics[width=3.5cm, height=3.2cm, trim=0 2cm 0 0]{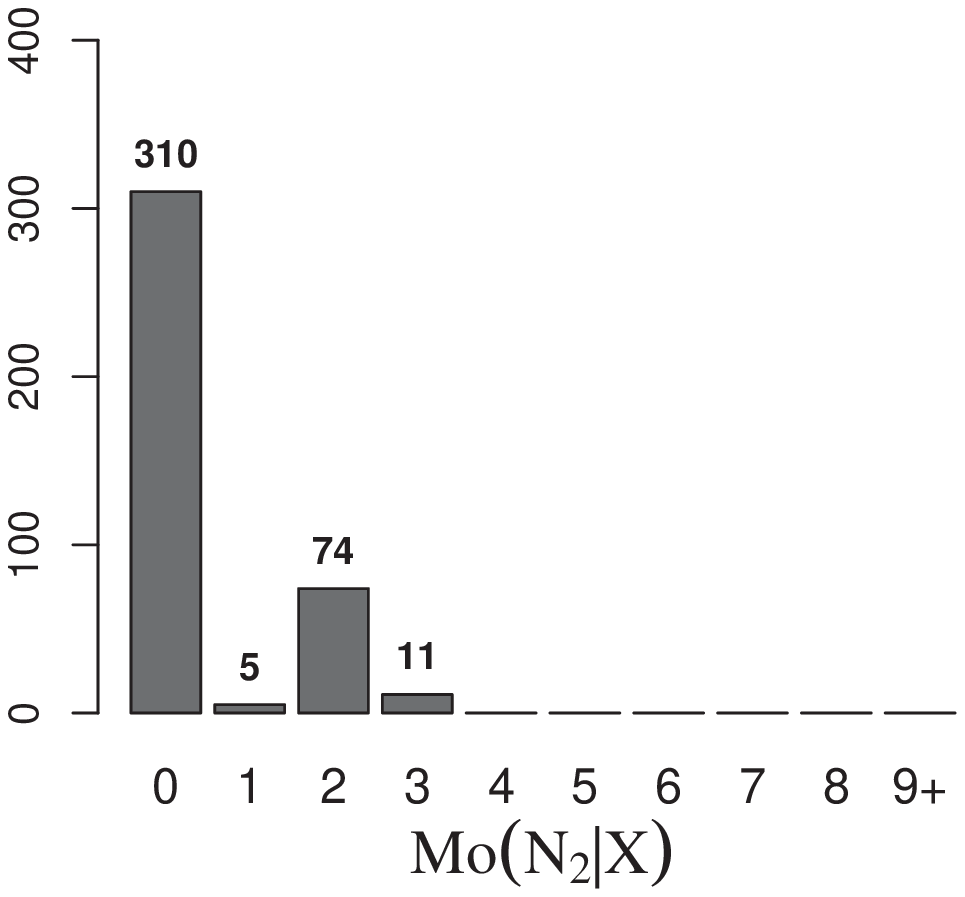}
	\label{fig:scene4_Nmode_P18_s2_c1}}\hspace{-.2in}
\subfigure[][DPM19 (C2)]{
	\includegraphics[width=3.5cm, height=3.2cm, trim=0 2cm 0 0]{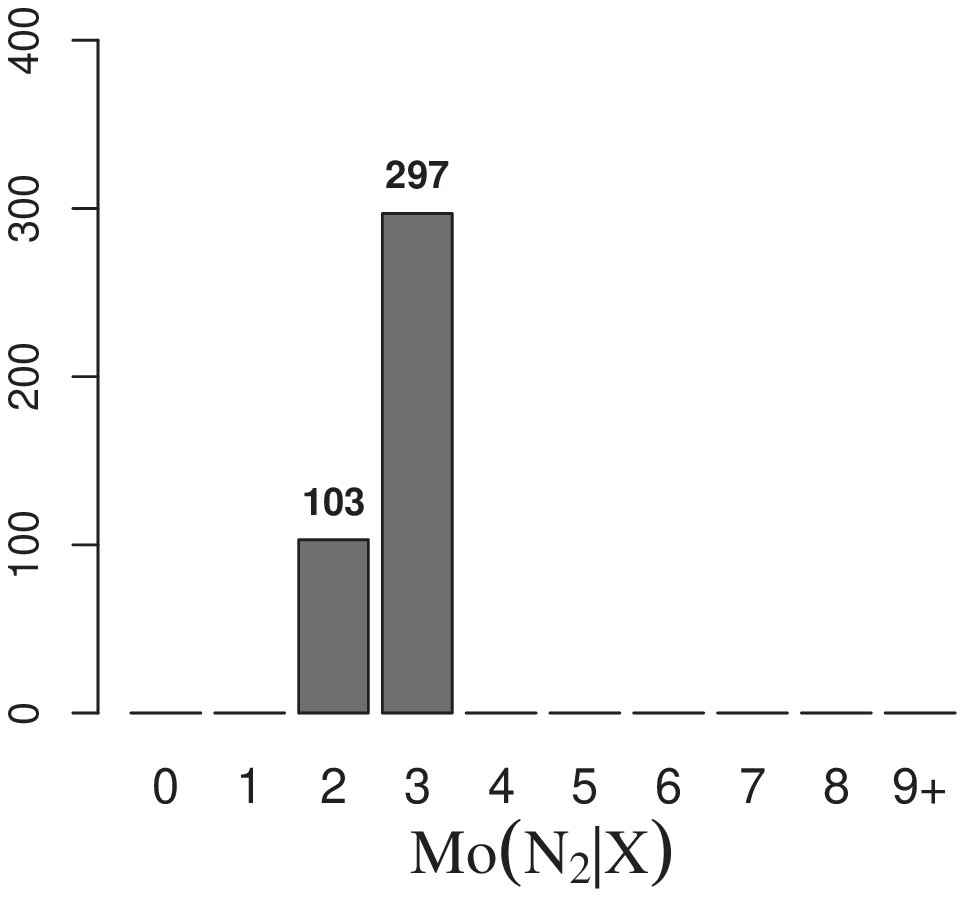}
	\label{fig:scene4_Nmode_P18_s2_c2}}\hspace{-.2in}
\subfigure[][DPM19 (C3)]{
	\includegraphics[width=3.5cm, height=3.2cm, trim=0 2cm 0 0]{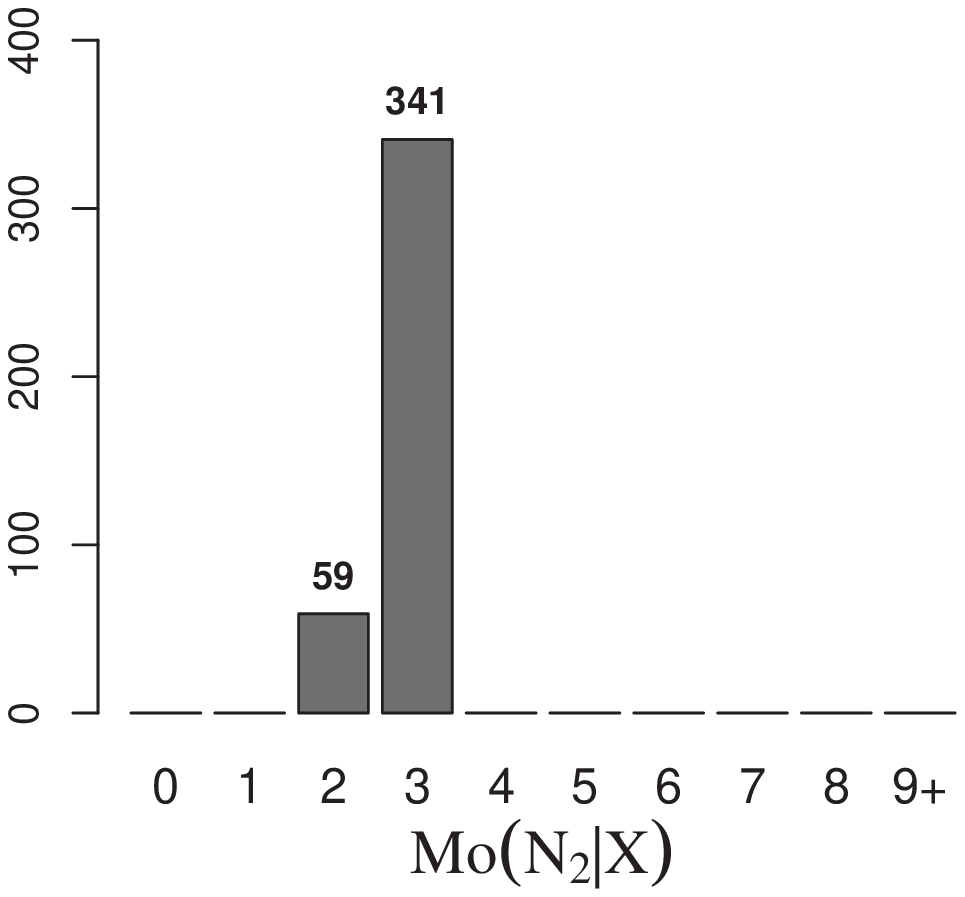}
	\label{fig:scene4_Nmode_P18_s2_c3}}\hspace{-.2in}
\subfigure[][DPM19 (C4)]{
	\includegraphics[width=3.5cm, height=3.2cm, trim=0 2cm 0 0]{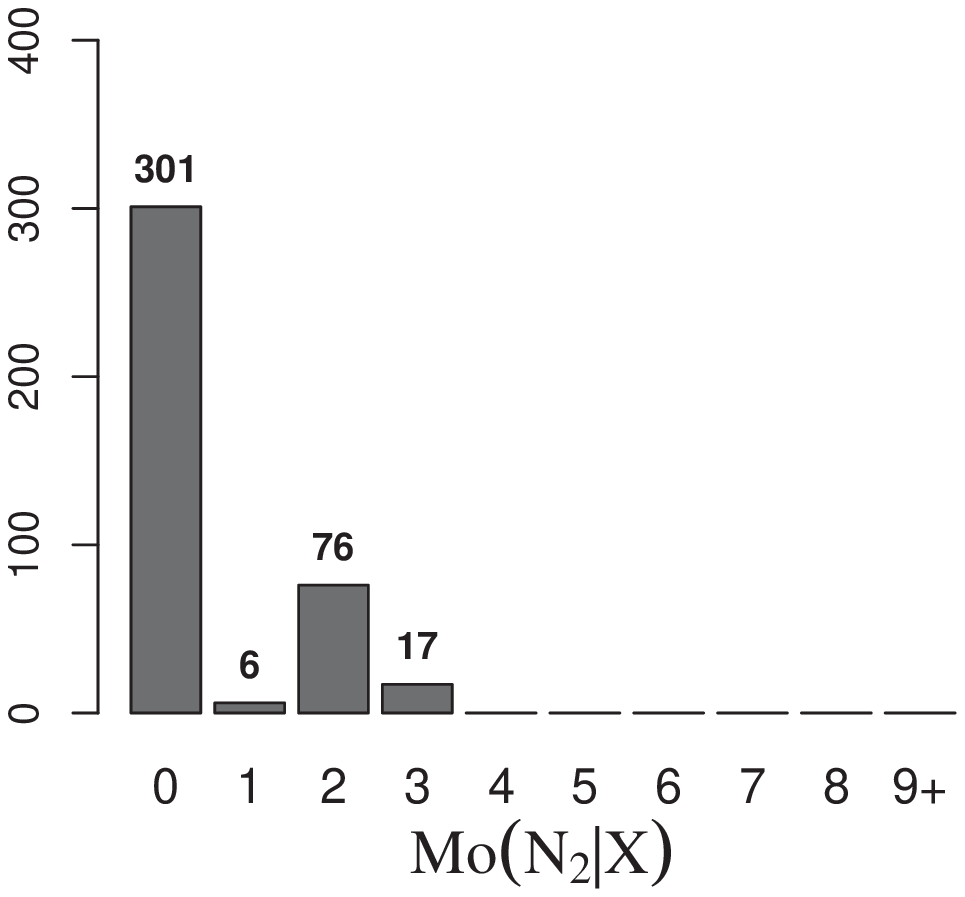}
	\label{fig:scene4_Nmode_P18_s2_c4}}\\
\caption{Counts distribution of the posterior modes of the number of changes in the variance ($N_2$) estimated for each Monte Carlo replication.}
\label{fig:scene4_Nmode_s2}
\end{figure}

BMCP  also provides more accurate estimates for the change point positions in the variance  (Table \ref{tab:scene4_mod_rho2}). As shown in Figure \ref{fig:scene4_Nmode_s2}, both models correctly estimate the true number of changes in the variance under configurations C2 and C3, with better performance of the BMCP. In scenarios C1 and C4,  DPM19  underestimates the number of changes in the variance indicating no change points in more than $75\%$ of the datasets while the BMCP overestimates this number for all data sets. Despite this, the most probable partitions (Table \ref{tab:scene4_mod_rho2}) under the BMCP correctly detected three changes in the variance and indicate that these changes occurred in the true positions or closer to them.

\begin{figure}[!htb]
%\begin{adjustwidth}{1cm}{1cm}
\centering
\subfigure[][BMCP (C1)]{
	\includegraphics[width=6cm, height=3.5cm, trim=0 2.5cm 0 0]{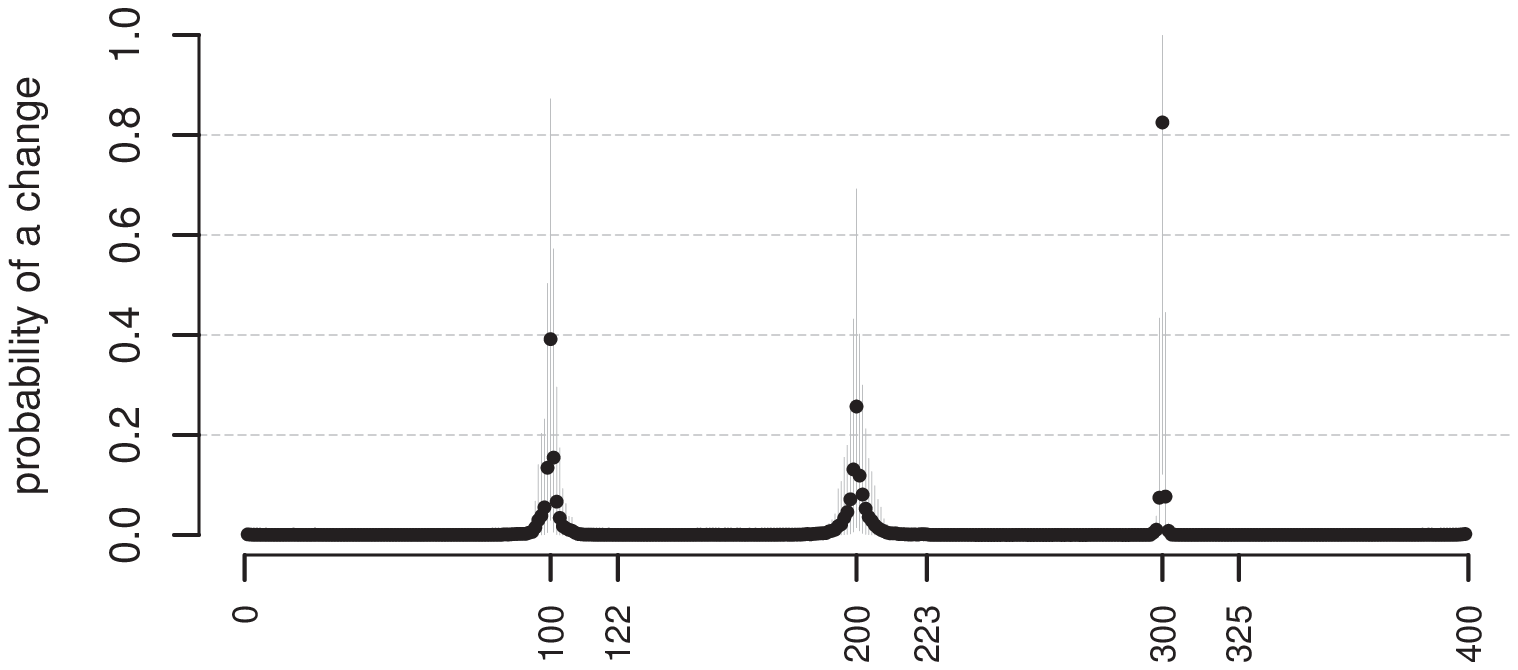}
	\label{fig:scene4_prob_IC_LP20_mu_c1}}\hspace{-.2in}
\subfigure[][DPM19 (C1)]{
	\includegraphics[width=6cm, height=3.5cm, trim=0 2.5cm 0 0]{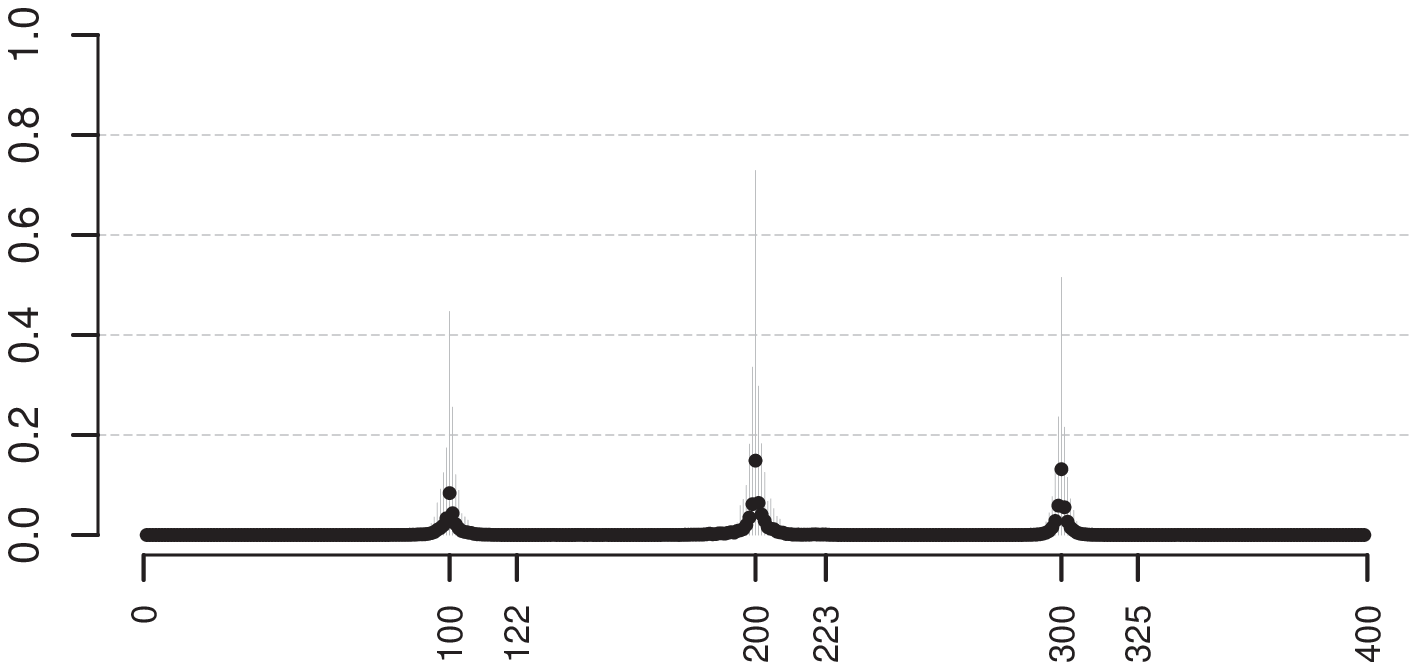}
	\label{fig:scene4_prob_IC_P18_mu_c1}}\\[-.35in]
\subfigure[][BMCP (C2)]{
	\includegraphics[width=6cm, height=3.5cm, trim=0 2.5cm 0 0]{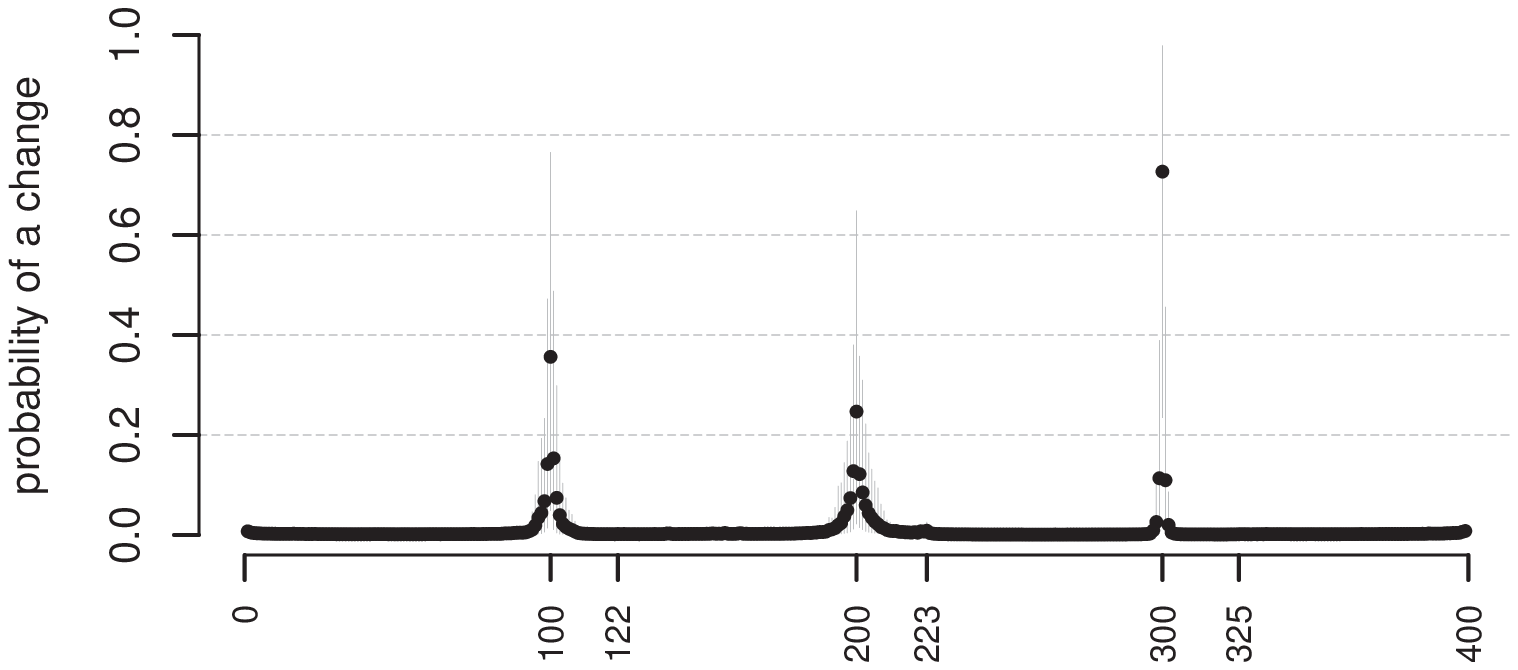}
	\label{fig:scene4_prob_IC_LP20_mu_c2}}\hspace{-.2in}
\subfigure[][DPM19 (C2)]{
	\includegraphics[width=6cm, height=3.5cm, trim=0 2.5cm 0 0]{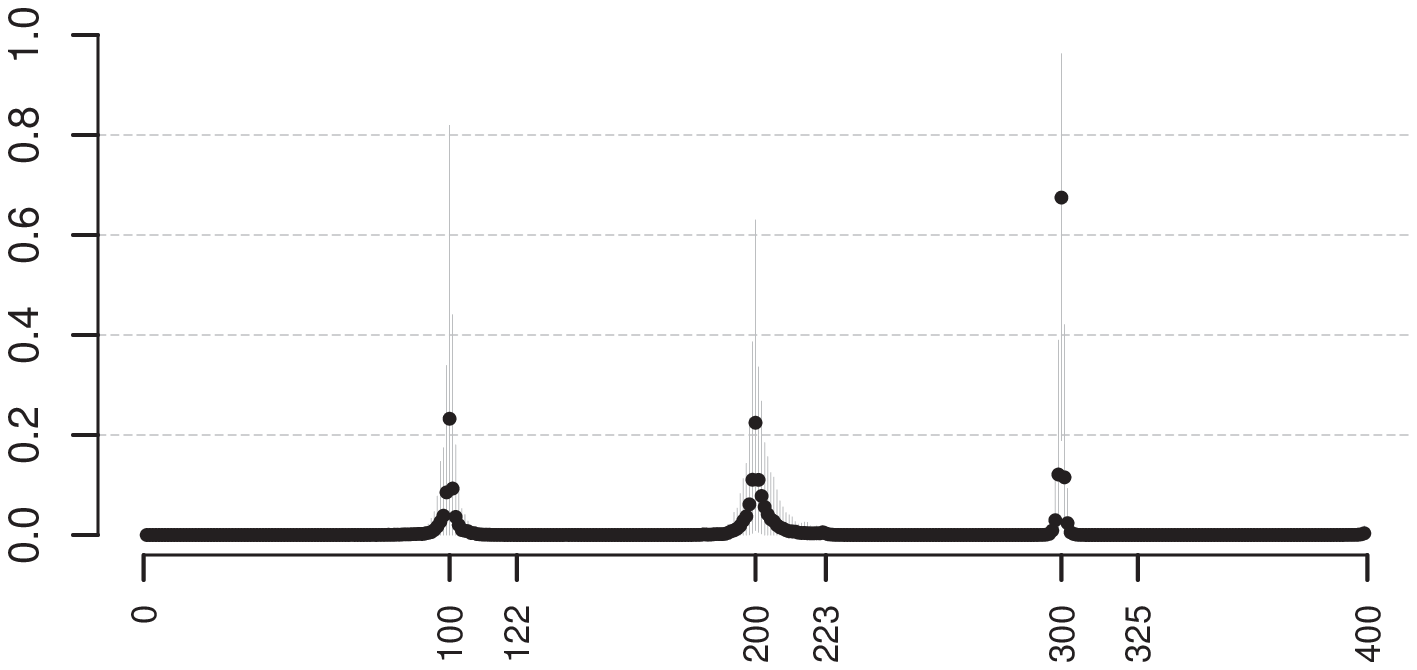}
	\label{fig:scene4_prob_IC_P18_mu_c2}}\\[-.35in]
\subfigure[][BMCP (C3)]{
 	\includegraphics[width=6cm, height=3.5cm, trim=0 2.5cm 0 0]{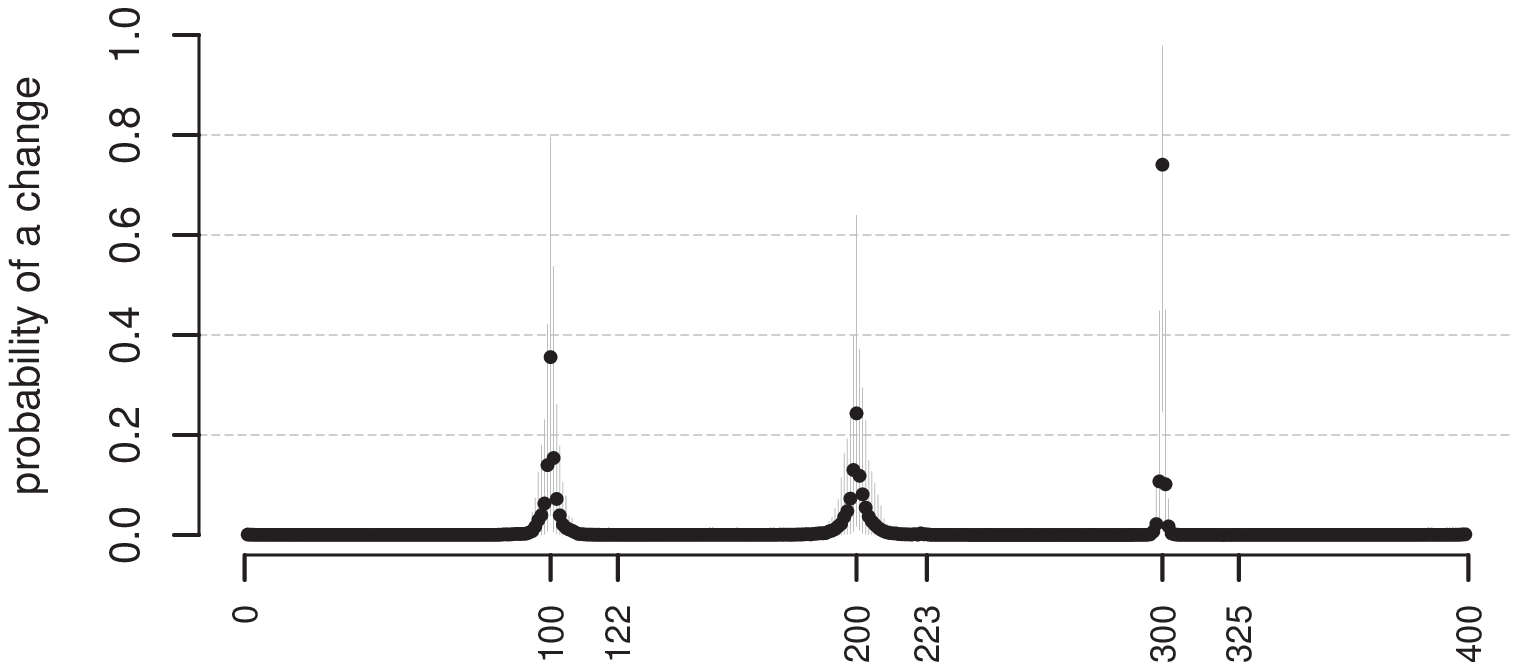}
 	\label{fig:scene4_prob_IC_LP20_mu_c3}}\hspace{-.2in}
\subfigure[][DPM19 (C3)]{
 	\includegraphics[width=6cm, height=3.5cm, trim=0 2.5cm 0 0]{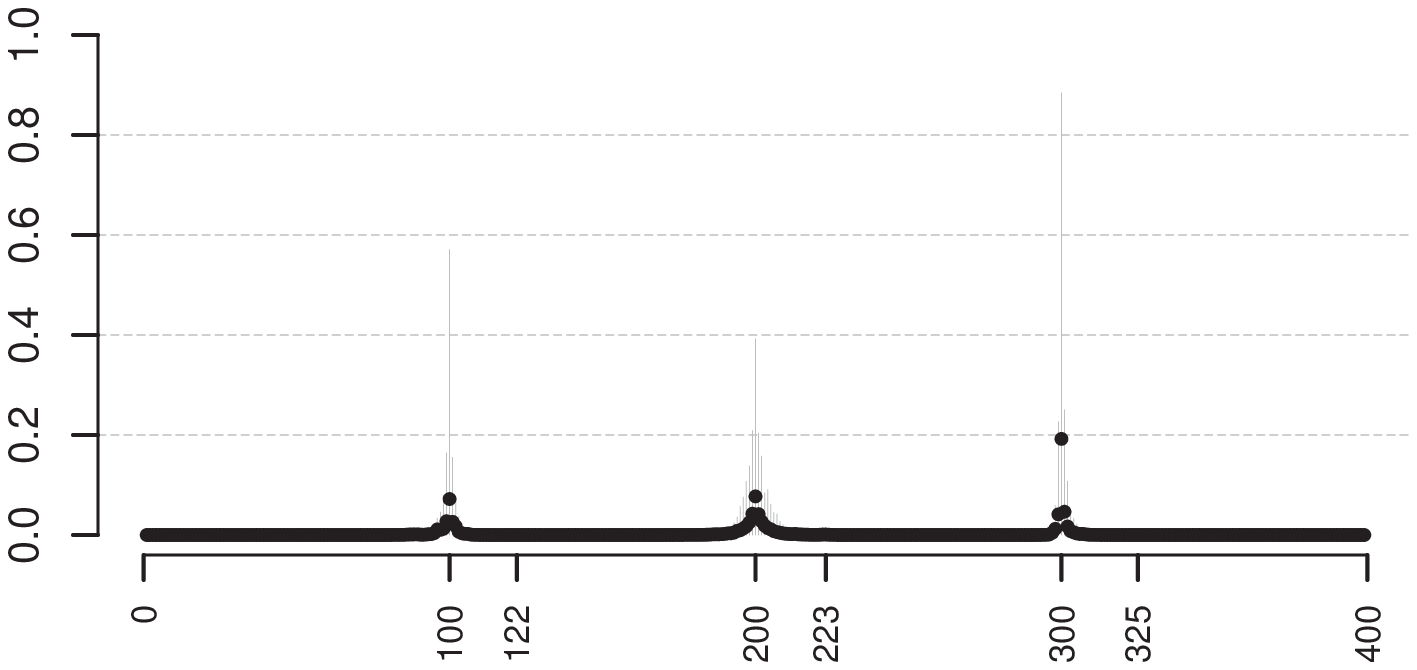}
 	\label{fig:scene4_prob_IC_P18_mu_c3}}\\[-.35in]
\subfigure[][BMCP (C4)]{
 	\includegraphics[width=6cm, height=3.5cm, trim=0 2.5cm 0 0]{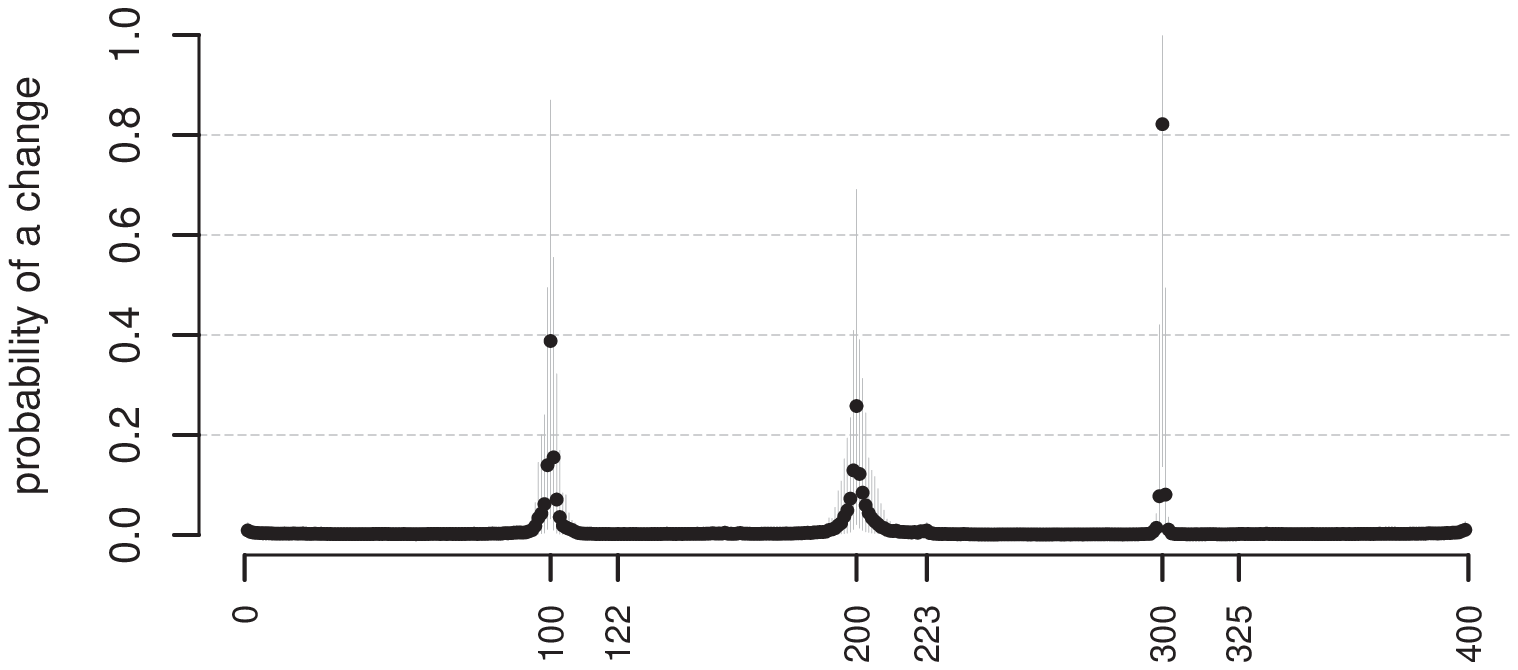}
 	\label{fig:scene4_prob_IC_LP20_mu_c4}}\hspace{-.2in}
\subfigure[][DPM19 (C4)]{
 	\includegraphics[width=6cm, height=3.5cm, trim=0 2.5cm 0 0]{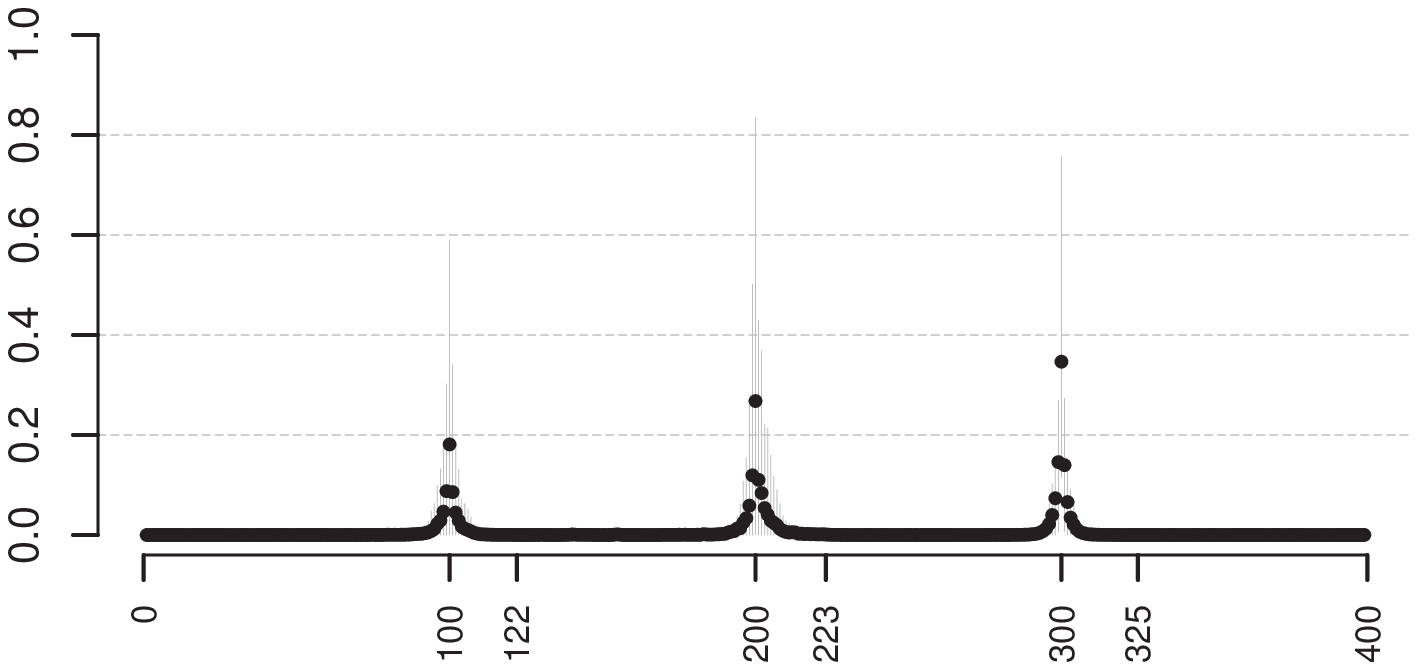}
 	\label{fig:scene4_prob_IC_P18_mu_c4}}\\
\caption{Average of the posterior probabilities of each instant to be an end point of $\rho_1$ (BMCP) and $E_1$ (DPM19) estimated for each Monte Carlo replication (dark bullets) and the $5\%$ and $95\%$ quantiles range of such probabilities.}
\label{fig:scene4_prob_IC_mu}
%\end{adjustwidth}
\end{figure}

\begin{figure}[!htb]
\centering
\subfigure[][BMCP (C1)]{
	\includegraphics[width=6cm, height=3.5cm, trim=0 2.5cm 0 0]{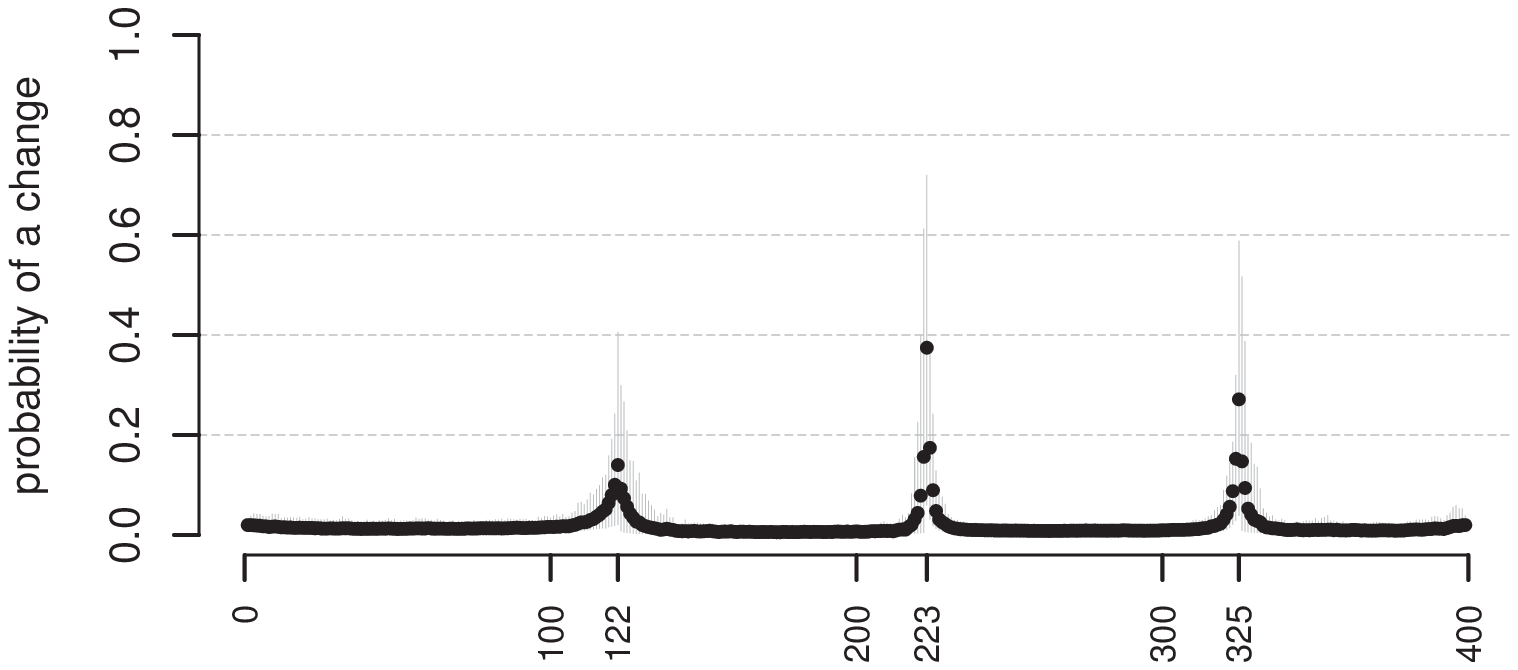}
	\label{fig:scene4_prob_IC_LP20_s2_c1}}\hspace{-.2in}
\subfigure[][DPM19 (C1)]{
	\includegraphics[width=6cm, height=3.5cm, trim=0 2.5cm 0 0]{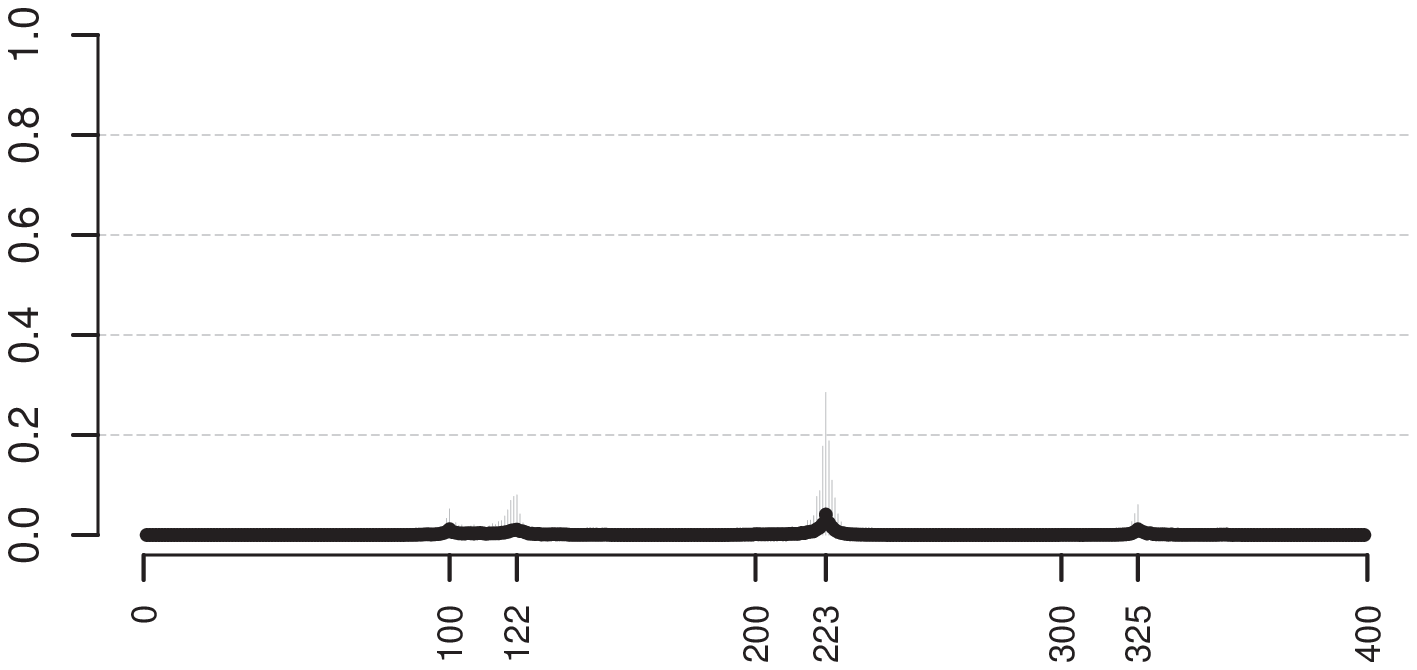}
	\label{fig:scene4_prob_IC_P18_s2_c1}}\\[-.35in]
\subfigure[][BMCP (C2)]{
	\includegraphics[width=6cm, height=3.5cm, trim=0 2.5cm 0 0]{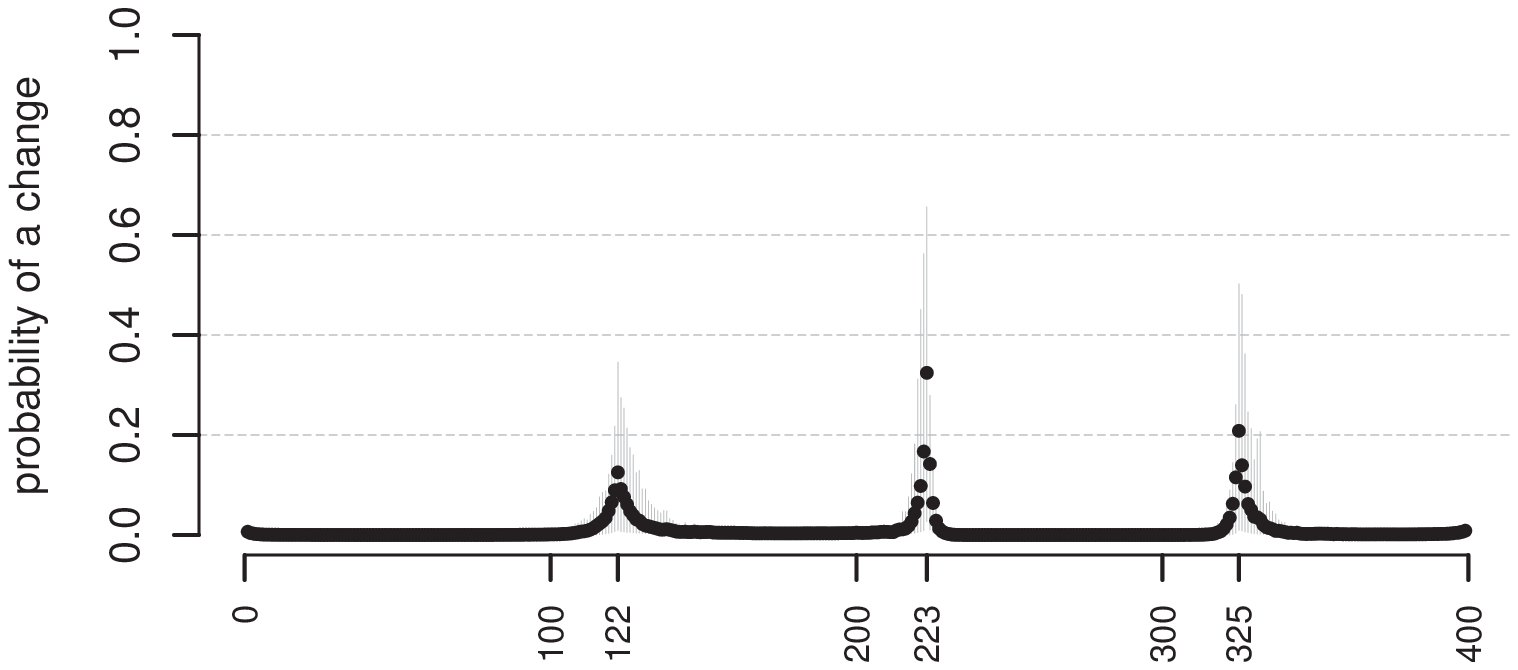}
	\label{fig:scene4_prob_IC_LP20_s2_c2}}\hspace{-.2in}
\subfigure[][DPM19 (C2)]{
	\includegraphics[width=6cm, height=3.5cm, trim=0 2.5cm 0 0]{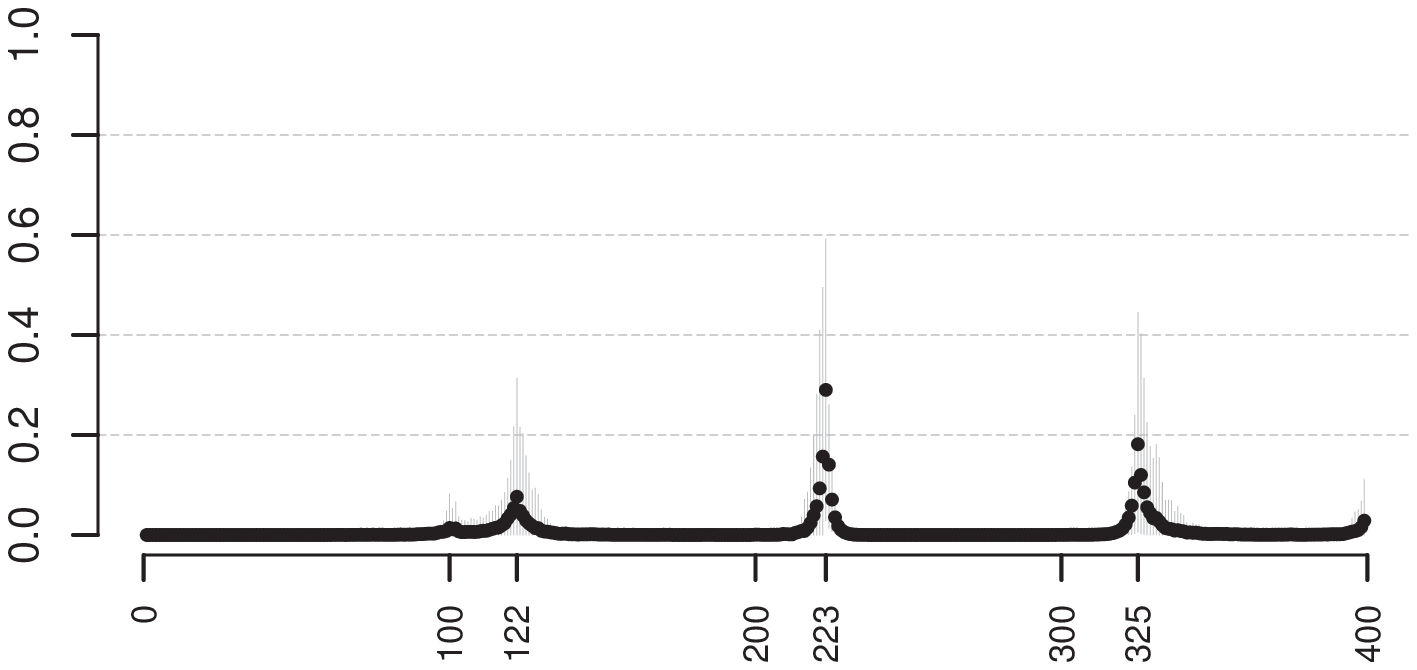}
	\label{fig:scene4_prob_IC_P18_s2_c2}}\\[-.35in]
\subfigure[][BMCP (C3)]{
	\includegraphics[width=6cm, height=3.5cm, trim=0 2.5cm 0 0]{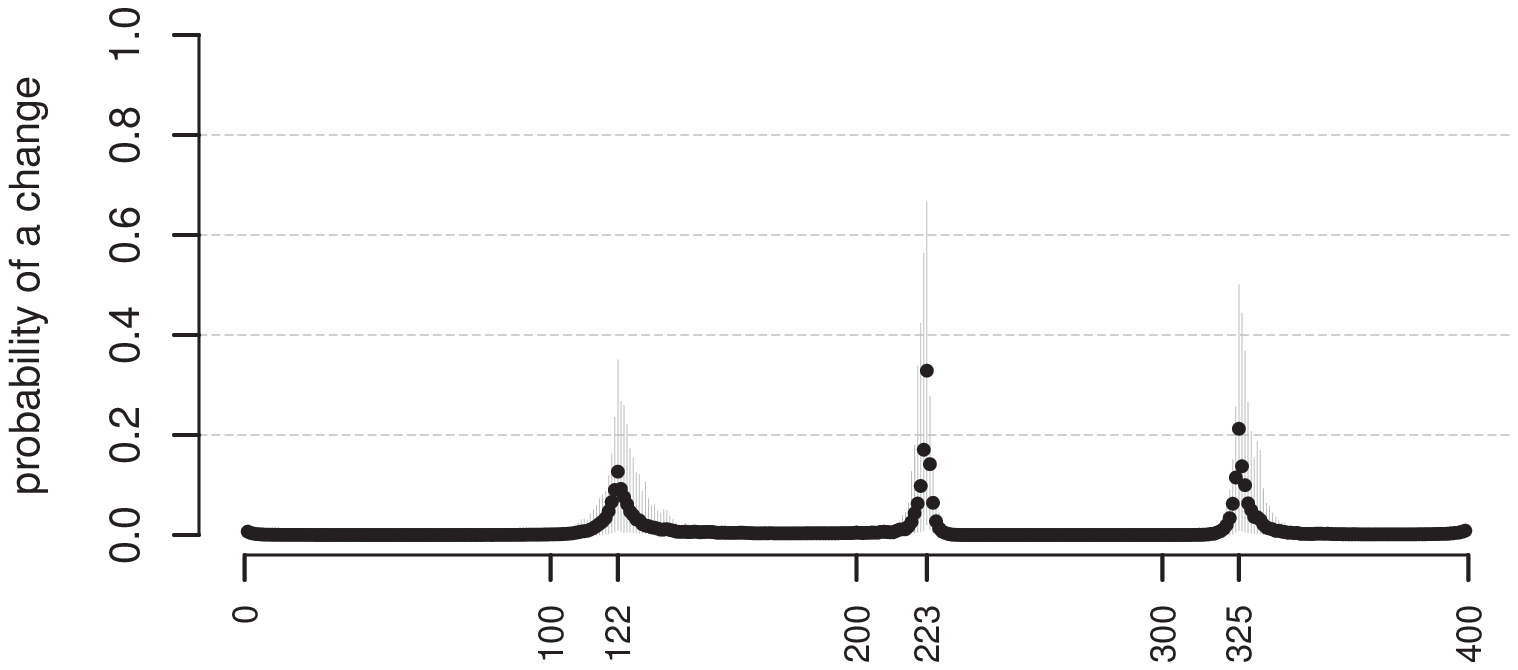}
	\label{fig:scene4_prob_IC_LP20_s2_c3}}\hspace{-.2in}
\subfigure[][DPM19 (C3)]{
	\includegraphics[width=6cm, height=3.5cm, trim=0 2.5cm 0 0]{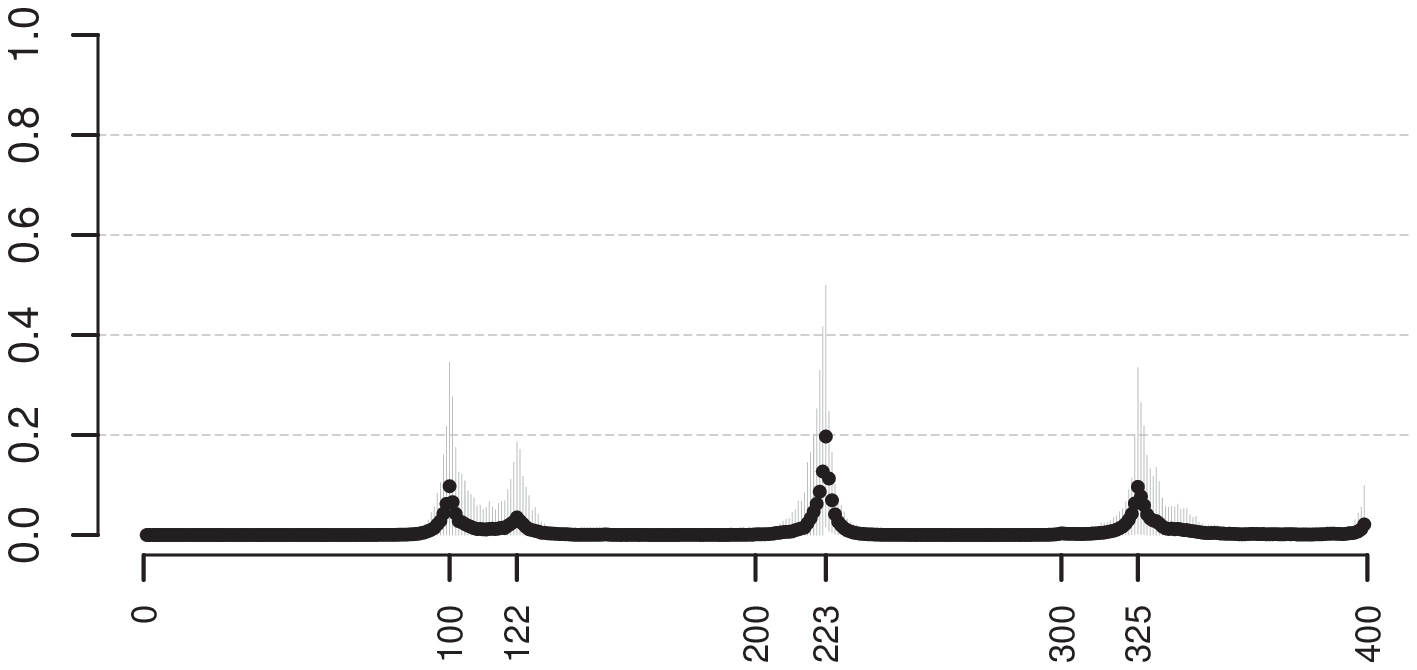}
	\label{fig:scene4_prob_IC_P18_s2_c3}}\\[-.35in]
\subfigure[][BMCP (C4)]{
	\includegraphics[width=6cm, height=3.5cm, trim=0 2.5cm 0 0]{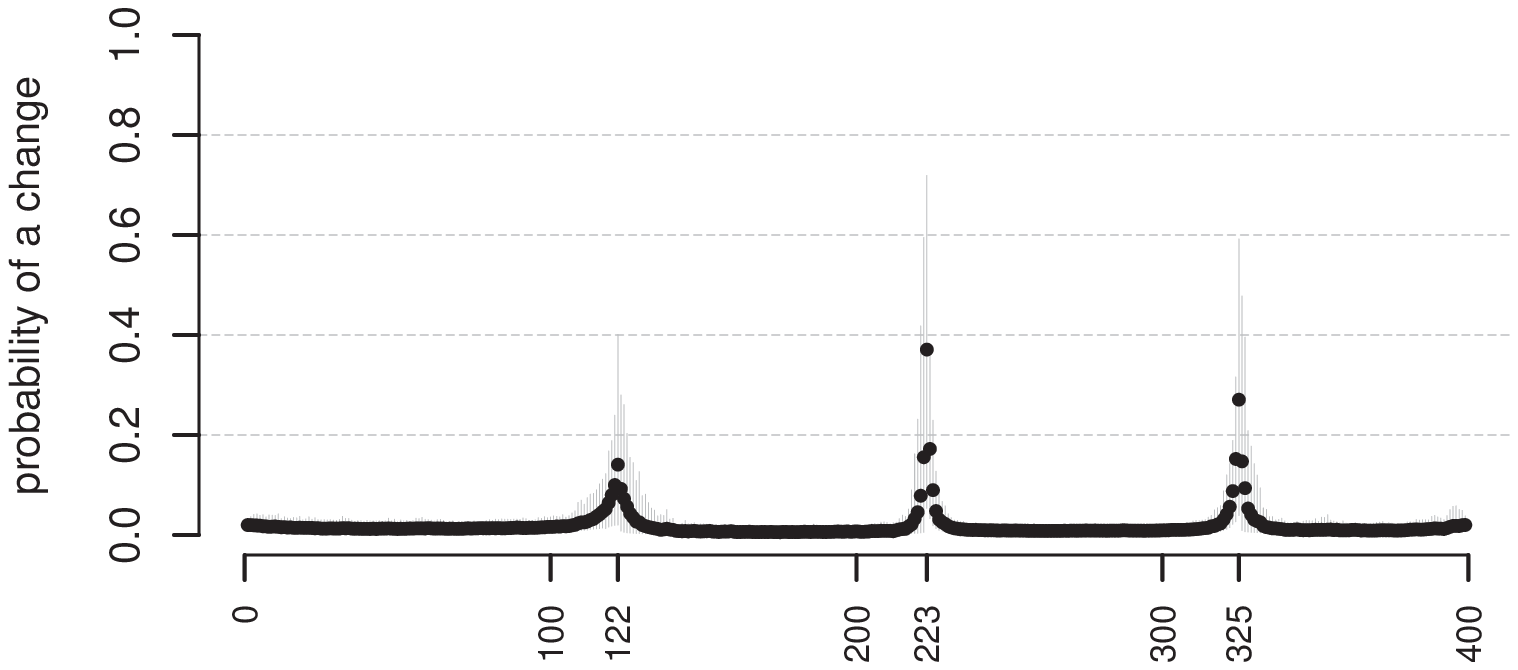}
	\label{fig:scene4_prob_IC_LP20_s2_c4}}\hspace{-.2in}
\subfigure[][DPM19 (C4)]{
	\includegraphics[width=6cm, height=3.5cm, trim=0 2.5cm 0 0]{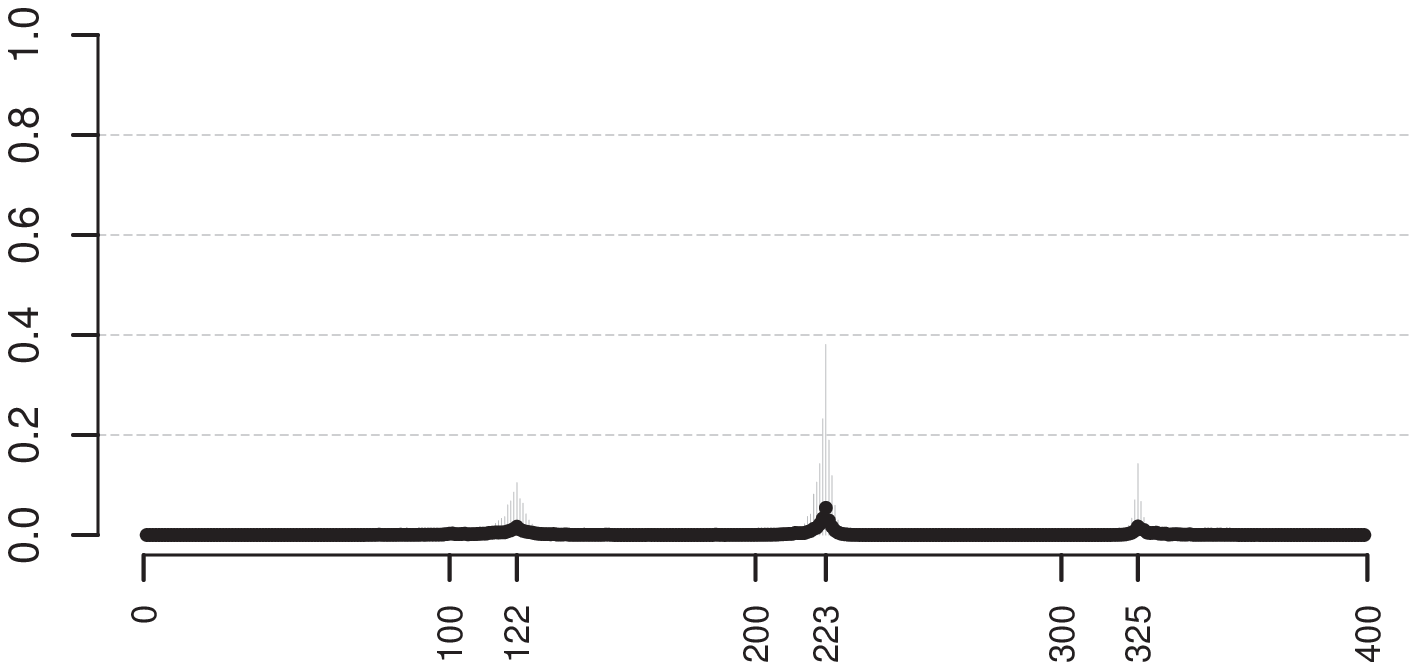}
	\label{fig:scene4_prob_IC_P18_s2_c4}}\\
\caption{Average of the posterior probabilities of each instant to be an end point of $\rho_2$ (BMCP) and $E_2$ (DPM19) estimated for each Monte Carlo replication (dark bullets) and the $5\%$ and $95\%$ quantiles range of such probabilities.}
\label{fig:scene4_prob_IC_s2}
\end{figure}

The probability of each position being a change displayed in Figures \ref{fig:scene4_prob_IC_mu} and \ref{fig:scene4_prob_IC_s2} also show a better performance of the BMCP to multiple change point identification in the mean and the variance for all prior specifications.

\begin{table}[!htb]
\centering 
\footnotesize
\begingroup
\renewcommand{\arraystretch}{.6} % Default value: 1
\begin{tabular}{p{0.3\textwidth}r}
	\\[-1ex]
	\hline \\[-1.8ex] 
	${Mo(\rho_1|\bm{X})}$ (C1) & \# \\ 
	\hline \\[-1.8ex]  
	$\{0,100,200,300,400\}$ & 71 \\ 
	$\{0,100,199,300,400\}$ & 30 \\ 
	$\{0,100,201,300,400\}$ & 22 \\ 
	$\{0,101,199,300,400\}$ & 17 \\ 
	$\{0,101,200,300,400\}$ & 16 \\ 
	\hline \\[-1.8ex] 
\end{tabular}\hspace{.05in}
\begin{tabular}{p{0.3\textwidth}r}
	\\[-1ex]
	\hline \\[-1.8ex] 
	${Mo(E_1|\bm{X})}$ (C1) & \# \\ 
	\hline \\[-1.8ex]  
	$\{0,400\}$             & 177 \\ 
	$\{0,200,300,400\}$     &  27 \\ 
	$\{0,100,200,400\}$     &  17 \\ 
	$\{0,100,200,300,400\}$ &  12 \\ 
	$\{0,199,300,400\}$     &   9 \\ 
	\hline \\[-1.8ex] 
\end{tabular}\\
\begin{tabular}{p{0.3\textwidth}r}
	\\[-1ex]
	\hline \\[-1.8ex] 
	${Mo(\rho_1|\bm{X})}$ (C2) & \# \\ 
	\hline \\[-1.8ex]  
	$\{0,100,200,300,400\}$ & 74 \\ 
	$\{0,100,199,300,400\}$ & 26 \\ 
	$\{0,101,200,300,400\}$ & 22 \\ 
	$\{0,100,201,300,400\}$ & 19 \\ 
	$\{0,100,202,300,400\}$ & 14 \\ 
	\hline \\[-1.8ex] 
\end{tabular}\hspace{.05in}
\begin{tabular}{p{0.3\textwidth}r}
	\\[-1ex]
	\hline \\[-1.8ex] 
	${Mo(E_1|\bm{X})}$ (C2) & \# \\ 
	\hline \\[-1.8ex]  
	$\{0,100,200,300,400\}$ & 53 \\ 
	$\{0,200,300,400\}$     & 37 \\ 
	$\{0,199,300,400\}$     & 21 \\ 
	$\{0,100,201,300,400\}$ & 17 \\ 
	$\{0,100,199,300,400\}$ & 13 \\ 
	\hline \\[-1.8ex] 
\end{tabular}\\
\begin{tabular}{p{0.3\textwidth}r}
	\\[-1ex]
	\hline \\[-1.8ex] 
	${Mo(\rho_1|\bm{X})}$ (C3) & \# \\ 
	\hline \\[-1.8ex]  
	$\{0,100,200,300,400\}$ & 77 \\ 
	$\{0,100,199,300,400\}$ & 21 \\ 
	$\{0,100,201,300,400\}$ & 18 \\ 
	$\{0,101,200,300,400\}$ & 17 \\ 
	$\{0,100,202,300,400\}$ & 12 \\ 
	\hline \\[-1.8ex] 
\end{tabular}\hspace{.05in}
\begin{tabular}{p{0.3\textwidth}r}
	\\[-1ex]
	\hline \\[-1.8ex] 
	${Mo(E_1|\bm{X})}$ (C3) & \# \\ 
	\hline \\[-1.8ex]  
	$\{0,400\}$         & 203 \\ 
	$\{0,300,400\}$     &  41 \\ 
	$\{0,200,300,400\}$ &  17 \\ 
	$\{0,100,200,400\}$ &  10 \\ 
	$\{0,301,400\}$     &   9 \\ 
	\hline \\[-1.8ex] 
\end{tabular}\\
\begin{tabular}{p{0.3\textwidth}r}
	\\[-1ex]
	\hline \\[-1.8ex] 
	${Mo(\rho_1|\bm{X})}$ (C4) & \# \\ 
	\hline \\[-1.8ex]  
	$\{0,100,200,300,400\}$ & 73 \\ 
	$\{0,100,199,300,400\}$ & 25 \\ 
	$\{0,101,200,300,400\}$ & 23 \\ 
	$\{0,100,201,300,400\}$ & 19 \\ 
	$\{0,100,202,300,400\}$ & 15 \\ 
	\hline \\[-1.8ex] 
\end{tabular}\hspace{.05in}
\begin{tabular}{p{0.3\textwidth}r}
	\\[-1ex]
	\hline \\[-1.8ex] 
	${Mo(E_1|\bm{X})}$ (C4) & \# \\ 
	\hline \\[-1.8ex]  
	$\{0,100,200,300,400\}$ & 43 \\ 
	$\{0,200,300,400\}$     & 39 \\ 
	$\{0,100,199,300,400\}$ & 17 \\ 
	$\{0,202,300,400\}$     & 15 \\ 
	$\{0,100,201,300,400\}$ & 12 \\ 
	\hline \\[-1.8ex] 
\end{tabular}
\endgroup
\caption{Top five posterior modes of $\rho_1$ (BMCP) and $E_1$ (DPM19) estimated for each Monte Carlo replication.}
\label{tab:scene4_mod_rho1}
\end{table}	

\begin{table}[!htb]
\centering 
\footnotesize
\begingroup
\renewcommand{\arraystretch}{.6} % Default value: 1
\begin{tabular}{p{0.3\textwidth}r}
	\\[-1ex]
	\hline \\[-1.8ex] 
	${Mo(\rho_2|\bm{X})}$ (C1) & \# \\ 
	\hline \\[-1.8ex]  
	$\{0,122,223,325,400\}$ & 25 \\ 
	$\{0,122,222,325,400\}$ & 13 \\ 
	$\{0,123,223,325,400\}$ & 10 \\ 
	$\{0,121,223,325,400\}$ &  9 \\ 
	$\{0,122,223,327,400\}$ &  8 \\ 
	\hline \\[-1.8ex] 
\end{tabular}\hspace{.05in}
\begin{tabular}{p{0.3\textwidth}r}
	\\[-1ex]
	\hline \\[-1.8ex] 
	${Mo(E_2|\bm{X})}$ (C1) & \# \\ 
	\hline \\[-1.8ex]  
	$\{0,400\}$         & 310 \\ 
	$\{0,123,223,400\}$ &   3 \\ 
	$\{0,223,325,400\}$ &   3 \\ 
	$\{0,100,221,400\}$ &   2 \\ 
	$\{0,100,223,400\}$ &   2 \\ 
	\hline \\[-1.8ex] 
\end{tabular}\\
\begin{tabular}{p{0.3\textwidth}r}
	\\[-1ex]
	\hline \\[-1.8ex] 
	${Mo(\rho_2|\bm{X})}$ (C2) & \# \\ 
	\hline \\[-1.8ex]  
	$\{0,122,223,325,400\}$ & 17 \\ 
	$\{0,122,222,325,400\}$ & 10 \\ 
	$\{0,123,223,325,400\}$ &  9 \\ 
	$\{0,124,223,325,400\}$ &  8 \\ 
	$\{0,125,223,325,400\}$ &  7 \\ 
	\hline \\[-1.8ex] 
\end{tabular}\hspace{.05in}
\begin{tabular}{p{0.3\textwidth}r}
	\\[-1ex]
	\hline \\[-1.8ex] 
	${Mo(E_2|\bm{X})}$ (C2) & \# \\ 
	\hline \\[-1.8ex]  
	$\{0,223,325,400\}$     & 22 \\ 
	$\{0,223,326,400\}$     & 14 \\ 
	$\{0,122,223,325,400\}$ &  8 \\ 
	$\{0,221,325,400\}$     &  8 \\ 
	$\{0,222,325,400\}$     &  8 \\ 
	\hline \\[-1.8ex] 
\end{tabular}\\
\begin{tabular}{p{0.3\textwidth}r}
	\\[-1ex]
	\hline \\[-1.8ex] 
	${Mo(\rho_2|\bm{X})}$ (C3) & \# \\ 
	\hline \\[-1.8ex]  
	$\{0,122,223,325,400\}$ & 16 \\ 
	$\{0,123,223,325,400\}$ &  9 \\ 
	$\{0,122,222,325,400\}$ &  8 \\ 
	$\{0,122,223,327,400\}$ &  7 \\ 
	$\{0,124,223,325,400\}$ &  7 \\ 
	\hline \\[-1.8ex] 
\end{tabular}\hspace{.05in}
\begin{tabular}{p{0.3\textwidth}r}
	\\[-1ex]
	\hline \\[-1.8ex] 
	${Mo(E_2|\bm{X})}$ (C3) & \# \\ 
	\hline \\[-1.8ex]  
	$\{0,223,325,400\}$ &  9 \\ 
	$\{0,100,223,325,400\}$ &  5 \\ 
	$\{0,223,327,400\}$ &  4 \\ 
	$\{0,100,223,327,400\}$ &  3 \\ 
	$\{0,100,223,331,400\}$ &  3 \\ 
	\hline \\[-1.8ex] 
\end{tabular}\\
\begin{tabular}{p{0.3\textwidth}r}
	\\[-1ex]
	\hline \\[-1.8ex] 
	${Mo(\rho_2|\bm{X})}$ (C4) & \# \\ 
	\hline \\[-1.8ex]  
	$\{0,122,223,325,400\}$ & 18 \\ 
	$\{0,123,223,325,400\}$ & 10 \\ 
	$\{0,121,223,325,400\}$ &  9 \\ 
	$\{0,122,222,325,400\}$ &  9 \\ 
	$\{0,122,223,327,400\}$ &  9 \\ 
	\hline \\[-1.8ex] 
\end{tabular}\hspace{.05in}
\begin{tabular}{p{0.3\textwidth}r}
	\\[-1ex]
	\hline \\[-1.8ex] 
	${Mo(E_2|\bm{X})}$ (C4) & \# \\ 
	\hline \\[-1.8ex]  
	$\{0,400\}$         & 301 \\ 
	$\{0,223,325,400\}$ &   4 \\ 
	$\{0,123,223,400\}$ &   3 \\ 
	$\{0,223,400\}$     &   3 \\ 
	$\{0,223,327,400\}$ &   3 \\ 
	\hline \\[-1.8ex] 
\end{tabular}
\endgroup
\caption{Top five posterior modes of $\rho_2$ (BMCP) and $E_2$ (DPM19) estimated for each Monte Carlo replication.}
\label{tab:scene4_mod_rho2}
\end{table}

\FloatBarrier
\newpage
\subsection{Additional results for the Case Study 1}\label{supp2_IR}

\begin{figure}[!htbp]
\begin{adjustwidth}{-.4cm}{-.4cm}
	\centering
	\subfigure[][$E_1$ ($m_1=m_2=3$)]{
		\includegraphics[width=4cm, height=3.5cm,trim=0 2cm 0 0]{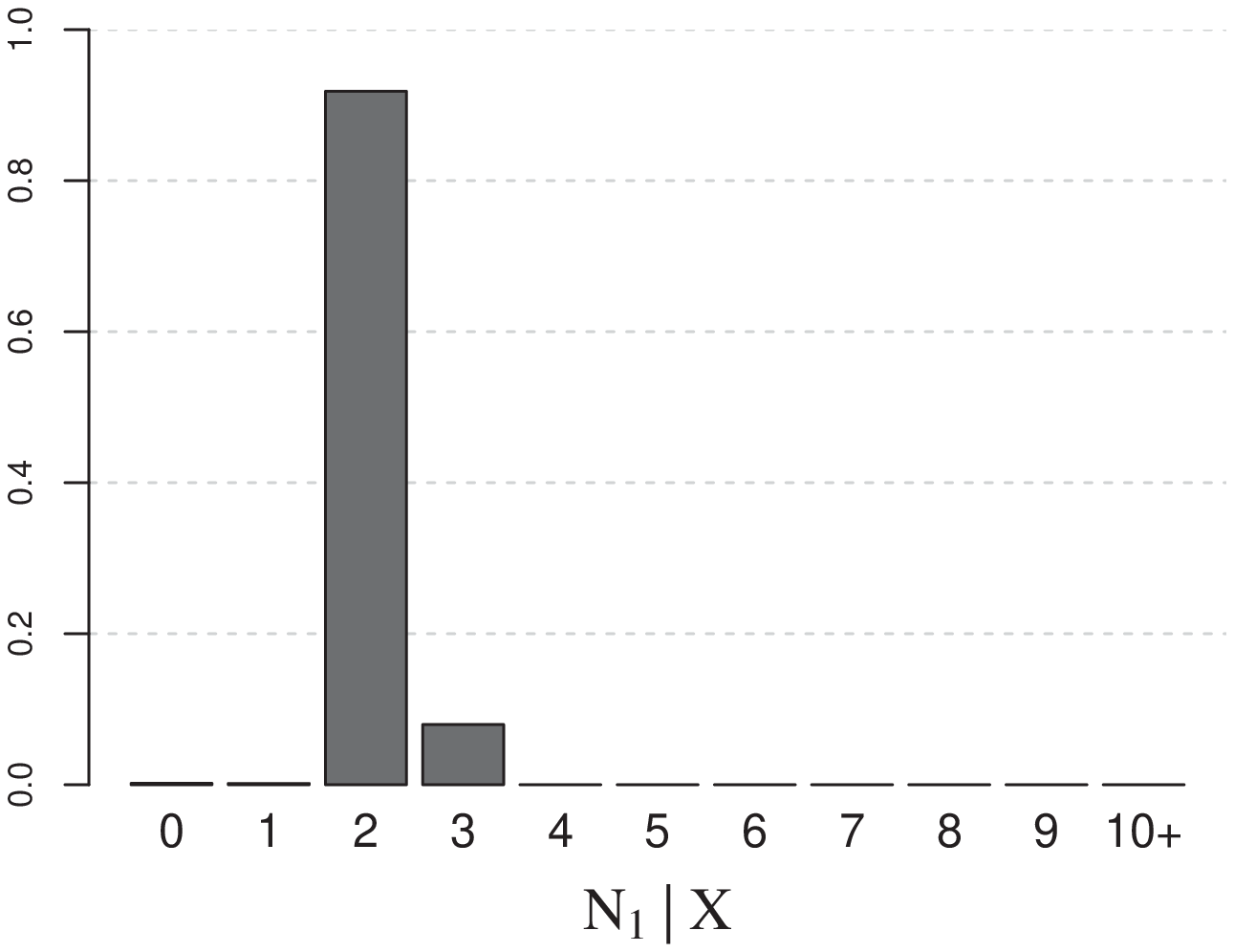}
		\label{fig:IR_N_P18_mu_m3}}%\hspace{.1in}
	\subfigure[][$E_2$ ($m_1=m_2=3$)]{
		\includegraphics[width=4cm, height=3.5cm, trim=0 2cm 0 0]{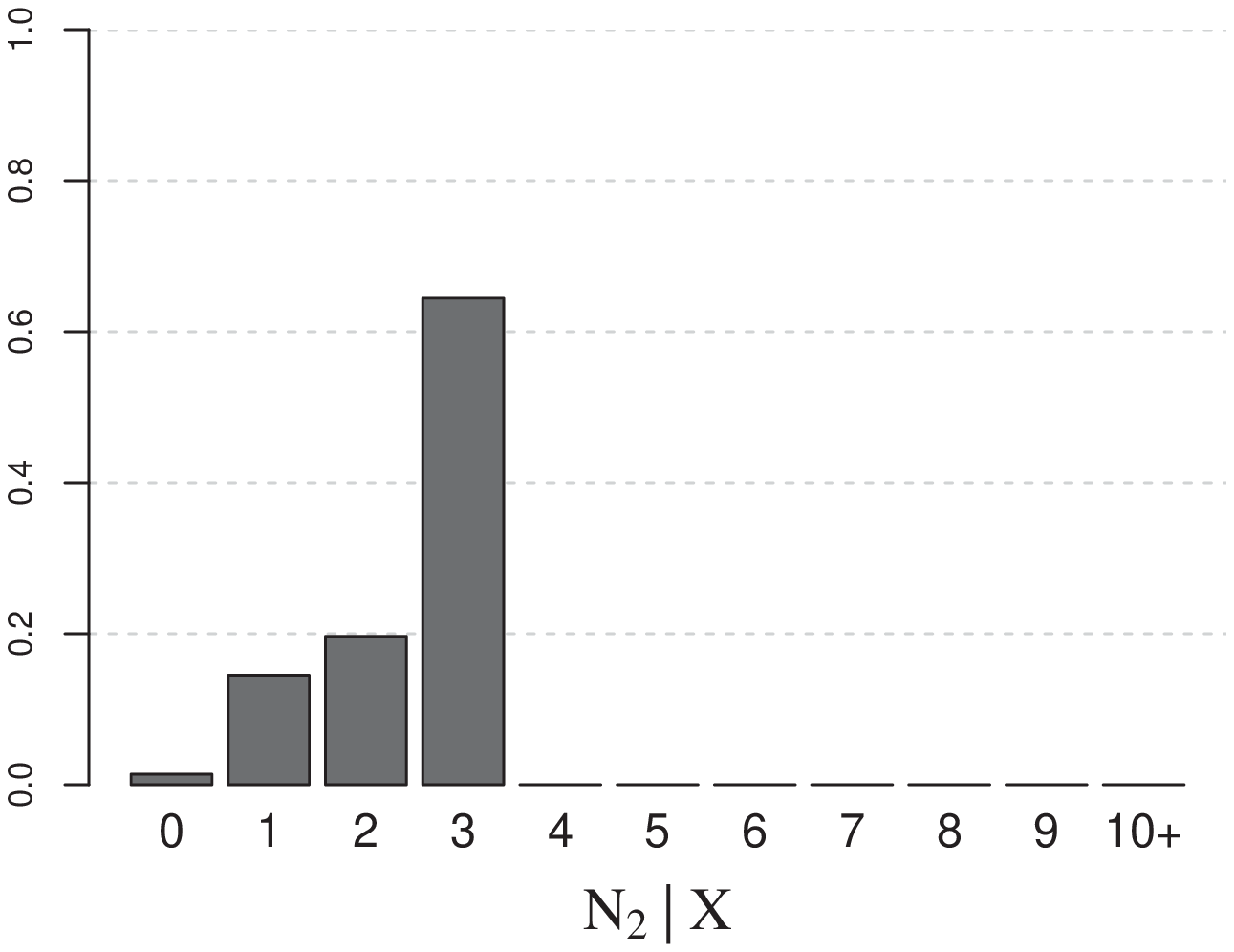}
		\label{fig:IR_N_P18_s2_m3}}%\\[-.1in]
	\subfigure[][$E_1$ ($m_1=m_2=100$)]{
		\includegraphics[width=4cm, height=3.5cm,trim=0 2cm 0 0]{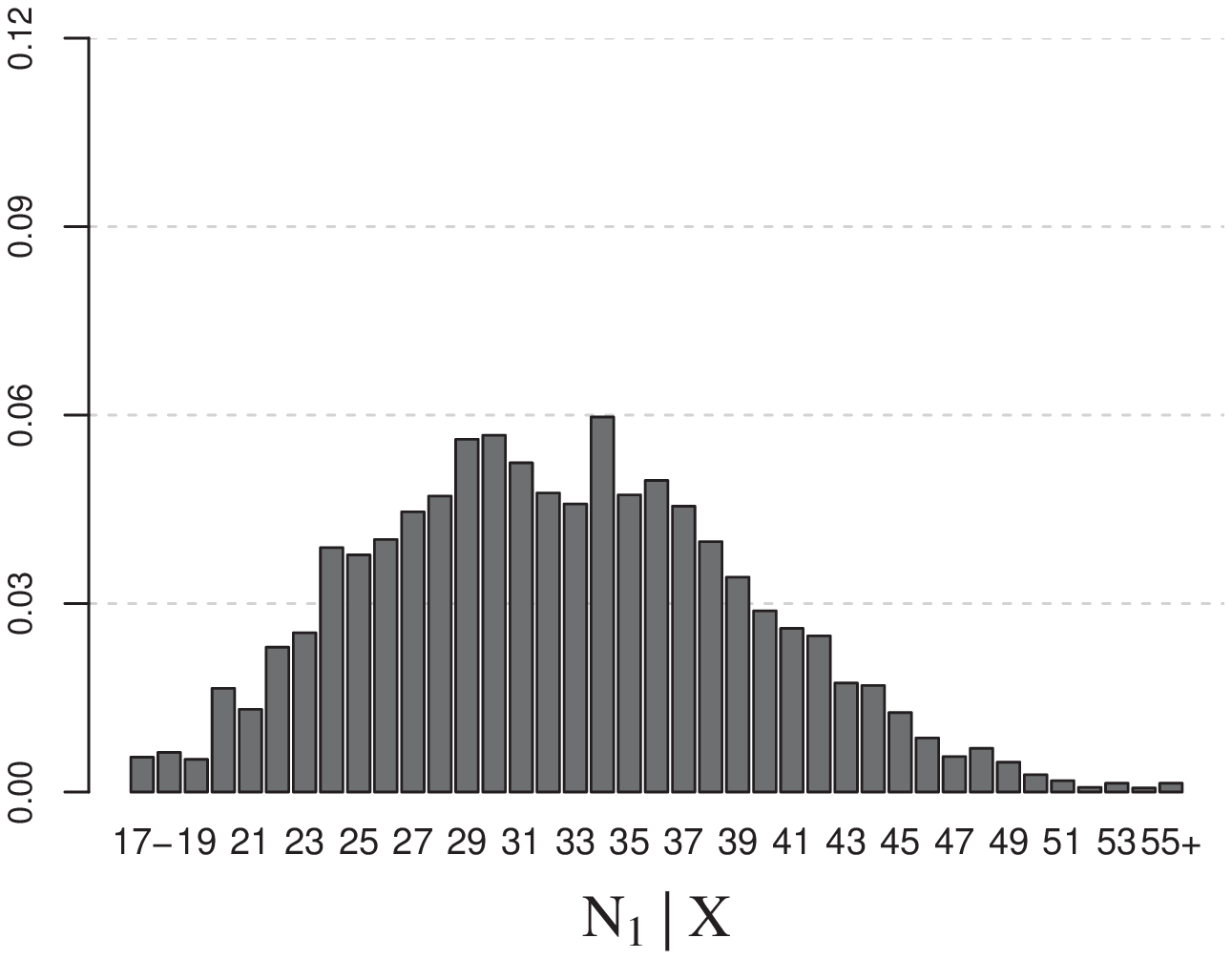}
		\label{fig:IR_N_P18_mu_m100}}%\hspace{.1in}
	\subfigure[][$E_2$ ($m_1=m_2=100$)]{
		\includegraphics[width=4cm, height=3.5cm, trim=0 2cm 0 0]{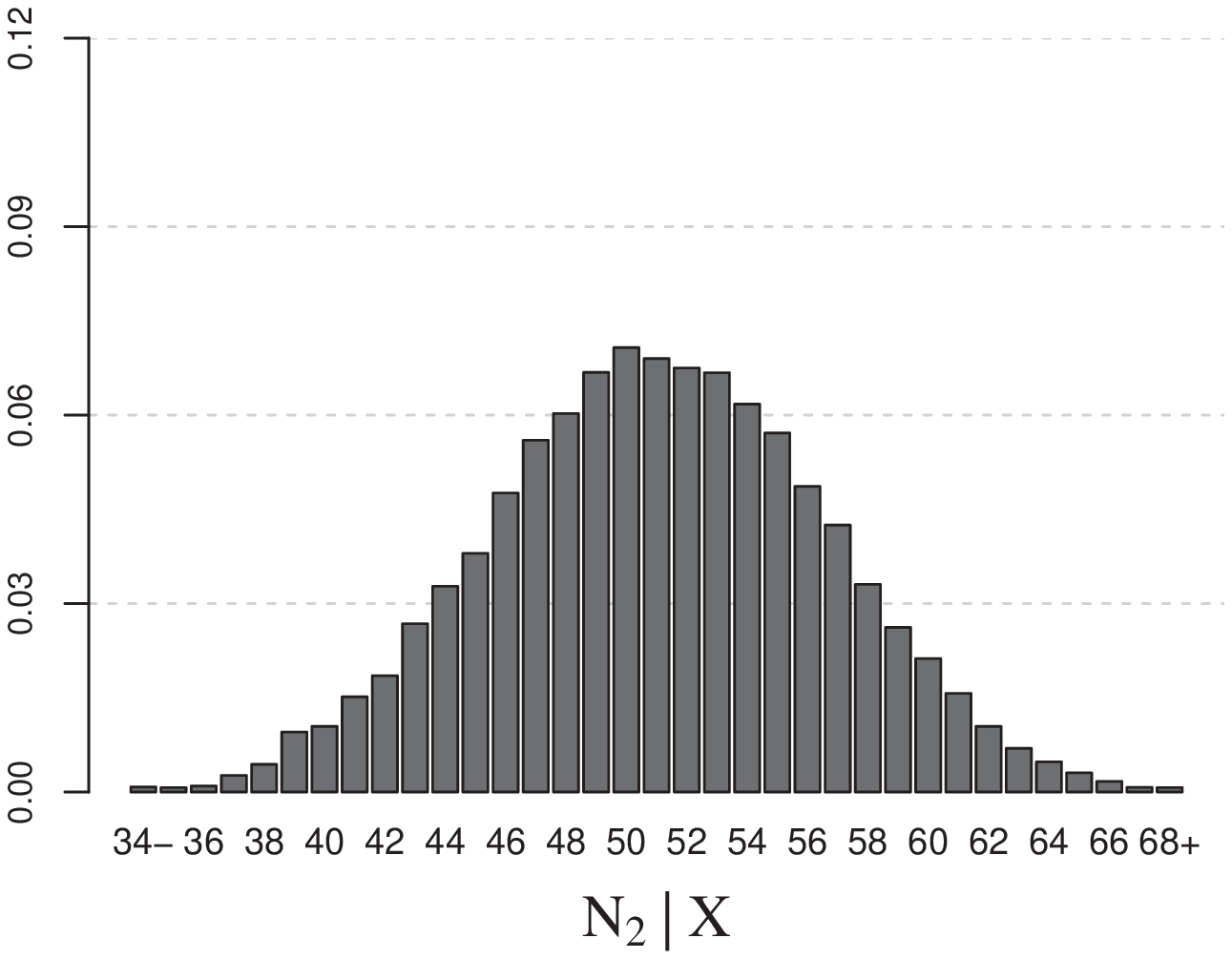}
		\label{fig:IR_N_P18_s2_m100}}
	\caption{\small Posterior distributions for the number of changes estimated by DPM19 model for the US ex-post Real Interest Rate dataset, considering different values for $m_1$ and $m_2$.}
	\label{fig:IR_N_m}
\end{adjustwidth}
\end{figure}

 We fitted the DPM19 model for the dataset considered in Section \ref{secReal_IR} assuming  the same prior distributions but different specifications for $m_1$ and $m_2$. Assuming $m_1=m_2=3$, results related to the estimates of the means are comparable to the ones obtained fitting the BMCP.
However, the posterior distribution of the number of changes in the variance is concentrated in the pre-fixed upper bound $m_2=3$ (Figure \ref{fig:IR_N_P18_s2_m3}), while BMCP estimates only one change in the variance.

\begin{figure}[!htbp]
\begin{adjustwidth}{-.4cm}{-.4cm}
	\centering		
	\subfigure[][$E_1$ ($m_1=m_2=3$)]{
		\includegraphics[width=4cm, height=3.5cm, trim=0 2cm 0 0]{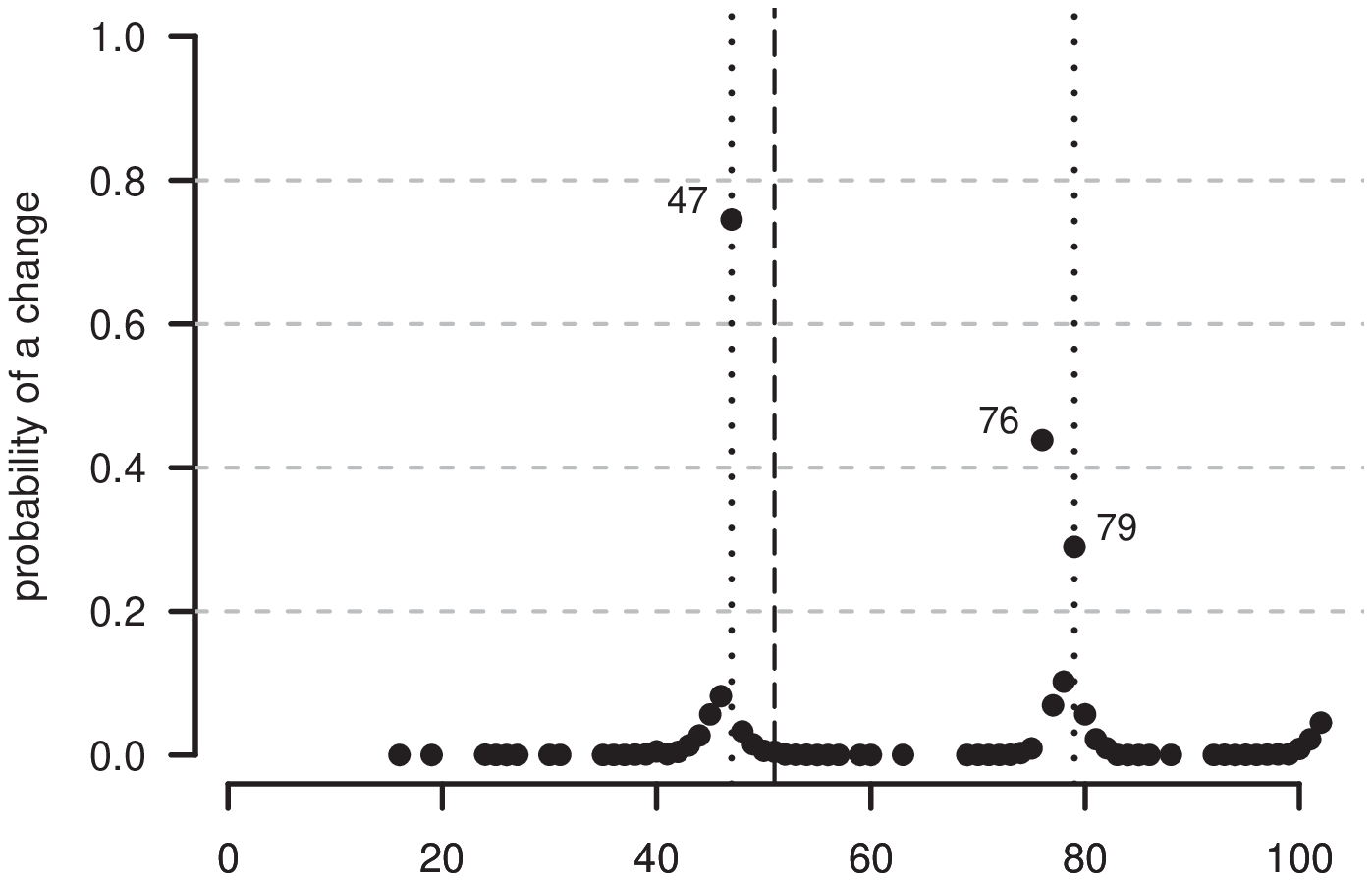}
		\label{fig:IR_prob_P18_mu_m3}}%\hspace{.1in}
	\subfigure[][$E_2$ ($m_1=m_2=3$)]{
		\includegraphics[width=4cm, height=3.5cm, trim=0 2cm 0 0]{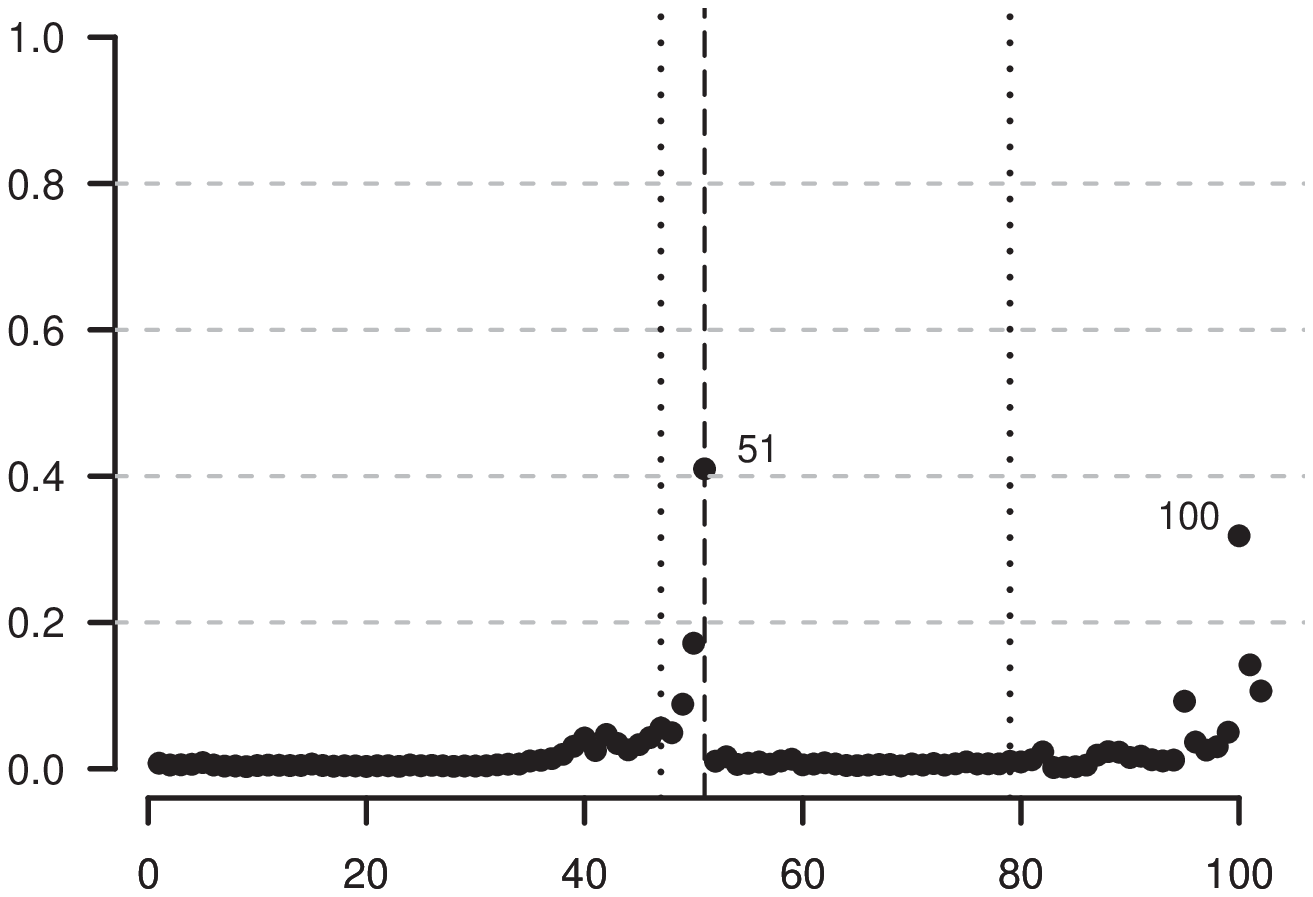}
		\label{fig:IR_prob_P18_s2_m3}}%\\[-.1in]
	\subfigure[][$E_1$ ($m_1=m_2=100$)]{
		\includegraphics[width=4cm, height=3.5cm, trim=0 2cm 0 0]{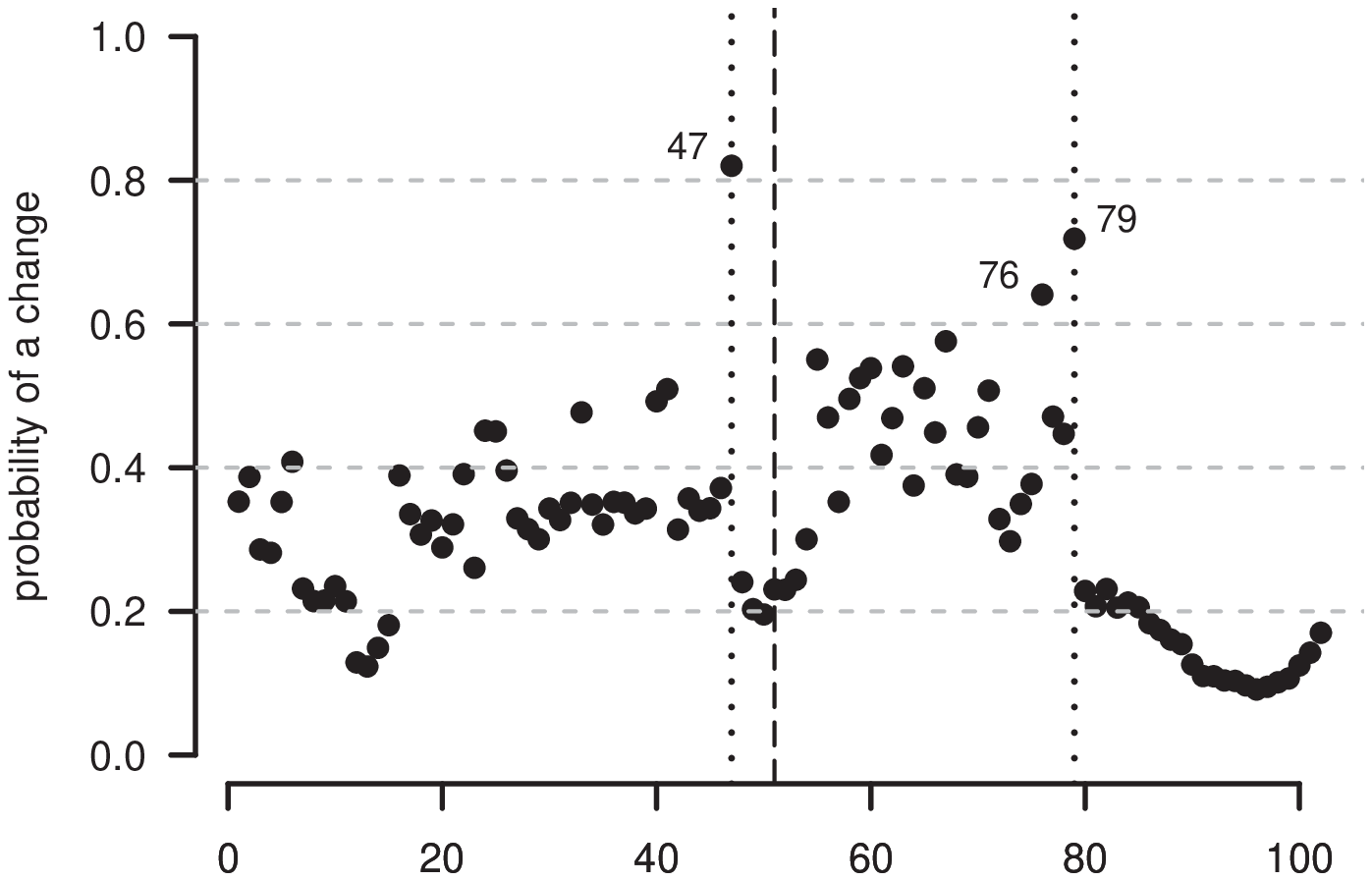}
		\label{fig:IR_prob_P18_mu_m100}}%\hspace{.1in}
	\subfigure[][$E_2$ ($m_1=m_2=100$)]{
		\includegraphics[width=4cm, height=3.5cm, trim=0 2cm 0 0]{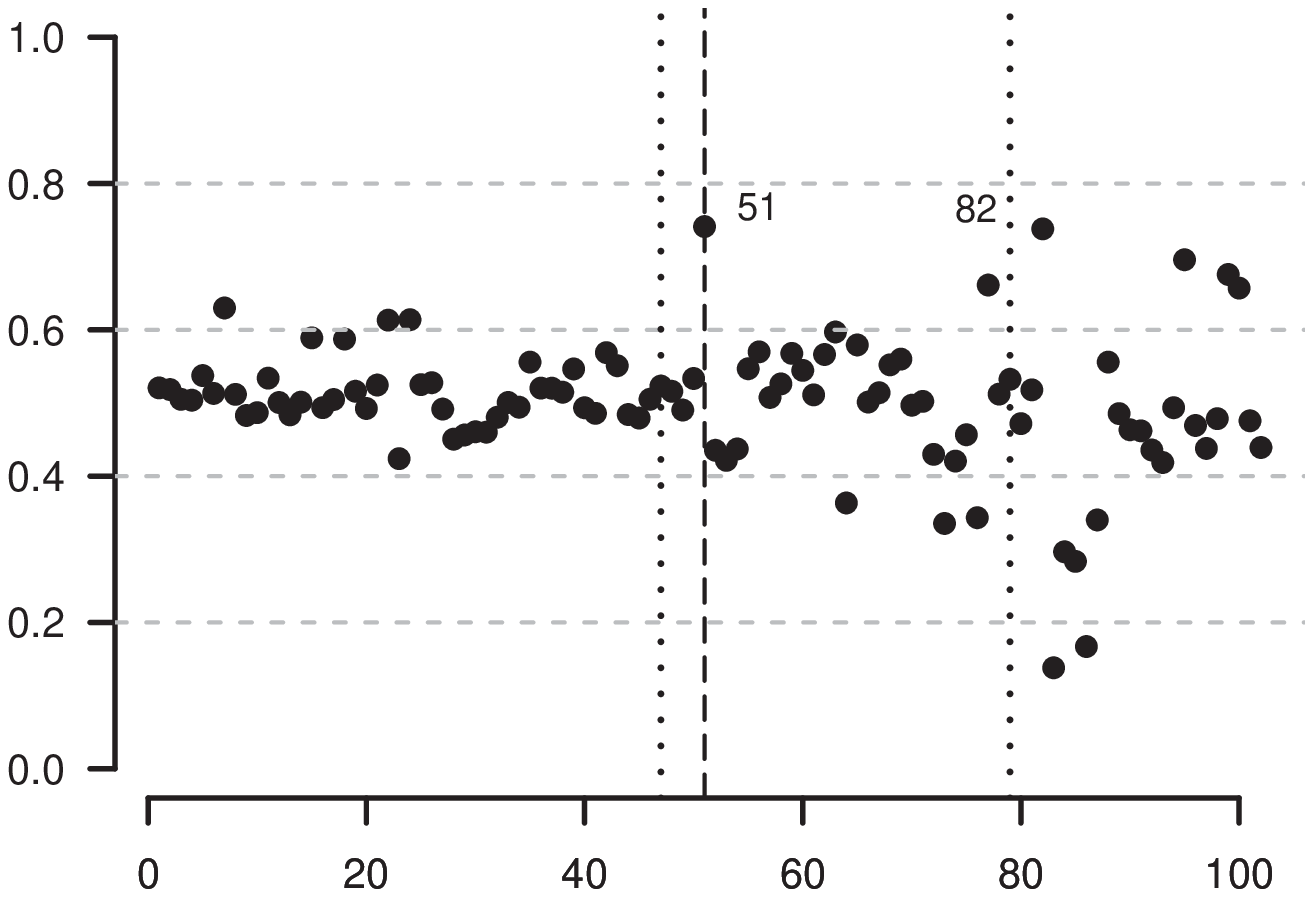}
		\label{fig:IR_prob_P18_s2_m100}}
	\caption{\small Posterior estimates for the probability a change at each instant (black bullets) under DPM19 model for the US ex-post Real Interest Rate dataset, considering different values for $m_1$ and $m_2$. The vertical dotted and dashed lines indicate the changes in $\bm{\mu}$ and $\bm{\sigma}$, respectively, according to the most likely partitions $\rho_1$ and $\rho_2$ estimated by the BMCP model in Section \ref{secReal_IR}.}
	\label{fig:IR_prob_m}
\end{adjustwidth}
\end{figure}

Considering $m_1=m_2=100$, besides having more uncertainty in the posterior estimates for the means and variances, the posterior distributions of the number of changes (Figure \ref{fig:IR_N_m}) and the posterior probabilities of a change (Figure \ref{fig:IR_prob_m}) indicated a high number of changes in both, the mean and variance. In this case, DPM19 presented poor results  to support the identification of possible regime changes if compared to what was  announced by \cite{garcia1996} and \cite{bai2003}.

The MCMC procedures took $155$ and $2150$ seconds to run $50,000$ iterations if $m_1=m_2=3$ and  $m_1=m_2=100$, respectively. We discarded the first $30,000$ iterations as the warm-up period.

\begin{figure}[!htbp]
\begin{adjustwidth}{-.4cm}{-.4cm}
\centering
\subfigure[][$m_1=m_2=3$]{
	\includegraphics[width=4cm, height=3cm, trim=0 1cm 0 0]{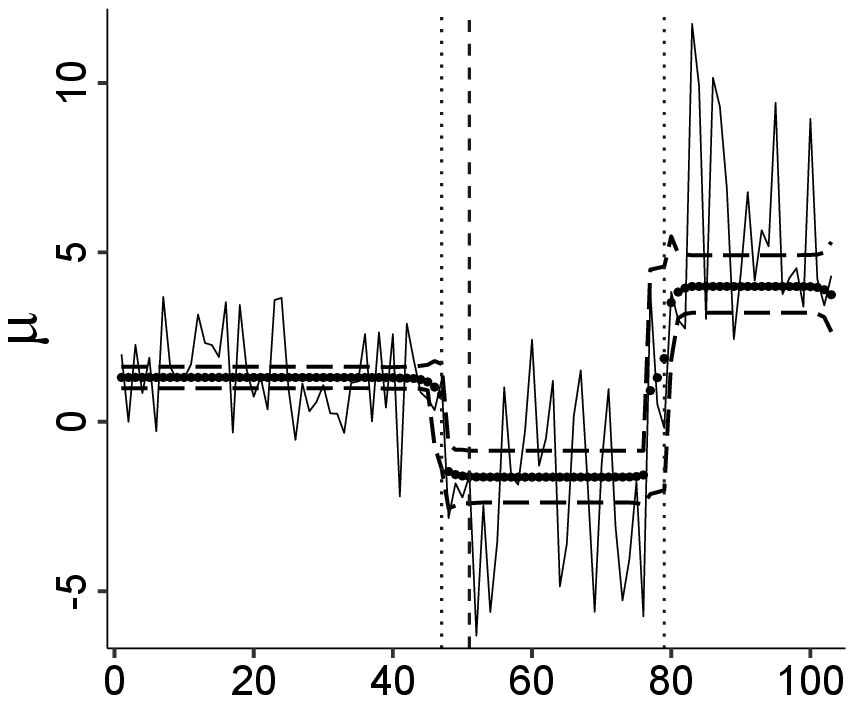}
	\label{fig:IR_mu_P18_m3}}%\hspace{.2in}
\subfigure[][$m_1=m_2=100$]{
	\includegraphics[width=4cm, height=3cm, trim=0 1cm 0 0]{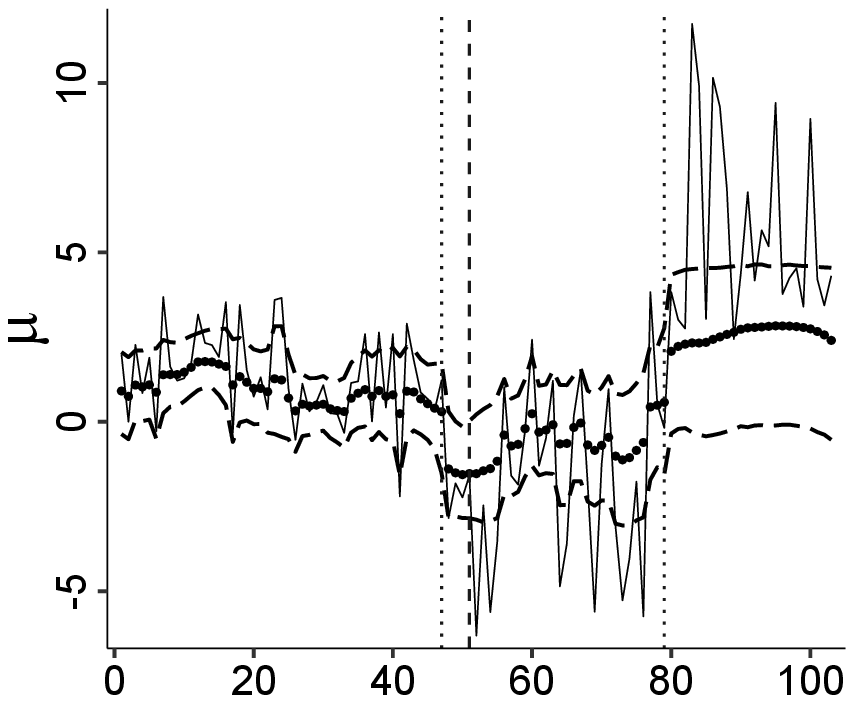}
	\label{fig:IR_mu_P18_m100}}%\\[.2in]
\subfigure[][$m_1=m_2=3$]{
	\includegraphics[width=4cm, height=3cm, trim=0 1cm 0 0]{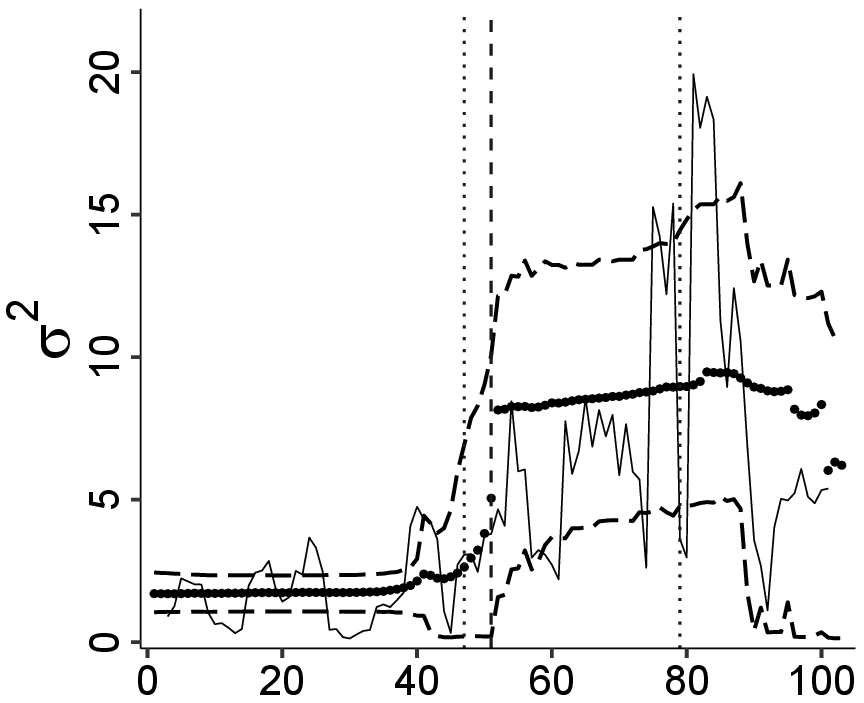}
	\label{fig:IR_s2_P18_m3}}%\hspace{.2in}
\subfigure[][$m_1=m_2=100$]{
	\includegraphics[width=4cm, height=3cm, trim=0 1cm 0 0]{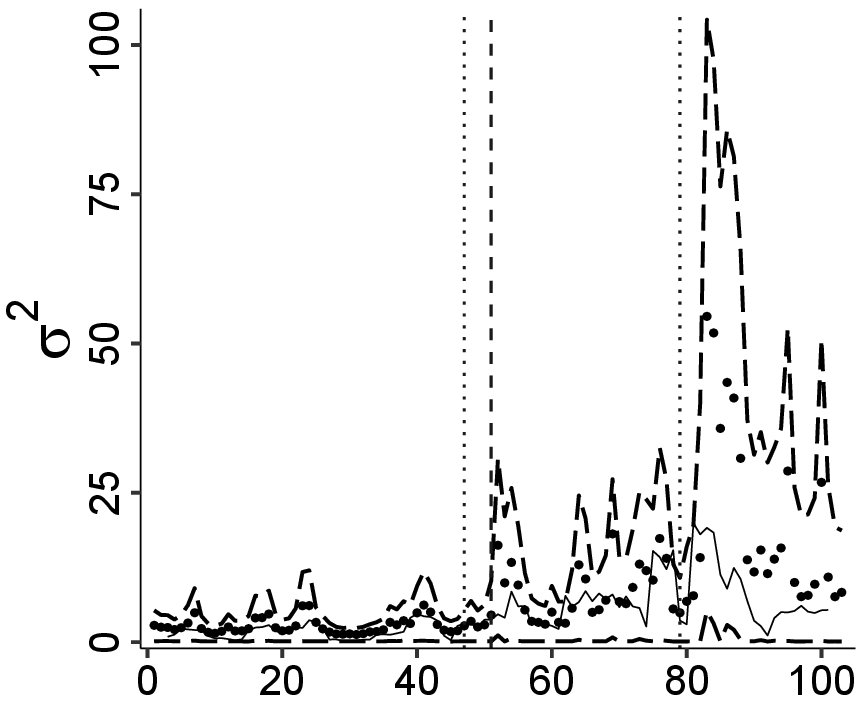}
	\label{fig:IR_s2_P18_m100}}%\hspace{-.1in}
\caption{\small Parameter estimates (black dots) and $90\%$ highest posterior density intervals (dashed lines) for the means (top) and variances (bottom) under DPM19 model considering different values for $m_1$ and $m_2$, for the US ex-post Real Interest Rate dataset. The black solid lines represent the observed data (top) and the moving sample variance calculated over ranges of length 5 (bottom). The vertical dotted and dashed lines indicate the changes in $\bm{\mu}$ and $\bm{\sigma}$, respectively, according to the most likely partitions $\rho_1$ and $\rho_2$ as estimated by the BMCP model in Section \ref{secReal_IR}.}
\label{fig:IR_PE_m}
\end{adjustwidth}
\end{figure}

%\begin{table}[H]%[!htbp]
%\centering
%\footnotesize
%\begin{tabular}{p{0.25\textwidth}p{0.075\textwidth}}
%	\\[-1ex]
%	\hline \\[-1.8ex]
%	DPM19 ($m_1=m_2=3$) & $p(E_1|\bm{X})$ \\
%	\hline \\[-1.8ex]
%	$\{0,47,76,103\}$ & 0.3160 \\
%	$\{0,47,79,103\}$ & 0.1931 \\
%	\hline \\[-1.8ex]
%\end{tabular}\\
%\begin{tabular}{p{0.25\textwidth}p{0.075\textwidth}}
%	\\[-1ex]
%	\hline \\[-1.8ex]
%	DPM19 ($m_1=m_2=3$) & $p(E_2|\bm{X})$ \\
%	\hline \\[-1.8ex]
%	$\{0,51,103\}$     & 0.0479 \\
%	$\{0,51,100,103\}$ & 0.0377 \\
%	\hline \\[-1.8ex]
%\end{tabular}\\
%\caption[Most likely partitions for case study 1 $(m_1=m_2=3)$]{Top most likely $E_1$ and $E_2$ based on the posterior probabilities, for case study 1, considering $(m_1=m_2=3)$.}
%\label{tab:IR_m}
%\end{table}
%

\FloatBarrier
\newpage
\subsection{Additional results for Case Study 2}\label{supp3_GC}

We analyze the dataset considered in Section \ref{secReal_GC} under the DPM19, LCIA05 and BH93 models. For the DPM19 model, we considered the same hyperparameters configuration described in the simulation study except for $m_1=100$, $m_2=10$, $\sigma_0^2=10^6$ and ${a=d=0.01}$. {The MCMC procedures took  $4.2$, $1.7$ and $1.4$ hours to run $50,000$ iterations under the  DPM19, LCIA05 and BH93 models, respectively. We discarded the first $50,000$ iterations as the warm-up period.}

\begin{figure}[!htbp]
\centering
\subfigure[][]{
	\includegraphics[width=12cm, height=3.7cm, trim=0 2.5cm 0 0]{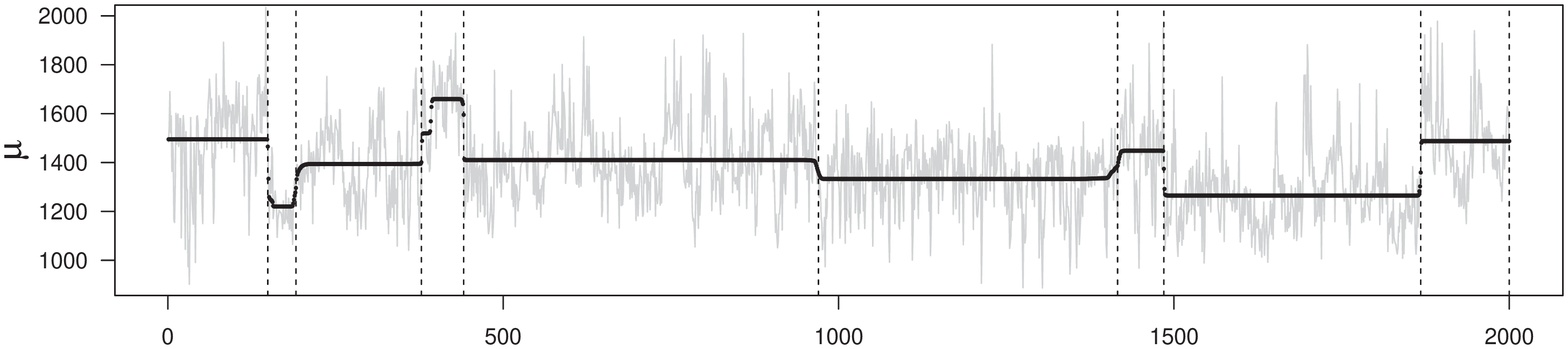}
	\label{fig:GC_mu_P18}}\vspace{-.4in}
\subfigure[][]{
	\includegraphics[width=12cm, height=3.7cm, trim=0 2.5cm 0 0]{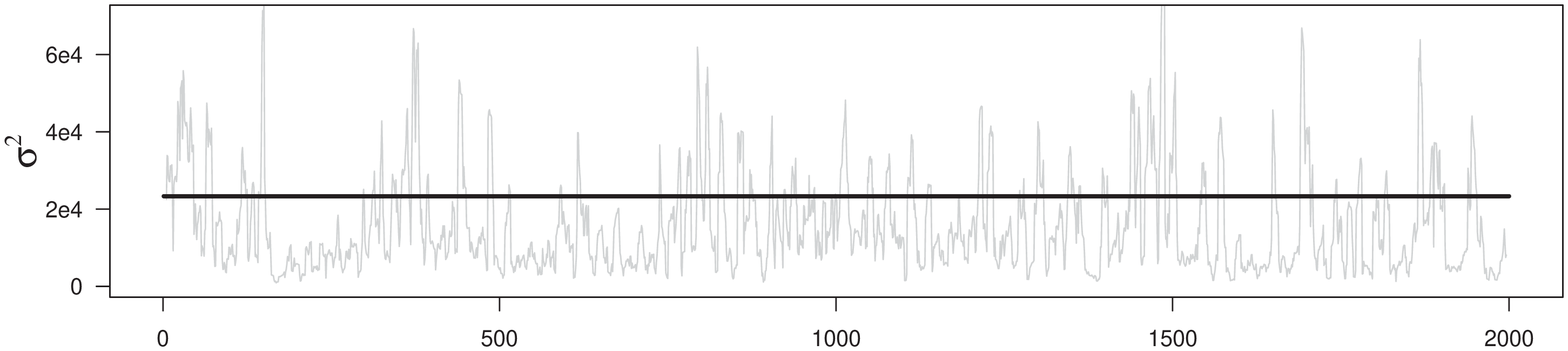}
	\label{fig:GC_s2_P18}}
\caption{\small Product estimates (black dots) for $\bm{\mu}$ (a) and $\bm{\sigma}$ (b), for the \texttt{HC1} dataset, under the DPM19 model. The gray lines represent the observed data in (a) and the moving sample variance, calculated over ranges of length nine, in (b). The vertical dashed lines in (a) represent the mean changes according to the most likely partition ${\rho_1=\{0,149,191,378,441,970,1416,1485,1868,2000\}}$.}
\label{fig:GC_PE_P18}
%\end{figure}
%
%\begin{figure}[!htbp]
	\centering
	\subfigure{
		\includegraphics[width=10.5cm,height=3.7cm, trim=0 2cm 0 0]{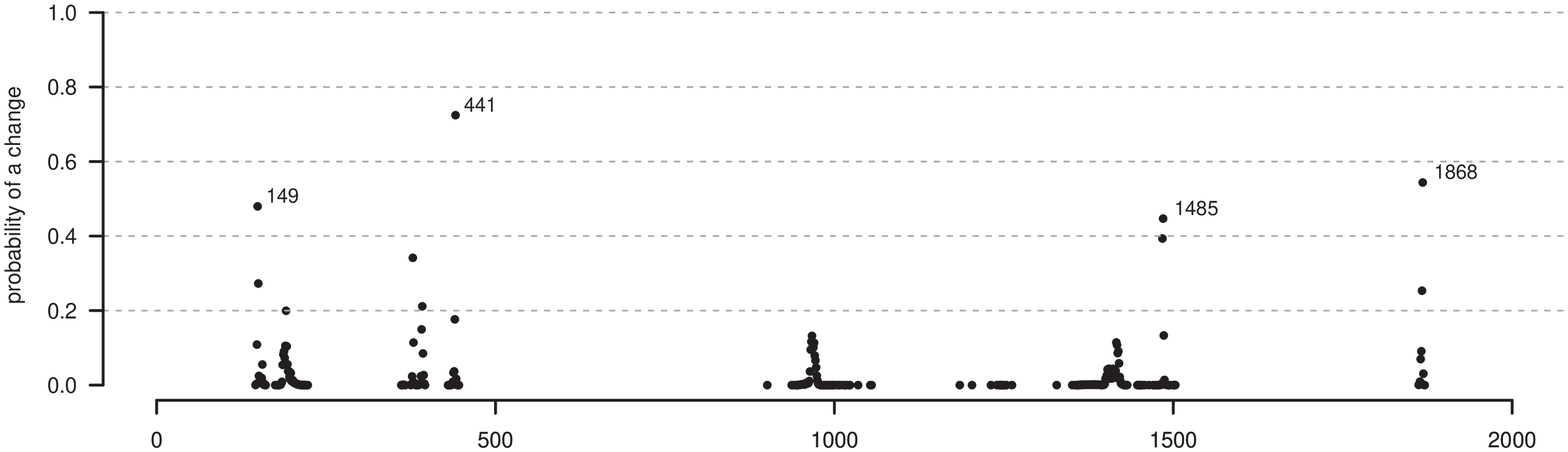}
		\label{fig:GC_prob_P18_mu}}
	\vspace{-.1in}
	\caption{\small Posterior probability of each position to be a change in $\bm{\mu}$, for the \texttt{HC1} dataset, under the DPM19 model. The labeled positions are those with probability greater than 0.4.}
	\label{fig:GC_prob_P18}
\end{figure}

Differently of our findings by fitting the BMCP model, DPM19 indicates no changes in the variance  and $8$ changes in the  mean with posterior probability equal to $1$, that is $P(N_2=0\mid\bm{X})=1 =P(N_1=8\mid\bm{X})=1$. For the mean  the posterior most probably partition is $\rho_1=\{149,191,378,441,970,1416,1485,1868,2000\}$ with posterior probability $0.0002$. The results for the DPM19 model are displayed in Figures \ref{fig:GC_PE_P18}  and \ref{fig:GC_prob_P18} .

\begin{figure}[!htbp]
\centering
\subfigure[][]{
	\includegraphics[width=12cm, height=3.7cm, trim=0 2.5cm 0 0]{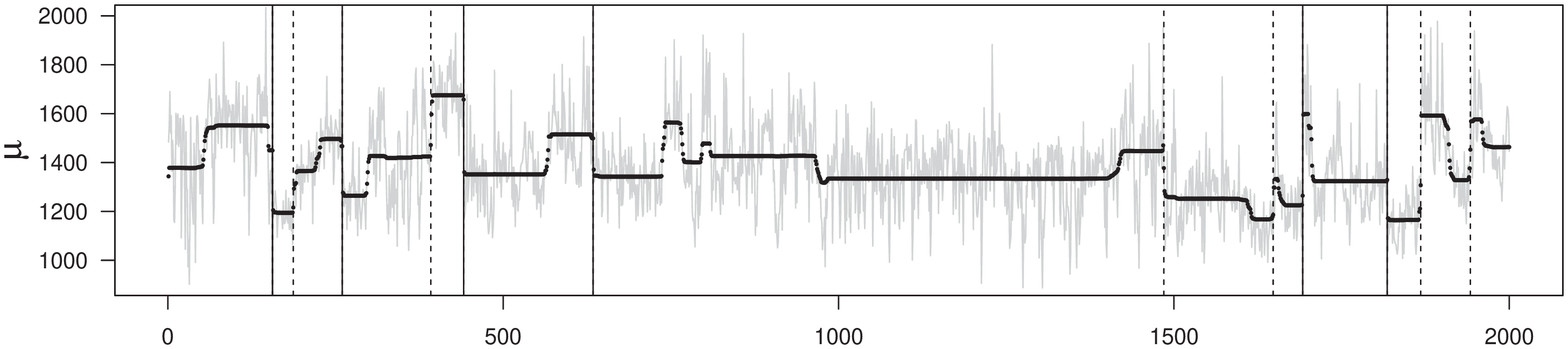}
	\label{fig:GC_mu_L99}}\\[-.4in]
\subfigure[][]{
	\includegraphics[width=12cm, height=3.7cm, trim=0 2.5cm 0 0]{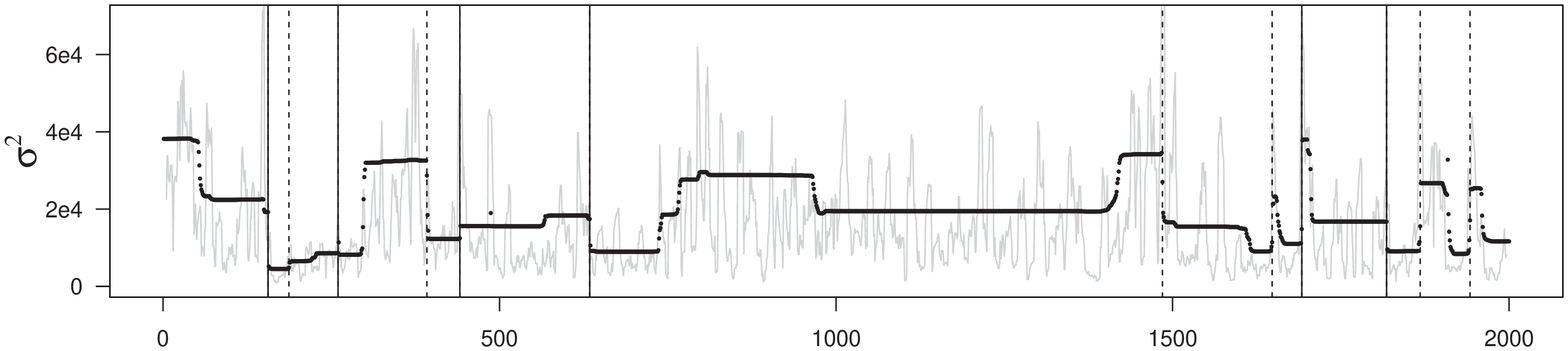}
	\label{fig:GC_s2_L99}}
\vspace{-.1in}
\caption{\small Product estimates (black dots) for $\bm{\mu}$ (a) and $\bm{\sigma}$ (b), for the \texttt{HC1} dataset, under LCIA05 model. The gray lines represent the observed data in (a) and the moving sample variance, calculated over ranges of length nine, in (b). The vertical lines represent the positions with probability of being a change greater than 0.4 (dashed lines) and greater than 0.7 (solid lines).}
\label{fig:GC_PE_L99}
%\end{figure}
%
%\begin{figure}[!htbp]
\centering
\vspace{-.1in}
\subfigure{
	\includegraphics[width=10.5cm, height=3.7cm, trim=0 2cm 0 0]{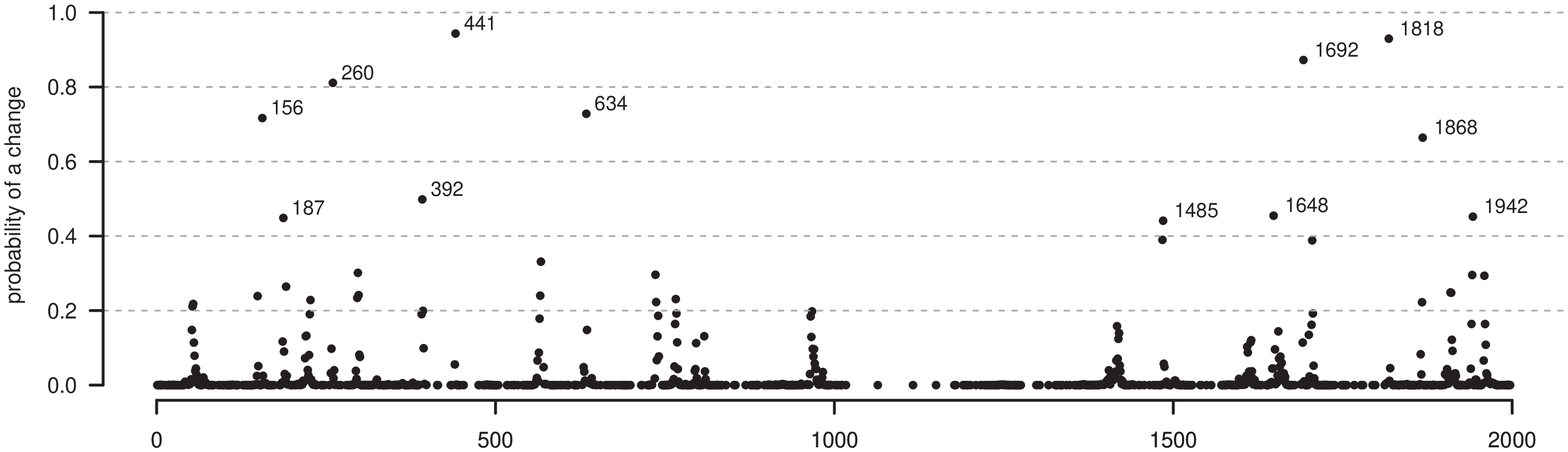}
	\label{fig:GC_prob_L99_mu}}
\vspace{-.2in}
\caption{\small Posterior probability of each position to be a change for the \texttt{HC1} dataset, under LCIA05 model. The labeled positions are those with probability greater than 0.4.}
\label{fig:GC_prob_L99}
\vspace{-.2in}
\subfigure{
	\includegraphics[width=6cm, height=4.2cm, trim=0 1cm 0 0]{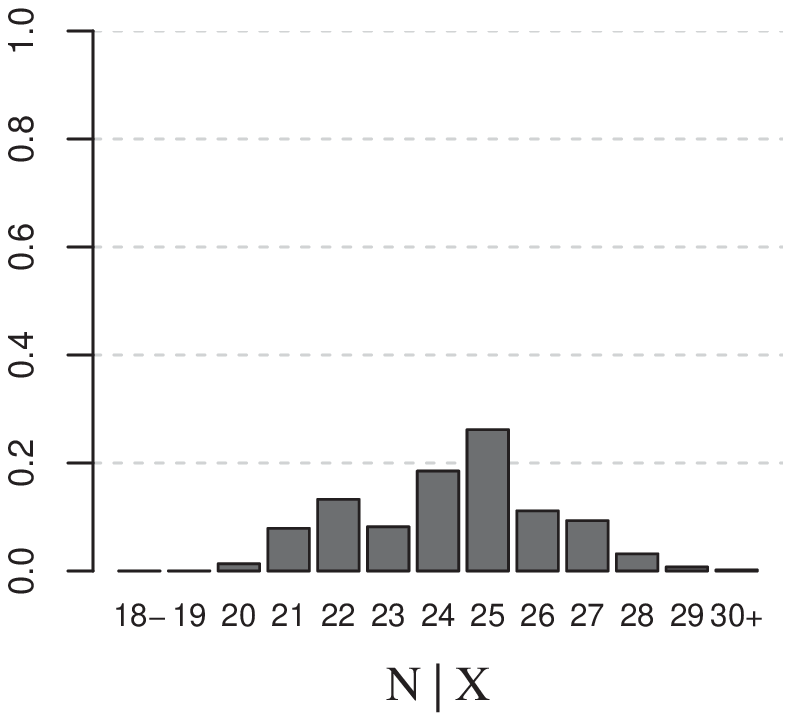}
	\label{fig:GC_N_L99_mu}}
\vspace{-.3in}
\caption{\small Posterior distribution of the number of changes in $(\bm{\mu,\sigma})$ for the \texttt{HC1} dataset, under LCIA05 model.}
\label{fig:GC_N_L99}
\end{figure}

For the LCIA05 model, we consider the same hyperparameters configuration described in the simulation study except for $v=100$ and ${a=d=0.02}$. The results for the LCIA05 model are displayed in Figures \ref{fig:GC_PE_L99}-\ref{fig:GC_N_L99}. This model identified several changes in the variance and the lower variance period indicated by the BMCP model can be also seen in Figure \ref{fig:GC_s2_L99}.

\begin{figure}[!htbp]
\centering	
%\end{figure}
%
%\begin{figure}[!htbp]
%\centering
\subfigure[][]{
	\includegraphics[width=12cm, height=3.7cm, trim=0 2.5cm 0 0]{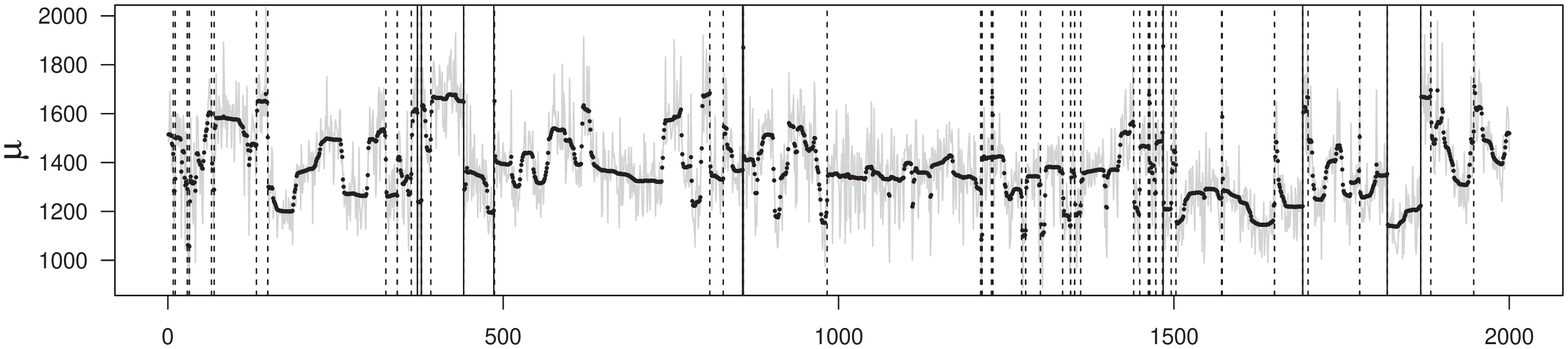}
	\label{fig:GC_mu_BH93_w1}}\\[-.4in]
\subfigure[][]{
	\includegraphics[width=12cm, height=3.7cm, trim=0 2.5cm 0 0]{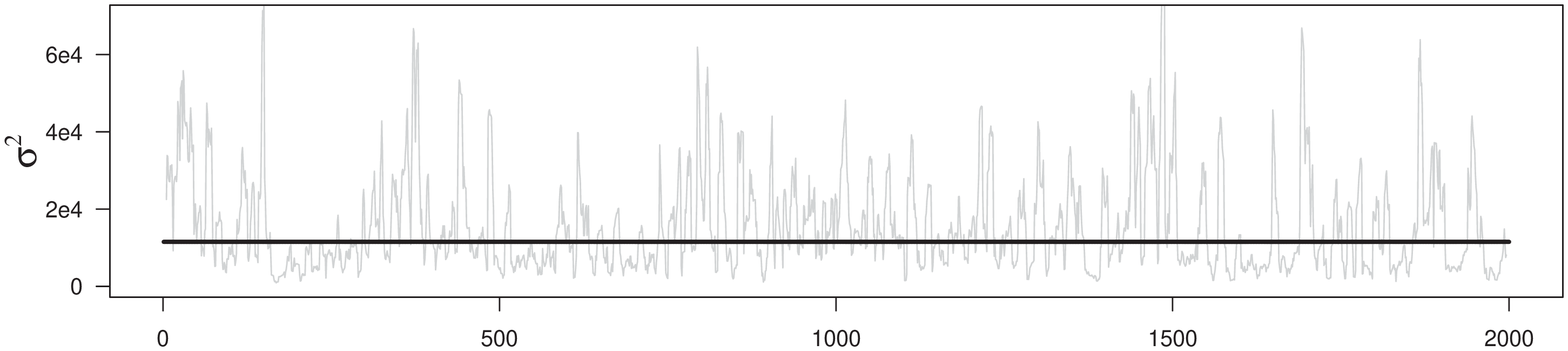}
	\label{fig:GC_s2_BH93_w1}}
\vspace{-.1in}
\caption{\small Product estimates (black dots) for $\bm{\mu}$ (a) and $\bm{\sigma}$ (b), for the \texttt{HC1} dataset, under BH93 model, considering $w_0=0.1$. The gray lines represent the observed data in (a) and the moving sample variance, calculated over ranges of length nine, in (b). The vertical lines in (a) represent the positions with probability of being a change in the mean greater than 0.5 (dashed lines) and greater than 0.9 (solid lines).}
\label{fig:GC_PE_BH93_w1}
\centering
\vspace{-.1in}
\subfigure{
	\includegraphics[width=10.5cm, height=3.7cm, trim=0 2cm 0 0]{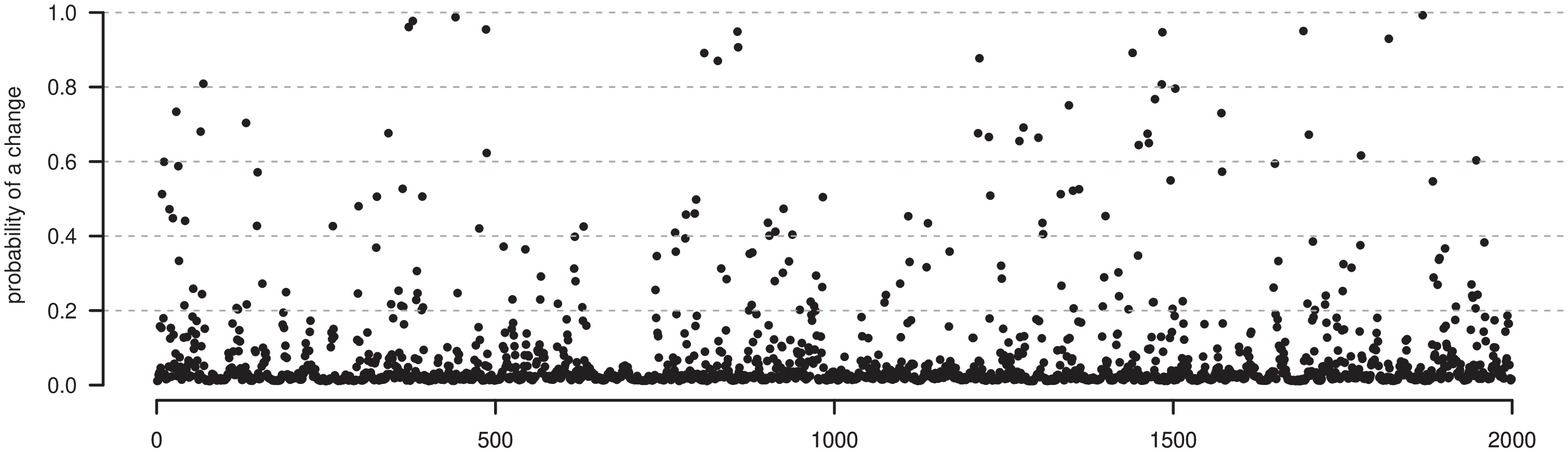}
	\label{fig:GC_prob_BH93_mu_w1}}
\vspace{-.2in}
\caption{\small Posterior probability of each position to be a change in $\bm{\mu}$, for the \texttt{HC1} dataset, under BH93 model, considering $w_0=0.1$.}
\label{fig:GC_prob_BH93_w1}
\vspace{-.2in}
\subfigure{
	\includegraphics[width=6cm, height=4.2cm, trim=0 1cm 0 0]{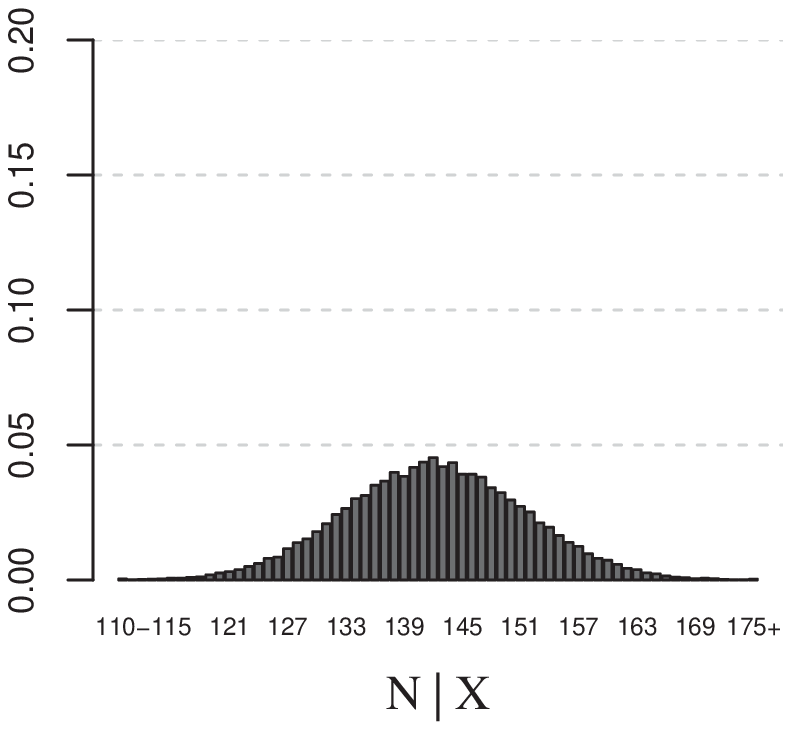}
	\label{fig:GC_N_BH93_mu_w1}}
\vspace{-.3in}
\caption{\small Posterior distribution of the number of changes in $\bm{\mu}$ for the \texttt{HC1} dataset, under BH93 model, considering $w_0=0.1$.}
\label{fig:GC_N_BH93_w1}
%\end{figure}
%
%\begin{figure}[!htbp]
\end{figure}

For the BH93 model, we considered the hyperparameters $(p_0,w_0)=(0.05,0.1)$. The results for the BH93 model are displayed in Figures \ref{fig:GC_PE_BH93_w1}-\ref{fig:GC_N_BH93_w1}. Under this hyperparameters configuration, the model indicates a high number of changes (posterior mode of $N=143$). As discussed in \cite{bh93}, lower values of $w_0$ imply smaller number of clusters. %Results for the BH93 model considering $w_0=0.0001$ are displayed in Figures \ref{fig:GC_N_BH93_w4}-\ref{fig:GC_PE_BH93_w4}. In this case, the posterior mode of $N$ is $46$.

\end{document}